\documentclass[10pt]{book}
\usepackage{amssymb,latexsym,amsmath}
\usepackage{cite}
\usepackage[dvips]{graphicx}
\usepackage[innercaption]{sidecap}
\usepackage[bf]{caption}
\usepackage{fancyhdr}
\usepackage{makeidx}
\usepackage[english]{babel}

\renewcommand{\chaptermark}[1]{\markboth{#1}{}}
\renewcommand{\sectionmark}[1]{\markright{\thesection\ #1}}
\fancypagestyle{plain}{
   \fancyhead{}
   \renewcommand{\headrulewidth}{0pt}
}
\renewcommand{\headrulewidth}{0pt}

\begin{document}
\ifx\href\undefined\else\hypersetup{linktocpage=true}\fi

\frontmatter
\begin{center}

\begin{LARGE}
Spin and orbital degrees of freedom \\
in transition metal oxides and oxide thin films\\
studied by soft x-ray absorption spectroscopy\\
\end{LARGE}

\begin{large}

$\qquad$

$\qquad$

$\qquad$

$\qquad$

$\qquad$

$\qquad$

$\qquad$

Inaugural - Dissertation\\
zur\\
Erlangung des Doktorgrades\\
der Mathematisch-Naturwissenschaftlichen Fakult\"at\\
der Universit\"at zu K\"oln\\

$\qquad$

$\qquad$

$\qquad$

$\qquad$

vorgelegt von\\
Maurits W. Haverkort\\
aus Swifterbant, Nederland\\

$\qquad$

$\qquad$

$\qquad$

$\qquad$

$\qquad$

$\qquad$

$\qquad$

$\qquad$

$\qquad$

$\qquad$

K\"oln 2005

\end{large}
\end{center}

\newpage
\begin{large}
$\qquad$

$\qquad$

$\qquad$

$\qquad$

$\qquad$

$\qquad$

$\qquad$

$\qquad$

$\qquad$

$\qquad$

$\qquad$

$\qquad$

$\qquad$

$\qquad$

$\qquad$

$\qquad$

$\qquad$

$\qquad$

$\qquad$

$\qquad$

$\qquad$

$\qquad$

$\qquad$

$\qquad$

$\qquad$\\
Berichterstatter:

Prof. Dr. L. H. Tjeng

Prof. Dr. T. Hibma

Prof. Dr. S. Bl\"ugel

$\qquad$

$\qquad$\\
Tag der m\"undlichen Pr\"ufung:

07.06.2005

$\qquad$

$\qquad$

\end{large}
\tableofcontents

\mainmatter
\renewcommand{\headrulewidth}{0.5pt}
\fancyhead[LE,RO]{\bfseries\thepage}
\fancyhead[LO]{\bfseries\rightmark}
\fancyhead[RE]{\bfseries\leftmark}

\chapter{Introduction}

The class of transition metal compounds shows an enormous richness of physical properties \cite{Tsuda91, Imada98}, such as metal-insulator transitions, colossal magneto-resistance, super-conductivity, magneto-optics and spin-depend transport. The theoretical description of these materials is still a challenge. Traditional methods using the independent electron approximation most of the time fail on even the simplest predictions. For example, many of the transition metal compounds, with NiO as the classical example, should be a metal according to band-structure calculations, but are in reality excellent insulators.

The single band Mott-Hubbard model \cite{Mott49, Hubbard63} explains very nicely why many correlated materials are insulating. But even the Mott-Hubbard model has some problems in describing the band-gap found for many of the transition metal compounds \cite{Sawatzky84}. With the recognition that transition metal compounds can be of the charge-transfer type or the Mott-Hubbard type \cite{Zaanen85}, depending on the ratio of $U$ and $\Delta$, also the band-gap can be understood. Whereby $U$ is defined as the repulsive Coulomb energy of two electrons on the same transition metal site and $\Delta$ is defined as the energy it costs to bring an electron from an oxygen site to a transition metal site.

The single band Mott-Hubbard model is, however, even when charge transfer effects are included, inadequate in describing the full richness found in many of the transition metal compounds \cite{Birgeneau00, Tokura00, Orenstein00}. It now becomes more and more clear that in order to describe transition metal compounds the charge, orbital, spin and lattice degrees of freedom should all be taken into account. Especially the orbital degrees of freedom have not been considered to the full extend until recently. In the manganates, for example, orbital and charge ordering of the Mn ions play an important role for the colossal magneto-resistance of these materials \cite{Ramirez97, Khomskii97, Mizokawa95, Mizokawa96, Mizokawa97}. An other example would be the metal-insulator transition in V$_{2}$O$_{3}$ \cite{Park00, Ezhov99, Mila00}. The orbital occupation of the V ion changes drastically at the phase transition \cite{Park00}. This change in orbital occupation will change the local spin-spin correlations which in-turn will change the effective band-width. This indicates that not only electron-electron Coulomb repulsion in a single band must be considered, but a full multi-band theory including all interactions must be considered in order to understand this prototypical Mott-Hubbard system.

\section{How to measure orbital occupations}

With the recognition that the local orbital occupation plays an important role in many of the transition metal compounds there is a need for experimental techniques that can measure the orbital occupation. This technique is soft x-ray absorption spectroscopy. For transition metal atoms one measures the local transition of a $2p$ core electron into the $3d$ valence shell. This type of spectroscopy has only developed into maturity over the last 20 years, both in terms of instrumentation as well as in terms of theoretical understanding of these spectra \cite{Thole97, Groot94, Groot05}. The pioneering work of Fink, Thole, Sawatzky and Fuggle, who used electron energy loss spectroscopy on narrow band and impurity systems has been very important for the development of soft x-ray absorption spectroscopy. They recognized, that the observed multiplet structures can provide an extremely detailed information about the local electronic structure of the ground and lower excited states of the system \cite{Thole85a, Fink85, Fink89}.

The $2p$ core-electron excitation into the $3d$ shell is dipole allowed. This has, first of all the advantage that good intensities are found. The locally dipole allowed transition has also the advantage of obeying the strict dipole selection rules. This means that the intensity found for a given final-state depends strongly on the symmetry of the initial state. In most cases the dependence on the initial state symmetry is so large that one does not need to have good resolution in order to determine which of the possible states is realized as the ground-state. Within chapter \ref{chapterXrayabs} we will show many examples taken from the literature of what can be measured with the use of x-ray absorption spectroscopy and how these dipole selection rules are active.

\section[Interpretation of x-ray absorption spectroscopy]{Interpretation of x-ray absorption spectros-copy}

For the interpretation of $2p$ x-ray absorption spectra cluster calculations are an essential tool. At the moment cluster calculations are still one of the best methods to describe both near ground-state properties and spectra of transition metal compounds. This might seem surprising while cluster calculations are not \textit{ab-inito} and the translational symmetry of the solid is not taken into account. There are two reasons why cluster calculations are so powerful. The first reason is that within cluster calculations the initial state and the final state are treated on an equal footing. This results in calculated spectra that can be compared to experiments in great detail. This is in strong contrast to density functional methods or Hartree-Fock methods which produce density of states and not spectra. One should realize that a density of states is not a spectrum. The second reason is that within cluster calculations the full electron-electron repulsion Hamiltonian can be included. The importance of full multiplet theory for the description of transition metal compounds will be stressed within the next section.

Part of the breakthrough in the understanding of $2p$ x-ray absorption spectroscopy on transition metal compounds was realized with the creation of good computation codes that can calculate the spectra of a cluster. Here the work done by Theo Thole \cite{Thole85a, Thole97} and Arata Tanaka \cite{Tanaka94} must be mentioned who both wrote a program able to do cluster calculations and calculate the x-ray absorption spectra with all its multiplet structure up to great detail.

  \begin{figure}[h]
   \begin{center}
    \includegraphics[width=120mm]{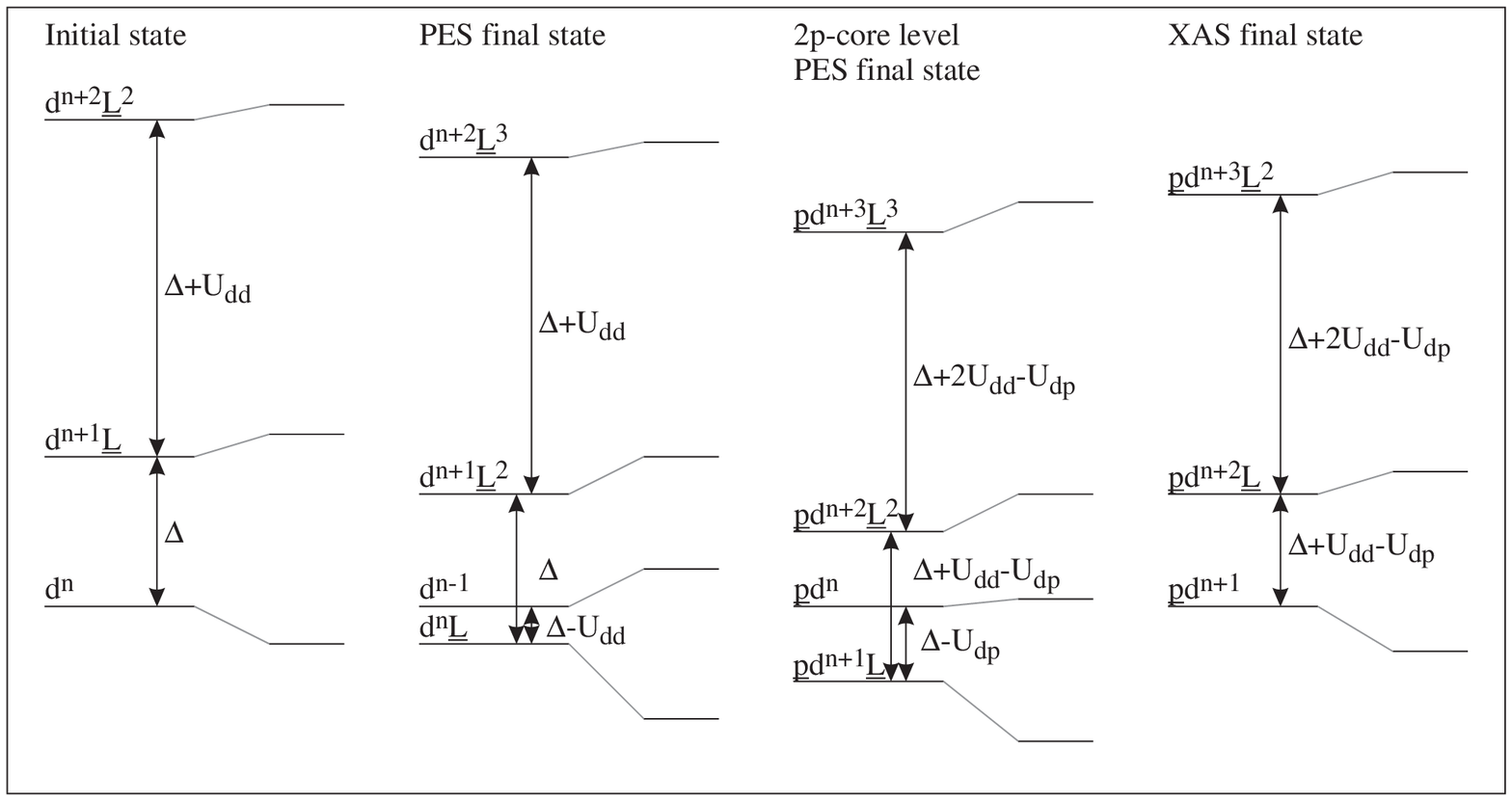}
    \caption{On-site energies of the \textit{bare} configurations for the initial state and different final states. $\Delta$ is the energy it costs to hop with one electron from the oxygen band to the transition metal $d$ shell. $U$ is the repulsive Coulomb energy between two electrons in the $3d$ shell ($U_{dd}$) or between a $3d$ electron and a $2p$ core electron ($U_{dp}$). $L$ denotes the oxygen or ligand states and $\underline{L}^n$ stands for $n$ holes in the ligands. The holes in the $\underline{p}$ shell for the $2p$-core level PES final states and for the XAS final states are holes in the $2p$-core level of the $3d$ transition metal ions. Scheme taken from J. Zaanen, G. A. Sawatzky and J. W. Allen \cite{Zaanen85}.}
    \label{introductionCIonsiteenergie}
    \end{center}
   \end{figure}

The cluster calculations are done within the configuration interaction scheme, allowing for the inclusion of hybridization of the transition metal $d$ orbitals with the oxygen $p$ orbitals \cite{Hubbard66, Fujimori84a, Fujimori84b}, denoted as the ligand orbitals ($L$). The on-site energies are parameterized with $U$ and $\Delta$ in the same way as done by J. Zaanen, G. A. Sawatzky and J. W. Allen \cite{Zaanen85}. In figure \ref{introductionCIonsiteenergie} we show the energy level diagram for the initial state and the final states of valence-band photo-electron spectroscopy, $2p$-core level photo-electron spectroscopy and $2p$-core level x-ray absorption spectroscopy. For the last two spectroscopies, the $2p$-core levels are core levels of the transition metal ions. For the initial state only the few lowest states are important. For the final state in principle all different configurations can be reached and are therefore important. The intensity of each final state does depend on the dipolar matrix elements between the initial state wave function and the final state wave-function under consideration. In order to reproduce the measured x-ray absorption spectra one has to have the correct final state as well as the correct initial state within the cluster calculation. This means that detailed information concerning the inital state can be obtained once the spectrum has been reproduced.

In figure \ref{introductionCIonsiteenergie} we only showed the on-site energies of the different configurations. The configurations will be split into different states. First of all the hybridization of different orbitals is not equal. For a transition metal in $O_h$ symmetry the $e_{g}$ orbitals hybridize more with the oxygen orbitals then the $t_{2g}$ orbitals. States with more holes in the $e_{g}$ shell will therefore be more covalent and have different energies than states where the $e_{g}$ orbitals are occupied. Second there will be a splitting due to the crystal field. Within mean-field theory this crystal field originates from the electric field made by the charges of the atoms that surround the atom under consideration. For transition metal oxides the oxygen atoms are charged negative and the electrons at the metal site do want to point away from the oxygen atom. Within $O_{h}$ symmetry this means that the $t_{2g}$ orbitals are lowered with respect to the $e_{g}$ orbitals. Third there will be a splitting between the different states within one configuration due to the electron-electron repulsion. The importance of electron-electron repulsion and the effect of screening for $3d$ elements will be discussed within the next section. Forth there are the spin-orbit coupling and magnetic interactions that split the different states within one configuration. Within a cluster calculation these interactions can be included and for many systems have to be included in order to reproduce the x-ray absorption spectra properly.

Besides cluster calculations there is an other way to deduce information from x-ray absorption spectra. B. T. Thole, P. Carra \textit{et al.} \cite{Thole92, Carra93a} have derived sum-rules that relate the total integrated intensity of polarized spectra to expectation values of some operators of the initial state. These sum-rules are very powerful due to there simplicity of use.

\section{The importance of full multiplet theory}

One might expect that within a solid the multiplet splitting due to electron-electron repulsion is largely screened. For the monopole part of the electron-electron repulsion this is indeed true. If one adds an extra charge to one atom within the solid the charges of the surrounding atoms (or other shells of the same atom) will change and thereby lower the total repulsion the added charge feels. This type of screening is also found experimentally. We define $U_{av}$ as the local average repulsion between two electrons. $U_{av}$ can be related to the Slater Integrals, $F^0$, $F^2$ and $F^4$. The expression then becomes; $U_{av}=F^0-\frac{14}{441}(F^2+F^4)$. This $U_{av}$ is within a solid much smaller than the Hartree-Fock value found for a free ion. Take Co$^{2+}$ in CoO for example. The Hartree-Fock value of $U_{av}$ for the free ion is about 25 eV. The $U_{av}$ experimentally found is about 6.5 eV \cite{Tanaka94}.

Surprisingly the multiplet splitting within a configuration in the solid is experimentally found not to be reduced from the free ion multiplet splitting. E. Antonides, E. C. Janse and G. A. Sawatzky \cite{Antonides77} did Auger spectroscopy on Cu, Zn, Ga and Ge metal and found that the $F^2$ and $F^4$ Slater integrals, the parameters describing the multiplet splitting between states with the same number of electrons are in reasonable agreement with Hartree-Fock calculations on a free ion. This implies that only $F^0$ can be efficiently screened. These findings have been confirmed on many transition metal compounds \cite{Ballhausen62, Fujimori84a, Fujimori84b, Graaf98, Groot90, Tjeng91a, Groot05}. This absence of screening for the multiplet part of the electron-electron interaction can be understood if one realizes that the multiplet splitting is due to the different shape of the local electron cloud and$\setminus$or due to different spin densities. Such differences are very difficult to screen by charges located externally. Moreover, there are also even states that have the same spin-resolved electron density matrix, $d_{m\sigma}^{\dagger}d_{m'\sigma'}$, but very different electron-electron repulsion interaction. Take a Ni$^{2+}$ ion for example. The $^3F_{M_L=\pm3, M_S=0}$ state, which belongs to the 21 fold degenerate ground-state, and the $^1G_{M_L=\pm3, M_S=0}$ state have the same electron density matrix $d_{m\sigma}^{\dagger}d_{m'\sigma'}$ but an energy difference of $\frac{12}{49}F^2+\frac{10}{441}F^4$. Within a mean-field approximation it is not possible to screen this splitting since these states have the same electron densities.

The full electron-electron interaction Hamiltonian is give by \cite{Ballhausen62, Sugano70, Cowan81}:
\begin{equation}
H_{e-e}=\sum_{\langle m m' m'' m'''\sigma\sigma'\rangle}  U_{mm'm''m'''\sigma\sigma'}l_{m\sigma}^{\dagger}{l'}^{\dagger}_{m'\sigma'}{l_{m''\sigma}''}{l_{m'''\sigma'}'''}\\
\end{equation}
with:
\begin{equation}\begin{split}
U_{mm'm''m'''\sigma\sigma'}=&\delta(m+m',m''+m''') \\
                                 &\sum_{k=0}^{\infty} \langle lm|C^k_{m-m''}|l''m''\rangle
                                  \langle l'''m'''|C^k_{m'''-m'}|l'm'\rangle R^k(ll'l''l''')\\
R^k(ll'l''l''')=&e^2\int_{0}^{\infty}\int_{0}^{\infty}\frac{r_{<}^{k}}{r_{>}^{k+1}}R_{l}^{n}(r_1)R_{l'}^{n'}(r_2)R_{l''}^{n''}(r_1)R_{l'''}^{n'''}(r_2) r_1^2 r_2^2 \delta r_1 \delta r_2
\end{split}\end{equation}
With $r_{<}=\textmd{Min}(r_1,r_2)$ and $r_{>}=\textmd{Max}(r_1,r_2)$. This Hamiltonian describes a scattering event of two electrons. These two electrons have before scattering the quantum numbers $m'',m''',\sigma$ and $\sigma'$. After the scatter event they have the quantum numbers $m,m',\sigma$ and $\sigma'$. The scatter intensity is given by $U_{mm'm''m'''\sigma\sigma'}$. The angular dependence can be solved analytically and is expressed in integrals over spherical harmonics. The radial part can not be solved analytically and is expressed in terms of Slater integrals over the radial wave equation; $R^k(ll'l''l''')$. For two $d$ electrons interacting with each other, the only important values of $k$ are $k=0,2$ or $4$. For these $k$ values the radial integrals are expressed as $F^0, F^2$ and $F^4$. 

Many calculations done on solids containing transition metal ions do not include the full electron-electron Hamiltonian, but make an approximation. The simplest approximation that one can make is to describe the electron-electron repulsion by two parameters $U_{0}$ and $J_{H}$, in which $U_{0}$ is the repulsive Coulomb energy between each pair of electrons and $J_{H}$ is the attractive Hund's rule exchange interaction between each pair of electrons with parallel spin. This leads to the following Hamiltonian:
\begin{equation}\begin{split}
H_{e-e}^{Simple}=&U_0\sum_{\langle mm'\sigma\sigma'\rangle} l^{\dagger}_{m\sigma}{l'}^{\dagger}_{m'\sigma'}l_{m\sigma}{l'}_{m'\sigma'}\\
-&J_H\sum_{\langle mm'\sigma \rangle}l^{\dagger}_{m\sigma}{l'}^{\dagger}_{m'\sigma}l_{m\sigma}{l'}_{m'\sigma}
\end{split}\end{equation}
This Hamiltonian is built up entirely by number operators $n_{m\sigma}=l^{\dagger}_{m\sigma}l_{m\sigma}$, and is therefore diagonal in the basis vectors that span the full electron-electron Hamiltonian.

\begin{table}
\begin{center}
\begin{footnotesize}
\begin{tabular}{|c||c|c|c|c|c|}
  \hline
    & \multicolumn{5}{c|}{\vphantom{\large{I}}Coulomb energy of Hund's rule ground-state}\\
  \hline
    \vphantom{\large{I}} & \multicolumn{2}{c|}{Full Hamiltonian}& Simple & Kanamori & Kanamori\\
                         & \multicolumn{2}{c|}{                }&        &          & mean field\\
  \hline
  $d^{0} $ &                                                    0 &                     0 &                   0 
         & 0 & 0\vphantom{\large{I}}\\
  $d^{1} $ &                                                    0 &                     0 &                   0 
         &           0 &           0 \\
  $d^{2} $ &    F$^0$-$\frac{ 8}{49}$F$^2$-$\frac{  9}{441}$F$^4$ &   U$_{0}$-  J$_{H}$-C &   U$_{0}$-  J$_{H}$ 
         &   U'   -  J &   U'   -  J \\
  $d^{3} $ &   3F$^0$-$\frac{15}{49}$F$^2$-$\frac{ 72}{441}$F$^4$ &  3U$_{0}$- 3J$_{H}$-C &  3U$_{0}$- 3J$_{H}$ 
         &  3U'   - 3J &  3U'   - 3J \\
  $d^{4} $ &   6F$^0$-$\frac{21}{49}$F$^2$-$\frac{189}{441}$F$^4$ &  6U$_{0}$- 6J$_{H}$   &  6U$_{0}$- 6J$_{H}$ 
         &  6U'   - 6J &  6U'   - 6J \\
  $d^{5} $ &  10F$^0$-$\frac{35}{49}$F$^2$-$\frac{315}{441}$F$^4$ & 10U$_{0}$-10J$_{H}$   & 10U$_{0}$-10J$_{H}$ 
         & 10U'   -10J & 10U'   -10J \\
  $d^{6} $ &  15F$^0$-$\frac{35}{49}$F$^2$-$\frac{315}{441}$F$^4$ & 15U$_{0}$-10J$_{H}$   & 15U$_{0}$-10J$_{H}$ 
         & 14U'+ U-10J & 14U'+ U-10J \\
  $d^{7} $ &  21F$^0$-$\frac{43}{49}$F$^2$-$\frac{324}{441}$F$^4$ & 21U$_{0}$-11J$_{H}$-C & 21U$_{0}$-11J$_{H}$ 
         & 19U'+2U-11J & 19U'+2U-11J \\
  $d^{8} $ &  28F$^0$-$\frac{50}{49}$F$^2$-$\frac{387}{441}$F$^4$ & 28U$_{0}$-13J$_{H}$-C & 28U$_{0}$-13J$_{H}$ 
         & 25U'+3U-13J & 25U'+3U-13J \\
  $d^{9} $ &  36F$^0$-$\frac{56}{49}$F$^2$-$\frac{504}{441}$F$^4$ & 36U$_{0}$-16J$_{H}$   & 36U$_{0}$-16J$_{H}$ 
         & 32U'+4U-16J & 32U'+4U-16J \\
  $d^{10}$ &  45F$^0$-$\frac{70}{49}$F$^2$-$\frac{630}{441}$F$^4$ & 45U$_{0}$-20J$_{H}$   & 45U$_{0}$-20J$_{H}$ 
         & 40U'+5U-20J & 40U'+5U-20J \\
  \hline
\end{tabular}
\end{footnotesize}
\end{center}
\caption{Energy of the Hund's rule ground-state of a $d^n$ electron configuration for different schemes. The relation between the Slater integrals $F^0$, $F^2$ and $F^4$ and the parameters $U_0$, $J_H$, $C$, $U'$, $U$, $J$, $J'$ is: $U_{0}=F^0$, $J_H=\frac{1}{14}(F^2+F^4)$, $C=\frac{1}{14}(\frac{9}{7}F^2-\frac{5}{7}F^4)$, $U'=U-2J$, $J'=J$, $U=F^0+\frac{4}{49}F^2+\frac{36}{441}F^4$ and $J=\frac{2.5}{49}F^2+\frac{22.5}{441}F^4$ \cite{Mizokawathesis, Marelthesis}. The center of each multiplet for the full Hamiltonian is $E_{av}(d^n)=(F^0-\frac{14}{441}(F^2+F^4))\frac{n(n-1)}{2}$.}
\label{TableE}
\end{table}

This extremely simple Hamiltonian leads for the Hund's rule high-spin ground-state to remarkably good results \cite{Marelthesis}. In table 1.1 we compare the total energy of a $d^n$ configuration ($n=0...10$) calculated using the full multiplet theory to that calculated using the simple scheme. To facilitate the comparison, we have rewritten the Slater integrals, $F^0$, $F^2$ and $F^4$ in terms of $U_0=F^0$, $J_H=\frac{1}{14}(F^2+F^4)$ and $C=\frac{1}{14}(\frac{9}{7}F^2-\frac{5}{7}F^4)$. We now see that the simple scheme is even exact in half of the n cases and has an error of $C$ for the other cases. Whereby the Hartree-Fock value of $C$ ranges from 0.5 eV for Ti$^{2+}$ to 0.8 eV for Cu$^{2+}$.

\begin{table}
\label{TableU}
\begin{center}
\begin{small}
\begin{tabular}{|c||c|c|c|c|c|}
  \hline
    & \multicolumn{5}{c|}{\vphantom{\Large{I}}Effective Coulomb interaction U$_{eff}$ for Hund's rule ground-state}\\
  \hline
    \vphantom{\Large{I}} & \multicolumn{2}{c|}{Full Hamiltonian}& Simple & Kanamori & Kanamori\\
                         & \multicolumn{2}{c|}{                }&        &          & mean field\\
  \hline
  $d^{1} $ &F$^0$-$\frac{ 8}{49}$F$^2$-$\frac{  9}{441}$F$^4$& U$_{0}$- J$_{H}$-C & U$_{0}$- J$_{H}$ 
         & U'- J & U'- J \vphantom{\Large{I}}\\
  $d^{2} $ &F$^0$+$\frac{ 1}{49}$F$^2$-$\frac{ 54}{441}$F$^4$& U$_{0}$- J$_{H}$+C & U$_{0}$- J$_{H}$ 
         & U'- J & U'- J \vphantom{\Large{I}}\\
  $d^{3} $ &F$^0$+$\frac{ 1}{49}$F$^2$-$\frac{ 54}{441}$F$^4$& U$_{0}$- J$_{H}$+C & U$_{0}$- J$_{H}$ 
         & U'- J & U'- J \vphantom{\Large{I}}\\
  $d^{4} $ &F$^0$-$\frac{ 8}{49}$F$^2$-$\frac{  9}{441}$F$^4$& U$_{0}$- J$_{H}$-C & U$_{0}$- J$_{H}$ 
         & U'- J & U'- J \vphantom{\Large{I}}\\
  $d^{5} $ &F$^0$+$\frac{14}{49}$F$^2$+$\frac{126}{441}$F$^4$& U$_{0}$+4J$_{H}$   & U$_{0}$+4J$_{H}$ 
         & U+4J & U+4J \vphantom{\Large{I}}\\
  $d^{6} $ &F$^0$-$\frac{ 8}{49}$F$^2$-$\frac{  9}{441}$F$^4$& U$_{0}$- J$_{H}$-C & U$_{0}$- J$_{H}$ 
         & U'- J & U'- J \vphantom{\Large{I}}\\
  $d^{7} $ &F$^0$+$\frac{ 1}{49}$F$^2$-$\frac{ 54}{441}$F$^4$& U$_{0}$- J$_{H}$+C & U$_{0}$- J$_{H}$ 
         & U'- J & U'- J \vphantom{\Large{I}}\\
  $d^{8} $ &F$^0$+$\frac{ 1}{49}$F$^2$-$\frac{ 54}{441}$F$^4$& U$_{0}$- J$_{H}$+C & U$_{0}$- J$_{H}$ 
         & U'- J & U'- J \vphantom{\Large{I}}\\
  $d^{9} $ &F$^0$-$\frac{ 8}{49}$F$^2$-$\frac{  9}{441}$F$^4$& U$_{0}$- J$_{H}$-C & U$_{0}$- J$_{H}$ 
         & U'- J & U'- J \vphantom{\Large{I}}\\
  \hline
\end{tabular}
\end{small}
\end{center}
\caption{The effective Hubbard U between lowest Hund's rule ground-state multiplets for different approximation schemes. The relation between the Slater integrals $F^0$, $F^2$ and $F^4$ and the parameters $U_0$, $J_H$, $C$, $U'$, $U$, $J$, $J'$ is: $U_{0}=F^0$, $J_H=\frac{1}{14}(F^2+F^4)$, $C=\frac{1}{14}(\frac{9}{7}F^2-\frac{5}{7}F^4)$, $U'=U-2J$, $J'=J$, $U=F^0+\frac{4}{49}F^2+\frac{36}{441}F^4$ and $J=\frac{2.5}{49}F^2+\frac{22.5}{441}F^4$ \cite{Mizokawathesis, Marelthesis}. The average U for the full Hamiltonian is: $U_{av}=F^0-\frac{14}{441}(F^2+F^4)$.}
\end{table}

This remarkable accuracy is important for the study of the conductivity gap in Mott-Hubbard insulators. The energy gap to move an electron from one site to another site far away is given by $U$:
\begin{equation}
U(d^n)=E(d^{n-1})+E(d^{n+1})-2E(d^{n})
\end{equation}
If multiplet effects are important, then there will be many different manners to move the electron. The gap is equal to the cheapest way to move one electron from one site to another site far away and thus given by the $U=U_{eff.}$, which involves the lowest multiplet states and not by $U_{av}$, which refers to the the multiplet average energies. In table 1.2 we have listed $U_{eff}$ for the various $d^n$ configurations $(n=1...9)$, having the high-spin Hund's rule ground-state. We can see here that the correspondence between the simple scheme and the full-multiplet theory is very good. The deviation is not more than $C\approx0.5 ... 0.8$ eV, which is quite acceptable since the accuracy in the determination of $U_{0}$ by experiment or calculation is of the same order.

  \begin{figure}
   \begin{center}
    \includegraphics[width=120mm]{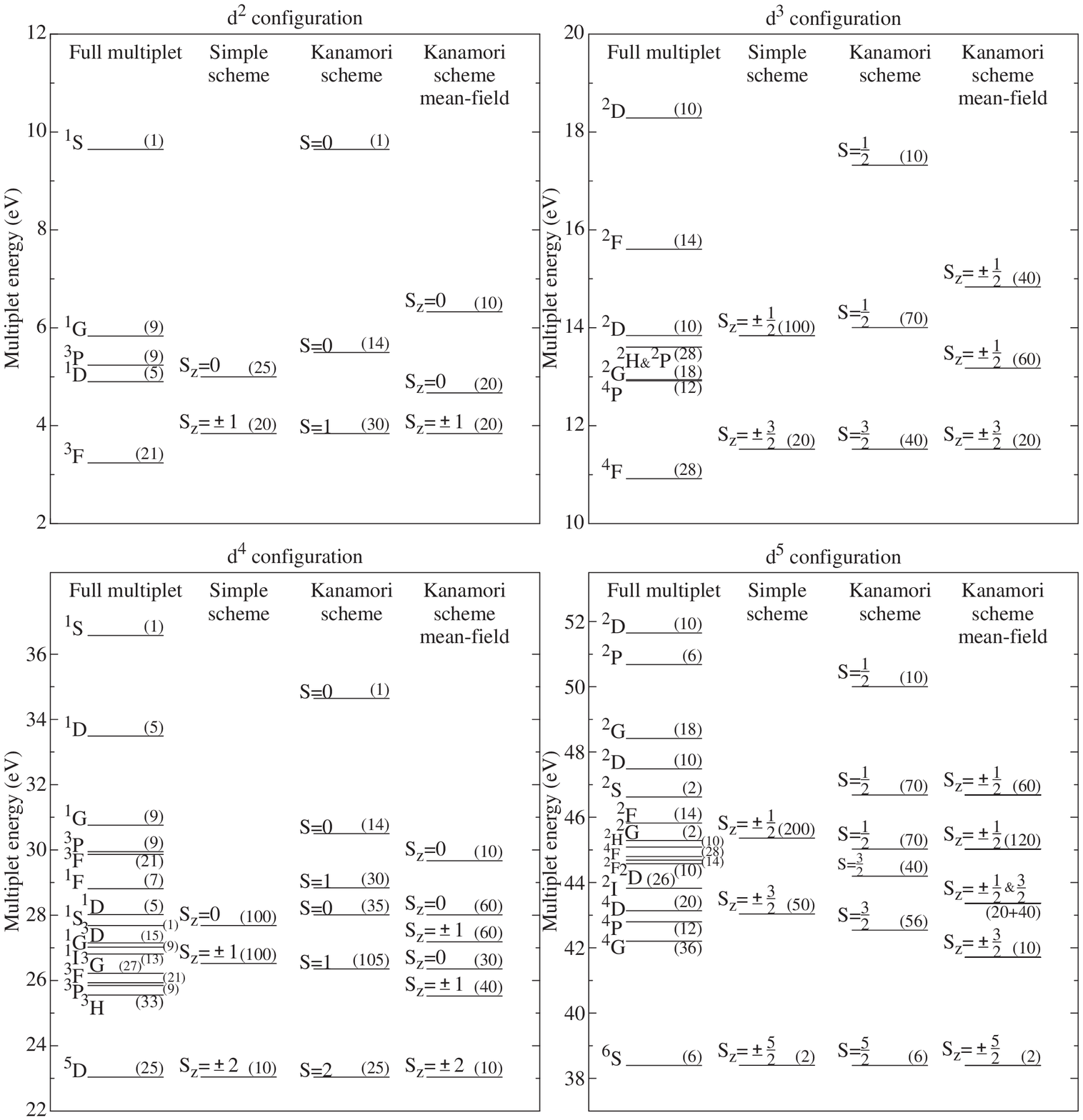}
    \caption{Energy levels of a $d^2$, $d^3$, $d^4$ and a $d^5$ configuration in spherical symmetry, split due to electron-electron repulsion within the full electron-electron repulsion Hamiltonian, a simplified scheme, as explained in the text, the Kanamori scheme and the Kanamori scheme calculated in mean-field theory. The values taken for the Slater integrals are $F^0$=5 eV, $F^2$=10 eV and $F^4$=6.25 eV. On the left of each term we wrote the term name. On the right of each term we show the multiplicity in brackets. The energies for the $d^6$, $d^7$ and $d^8$ configurations can be found by the equivalence of electrons and holes. In spherical symmetry the $d^n$ configuration has the same multiplet splitting as the $d^{10-n}$ configuration, only shifted in total energy.}
    \label{introductionMultipletEnergies}
    \end{center}
   \end{figure}

The shortcomings of the simple scheme do, however, show up dramatically when we have to consider the presence of states other than the Hund's rule ground-state. In figure \ref{introductionMultipletEnergies} we show the energy level diagram for the $d^n$ configuration $(n=2...5)$, in spherical symmetry as an example. We can see that the multiplet splitting within the simple scheme is very different from that within the full multiplet theory. This first of all means that the simple scheme is completely useless for calculating high energy excitations, like soft-x-ray absorption spectra. One truly needs the full multiplet theory to explain the many sharp structures observed in experiments \cite{Antonides77, Ballhausen62, Fujimori84a, Fujimori84b, Graaf98, Groot90, Tjeng91a, Groot05}. Not only for spectroscopy the simple scheme is inadequate, also for calculating ground-state properties, such as the magnetic susceptibility: the multiplicities are quite different. Within the simplified scheme only the magnetic angular spin momentum $S_z$ is a good quantum number, whereas in the full electron-electron repulsion Hamiltonian also the angular momenta of spin and orbital moment, $S$ and $L$, are good quantum numbers. In other words the simple scheme brakes symmetry. It is almost needles to state that the simple scheme is by far to inaccurate to calculate, the relative stability of, for instance, the different spin-states within a particular configuration. Figure \ref{introductionMultipletEnergies} shows clearly that the energy levels in the two schemes differ by many electron volts. 

In order to improve on the simple scheme, but still simplify the full electron-electron Hamiltonian we will have a look at the simplification as proposed by J. Kanamori. He proposed a simplification that tries to preserve the multiplet character of the electron-electron interaction as much as possible. The approximation he made is that the electron-electron scattering events can be expressed in terms that only depend on wether the scattered electrons are in the same band (orbital) or in different bands (orbitals). The Hamilton found by these approximations is \cite{Kanamori63}:
\begin{equation}\begin{split}
H_{e-e}^{Kanamori}=&U\sum_{m}l^{\dagger}_{m\uparrow}{l}^{\dagger}_{m\downarrow}l_{m\uparrow}{l}_{m\downarrow}\\
+&U'\sum_{m\neq m'}l^{\dagger}_{m\uparrow}{l'}^{\dagger}_{m'\downarrow}l_{m\uparrow}{l'}_{m'\downarrow}\\
+&U'\sum_{m>m'\sigma}l^{\dagger}_{m\sigma}{l'}^{\dagger}_{m'\sigma}l_{m\sigma}{l'}_{m'\sigma}\\
+&J\sum_{m>m'\sigma}l^{\dagger}_{m\sigma}{l'}^{\dagger}_{m'\sigma}{l'}_{m'\sigma}l_{m\sigma}\\
+&J\sum_{m\neq m'}l^{\dagger}_{m\uparrow}{l'}^{\dagger}_{m'\downarrow}l'_{m'\uparrow}{l}_{m\downarrow}\\
+&J'\sum_{m\neq m'}l^{\dagger}_{m\uparrow}{l}^{\dagger}_{m\downarrow}l'_{m'\uparrow}{l'}_{m'\downarrow}\\
\end{split}\end{equation}

$U$ is the parameter that describes the direct repulsion between two electrons in the same orbital; $U=\int \int \phi^*(r_1)\phi^*(r_2)$ $\frac{e^2}{|r_1-r_2|}$ $\phi(r_1)\phi(r_2)\delta r_1 \delta r_2$. $U'$ describes the direct repulsion between two electrons in different orbitals; $U'=\int \int \phi^*(r_1)\phi'^*(r_2)$ $\frac{e^2}{|r_1-r_2|}$ $\phi(r_1)\phi'(r_2)\delta r_1 \delta r_2$. It is assumed that this repulsion is equal between all different orbitals, $\phi$ and $\phi'$. One can easily understand that this is not true. For example, the repulsion between the $d_{xy}$ and the $d_{x^2-y^2}$ orbital is larger than between the $d_{z^2}$ and the $d_{x^2-y^2}$ orbital. The value of $U'$ taken when we compare the Kanamori scheme with the full multiplet calculation is the average of the repulsion between all different orbitals \cite{Mizokawathesis}. The process where two electrons are interchanged is described by the exchange integral; $J=\int \int \phi^*(r_1)\phi'^*(r_2)$ $\frac{e^2}{|r_1-r_2|}$ $\phi'(r_1)\phi(r_2)\delta r_1 \delta r_2$. The process where two electrons residing in the same orbital scatter on each other and are transferred from one orbital into another is characterized by the integral; $J'=\int \int \phi^*(r_1)\phi^*(r_2)$ $\frac{e^2}{|r_1-r_2|}$ $\phi'(r_1)\phi'(r_2)\delta r_1 \delta r_2$.

In table 1.1 and 1.2 we show the energy of the lowest Hund's rule multiplet and the effective $U$ within the Kanamori scheme. Like in the case of the simple scheme, one find that the Kanamori scheme gives very similar and sometimes identical values as the full multiplet theory, as long as one limits oneself to the high-spin Hunds's rule ground-state.

The energetics of the excited states and the multiplicities of the states as calculated using the Kanamori scheme differ nevertheless still quite appreciably from the full multiplet theory. Although, the spread of the multiplets is now much improved as compared to that coming from the simple scheme, the Kanamori multiplet structure still will not be of use to explain the high energy soft-x-ray absorption data, which usually contain many very detailed and sharp fine structures. Also, like in the case for the simple scheme, the Kanamori approximation will be quite unreliable when one wants to calculate the relative energies of the different spin-states within a particular $d^n$ configuration.

Nevertheless, the Kanamori scheme is better than the simple scheme: (1) The spread of the multiplet splitting is much better reproduced; (2) The multiplicity for the Hund's rule ground-state is less wrong; within the Kanamori scheme all states with maximum $S$ are degenerate, while in the single scheme it is only all states with maximum $|S_z|$; and (3) within the Kanamori scheme $S$ is a good quantum number, whereas in the simple scheme it is not.

Sofar we have discussed approximations made in the scattering amplitude $U_{mm'm''m'''\sigma\sigma'\sigma''\sigma'''}$. There are however, more fundamental approximations made when calculating the electron-electron repulsion. Within the electron-electron repulsion Hamiltonian $U_{mm'm''m'''\sigma\sigma'\sigma''\sigma'''}$ is multiplied by two creation and two annihilation operators, $l_{m\sigma}^{\dagger}{l'}^{\dagger}_{m'\sigma'}{l_{m''\sigma}''}{l_{m'''\sigma'}'''}$. In order to calculate eigenvalues of a particular state one has to evaluate the integral $\langle l_{m\sigma}^{\dagger}{l'}^{\dagger}_{m'\sigma'}{l_{m''\sigma}''}{l_{m'''\sigma'}'''} \rangle$. If one uses a mean-field approximation, the Hartree-Fock approximation or density functional theory, this integral over four operators reduces to two integrals over two operators, $\langle l_{m\sigma}^{\dagger}{l_{m''\sigma}''} \rangle \langle {l'}^{\dagger}_{m'\sigma'}{l_{m'''\sigma'}'''} \rangle$. We will discuss the effect of this mean field approximation within the Kanamori scheme, as it shows very instructive what happens.

Within mean-field theory the expectation values of the Kanamori Hamiltonian reduce to:
\begin{equation}\begin{split}
\langle H_{e-e}^{Kanamori} \rangle = &  U\sum_{m} \langle l^{\dagger}_{m\uparrow}l_{m\uparrow} \rangle \langle {l}^{\dagger}_{m\downarrow}{l}_{m\downarrow} \rangle\\
+ & U'\sum_{m\neq m'} \langle l^{\dagger}_{m\uparrow}l_{m\uparrow} \rangle \langle {l'}^{\dagger}_{m'\downarrow} {l'}_{m'\downarrow} \rangle\\
+ & U'\sum_{m>m'\sigma} \langle l^{\dagger}_{m\sigma}l_{m\sigma} \rangle \langle {l'}^{\dagger}_{m'\sigma}{l'}_{m'\sigma} \rangle\\
- & J\sum_{m>m'\sigma} \langle l^{\dagger}_{m\sigma}l_{m\sigma} \rangle \langle {l'}^{\dagger}_{m'\sigma}{l'}_{m'\sigma} \rangle\\
\end{split}\end{equation}
The two terms, $J\sum_{m\neq m'}l^{\dagger}_{m\uparrow}{l'}^{\dagger}_{m'\downarrow}l'_{m'\uparrow}{l}_{m\downarrow}$ and $J'\sum_{m\neq m'}l^{\dagger}_{m\uparrow}{l}^{\dagger}_{m\downarrow}l'_{m'\uparrow}{l'}_{m'\downarrow}$ drop since they are only non-zero between two wave-functions that differ by two single electron wave functions from each other. We now see that the Kanamori Hamiltonian within mean-field theory only depends on number operators and therefore that it is diagonal within the basis vectors that span the full electron-electron Hamiltonian. Within mean-field theory there are remarkable similarities between the Kanamori scheme and the simple approximation we started with. The Kanamori Hamiltonian can in mean-field approximation be described by three parameters, $U$, $U'$ and $J$, in which $U$ is the repulsive Coulomb energy between two electrons in the same orbital, with opposite spin, $U'$ is the repulsive Coulomb energy between two electrons in different orbitals and $J$ is the attractive Hund's exchange interaction between each pair of electrons with parallel spin. The only difference with the simple scheme is that we introduced a $U$ and $U'$ that accounts for a difference between repulsion when two electrons are in the same orbital, or if they are in a different orbital.

In table 1.1 and 1.2 we show the energy of the lowest Hund's rule multiplet and the effective $U$ in the mean-field approximation of the Kanamori scheme. It should not be a surprise that these values are very similar to the full-multiplet theory. In fact the values are identical to the values found for the Kanamori scheme. So for the energy of the Hunds's rule ground-state it does not matter if one calculates the Kanamori scheme in mean-field theory or not. The multiplicity of the ground-state and the energies of the excited state as calculated within the Kanamori scheme within mean-field approximation differ considerably from the Kanamori scheme as can be seen in figure \ref{introductionMultipletEnergies}. The spread of the multiplets is, like in the simple scheme, much smaller than in the full multiplet calculation. The multiplicity of the ground-state is incorrect and similar as in the simple scheme, only $S_z$ is a good quantum number and $S$ is not.

The importance of the full electron-electron repulsion Hamiltonian should not be underestimated. From high energy spectroscopy it has been shown that the full electron-electron repulsion Hamiltonian has to be included to explain the spectra. This is also true for ground-state properties. For example there is a lively discussion about the spin-states of many of the cobaltates materials synthesized. A meaningful comparison of different multiplet terms can only be done if the electron-electron repulsion is correctly included. The differences of total energies between different spin states are often small and the differences between the full electron-electron repulsion Hamiltonian and the approximations shown here are quite large. Another example is CoO. There one finds an ordered orbital momentum that is quite large and experiment hints that it even might be larger than $1\mu_B$ \cite{Jauch02}. For a $d^7$ system with one hole in the $t_{2g}$ orbitals this is not so easy to understand as the maximum orbital momentum of a $t_{2g}$ orbital is $1\mu_B$. The only way to understand this large orbital momentum is by inclusion of the correct full electron-electron repulsion Hamiltonian. A detailed explanation of this effect can be found in chapter \ref{ChapterCoO}.

Finally, we would like to remind the reader again that there are different definitions of $U$ and $J$ around. The $U$ as defined in the simple scheme is equal to $F^0$, the $J$ to $\frac{1}{14}(F^2+F^4)$. The $U$ in the Kanamori scheme is equal to $F^0+\frac{4}{49}F^2+\frac{36}{441}F^4$ and the $J$ to $\frac{2.5}{49}F^2+\frac{22.5}{441}F^4$. Within full multiplet theory $U_{av}$ is equal to $F^0+\frac{14}{441}(F^2+F^4)$. The effective $U_{eff.}$ depends on the configuration one looks at. Within this thesis when we talk about $U$ we mean the $U$ defined within the full multiplet theory with respect to the multiplet average.

\section{Scope}

This thesis can be divided into three parts. Within the first part, ranging from chapter \ref{chapterXrayabs} to \ref{ChapterSumrules} we will discuss some properties of x-ray absorption spectroscopy and show how these spectra can be interpreted with the use of cluster calculations and sum-rules. The second part, ranging from chapter \ref{ChapterNiO} to chapter \ref{ChapterMnO} will be about applications in thin-film research. In these chapters we will have a closer look on strain induced low symmetry crystal fields and orbital and spin magnetic moments. The third part, ranging from chapter \ref{ChapterCobaltates} to \ref{ChapterVO2}, will deal with spin and orbital degrees of freedom in bulk transition metal compounds.

In chapter \ref{chapterXrayabs} an overview from literature is given of the different material properties, or operator values, one can measure with the use of x-ray absorption spectroscopy. With the use of cluster calculations and dipole selection rules it is shown why x-ray absorption spectroscopy is extremely sensitive to many of the initial state operator values we are interested in.

In chapter \ref{ChapterCluster} some concepts of cluster calculations are given. This overview is by no means complete, as already many very good textbooks exist on this topic. We do not concentrate on simplifications one can make in diagonalizing the Hamiltonian, but assume this is done with the use of a PC and brute force.

In chapter \ref{ChapterSumrules} we discuss the sum-rules present in x-ray absorption spectroscopy as derived by Thole and Carra \textit{et al.} \cite{Thole92, Carra93a}. These sum-rules have become very important in the interpretation of x-ray absorption spectroscopy due to their simplicity of use. In chapter \ref{ChapterSumrules} we will spent some time on how these sum-rules can be derived based on second quantization on the same line of thought as proposed by M. Altarelli \cite{Altarelli93}.

In chapter \ref{ChapterNiO} we present linear dichroism in the Ni $L_{2,3}$ x-ray absorption spectra of a monolayer NiO(001) on Ag(001) capped with MgO (001). The dichroic signal appears to be very similar to the magnetic linear dichroism observed for thicker antiferromagnetic NiO films. A detailed experimental and theoretical analysis reveals, however, that the dichroism is caused by crystal field effects. We present a practical experimental method for identifying the independent magnetic and crystal field contributions to the linear dichroic signal in spectra of NiO films with arbitrary thickness and lattice strain.

In chapter \ref{ChapterCoO} we first used XAS to study the properties of CoO bulk, as well as thin films. We confirm that the Co ion in CoO has a free orbital momentum in cubic symmetry. We confirm that spin-orbit coupling is very important for understanding the properties of CoO and show that it is not reduced from the Hartree-Fock value for a free Co$^{2+}$ ion. With the use of cluster calculations we can get a full consistent understanding of the XAS spectra and the polarization dependence of CoO. For CoO thin films we used XAS to show that we can control the orbital momentum and spin direction with the application of strain to thin CoO films. This finding opens up great opportunities for the use of exchange-bias, where people put an antiferromagnet adjoined to a ferromagnet in order to shift the magnetization hysteresis loop in one of the magnetic field directions \cite{Meiklejohn56, Meiklejohn57}. With the application of strain in the antiferromagnet one can chose if the system will be exchange-biased in the plane of the thin film, or perpendicular to the thin film surface. It also has great implications for the understanding of the exchange-bias phenomenon so far. The exchange-bias effect takes place at the interface between the ferromagnet and antiferromagnet \cite{Nogues99, Berkowitz99}. At the interface there will be strain in the antiferromagnet and one can not assume that the antiferromagnet has the same spin structure at the interface as it has in the bulk.

In chapter \ref{ChapterMnO} we show how one can orientate spins in antiferromagnetic thin films with low magnetocrystalline anisotropy ($d^3$, $d^5$ and $d^8$ systems in $O_h$ symmetry) via the exchange coupling to adjacent antiferromagnetic films with high magnetocrystaline anisotropy ($d^6$ and $d^7$ systems in nearly $O_h$ symmetry). We have grown MnO thin films on CoO thin films with different predetermined spin orientation. With the use of Mn $L_{2,3}$ soft x-ray absorption spectroscopy we show that the Mn spin 'follows' the Co spin direction.

In chapter \ref{ChapterCobaltates} we study the spin-state problem within the cobaltates. Normally one is used to discuss the spin direction, or the magnetic spin angular momentum, $S_z$. Within the cobaltates, $d^6$ compounds, there is however a discussion about the size of the spin, or the spin angular momentum, $S^2=S(S+1)$. $S^2$ can be 0, 2, or 6 (S=0,1,2), referred to as a low-spin state, an intermediate-spin state and a high-spin state. Within the literature there is a lot of confusion about the spin state as deduced from magnetic, neutron and x-ray diffraction measurements in the newly synthesized layered cobalt perovskits \cite{Martin97, Maignan99, Yamaura99a, Yamaura99b, Loureiro01, Vogt00, Suard00, Fauth01, Burley03, Mitchell03, Moritomo00, Respaud01, Kusuya01, Frontera02, Fauth02, Taskin03, Soda03, Loureiro00, Knee03, Wu00, Kwon00, Wang01, Wu01, Wu02, Wu03}. These measurements determine the size of the spin angular momentum ($S^2$) from the maximum size of the magnetic spin momentum $S_z$. XAS is directly sensitive to the expectation value of $S^2$, the spin angular momentum. We carried out a test experiment using a relatively simple model compound, namely Sr$_2$CoO$_3$Cl, in which there are no spin state transitions present and in which there is only one kind of Co$^{3+}$ ion coordination \cite{Loureiro00,Knee03}. Important is that this coordination is identical to the pyramidal CoO$_{5}$ present in the heavily debated layered perovskites \cite{Martin97, Maignan99, Yamaura99a, Yamaura99b, Loureiro01, Vogt00, Suard00, Fauth01, Burley03, Mitchell03, Moritomo00, Respaud01, Kusuya01, Frontera02, Fauth02, Taskin03, Soda03}. Using a \textit{spectroscopic} tool, that is soft x-ray absorption spectroscopy (XAS), we demonstrate that pyramidal Co$^{3+}$ ions are not in the often claimed intermediate-spin state but unambiguously in a high-spin state. This outcome suggests that the spin states and their temperature dependence in layered cobalt perovskites may be rather different in nature from those proposed in the recent literature.

In chapter \ref{ChapterLaTiO3} we study LaTiO$_{3}$. There has been a strong debate about the role of orbital degrees of freedom within the titanates and in LaTiO$_{3}$ especially \cite{Cwik03, Goral83, Meijer99, Mizokawa96, Keimer00, Khaliullin00, Mochizuki03, Pavarini04, Solovyev04, Craco04}. With the use of spin resolved circular polarized photo electron spectroscopy we confirmed that the orbital momentum in LaTiO$_{3}$ is indeed quenched \cite{Keimer00}. With the use of XAS we show that this is due to a relative large crystal field in the order of 120 to 300 meV. For a realistic description of materials one should not forget that there is a strong coupling between the orbitals and the lattice. 

In chapter \ref{ChapterVO2} we look at the metal-insulator transition in VO$_{2}$. With the use of XAS we show that the metal-insulator transition within this material is accompanied by a change in orbital occupation. The orbital occupation changes from almost isotropic in the metallic phase to the almost completely $\sigma$-polarized in the insulating phase, in close agreement with the two-site cluster model \cite{Tanaka04}. This very strong orbital polarization leads in fact to a change of the electronic structure of VO$_{2}$ from a 3-dimensional to effectively a 1-dimensional system \cite{Khomskii05}. The V ions in the chain along the c-axis are then very susceptible to a Peierls transition. In this respect, the MIT in VO$_{2}$ can indeed be regarded as a Peierls transition \cite{Wentzcovitch94}. However, to achieve the required dramatic change of the orbital occupation one also need the condition that strong electron correlations bring this narrow band system close to the Mott regime \cite{Rice94}. The MIT in VO$_{2}$ may therefore be labelled as a "collaborative" Mott-Peierls transition.

\chapter{X-ray absorption spectroscopy}
\label{chapterXrayabs}

X-ray absorption spectroscopy (XAS) is based on core-level absorption. This means that one makes excitations with light of high energy with respect to the chemical binding energies of the system. These excitations are dipole allowed and therefore do have good absorption cross sections. For $2p$ elements like C,N and O one can make a $1s$ to $2p$ excitation, found in the 300 - 700 eV range. For the $3d$ transition metals one can make a $2p$ to $3d$ excitation found in the 350 to 950 eV energy range. The $4d$ transition metals have their $2p$ to $4d$ excitations at an energy from 2000 to 3500 eV. For the rare earth Lanthanides one can make a $3d$ to $4f$ excitation found in the 800 to 1700 eV energy range. These are the most useful excitations, but more excitations are possible. For the $3d$ elements one could make excitations into the $4p$ shell, from 1s, but these excitations do not probe directly the valence shell and are therefore less informative. In this thesis we will concentrate on the $3d$ transition metals and excitations from the $2p$ to $3d$ shell therein. Historically this absorption edge is called the $L$ edge.

The light source used to create light within this energy range is synchrotron radiation. Synchrotron radiation of modern synchrotrons is bright enough to do most XAS experiments wanted. Important is, however, to be able to scan through the energy range of the spectrum in a fast and reproducible way. Modern beam-lines become more and more stable and have a very good accuracy and reproducibility, necessary in many of the dichroic experiments done in this thesis. Stability really becomes a big issue when one wants to compare spectra taken at different temperatures. The spectra in this thesis have been taken at the banding magnet Dragon beam-line at the NSRRC of Taiwan and the undulator Dragon beam-line ID08 at the ESRF in Grenoble. Although banding magnet beam-lines are less brilliant than undulator beam-lines, they are still a good choice for absorption spectroscopy. This while not the brilliancy of the light source is the bottle neck in many experiments, but stability, scanning speed and reproducibility is more important. The absorption measurements at ID08 have been done with a fixed gap of the undulator.

  \begin{figure}[!h]
   \begin{center}
    \includegraphics[width=90mm]{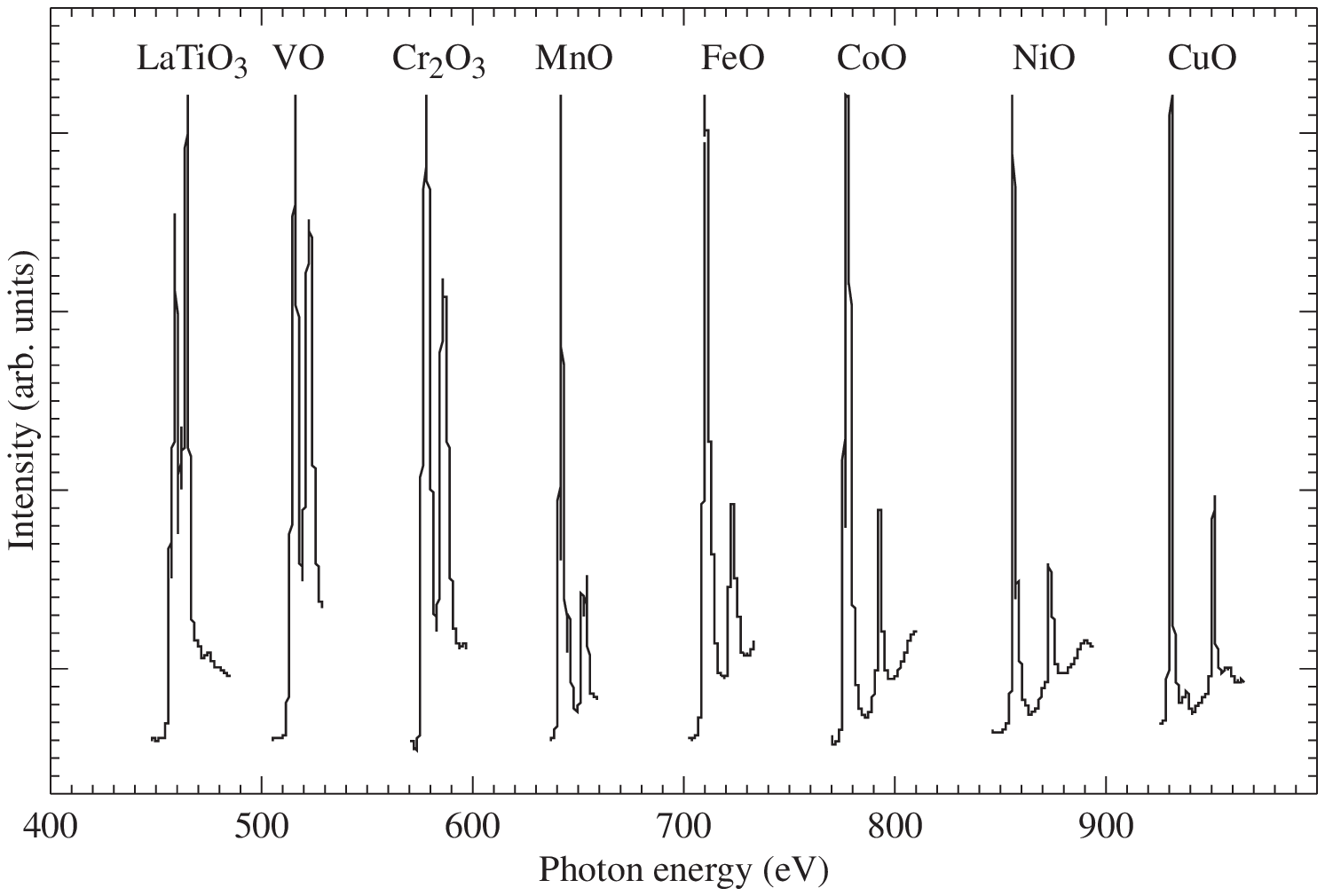}
    \caption{2p-Core level spectroscopy of different transition metal compounds. Each element has its absorption edge at a different energy.}
    \label{introductionEnergyXAS}
    \end{center}
   \end{figure}
  
In figure \ref{introductionEnergyXAS} we show the XAS spectra for some selected $3d$ transition metal oxides. The first thing one should notice is that each element has its absorption edge at a different energy. The reason for that is actually very simple. The more protons in the nucleus the more the core electrons are bound to the ion and the more energy it will cost to bring a core electron into the valence shell. The fact that each element has its own energy for the core level absorption edge makes XAS very powerful. By selecting the energy of a specific element one will get information about the properties of only that element in the material. This allows one to do element specific measurements.
  
  \begin{SCfigure}[][!h]
    \includegraphics[width=60mm]{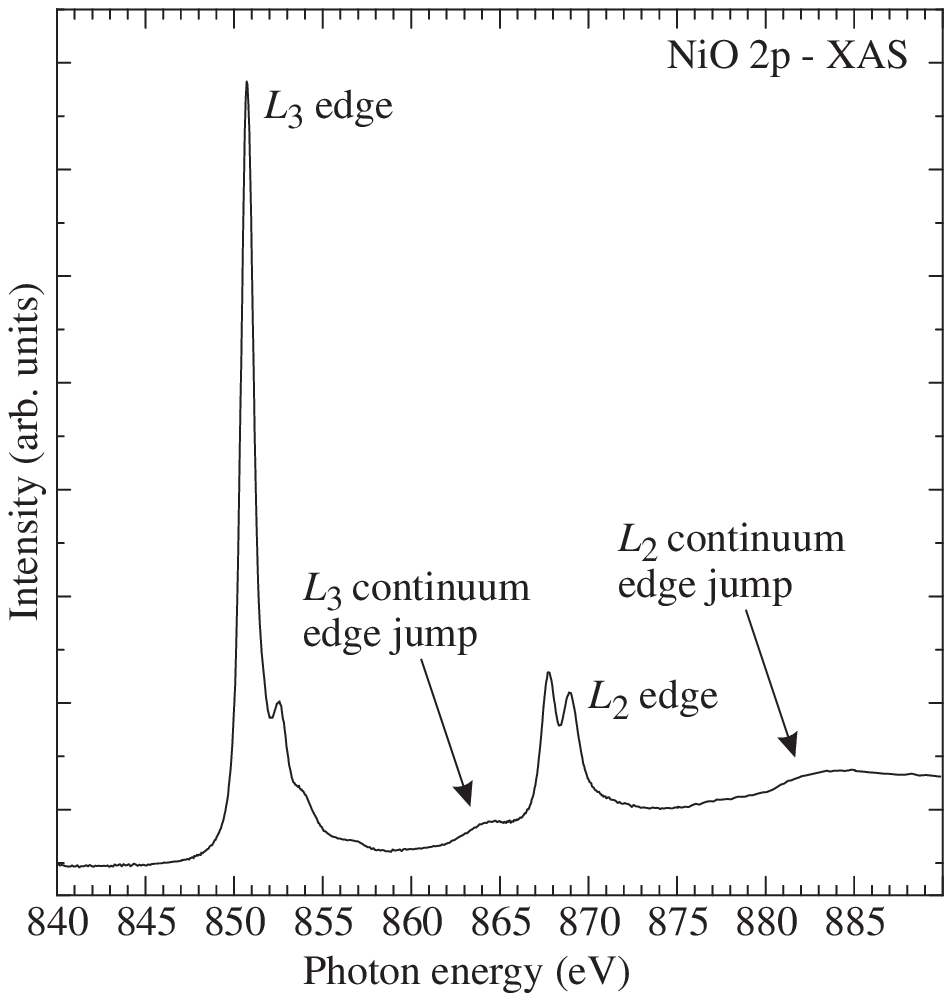}
    \caption{A general example of $2p$-XAS. One can clearly see the general features present. The $L_2$ and the $L_3$ edges and the corresponding continuum edge-jumps about 15 eV higher.\newline}
    \label{introductionNiO}
  \end{SCfigure}

There are some features that all $2p$ core level XAS spectra have in common. In figure \ref{introductionNiO} we show the $2p$-XAS spectra of NiO. One can clearly see two sets of peaks, one around 850 eV, the other around 867 eV. The spectrum is split in two parts due to the  $2p$-core level spin-orbit coupling. This energy is not small, because the $2p$ electrons are close to the core and therefore should be treated relativistically. Historical the two different peak structures are called the $L_{2}$ edge and the $L_{3}$ edge. In figure \ref{introductionEnergyXAS} one can see that the splitting between the $L_3$ and the $L_2$ edge is larger for the late transition metal compounds than for the early transition metal compounds. The late transition metal atoms have more charge at the nucleus and therefore a tighter bound $2p$ shell, which enlarges the $2p$ spin-orbit coupling constant. In figure \ref{introductionNiO} one can also see that the intensity above 885 eV is larger than the intensity below 845 eV. This is called the edge-jump. The increase in overall spectral intensity, often called the background, takes largely place in two distinct small energy areas. In accordance to G. van der Laan et al. \cite{Laan86}, these have been labelled as the continuum edge-jump belonging to the $L_2$ and the $L_3$ edge. We will show below that the $L_2$ and the $L_3$ edges are excitons. The $2p$-core hole is bound to the additional $3d$ electron whereas the continuum-edge jump is due to non-bonding final states.

  \begin{figure}[!h]
   \begin{center}
    \includegraphics[width=90mm]{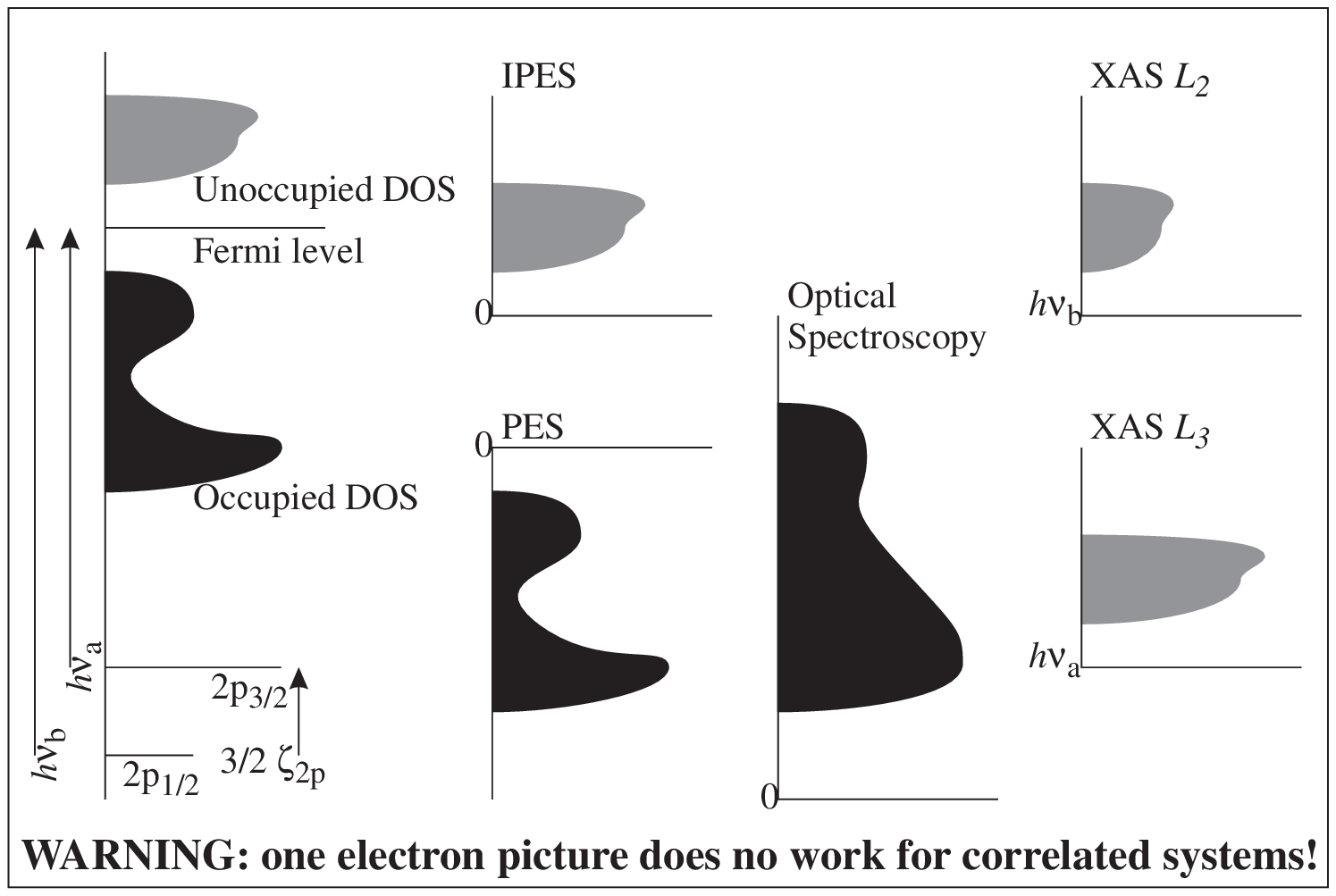}
    \caption{Interpretation of spectra within a one-electron theory. The shape of the spectra represents the band-structure.}
    \label{introductionOneElectron}
   \end{center}
  \end{figure}
  
\section{One-electron theory}  

In order to understand why these features are present we start by describing what one should expect if no electron-electron correlations were present. Without correlations materials and there spectra can very well be explained in an one-electron picture. In figure \ref{introductionOneElectron} we show, on the left side a density of states (DOS) for an arbitrary insulator, with a $3d$ valence shell. Below the fermi energy the levels are occupied and above the fermi energy they are empty. We also show the $2p$ core level at deep energies. Other core levels, the $1s$, $2s$, $3s$, and $3p$ are omitted for clarity. The $2p$ core level is a delta function, since there is no overlap between core levels of neighboring atoms, resulting in a flat band and a delta function as DOS. It is important to notice that we drew two delta functions for the $2p$ core level. This is done on purpose while the $2p$ core level is split by spin-orbit coupling. Since we are talking about a core level the spin-orbit coupling constant is not small, but it varies from 1.88 eV for K to 15.7 eV for Zn. The $2p$ shell has an orbital momentum of 1 and spin of $\frac{1}{2}$. This results for a $2p$ electron in a total angular momentum $j=\frac{1}{2}$ or $j=\frac{3}{2}$. There are 2 times as much $2p$ orbitals with $j=\frac{3}{2}$ than orbitals with $j=\frac{1}{2}$. Therefore we drew one delta function twice as high as the other one. Excitations made from the 2p orbital with $j=\frac{1}{2}$ to the $3d$ valence band are called the $L_2$ edge and excitations made from the $2p$ orbital with $j=\frac{3}{2}$ the $L_3$ edge.

In the middle left panel of figure \ref{introductionOneElectron} we drew the photo emission spectra (PES) and inverse photo emission spectra (IPES). Within a single electron theory these types of spectroscopy measure the occupied and unoccupied DOS respectively. For optical spectroscopy one makes excitations from the occupied DOS to the unoccupied DOS. This means the spectra measured is a convolution of both, the occupied and the unoccupied DOS. There is a minimum energy required to make an excitation at all, the optical gap. This is the distance between the highest occupied state to the lowest unoccupied state. In the right panel of figure \ref{introductionOneElectron} we show the XAS spectra expected in a single electron theory. The principle of x-ray spectroscopy is similar to optical spectroscopy, we make an excitation from the occupied core DOS to the unoccupied DOS. Again the spectrum found is a convolution of the core DOS with the unoccupied valence DOS. But in this case the $2p$ core DOS consists of 2 delta peaks. Therefore the $2p$ XAS consists of 2 times the unoccupied DOS shifted in energy by $\frac{3}{2}$ times the $2p$ core level spin orbit constant. Thereby one should notice that, since there are 2 times as much $2p$ core orbitals with $j=\frac{3}{2}$ than with $j=\frac{1}{2}$, there is a difference in intensity of a factor of two between the $L_2$ and $L_3$ edge.
  
  \begin{SCfigure}[][!h]
    \includegraphics[width=60mm]{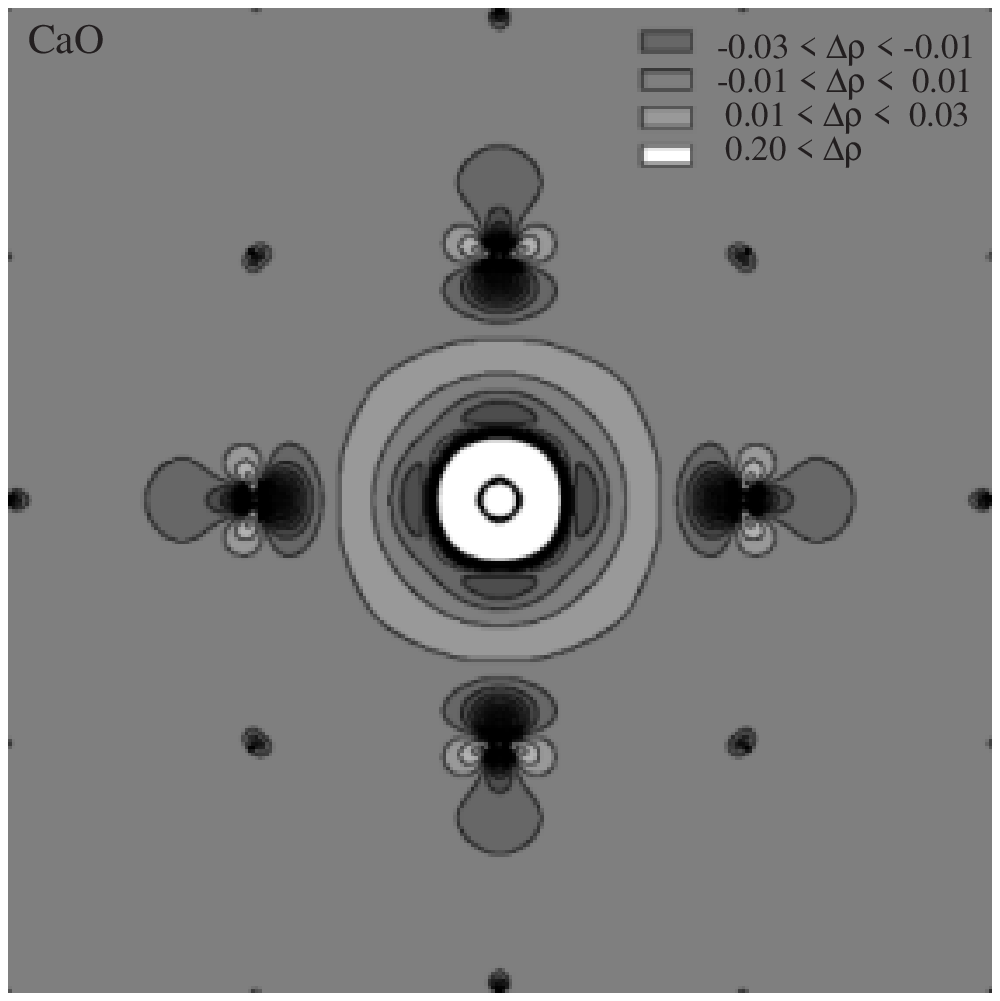}
    \caption{LDA electron density of the $2p$ core-hole $3d$ electron $2p$-XAS final-state exciton in CaO. Shown is the difference charge density of CaO with a core hole on the center Ca atom and one extra valence electron, and the charge density of normal CaO. On the corners and the middle of the sides Ca atoms are placed. On $\frac{1}{4}$ and $\frac{3}{4}$ of each side O atoms are placed. The rest of the square is filled in a checkerboard manner.}
    \label{introductionRho}
  \end{SCfigure}

\section{Excitons}

Within one-electron theory the $L_2$ edge is equal in line-shape to the $L_3$ edge and both edges are equal to the IPES spectra. If we look back at the spectra of NiO in figure \ref{ChapterNiO}, or further to any spectrum in this thesis, we see that the $L_2$ edge is generally quite different from the $L_3$ edge. This indicates that something is wrong with our one-electron explanation. We neglected the interaction between electrons and especially the interaction between the $2p$ core hole created and the electrons that are in the valence band, including the electron that is added to the valence band due to our spectroscopic process. These interactions are not small and should be taken serious. It is true that within a solid part of the electron-electron interactions are screened. The monopole part of the electron-electron interaction ($U$ or $F^0$) is within a solid reduced from the atomic value, however still important for the understanding of many transition metal compounds \cite{Mott49, Hubbard63, Sawatzky84, Zaanen85, Tsuda91, Imada98}. The multipole part of the electron-electron repulsion ($F^2$ and $F^4$, for $d-d$ interactions) is however not screened as shown, for example, by E. Antonides \textit{et al.} \cite{Antonides77}. For $2p$-XAS also the monopole part of the $2p$-$3d$ electron-electron interaction is very important \cite{Groot93b}.

It is well know that optical transitions can form excitons and exactly that is what happens in $2p$-XAS. In order to get a feeling of how tight the electron that is added to the valence band is bound to the hole it leaves behind we did a LDA calculation on CaO. CaO is a simple cubic rock-salt with a $4s^0 3d^0$ configuration at the Ca site. CaO is a band-insulator that should be reproduced quite nicely in LDA. In figure \ref{introductionRho} we plotted the final state charge-density minus the initial state charge density of CaO. The initial state charge density is the normal charge density of CaO, but now not calculated for a single unit cell, but for a cubic supper-cell consisting of 32 Ca atoms and 32 O atoms. The final state charge density has a $2p$ core hole on one of the Ca sites and an extra electron in the valence band. The charge density shown in figure \ref{introductionRho} includes all occupied bands, including the semi core states, but excluding the core states. The Ca-$1s$, $2s$ and $2p$ are treated as core-states. For the O atoms only the $1s$ states are treated as core states. The result is that the integral over the difference charge density, as plotted, is equal to one. The size of figure \ref{introductionRho} is 4 times the Ca-O distance on each side. The atom on which the $2p$ core hole has been made can be found exactly in the middle. Above, below and on the left and right of the excited Ca atom are O atoms. Around that are eight Ca atoms. This checkerboard like pattern of Ca and O atoms continues, until the edge of the figure. The calculations have been done with the use of WIEN2k \cite{Wien2k}. The colors are such that the large grey area means that the charge density for the final-state exciton is equal to the initial state. Lighter areas mean that the final state has more charge, dark areas mean less charge. One should note that the completely black areas are there because lines separating different charge densities got to close to each other to resolve, this does not mean that there is much less charge in these areas.

  \begin{SCfigure}[][h]
    \includegraphics[width=60mm]{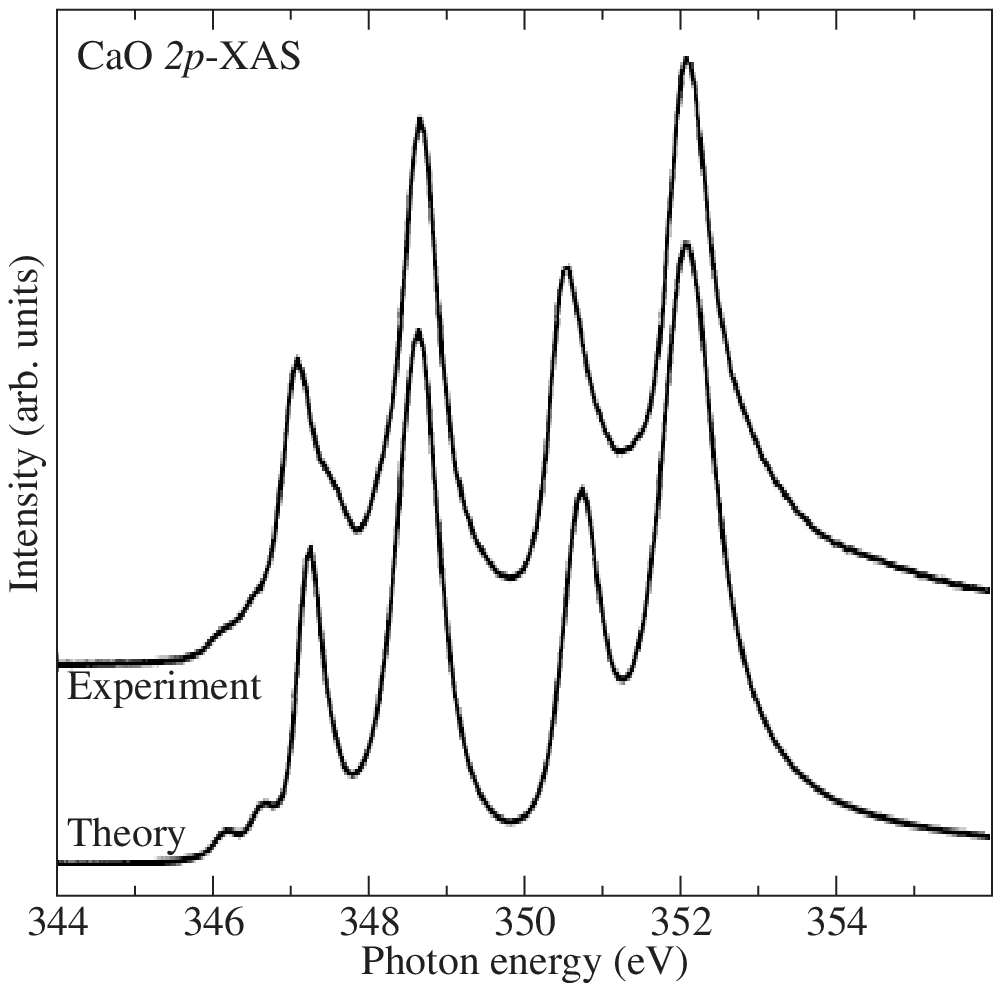}
    \caption{2p-XAS of CaO. The spectra can be reproduced within a Ca single atom calculation, including cubic crystal fields, a direct consequence of the localized final-state of 2p-XAS. The spectra and calculation are copied from F. J. Himpsel \textit{et al.} \cite{Himpsel91}.\newline}
    \label{introductionCaOXAS}
  \end{SCfigure}

In figure \ref{introductionRho} we see that the entire electron added to the valence shell is gathered around the atom with the core hole. Close to the core there is a large increase of charge. The semi-core shells contract due to the missing $2p$ electron. Then there is a small area where there is a small depletion of charge, followed by a wider 'corona' of extra charge. All of this happens however within the neighborhood of the Ca atom on which the core hole is made. If we look at the total change of charge around each atom we find that within our super-cell all atoms stay neutral, except for the O atoms close to the exciton. These atoms \underline{lose} about 0.02 electrons each. The Ca atom on which the core hole is made loses a $2p$ electron and gains almost one (about 0.85, depending on the muffin-tin sphere radius) electron within the muffin tin sphere. The additional charge is located in the inter-muffin-tin sphere area, close to the Ca atom with a $2p$ core hole. The $2p$ core hole completely binds the additional valence electron plus the electrons lost on the neighbor O atoms within a radius of about 1.5 \AA. Additional there are some multipole induced changes around the exciton. The O atoms close to the exciton have less charge in the $p$ orbital pointing to the exciton and more charge in the orbitals perpendicular to the direction to the exciton. The multipole changes on the next atom, a Ca atom, are very small.

We see that the final state of $2p$-XAS is a tightly bound exciton, that can be largely understood when considered only one atom. However the O atoms that are close to the exciton do react and for a full description should not be forgotten. Within a $TMO_6$ cluster however the entire exciton can be understood. Good news, since this is exactly the cluster that we can calculate without running into calculations that become to big for a PC.

In figure \ref{introductionCaOXAS} we show the $2p$ XAS of CaO as measured and calculated by F. J. Himpsel \textit{et al.} \cite{Himpsel91}. The top line is the experimental spectra, the bottom a theoretical calculation done for a single atom in cubic symmetry. We see that the spectra is excellent reproduced. A direct consequence of the closely bound exciton representing the final state of $2p$-XAS.
   
  \begin{SCfigure}[][h]
    \includegraphics[width=60mm]{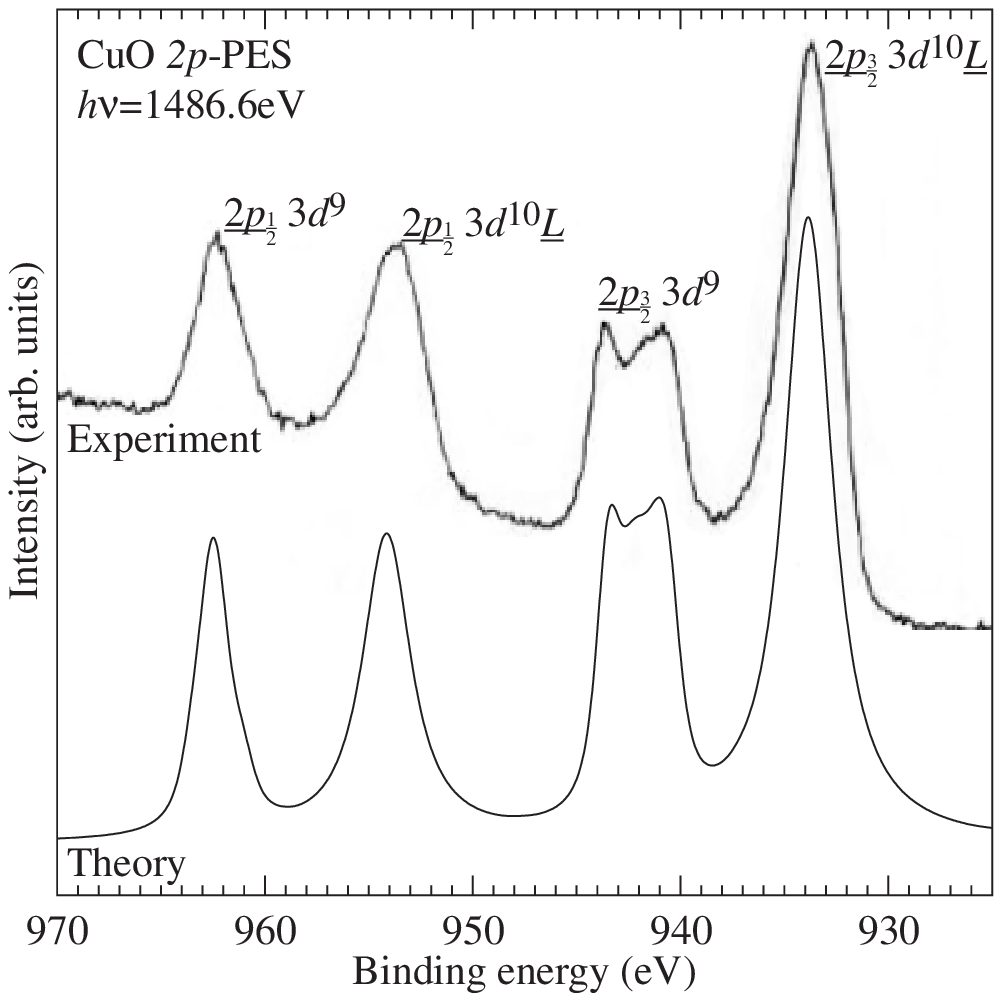}
    \caption{Cu $2p$ core level photo emission. The spectra are copied from. J. Ghijsen \textit{et al.} \cite{Ghijsen88}. \newline}
    \label{introductionCuO2pPES}
  \end{SCfigure}

The $2p$-XAS spectra are excitonic due to the large $2p$-$3d$ electron-electron interaction. That this interaction is really present for real materials can be nicely seen in $2p$ core level x-ray photo emission (XPS). In one-electron theory $2p$-XPS is quite boring to look at. Core levels are atomic like, have no k dependence and have a delta peak as DOS. In one-electron theory $2p$-XPS should be two delta peaks with an intensity ratio of 2 to 1 due to the $2p$ spin-orbit coupling. If there is however an interaction between $2p$ core levels and the $3d$ valence shell one would expect more peaks in the $2p$-XPS. J. Ghijsen \textit{et al.} measured $2p$-XAS on CuO and found four peaks. They explained these four peaks quite naturally in a many-electron picture. One should realize that Cu$^{2+}$ has 9 electrons in the $d$ shell. Therefore it is possible to remove a $2p$ electron and do nothing, this will change the Cu configuration from $3d^9$ to $2p^53d^9$. The second option is to remove a $2p$ core electron and move one electron from the oxygen to the Cu atom. If we denote the oxygen band by $L$ we could write the electron configuration in this case as $2p^53d^{10}\underline{L}$. Where $\underline{L}$ means a hole in the oxygen band. The $2p^53d^{10}\underline{L}$ configuration should be one peak, whereas for the $2p^53d^9$ configuration we would expect multiplet effects. 

Ghijsen \textit{et al.} \cite{Ghijsen88} measured the $2p$-XPS of CuO and found two peaks per edge, one with multiplet splitting and one without multiplet splitting. In figure \ref{introductionCuO2pPES} we show their measurements on CuO. The peak with the lowest energy does not show multiplet effects and originates from the $2p^53d^{10}\underline{L}$ configuration. The structured peak about 7 eV higher in energy is the multiplet of the $2p^53d^9$ configuration. This shows that the $2p$ hole $3d$ electron attraction is so large that it is energetically more favorable to make a hole in the oxygen band and put this electron close to the created $2p$ core hole than to leave the Cu atom with an extra positive charge. Below the measured spectra we also show a cluster calculation, that describes the spectra quite well. From cluster fits to the $2p$ XPS spectra as well as to valence band PES spectra Ghijsen \textit{et al.} \cite{Ghijsen88} deduce that there is an effective attractive potential of about 8.8 eV between a $2p$ core hole and $3d$ valence electron in CuO. This hole-electron attraction between $2p$ core holes and $3d$ electrons is also responsible for making $2p$ x-ray absorption spectroscopy excitonic like.
   
  \begin{figure}[!h]
   \begin{center}
    \includegraphics[width=75mm]{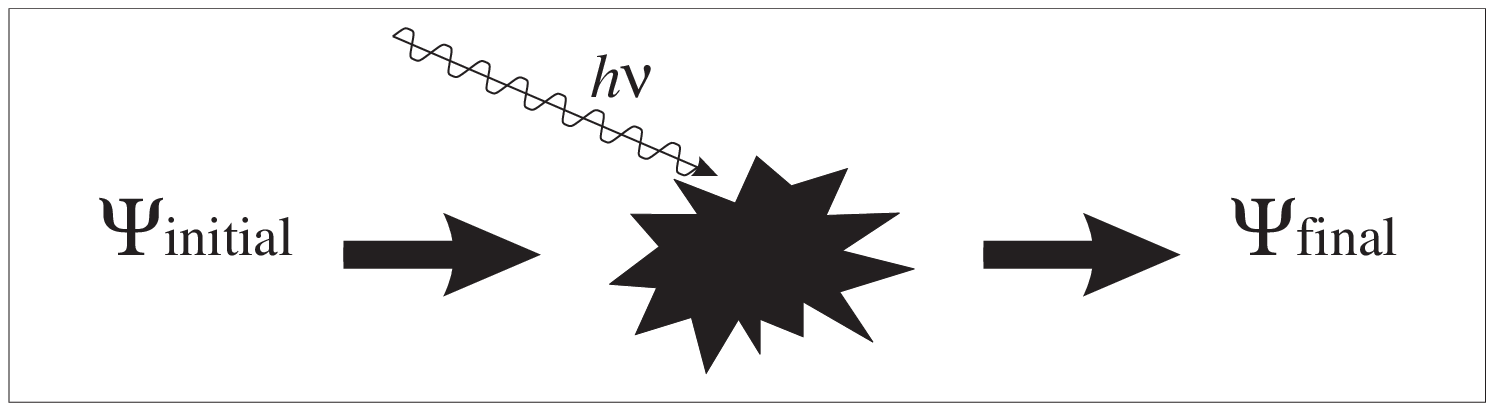}
    \caption{A spectroscopic proses brings the system from the initial state, in which properties we are interested, to a final state.}
    \label{introductionIn-O-Out}
   \end{center}
  \end{figure}
  
Excitons are very well known in optical spectroscopy \cite{Ashcroft76,Kittel53} and often hard to understand. If correlations are important the one-electron picture brakes down and one can not talk about bands anymore. An exciton can only be understood as a many-electron wave function. That brings us to figure \ref{introductionIn-O-Out}. Here we show in a cartoon how to look at $2p$ XAS. There is a many-electron initial state wave function, describing all electrons in the system. This is the ground-state wave function. By shining light on the sample this initial state wave function is transformed into a final state wave function. Here we should warn the reader. In this context the initial state wave function describes the valence band. All core shells in the initial state wave function are filled and $2p$ core level spin orbit coupling is not important for the initial state wave function, as the $2p$ core shell is fully filled. For the final state wave functions we have excitons. They have nothing to do with the bands calculated for the material and $2p$ spin orbit coupling for the final state wave function is important, since here we do have a hole in the $2p$ core shell.

In optical spectroscopy calculating the excitons is far from easy as many more, free electron like states are present at equal energies. For $2p$ XAS the excitonic binding energy is very large. Actually almost all intensity is in the excitonic excitation and only a small part of the intensity is due to excitations where the $2p$ core hole and the additional $3d$ electron are not bound. The non-excitonic part of the spectrum shows up as continuum jump at 8.0-15.0 eV above the white line of the spectrum. Excitons alone can be understood quite well, since an exciton lives largely on a single atom, so atomic multiplet theory will explain the spectra quite well. For the less tight bound excitons it is important to consider a small cluster calculation of one TM with the neighboring O atoms. Exactly that is what we have used to understand $2p$ core level spectroscopy. A small cluster with full inclusion of all electron-electron interaction, leading to many multiplets. It should be noted that in order to calculate a spectrum correctly the final state as well as the initial state have to be calculated correctly. Therefore these high-energy spectra still contain a lot of information about the initial state, or ground-state wave function.

Atomic multiplet theory has been known very long and for atoms imbedded in a crystal the crystal field theory was developed about 60 years ago and many textbooks have been written on one single atom in a crystal field \cite{Ballhausen62, Sugano70}. In the last 15 years computer codes have been developed \cite{Cowan81,Thole97,Groot05,Tanaka94} to do crystal field calculations. With the modern desktop computers these calculations can be run on any pc available, thereby opening up the very detailed interpretation of XAS spectra as we have today. That a single atom calculation embedded in an electric field mimicking the rest of the crystal (crystal field theory) already works quite remarkable in describing $2p$ XAS spectra has been shown by fits to measured spectra on very many different samples \cite{Groot05,Thole97}. In this thesis we do more than single atom calculations, since charge transfer from the O atoms to the TM atoms is often important in understanding the physics of oxides. Our calculations are done on a small cluster. Typical clusters taken into account are TMO$_{6}$ clusters. This theory is about as old as the crystal field theory and the program of Thole \cite{Cowan81,Thole97,Groot05} as well as the program of Tanaka \cite{Tanaka94} can do these cluster calculations on any desktop pc within reasonable time. In this thesis Tanaka's program XTLS8.0 has been used.

In short $2p$ XAS can be very well calculated while there is a core hole at very high energies (350-950 eV) resulting in a flat band. This core hole is bound very strong with the 3d valence electrons due to the large $2p$ core hole - $3d$ electron attraction. Thereby creating strongly bound excitons. This results in atomic like physics which is very well known.

\section{Multiplets and selection rules}

Now lets have a closer look at the exciton itself. The exciton is the final state of $2p$ XAS. The final state has a $2p^{5}3d^{n+1}$ configuration. The $2p$ core hole can be in six different orbitals which are split in a group of four orbitals with $j=\frac{3}{2}$ and a group of two orbitals with $j=\frac{1}{2}$ due to the $2p$ spin orbit coupling. So for $n=9$, a $2p^{5}3d^{10}$ configuration there are 6 final states split in two groups. For $n=8$ however one has to consider the hole in the $3d$ shell. This hole can be in 10 different orbitals and therefore there are 60 different final states. Electron electron repulsion between a $3d_{z^{2}}$ orbital and a $2p_{z}$ orbital is larger than between a $3d_{x^{2}-y^{2}}$ and a $2p_{z}$ orbital, which can be understood by looking at the electron-electron density overlap of these two orbitals. This effect is known as a multiplet effect and causes the sixty final states present in a $2p^{5}3d^{9}$ configuration to be spread out over an energy range. For $n=7$, a $2p^{5}3d^{8}$ configuration there are $\frac{6\times10\times9}{2}=270$ states. 6 for the $p$ hole, 10 for the first $d$ hole, 9 for the second and divided by 2 since electrons are equivalent and it does not matter if the first hole in the $d$ shell was made on orbital a and the second on orbital b or the first hole was made on orbital b and the second on orbital a. 

  \begin{SCfigure}[][!h]
    \includegraphics[width=60mm]{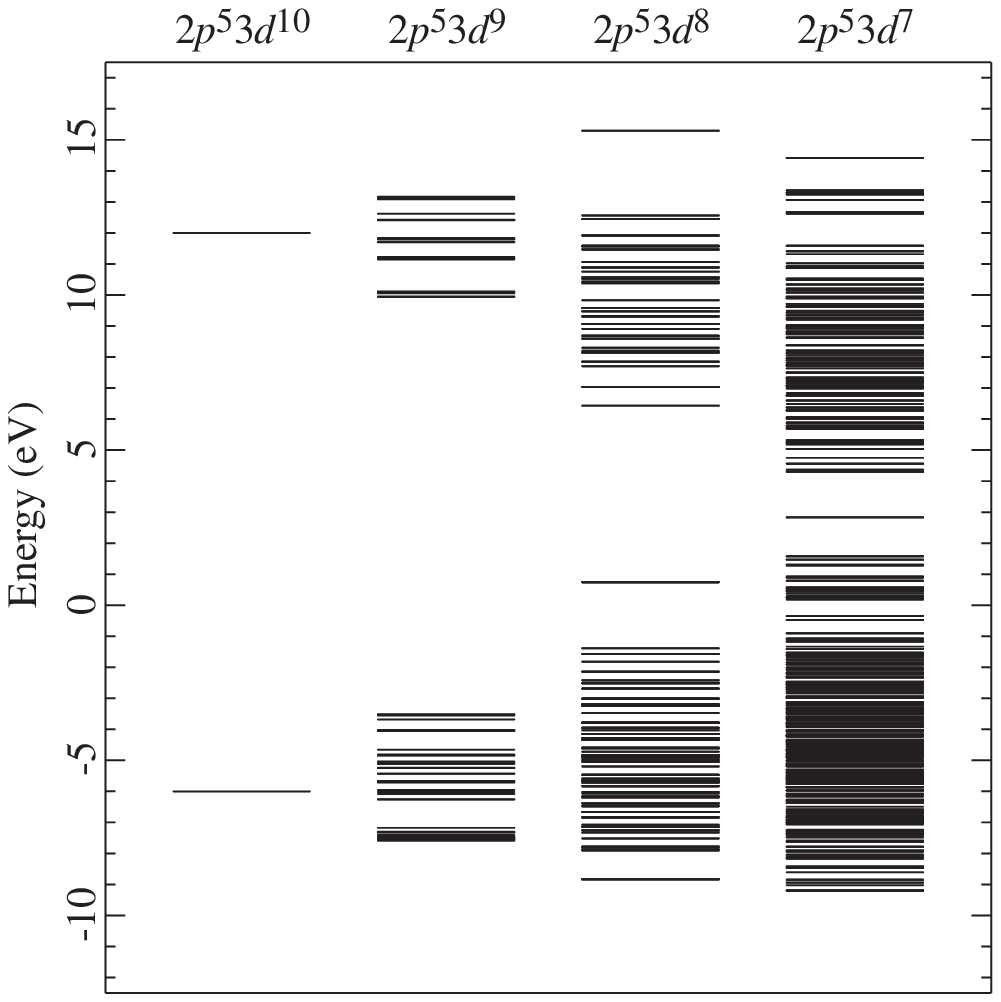}
    \caption{Multiplets of the $2p^53d^{n+1}$ final states of $2p$-XAS. \newline}
    \label{introductionFinalStateEnergyLevelDiagram}
  \end{SCfigure}

In figure \ref{introductionFinalStateEnergyLevelDiagram} we show the atomic multiplet calculations for $n=9$ to $n=6$. Each configuration has its own amount of final states and its own multiplet. Naturally if one changes from one element to an other some parameters change. The Slater integrals, determining the size of the electron electron repulsion differ from element to element and the crystal fields present are different from material to material, but the general structure of a multiplet is strongly dependent on the configuration one has. 

  \begin{figure}[!h]
   \begin{center}
    \includegraphics[width=120mm]{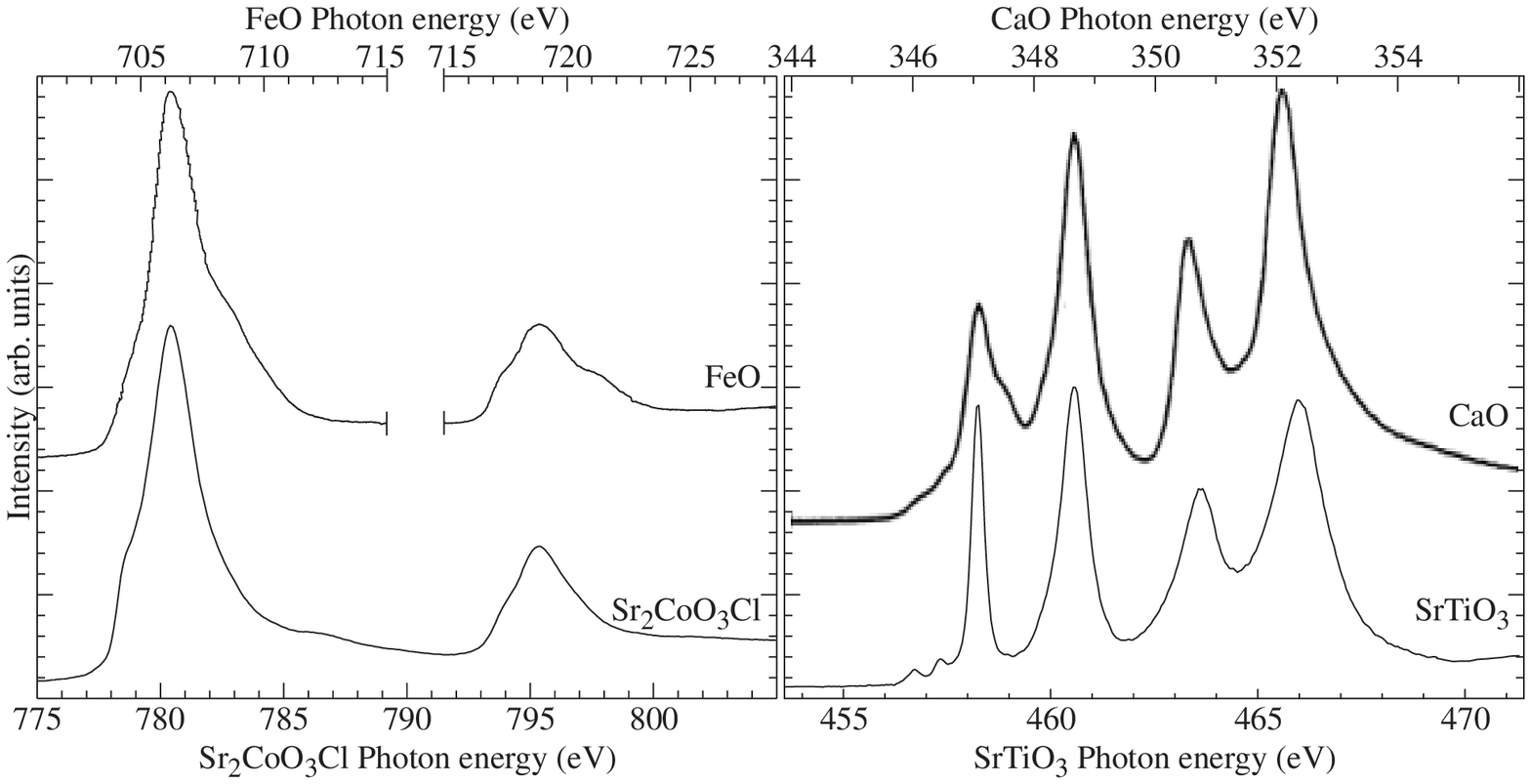}
    \caption{Experimental spectra of FeO and Sr$_{2}$CoO$_{3}$Cl on the left side and CaO and SrTiO$_{3}$ on the right side. FeO and Sr$_{2}$CoO$_{3}$Cl are $d^6$ systems and have the same $2p^53d^7$ final state configuration, CaO and SrTiO$_{3}$ are $d^0$ systems and have the same $2p^53d^1$ final state configuration. The configuration (valence) is very important for the line-shape of the spectra. The FeO spectrum has been copied from J.-H. Park \textit{et al.} \cite{Park94}. The CaO spectrum has been copied from F. J. Himpsel \textit{et al.} \cite{Himpsel91}.}
    \label{introductionValenceDependence}
   \end{center}
  \end{figure}

In figure \ref{introductionValenceDependence} we compare the $2p$ Fe XAS of FeO taken from J.-H. Park \textit{et al.} \cite{Park94} with the $2p$ Co XAS of Sr$_{2}$CoO$_{3}$Cl as well as the spectra of CaO taken from from F. J. Himpsel \textit{et al.} \cite{Himpsel91} with the spectra of SrTiO$_{3}$. Fe in FeO is 2+ and has therefore a $d^{6}$ configuration. Co in Sr$_{2}$CoO$_{3}$Cl is 3+ and has also a $d^{6}$ configuration. Ca in CaO is 2+ and has a $d^{0}$ configuration and Ti in SrTiO$_{3}$ is 3+ and has also a $d^{0}$ configuration. As one can see are the spectra of FeO and Sr$_{2}$CoO$_{3}$Cl alike. The same is true for the spectra of CaO and SrTiO$_{3}$. Naturally there are differences. For lighter transition elements the $2p$ core level spin orbit coupling constant is smaller. So to make a good comparison we added some space in the FeO spectra between the $L_{2}$ and the $L_{3}$ edge. Thereby the crystal fields present are not equal, resulting in slightly different positions of the peaks and differences in intensities of some peaks.

If we now compare the spectra in figure \ref{introductionValenceDependence} with the multiplets calculated in figure \ref{introductionFinalStateEnergyLevelDiagram} we see that there are far more multiplets present than peaks in the spectra. For a d$^6$ compound we see that there are almost everywhere final states multiplets in the energy region of the spectra, but we do see very distinct peaks with clear structure. This can be explained very nicely with the use of dipole selection rules.

The intensity of a $2p$ XAS spectrum is proportional to the square of $\langle\Psi_{i}|O|\Psi_{f}\rangle$. Here $\Psi_{i}$ stands for the initial state wave function, $\Psi_{f}$ for one of the many final state wave functions and O can be approximated as a dipole operator. We know that $\Psi_{i}|O|\Psi_{f}$ has to be even, otherwise the integral $\langle\Psi_{i}|O|\Psi_{f}\rangle$ will be zero. O, the dipole operator is odd and we further know that $p$ and $f$ orbitals are odd and $s$ and $p$ orbitals are even. Using these symmetries we find the following selection rules for excitations to be allowed.
\begin{eqnarray}
\Delta l=\pm 1\\
\Delta j=0,\pm 1\nonumber\\
\Delta m_{l}=0,\pm 1\nonumber\\
\Delta m_{s}=0\nonumber\\
\Delta L=0,\pm 1\nonumber\\
\Delta S=0\nonumber\\
\Delta J=0,\pm 1\nonumber
\end{eqnarray}

  \begin{figure}[!h]
   \begin{center}
    \includegraphics[width=90mm]{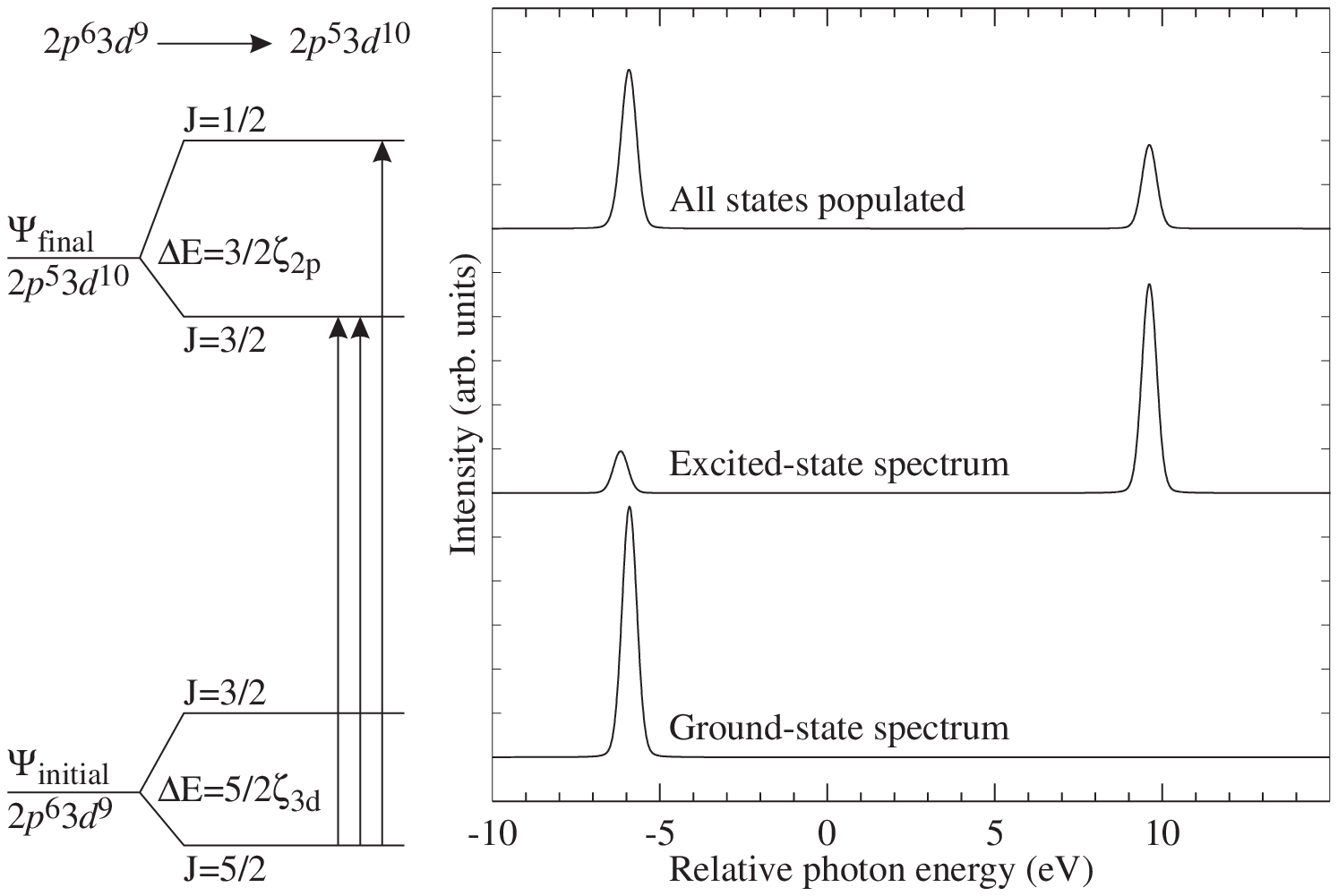}
    \caption{Left: Initial state of a $d^9$ system split by $3d$ spin orbit coupling and final state split by, much stronger, $2p$ spin orbit coupling. The arrows indicate the dipole allowed transitions. Right: the spectra for the ground-state, first excited state and for a system where all states are equally occupied.}
    \label{introductionJDependence}
   \end{center}
  \end{figure}

Lets make it more clear on a simple example. We start with studying the spectra of a $d^9$ compound, Ni metal for example, since for Ni metal there exist a very instructive measurement by P. Gambardella \textit{et al.} \cite{Gambardella02}, illustrating the effects of selection rules. The final state of $2p$ XAS on a $d^9$ compound has a $2p^53d^{10}$ configuration. There are six states split by the $2p$ spin orbit coupling into 4 states with $J=\frac{3}{2}$ and 2 states with $J=\frac{1}{2}$. The initial state has the $d^9$ configuration. There are 10 states, also split by spin orbit coupling into 6 states with $J=\frac{5}{2}$ and four states with $J=\frac{3}{2}$.

On the left side of figure \ref{introductionJDependence} we show the initial and final state configurations and the splitting within them present due to spin orbit coupling. For the lowest state of the initial configuration we have $J=\frac{5}{2}$. We also have the selection rule $\Delta J=0,\pm 1$. So from the initial states with $J=\frac{5}{2}$ only the final states with $J=\frac{3}{2}$ can be reached and the final states with $\frac{1}{2}$ can not be reached. For the initial states with $J=\frac{3}{2}$ both final states can be reached. On the right side of figure \ref{introductionJDependence} we show the spectra calculated for the ground-state with $J=\frac{5}{2}$ and for the first excited state with $J=\frac{3}{2}$. We also show the spectra assuming all initial states are populated equally.

  \begin{SCfigure}[][!h]
    \includegraphics[width=60mm]{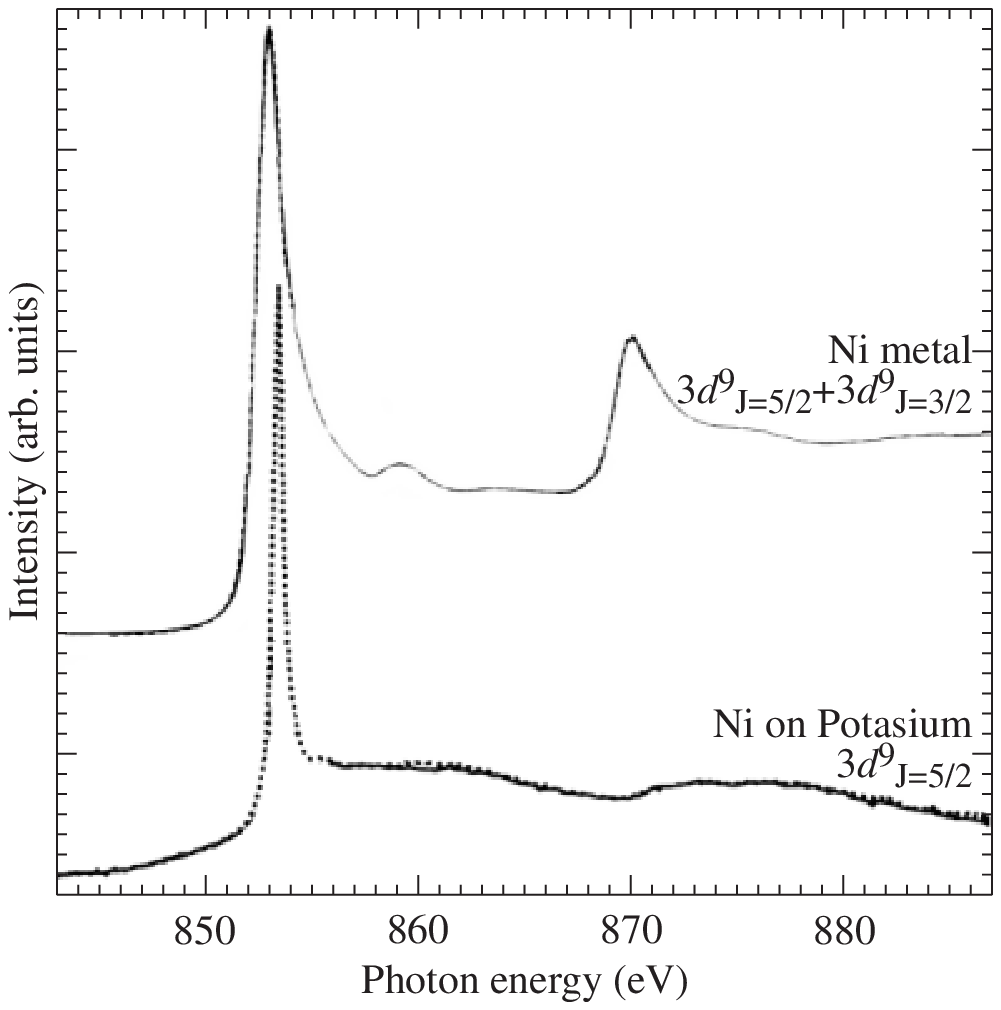}
    \caption{$2p$ XAS spectra of Ni metal and Ni on Potassium. The differences in Ni spectra can be explained by the differences in the Ni ground-state and the selection rules $\Delta J=0, \pm 1$. The spectra have been copied from P. Gambardella \textit{et al.} \cite{Gambardella02}. \newline}
    \label{introductionNiMetalXAS}
  \end{SCfigure}

However, in Ni metal $3d$ spin-orbit coupling is not the most important interaction present. Covalency between the Ni atoms and band formation is more important and splits the $3d$ levels resulting in an almost equal population of states with $J=\frac{3}{2}$ and states with $J=\frac{5}{2}$. The spectra for Ni metal measured by P. Gambardella \textit{et al.} \cite{Gambardella02} are presented in figure \ref{introductionNiMetalXAS}. They look most like the spectra calculated with all states populated. P. Gambardella \textit{et al.} \cite{Gambardella02} showed that Ni atoms placed on Potassium (K) retrieve their atomic character and spin orbit coupling becomes the most important low energy interaction again. $J$ is a good quantum number and the spectra is equal to a cluster calculation where the ground-state is a state with $J=\frac{5}{2}$.

  \begin{SCfigure}[][!h]
    \includegraphics[width=60mm]{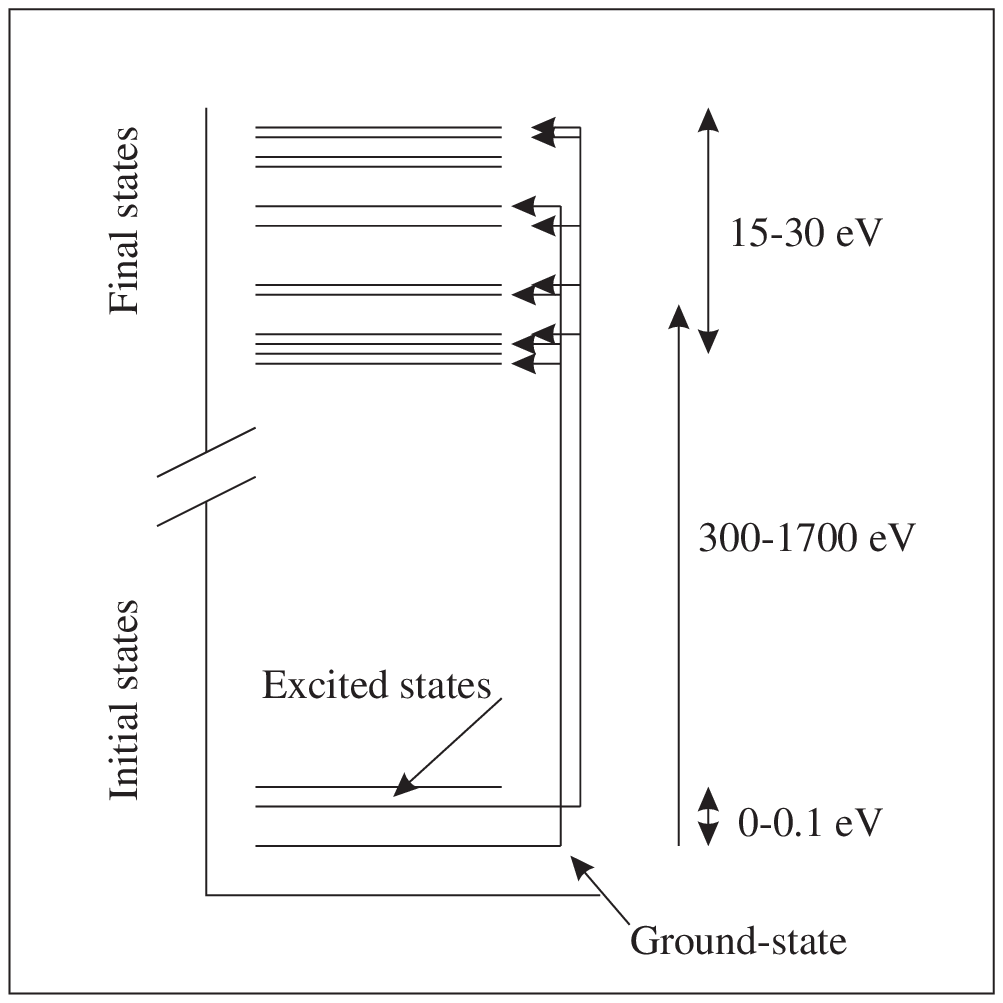}
    \caption{General idea behind $2p$-XAS. There is a ground-state, that might be degenerate, and have important lower excited states. There are, much higher in energy, many final states. Strong dipole selection rules determine which of the final states can be reached. For each of the initial state there the set of final states that can be reached is different, making $2p$-XAS a very sensitive probe.}
    \label{introductionGeneralIdea}
  \end{SCfigure}

The basic idea behind $2p$ XAS can be stated as follows. There is a ground-state belonging to a $3d^n$ configuration, there are many final states, belonging to the $2p^53d^{n+1}$ configuration. Of these final states only a few can be reached in $2p$ XAS experiments due to the strict selection rules. There might be excited states close by the ground-state that are important to consider as well. If the energy splitting is small temperature might cause these states to be populated, assuming Boltzmann statistics. A small perturbing Hamiltonian like exchange interaction might shift these states in and out population range with different temperatures. Each excited state belonging to the initial state will in principle have a different set of final states it can reach. Therefore $2p$ XAS is very sensitive to small changes in the ground-state. Small shifts in the initial state are able to make very large shift in spectral intensity, since a different set of final states can be reached for each initial state. In figure \ref{introductionGeneralIdea} we show this graphically.
   
  \begin{figure}[!h]
   \begin{center}
    \includegraphics[width=90mm]{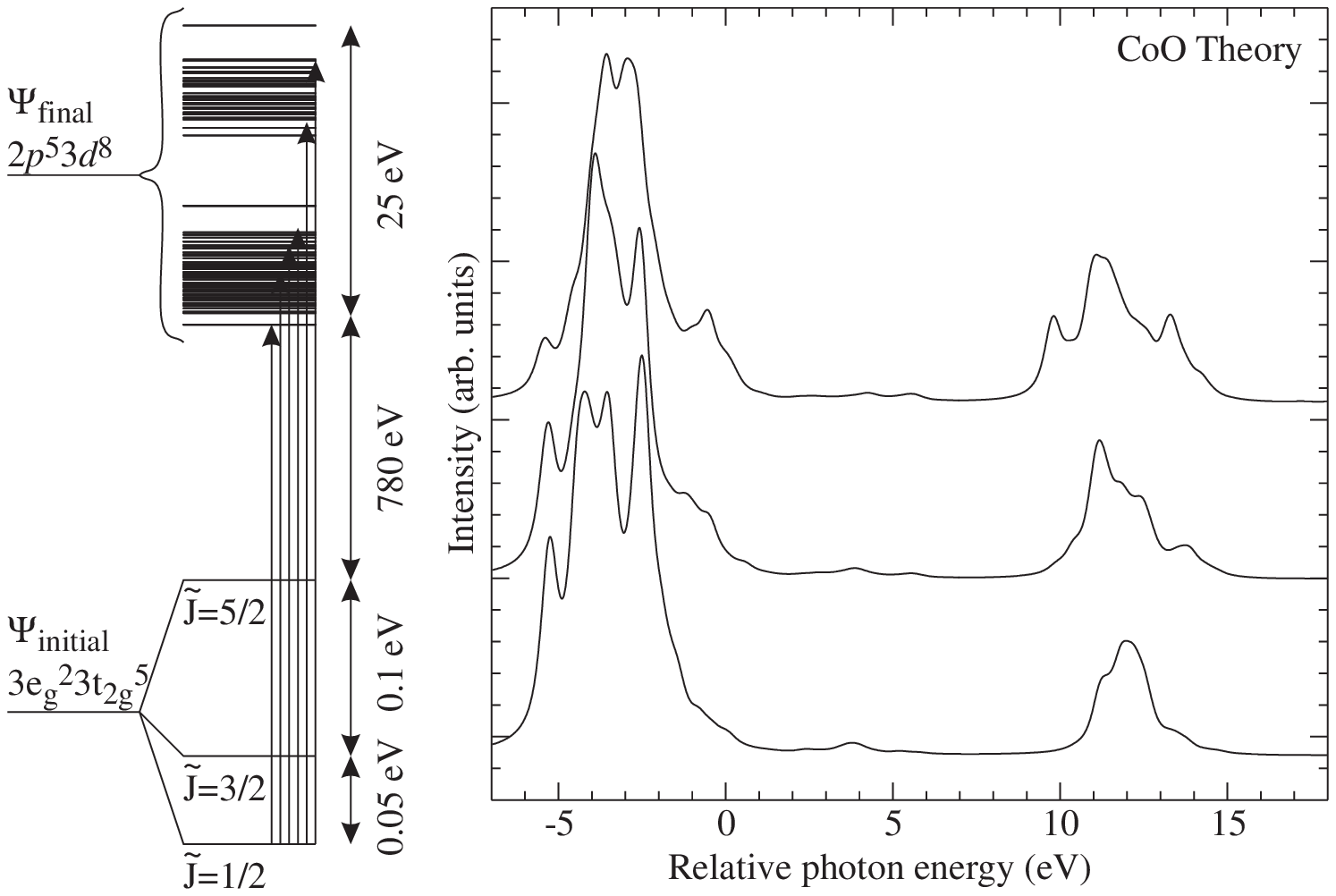}
    \caption{Left: energy level diagram of CoO, the $e_g^2t_{2g}^5$ initial state is split by spin orbit coupling into a lowest triplet, first excited quintet and highest septet. Right: the different theoretical spectra for the triplet, quintet and septet in CoO.}
    \label{introductionCoOTheory}
   \end{center}
  \end{figure}
   
In order to show this temperature dependence due to population of different initial states we turn our attention to CoO. CoO is a material where the Co atoms are in a $d^7$ configuration and it is know for CoO that there is a large orbital momentum present at the Co site \cite{Neubeck99, Jauch02}. Co in CoO is surrounded by six O atoms in Octahedral ($O_h$) symmetry. In $O_h$ symmetry the $d$ orbitals split in three $t_{2g}$ orbitals and two $e_{g}$ orbitals. The $e_{g}$ orbitals being about 1.1 eV higher in energy. In order to keep things simple we assume an occupation of $e_{g}^2t_{2g}^5$. The $t_{2g}$ shell has a pseudo orbital momentum of \~{l}=1. With one hole in the $t_{2g}$ shell and a spin momentum of $S=\frac{3}{2}$ we find that the ground state is 12 fold degenerate. This degeneracy is lifted by spin orbit coupling and splits into a ground-state doublet with \~{J}=$\frac{1}{2}$, a first excited quartet, about 500 K higher in energy with \~{J}=$\frac{3}{2}$, and a sixted, about 1500 K higher in energy than the ground-state. This sixted has \~{J}=$\frac{5}{2}$. The final states of $2p$ XAS on CoO all belong to the $2p^{5}3d^{8}$ configuration and are about 780 eV higher in energy.

In figure \ref{introductionCoOTheory} we show the energy level diagram of the initial states belonging to the $e_{g}^2t_{2g}^5$ configuration and of the final states belonging to the $2p^{5}3d^{8}$ configuration. Each initial state with a different \~{J} can make excitations to a different set of final states and therefore each spectrum of each initial state will be different. The spectra for the 3 different initial states are shown on the right side. At T=0 K we should see the spectrum belonging to the state with \~{J}=$\frac{1}{2}$ and with increase in temperatures we should see a change towards the spectrum with \~{J}=$\frac{3}{2}$ and \~{J}=$\frac{5}{2}$. Population of excited states is according to Boltzmann statistics.
   
  \begin{SCfigure}[][!h]
    \includegraphics[width=60mm]{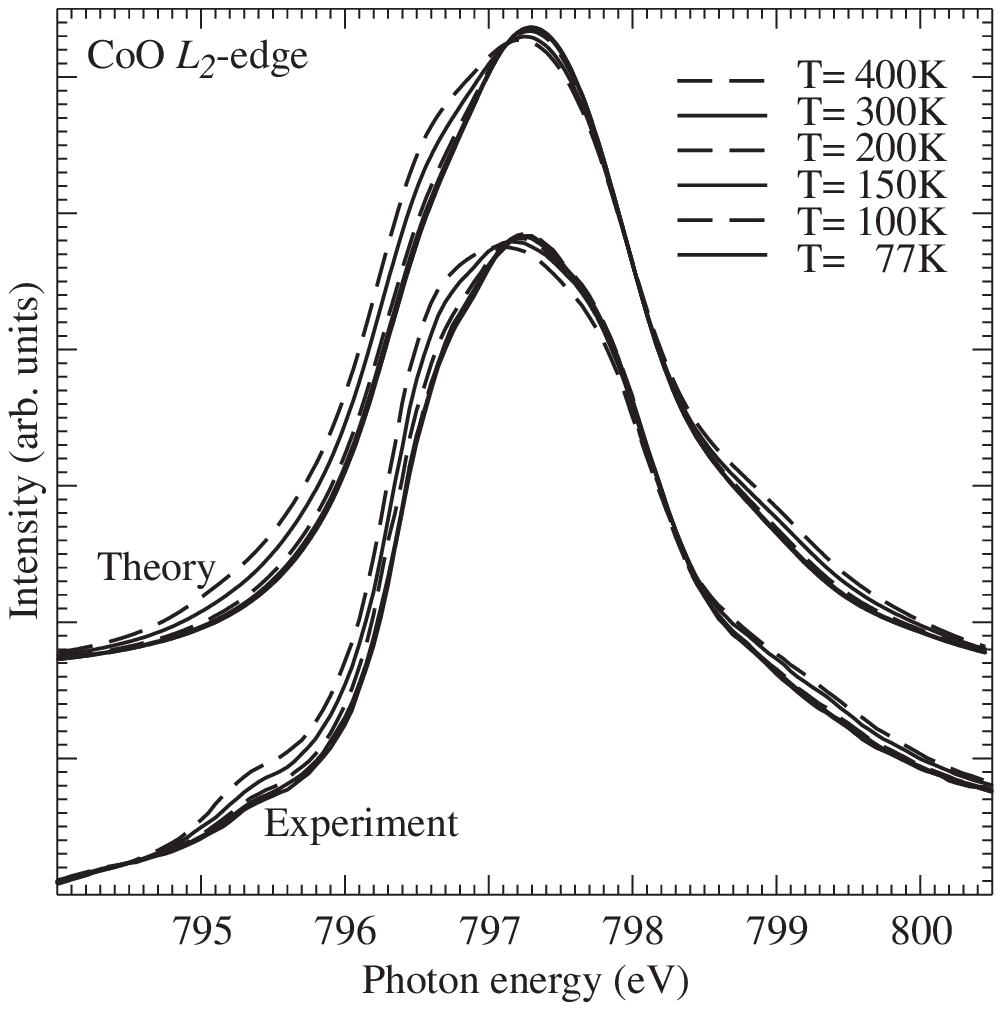}
    \caption{Experimental and theoretical temperature dependence in the CoO $2p$-XAS $L_{2}$ edge spectra due to population of excited states. \newline}
    \label{introductionCoOL2XAS}
  \end{SCfigure}   
   
In figure \ref{introductionCoOL2XAS} we show the $L_2$ edge of the measured spectra of poly crystalline CoO on Al. We also show a cluster calculation including the temperature effect discussed in the previous paragraph. In this case we included the O atoms and did not do a single atom calculation, but calculated the spectrum for a CoO$_6$ cluster. As one can see we do reproduce the temperature dependence and general line shape with our cluster calculations. In chapter \ref{ChapterCoO} we will discuss more about CoO thin films.

There are more selection rules then $\Delta J=0,\pm 1$ that can be used to see large differences in spectra when small changes in the ground state or population of the low energy excited states happen. One of them used in chapter \ref{ChapterCobaltates} is the selection rule $\Delta S=0$. If a transition metal is placed in a cubic crystal field the $d$ orbitals split into two $e_{g}$ and three $t_{2g}$ orbitals. If the splitting is small first the five orbitals with spin up will be occupied upon filling and only when they are full the 5 orbitals with spin down will be filled. This in order to minimize the electron-electron repulsion according to Hund's rules. If the splitting between the $e_{g}$ and $t_{2g}$ orbitals becomes big however Hund's rules brake down and first the three $t_{2g}$ orbitals with spin up are filled, then the three $t_{2g}$ orbitals with spin down are filled and after that the $e_{g}$ orbitals with spin up are filled. For a $d^{6}$ configuration many materials are found that are on the border of being in the high spin configuration $t_{2g}^4e_{g}^2$ with $S=2$, or low spin, configuration $t_{2g}^6$ with $S=0$.
   
  \begin{figure}[!h]
   \begin{center}
    \includegraphics[width=120mm]{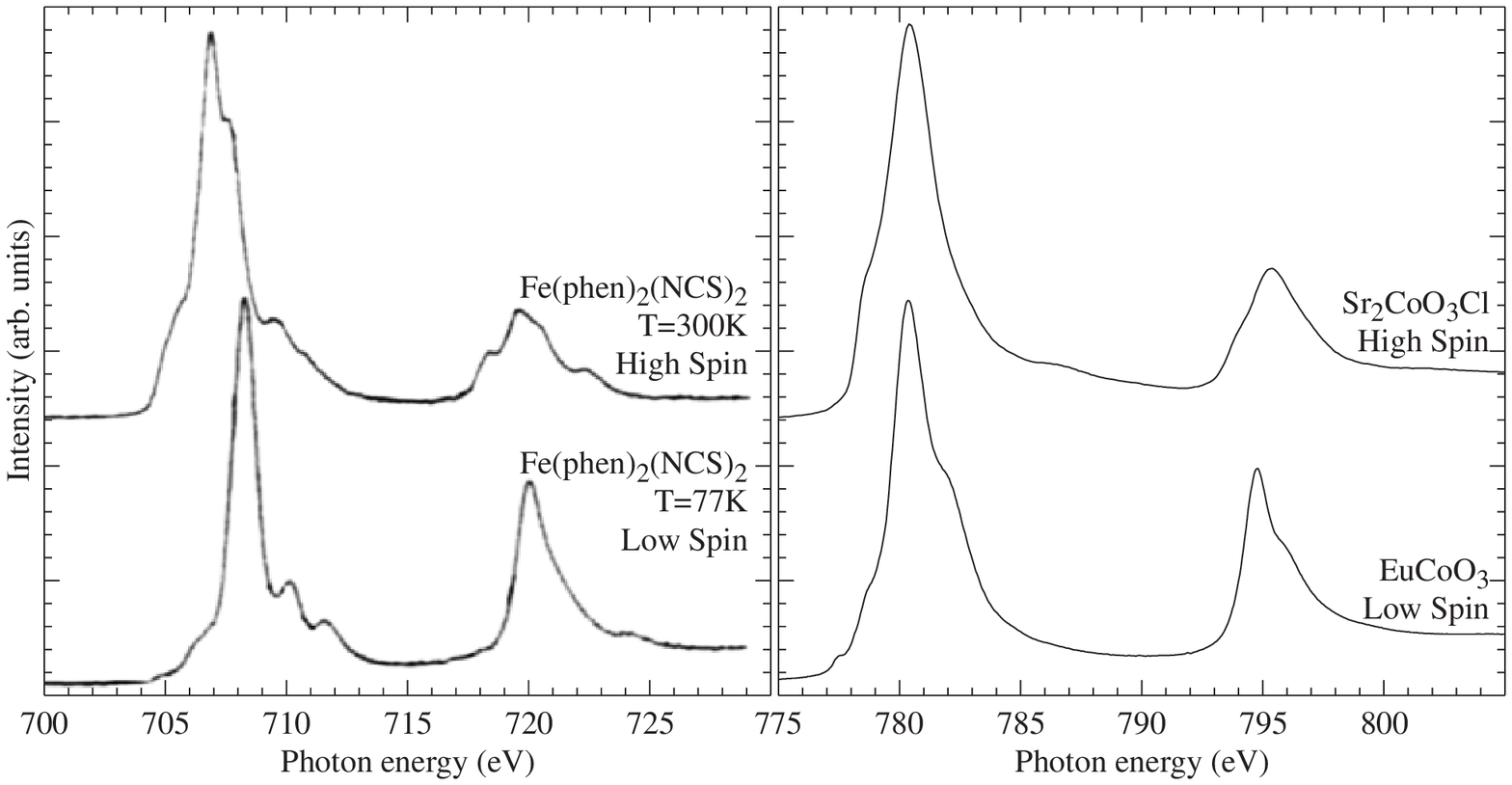}
    \caption{Difference in spectra for $d^6$ high-spin and low-spin electron arrangements. Fe$^{2+}$ and Co$^{3+}$ have similar spectra, but there is a clear difference between the spectra of high-spin and low-spin compounds. The Fe(phen)$_{2}$(NCS)$_{2}$ spectra have been copied from C. C. dit Moulin \textit{et al.} \cite{Moulin92}.}
    \label{introductionSpinDependence}
   \end{center}
  \end{figure}

Using XAS C. C. dit Moulin \textit{et al.} \cite{Moulin92} confirmed that for Fe(phen)$_{2}$(NCS)$_{2}$ there is a spin state transition at T=177 K. In figure \ref{introductionSpinDependence} we show their $2p$ XAS spectra of Fe(phen)$_{2}$(NCS)$_{2}$ at 77 K and at 300 K. The spectral changes are remarkable making XAS a very good tool to study the spin state of materials. Especially valuable for the newly synthesized layered cobaltates where for many materials there is a dispute about the spin state of the Co ion \cite{Martin97, Maignan99, Yamaura99a, Yamaura99b, Loureiro01, Vogt00, Suard00, Fauth01, Burley03, Mitchell03, Moritomo00, Respaud01, Kusuya01, Frontera02, Fauth02, Taskin03, Soda03, Loureiro00, Knee03, Wu00, Kwon00, Wang01, Wu01, Wu02, Wu03}. In figure \ref{introductionSpinDependence} we also show the spectra of EuCoO$_{3}$ and Sr$_{2}$CoO$_{3}$Cl, a low and high spin system respectively. Showing how easy it can be to determine the spin state of different materials.

We would like to note that $2p$ XAS spectroscopy is one of the few, if not the only method that is sensitive to the value of $S$, the spin momentum, and does not deduces the value of $S$ from the maximum value of the magnetic spin momentum $S_z$, or the degeneracy of the state.

\section{Polarization dependence}

Sofar we have only discussed isotropic XAS spectra. But light has an electric vector and can be linear or circular polarized. Magnetic dichroism spectra at the $L_{2,3}$ edges of transition metals has for the first time been measured in the beginning of the nineteens \cite{Chen90a, Chen90b, Tjeng91b, Tobin92}. Shortly thereafter it has been calculated by G. van der Laan and B. T. Thole \cite{Laan91}, that strong magnetic dichroism is present in the $2p$ x-ray absorption spectra of $3d$ transition-metal ions. X-ray magnetic circular dichroism became a widely used tool after the derivation of sum-rules. It has been shown \cite{Thole92, Carra93a} that the total integrated intensity of circular polarized spectra can be related to the average aligned spin and orbital magnetic momentum. The multiplet structure of the polarized spectra can contain even more information. Not only large circular dichroism is found in the $2p$ absorption edges of transition metals. Also large magnetic linear dichroism has been predicted \cite{Laan91} and measured \cite{Kuiper93, Alders95, Alders98}. Besides a magnetic origin, linear dichroism can also be induced by a non-cubic orbital occupation \cite{Chen92, Regueiro95, Park00} or local, non-cubic, crystal fields \cite{Haverkort04}. A review about dichroism effects in $2p$ x-ray absorption spectroscopy has been written by J. St\"ohr and R. Nakajima \cite{Stohr98b} and by F. M. F. de Groot \cite{Groot94}.

In the next paragraphs we will first discuss the polarization dependence for systems with a non-cubic orbital occupation. We will show the spectra of C. T. Chen \textit{et al.} \cite{Chen92} measured on CuO planes of an High-T$_{C}$ super conductor as an example. Then we will show how crystal fields can be determined for thin NiO films \cite{Haverkort04}. After that we will continue with NiO and explain the magnetic linear dichroism therein. We will use the measurements of D. Alders \textit{et al.} \cite{Alders95, Alders98} who showed, among other properties of the magnetic linear dichroic spectra, how with the use of linear dichroism, the average aligned magnetic moments of NiO thin films can be determined. The last feature of XAS we will show will be about magnetism in CoO, where we will discus the circular polarization dependence and show that the spectra contain information about $L_{z}$ and $S_{z}$.
   
  \begin{SCfigure}[][!h]
    \includegraphics[width=60mm]{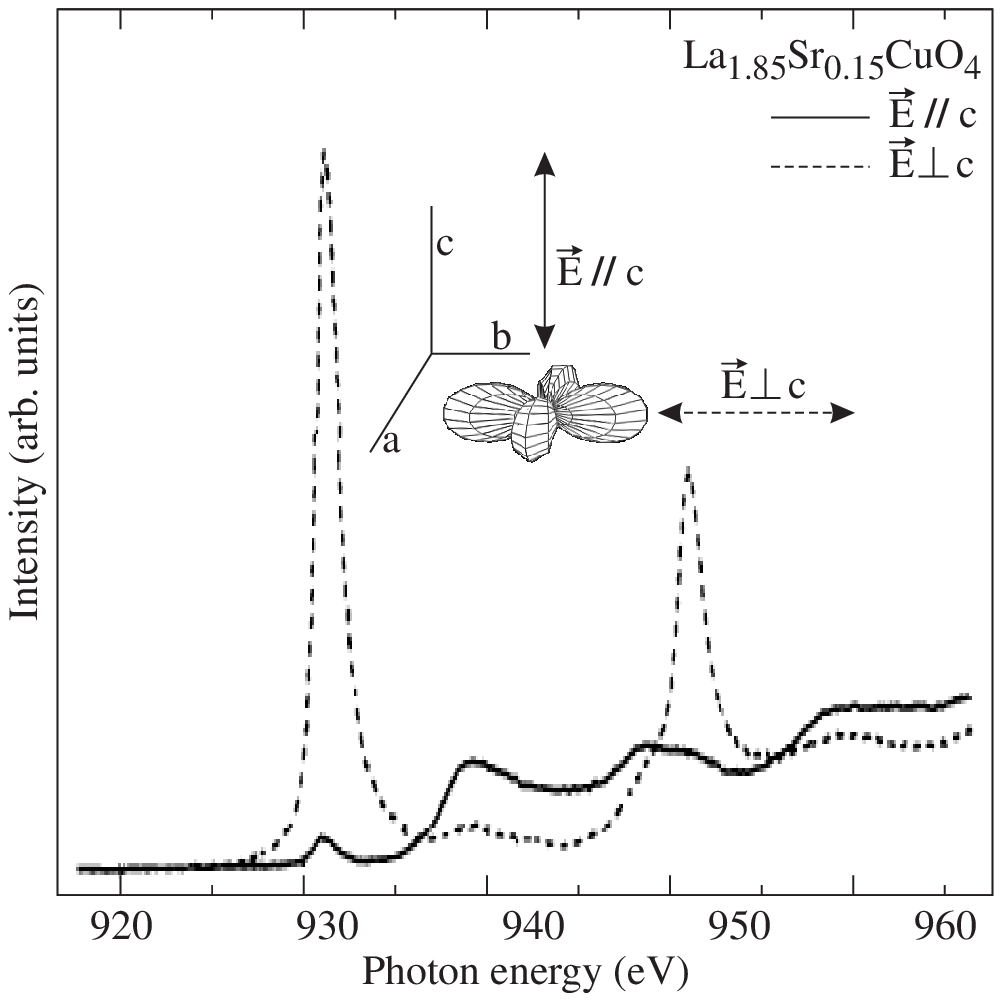}
    \caption{Linear polarization dependence in La$_{1.85}$Sr$_{0.15}$CuO$_{4}$ due to orbital polarization of the ground-state. The spectra are copied from C. T. Chen \textit{et al.} \cite{Chen92} \newline}
    \label{introductionLaSrCuO4}
  \end{SCfigure} 
  
The simplest form of dichroic spectroscopy in the x-ray regime is that of linear dichroism due to a difference in orbital occupation. The difference in total intensity caused by a non-cubic orbital occupation can be understood in a one-electron picture. Lets start with a simple example where we make an excitation from a $s$ orbital to a $p$ orbital. There are three different orbitals ($p_x$, $p_y$, and $p_z$) one can excite to. If we use z polarized light then the intensity found for exciting the $s$ orbital to a $p_x$ orbital is proportional to the square of $\langle s|z|p_x\rangle$. If we evaluate the integral $\langle s|z|p_x\rangle$ we notice that $s$ is even in z, z is odd in z and $p_x$ is again even in z. The total integrand is thus odd in z. The integral over an odd function is zero. With z polarized light one can only excite an $s$ orbital to the $p_z$ orbital. If we now take a system where we have a $p^5$ shell, then the orbital occupied can be measured by XAS. Just look at the total intensity of the spectra with respect to the intensity of the x,y, and z polarized spectra. The ratio of intensity with x, y, or z polarized light is equal to the ratio of the number of holes in the $p_x$, $p_y$, or $p_z$ orbital.

For excitations from a $p$ to $d$ shell things become a bit more complicated, since the $p$ shell has 3 and the $d$ shell 5 orbitals, but the basic principle is the same. With x polarized light one can only excite a $p_y$ orbital to a $d_{xy}$, a $p_z$ orbital to a $d_{xz}$ orbital and an $p_{x}$ orbital to an $d_{x^2}=-\sqrt{\frac{1}{4}}d_{z^2}+\sqrt{\frac{3}{4}}d_{x^2-y^2}$ orbital. The other excitations are zero because the integrand is odd. For the other polarization directions a cyclic permutation of the coordinates will do. In the end we can write for the total intensities measured with linear polarized light for $p$ to $d$ excitations:
\begin{eqnarray}
I_x=\frac{1}{2}\underline{n}_{xy}+\frac{1}{2}\underline{n}_{xz}+\frac{2}{3}\underline{n}_{x^2}\\\nonumber
I_y=\frac{1}{2}\underline{n}_{xy}+\frac{1}{2}\underline{n}_{yz}+\frac{2}{3}\underline{n}_{y^2}\\\nonumber
I_z=\frac{1}{2}\underline{n}_{xz}+\frac{1}{2}\underline{n}_{yz}+\frac{2}{3}\underline{n}_{z^2}
\end{eqnarray}
Where $\underline{n}_{x^2}=d_{x^2}d_{x^2}^{\dag}=\frac{1}{4}\underline{n}_{z^2}+\frac{3}{4}\underline{n}_{x^2-y^2}-\sqrt{\frac{3}{16}}(d_{z^2}d_{x^2-y^2}^{\dag}+d_{x^2-y^2}d_{z^2}^{\dag})$ and $\underline{n}_{y^2}=d_{y^2}d_{y^2}^{\dag}=\frac{1}{4}\underline{n}_{z^2}+\frac{3}{4}\underline{n}_{x^2-y^2}+\sqrt{\frac{3}{16}}(d_{z^2}d_{x^2-y^2}^{\dag}+d_{x^2-y^2}d_{z^2}^{\dag})$. 
The polarization dependence due to orbital occupation has been demonstrated very nicely in the XAS of High T$_C$ superconductors \cite{Chen92, Regueiro95}. High T$_C$ superconductors have a CuO plane where the valence of the Cu is $d^9$. The one hole at the Cu site is in the d$_{x^2-y^2}$ orbital. This means that for x and y polarized light we would expect to see some intensity and for z polarized light we would expect to see no intensity. This effect has been demonstrated by C. T. Chen \textit{et al.} \cite{Chen92} on La$_{1.85}$Sr$_{0.15}$CuO$_{4}$ their measurements can be seen in figure \ref{introductionLaSrCuO4}. We would like to note that the contrast seen has nothing to do with the energy splitting between the $d_{z^2}$ and $d_{x^2-y^2}$ orbital. XAS would be equally sensitive if the energy splitting was just barely larger than the temperature of the system. Second we would like to note the huge contrast there exists between the different polarizations, making XAS a very sensitive tool to study orbital occupations.

In chapter \ref{ChapterVO2} we have used this technique on VO$_{2}$ and in chapter \ref{ChapterCoO} on CoO. For VO$_{2}$ the question of which orbital is occupied is one of the basic assumptions to about all theories of the metal-insulator transition present in this compound. To test this against experiment is most valuable. For CoO orbital occupation is closely related to the orbital momentum present. From the information which orbital is occupied a detailed understanding of the materials can be achieved.

  \begin{SCfigure}[][!h]
    \includegraphics[width=50mm]{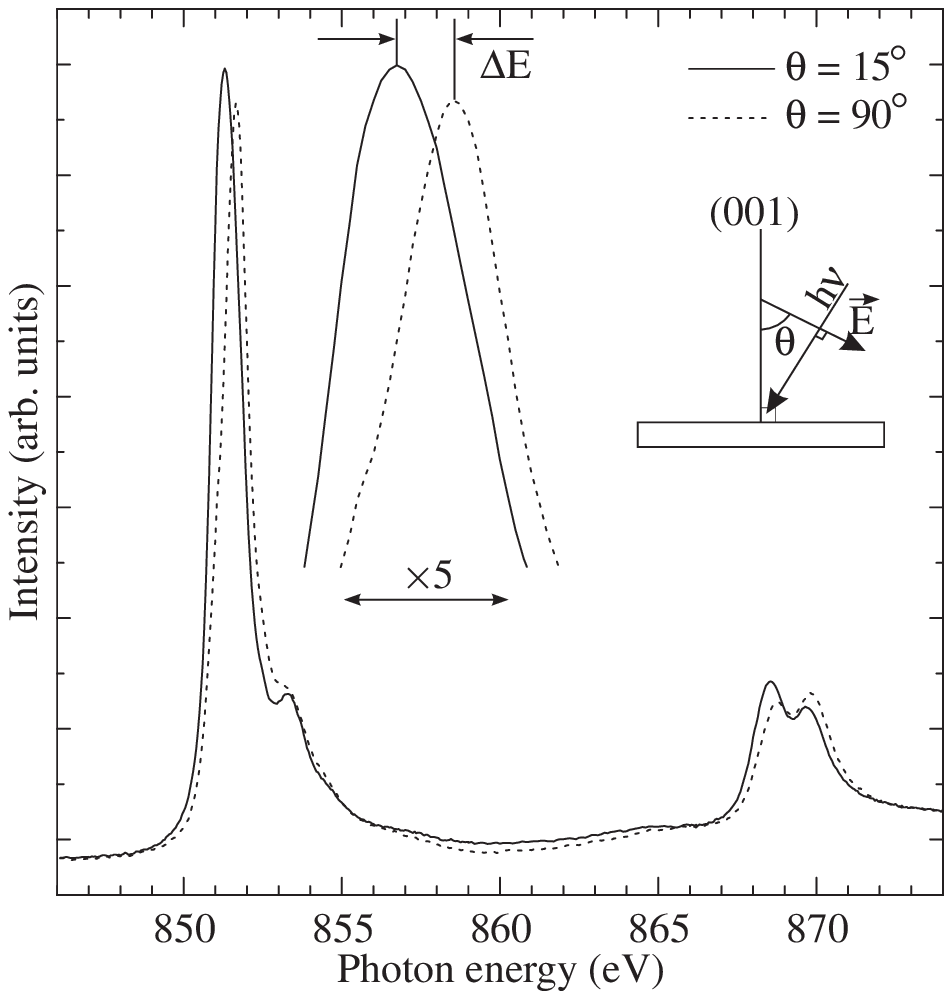}
    \caption{Linear dichroism in NiO due to crystal field splitting. The energy difference between the $L_3$ edge maximum taken at different polarizations is about equal to the crystal field splitting between the $d_{z^2}$ and $d_{x^2-y^2}$ orbital. \newline}
    \label{introductionNiOCrystalField}
  \end{SCfigure} 

We are not only interested in orbital occupations, but also in the energy scale of the splitting between the orbitals. The energy splitting between the orbitals, or crystal fields can be measured with polarization dependent XAS. For that we have a look at figure \ref{introductionNiOCrystalField} where we show the $2p$ XAS spectra of one monolayer of NiO on Ag, kept with MgO. A thin film of NiO on Ag is compressed in plane and therefore has a $d_{x^2-y^2}$ orbital that has a higher energy than the $d_{z^2}$ orbital. The energy splitting can be read of very easily in the polarized XAS spectra. One should just remember that for z polarized light one can excite to the $d_{z^2}$ orbital, but not to the $d_{x^2-y^2}$ orbital. Whereas with x polarized light one excites with an intensity of $\frac{3}{4}$ to the $d_{x^2-y^2}$ orbital and only with an intensity of $\frac{1}{4}$ to the $d_{z^2}$ orbital. To a good approximation the energy splitting between the $L_{3}$ edge for different polarizations is therefore equal to the crystal field splitting between the $d_{x^2-y^2}$ and the $d_{z^2}$ orbital. In chapter \ref{ChapterNiO} we will discuss this in more detail and explain also the changes in intensity found due to the crystal fields. The determination of crystal fields is often extremely important. In chapter \ref{ChapterCoO} we determine the crystal fields of thin CoO films, grown on different substrates. The crystal fields in CoO are important since they determine the single ion anisotropy, which is closely related to the effectiveness of exchange-bias. In chapter \ref{ChapterLaTiO3} we determine the size of the crystal fields in LaTiO$_3$. We do not have polarization dependent data on LaTiO$_{3}$, since the crystals are not single domain, but also from the isotropic line-shape information about the crystal fields can be deduced. There has been a lively debate about the importance of orbital fluctuations in this compound. Orbital fluctuations are only possible if the crystal fields splitting these orbitals are small. By determining the local crystal fields present in LaTiO$_{3}$ we are able to determine if the orbitals are spatially locked, or if they can fluctuate.

Not only orbital ordering and crystal fields lead to linear dichroism. Also magnetic ordering does so. On first account this might sound strange since the dipole operator does not act on spin space. If spin orbit interaction is neglected the initial (final) wave function can be written as $\Psi_{e}^{i(f)} \chi_{s}^{i(f)}$, where $\Psi_{e}^{i(f)}$ is the electron part of the wave function and $\chi_{s}^{i(f)}$ the spin part. The $2p$-XAS intensity is proportional to the square of $\langle\Psi_{e}^{i} \chi_{s}^{i}|q|\Psi_{e}^f \chi_{s}^{f}\rangle= \langle\Psi_{e}^{i} |q|\Psi_{e}^f\rangle\langle\chi_{s}^{i}|\chi_{s}^{f}\rangle$. Where q is the dipole operator. For cubic electron distributions this is independent of polarization and this intensity is only nonzero if $\chi^i=\chi^f$. No room for magnetic linear dichroism without spin orbit coupling thus.

We know however that for $2p$ core levels the spin orbit coupling is very large. In order to become sensitive to the spin ordering in the valence shell we also need a coupling between the core hole and the valence electrons. This is realized while the final state is strongly excitonic. The $p$ hole attracts the $d$ electrons quite hard. Coulomb attraction is orbital and spin dependent. The spin dependent part has become know as the exchange energy. From atomic physics Hunds rules are know to describe the lowest arrangement for electrons and one of Hunds rules states that electron electron repulsion tries to maximize the total spin momentum S. In other words the repulsion of two electrons with the same spin is smaller than the repulsion of two electrons with opposite spins. Hunds rules however are not exact, but rules of thumb, and sometimes, especially when two different shells are involved, fail. In order to understand the magnetic linear dichroism it is important to calculate the full multiplet structure of the final state excitons involved. Based on atomic, full multiplet calculations, including a cubic crystal field, strong magnetic linear dichroism has been predicted by G. van der Laan and B. T. Thole \cite{Laan91}. These predictions have been confirmed by P. Kuiper \textit{et al.} \cite{Kuiper93} on Fe$_{2}$O$_{3}$. They also give a relative simple, intuitive explanation of the reason why magnetic linear dichroism should be present in Fe$^{2+}$, a $d^5$ system. We will not show their spectra as an example, but use the example of magnetic linear dichroism in NiO. This while the final state of NiO is easier to understand and one can calculate the $L_2$ edge and the magnetic linear dichroism therein by hand.
   
  \begin{SCfigure}[][!h]
    \includegraphics[width=60mm]{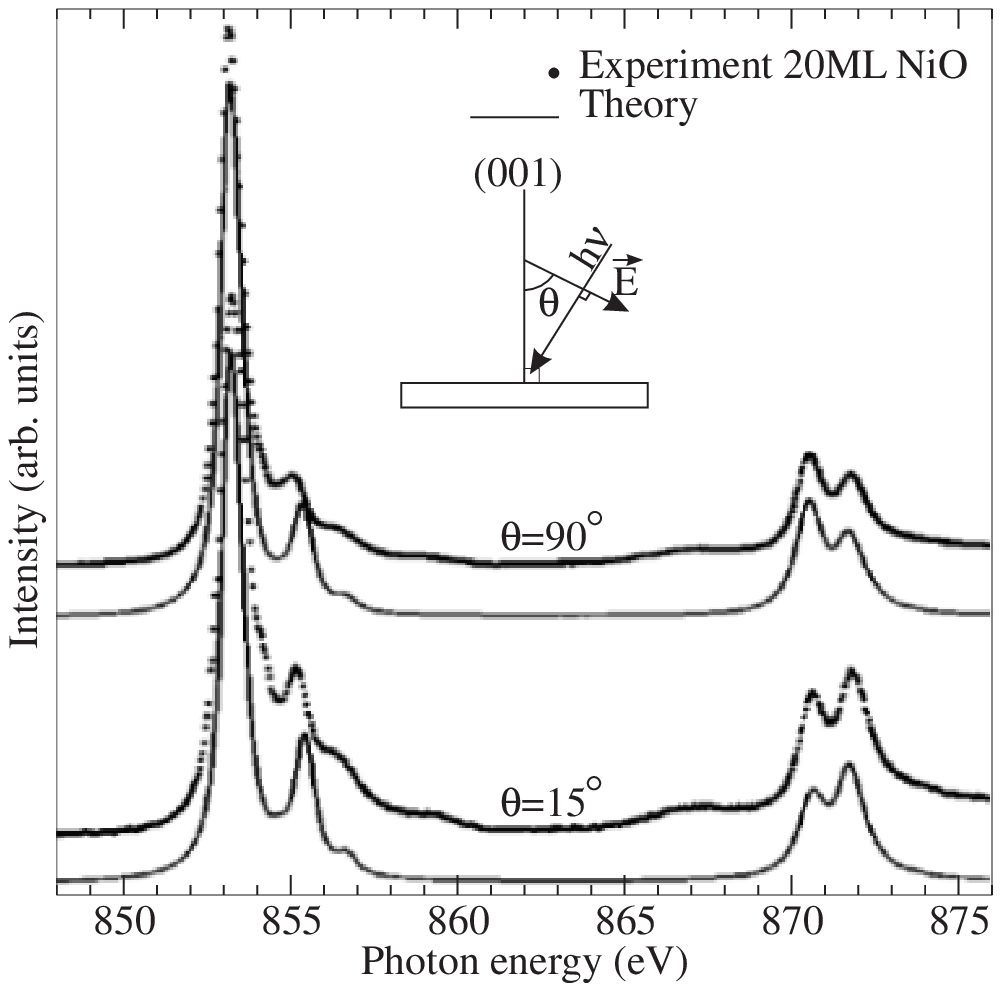}
    \caption{Magnetic linear dichroism in thin films of NiO on MgO. The linear dichroism arises due to the antiferromagnetic ordering within the NiO thin film. The spectra have been copied from D. Alders \textit{et al.} \cite{Alders98}. \newline}
    \label{introductionMagneticOrdering}
  \end{SCfigure} 

D. Alders \textit{et al.} \cite{Alders98} measured the magnetic linear dichroism of NiO thin films and in figure \ref{introductionMagneticOrdering} we show their spectra of 20 ML NiO on MgO. NiO is an antiferromagnetic rock salt with a (112) easy axes for the spins. In bulk NiO all 24 domains are available $(\pm1\pm1\pm2),(\pm1\pm2\pm1),(\pm2\pm1\pm1)$ and on average no dichroic signal is found. For thin films of NiO on MgO D. Alders \textit{et al.} \cite{Alders98} showed that the $(\pm1\pm1\pm2)$ domains are stabilized and a nett magnetic linear dichroic effect can be found. This effect has been widely used in order to study antiferromagnetic systems and to map out different domains.

In order to show the details why magnetic linear dichroism works one needs to use a multi electron approach, since besides $2p$ spin-orbit coupling, $2p$--$3d$ electron correlations are needed in order to create a linear dichroic effect. We will discuss the example of a $3d^8$ spectrum, like NiO. For a $3d^8$ system a $2p$ electron is excited into the valence $3d$ band. In a multi electron notation the $3d^8$ state is excited to $2p^53d^9$. For the $3d^8$ initial state we assume an occupation of $\underline{d}_{x^2-y^2}^{\uparrow}\underline{d}_{z^2}^{\uparrow}$ This state is orbitally $O_h$ symmetric and fully spin polarized in the z direction. There should not be any orbital induced linear dichroism (as long as the crystal field is of $O_h$ symmetry), but there should be magnetic linear dichroism. For the final state one should realize that the $2p$ spin orbit coupling is by far the largest interaction. We will assume that the $2p$ final state orbitals are eigenstates of the $2p$ spin-orbit coupling Hamiltonian. There are two different eigenstates for the $2p$ spin-orbit coupling Hamiltonian, $j=\frac{1}{2}$ ($l.s=-1$) and $j=\frac{3}{2}$ ($l.s=\frac{1}{2}$). These eigenstates can be seen in the spectrum as two distinct edges. The $L_2$ edge for $j=\frac{1}{2}$ and the $L_3$ edge for $j=\frac{3}{2}$. We will concentrate us here on the $L_2$ edge, since this edge shows the nicest magnetic linear dichroism. For the $L_2$ edge $j=\frac{1}{2}$. This means the $2p$ orbitals available are  $|\frac{1}{2} \quad\frac{1}{2}\rangle =\sqrt{\frac{1}{3}}p_{x}^\downarrow -{\rm{i}}\sqrt{\frac{1}{3}}p_{y}^\downarrow -\sqrt{\frac{1}{3}}p_{z}^\uparrow$ and $|\frac{1}{2} - \frac{1}{2} \rangle =\sqrt{\frac{1}{3}}p_{x}^\uparrow +{\rm{i}}\sqrt{\frac{1}{3}}p_{y}^\uparrow +\sqrt{\frac{1}{3}}p_{z}^\downarrow$. The $3d$ orbitals available for the final state are in principal all five $d$ orbitals with 2 fold spin degeneracy. We will however restrict ourselves in this example to the two $e_g$ orbitals with two fold spin degeneracy. This is strictly only correct if $10Dq >> G_1^{pd}$. I.e the crystal field is stronger than the $2p$--$3d$ electron-electron repulsion. Although this is not fulfilled in NiO one can sill use this example to get qualitative good agreement between theory and experiment. In total we have a basis of 8 wave functions for the final state. The $2p$ orbital can be $|\frac{1}{2} \quad\frac{1}{2}\rangle$, or $|\frac{1}{2} -\frac{1}{2}\rangle$. The $3d$ orbital can be $d_{x^2-y^2}$ or d$_{z^2}$. The $3d$ spin can be up or down. 

These 8 states do not have the same energy, but are split by the $2p$--$3d$ electron-electron Hamiltonian. As a first approximation one could take only the exchange part of the electron-electron repulsion Hamiltonian and assume an energy difference between the states that couple to $K=1$ and $K=0$, where $K=J_{2p}+S_{3d}$ . This however is not sufficient since there is a large difference in electron-electron repulsion between different orbital occupations. If one looks at the electron densities, or the part of the Hamiltonian that depends on $n_{m\sigma}n_{m'\sigma'}$ one will find that the $p_z$ orbital has a much larger density overlap with a $d_{z^2}$ orbital than with a $d_{x^2-y^2}$ orbital. Group theory teaches us that there are 3 different states, one doublet and two triplets. The doublet has $K=0$, the two triplets have $K=1$. If one diagonalizes the $2p$--$3d$ electron-electron repulsion Hamiltonian on the basis of the eight wave functions we limited ourselves too one recovers this doublet and two triplets. The doublet has an energy of $-\frac{2}{15}G^1-\frac{3}{35}G^3$. The lowest triplet has an energy of $-\frac{2}{15}G^1-\frac{11}{245}G^2$ and the highes triplet has an energy of $\frac{2}{45}G^1-\frac{3}{245}G^3$. If one fills in the atomic Hartree-Fock values for the Sleighter integrals one finds the doublet to be the lowest-state. The first triplet is 140 meV higher in energy. The second triplet is 1.27 eV higher in energy. This can also be seen in the NiO $L_2$ spectra where one sees two peaks at $\pm$ 1eV difference. The left peak belongs to a doublet and a triplet, the right peak belongs to a triplet.

\begin{table}\label{NiOMCDfinalstates}
\begin{equation}\nonumber
\begin{array}{|c|c|cc|}
\hline
   \Psi_{final} & E & I_x & I_z \\
\hline
  +\sqrt{\frac{1}{2}}|\frac{1}{2}-\frac{1}{2}\rangle \underline{d}_{x^2-y^2}^{\uparrow}- \sqrt{\frac{1}{2}}|\frac{1}{2}\quad\frac{1}{2}\rangle \underline{d}_{x^2-y^2}^{\downarrow} & -\frac{6}{45}G^1-\frac{21}{245}G^3 & \frac{1}{18} & 0 \\
 -\sqrt{\frac{1}{2}}|\frac{1}{2}-\frac{1}{2}\rangle \underline{d}_{z^2}^{\uparrow}+ \sqrt{\frac{1}{2}}|\frac{1}{2}\quad\frac{1}{2}\rangle \underline{d}_{z^2}^{\downarrow}  &                                    & \frac{3}{18} & 0 \\
\hline
  +\sqrt{\frac{1}{2}}|\frac{1}{2}-\frac{1}{2}\rangle \underline{d}_{x^2-y^2}^{\uparrow}+ \sqrt{\frac{1}{2}}|\frac{1}{2}\quad\frac{1}{2}\rangle \underline{d}_{x^2-y^2}^{\downarrow} & -\frac{6}{45}G^1-\frac{11}{245}G^2 & \frac{1}{18} & 0 \\
  -\sqrt{\frac{3}{4}}|\frac{1}{2}\quad\frac{1}{2}\rangle \underline{d}_{z^2}^{\uparrow}+ \sqrt{\frac{1}{4}}|\frac{1}{2}-\frac{1}{2}\rangle \underline{d}_{x^2-y^2}^{\downarrow} &                                    & 0 & 0 \\
  +\sqrt{\frac{1}{4}}|\frac{1}{2}\quad\frac{1}{2}\rangle \underline{d}_{x^2-y^2}^{\uparrow}- \sqrt{\frac{3}{4}}|\frac{1}{2}-\frac{1}{2}\rangle \underline{d}_{z^2}^{\downarrow} &                                    & 0 & \frac{2}{18} \\
\hline
\hline
  +\sqrt{\frac{1}{2}}|\frac{1}{2}-\frac{1}{2}\rangle \underline{d}_{z^2}^{\uparrow}+ \sqrt{\frac{1}{2}}|\frac{1}{2}\quad\frac{1}{2}\rangle \underline{d}_{z^2}^{\downarrow} &  \frac{2}{45}G^1-\frac{ 3}{245}G^3 & \frac{3}{18} & 0 \\
  +\sqrt{\frac{1}{4}}|\frac{1}{2}\quad\frac{1}{2}\rangle \underline{d}_{z^2}^{\uparrow}+ \sqrt{\frac{3}{4}}|\frac{1}{2}-\frac{1}{2}\rangle \underline{d}_{x^2-y^2}^{\downarrow} &                                    & 0 & 0 \\
  +\sqrt{\frac{3}{4}}|\frac{1}{2}\quad\frac{1}{2}\rangle \underline{d}_{x^2-y^2}^{\uparrow}+ \sqrt{\frac{1}{4}}|\frac{1}{2}-\frac{1}{2}\rangle \underline{d}_{z^2}^{\downarrow} &                                    & 0 & \frac{6}{18} \\
\hline
\end{array}
\end{equation}
\caption{$L_2$ final states of $2p$-XAS on NiO. The $|\frac{1}{2}\pm\frac{1}{2}\rangle$ states denote a $2p$ core hole with $j=\frac{1}{2}$ and $m_j=\pm\frac{1}{2}$. Whereby $|\frac{1}{2}-\frac{1}{2}\rangle= \sqrt{\frac{1}{3}}\underline{p}_x^{\downarrow}-{\rm{i}}\sqrt{\frac{1}{3}}\underline{p}_y^{\downarrow}- \sqrt{\frac{1}{3}}\underline{p}_z^{\uparrow}$ and $|\frac{1}{2}\quad\frac{1}{2}\rangle= \sqrt{\frac{1}{3}}\underline{p}_x^{\uparrow}+{\rm{i}}\sqrt{\frac{1}{3}}\underline{p}_y^{\uparrow}+ \sqrt{\frac{1}{3}}\underline{p}_z^{\downarrow}$. The intensities for x and z polarized light are calculated assuming a $\underline{d}_{x^2-y^2}^{\uparrow}\underline{d}_{z^2}^{\uparrow}$ initial state.}
\end{table}

In order to understand the magnetic linear dichroism we have to take a look at the final states. In table \ref{NiOMCDfinalstates} we show the eight different final states, their energies and the intensity these states contribute to a spectrum with x or z polarized light. Now one can see that the doublet, with $K=0$, can only be excited by x polarized light, whereas the two triplets can be excited by both polarizations and have more intensity for z polarized light than for x polarized light. We see that for x polarized light the peak at the lower energy is higher in intensity and that for z polarized light the latest peak is higher in intensity.

This more or less complete calculation shows what goes into the explanation of magnetic linear dichroism in cubic symmetry. One can also explain these spectra with the use of coupling of orbital momenta. For the final state $J_{2p}$ and $S_{3d}$ couple to $K=J_{2p}+S_{3d}$. $K$ can be 2, 1 or 0, while $S_{3d}=\frac{1}{2}$ and $J_{2p}=\frac{3}{2}$ or $\frac{1}{2}$. The initial state has $J_{2p}=0$ and $S=1$ and therefore $K=1$. If we fully magnetize the sample we have $M_{K}=1$ in the initial state. For angular momenta we have the polarization dependent selection rules $\Delta M_K=0$ for z polarized light and $\Delta M_K=\pm 1$ for x,y polarized light. Starting from $M_K=1$ one reaches $M_K=1$ final states with z polarized light and $M_K=0,2$ states with x,y polarized light. The final states with $K=0$ do not contain $M_K=1$ levels and can therefore not be reached with z polarized light. For the $L_{2}$ edge ($J_{2p}=\frac{1}{2}$) this shows very clearly since there are only final states with $K=1$ or $K=0$.

  \begin{SCfigure}[][!h]
    \includegraphics[width=60mm]{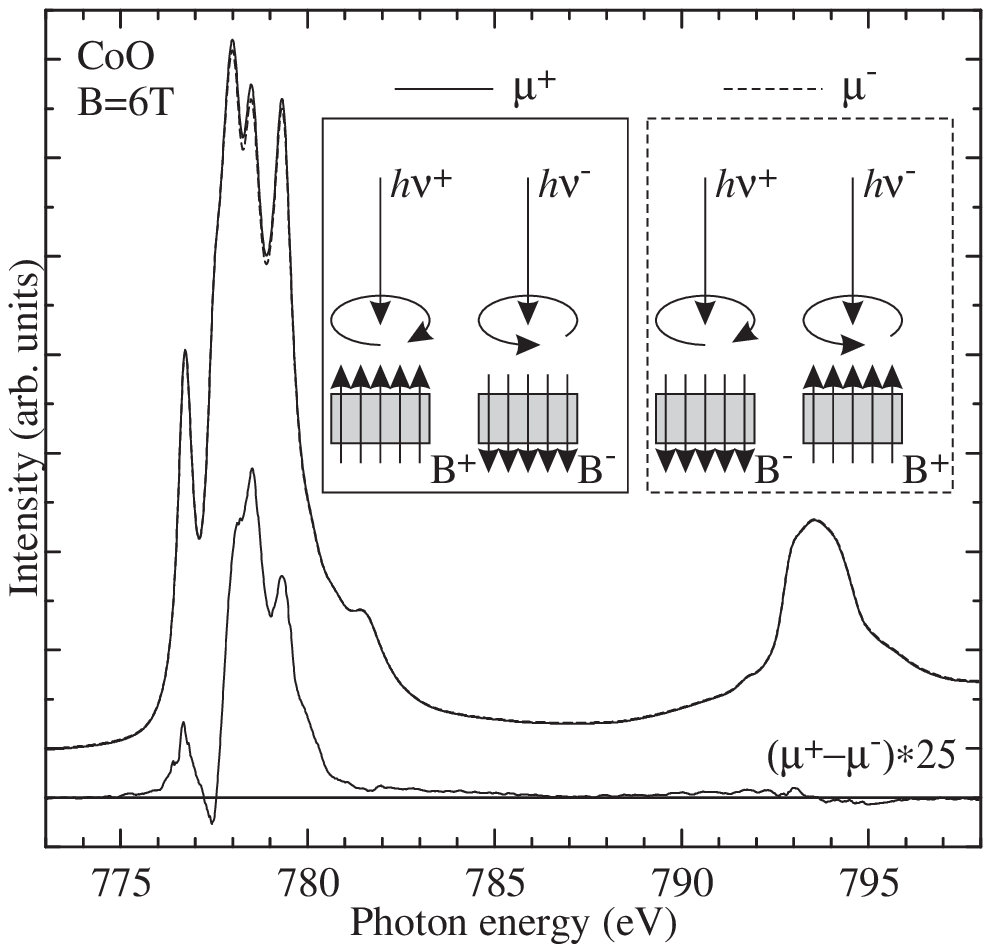}
    \caption{Circular dichroism in CoO in the paramagnetic state at 291K in a field of 6 Tesla. The circular dichroism arises due to the induced magnetic moment by the magnetic field. \newline}
    \label{introductionCoOXMCD}
  \end{SCfigure} 
  
The last option of $2p$-XAS spectroscopy we want to discus is x-ray magnetic circular dichroism (XMCD). For systems with net aligned magnetic moments the spectra taken with right and left circular polarized light are different. Magnetic moments can be due to ferromagnetic ordering, but can equally well be induced by a magnetic field. In figure \ref{introductionCoOXMCD} we show these measurements for CoO in a 6T magnetic field. One can use cluster calculations to reproduce the spectra. If the spectra are reproduced information about the magnetic state can be obtained. There is however an even simpler way to deduce information from circular dichroic spectra. Thole \textit{et al.} \cite{Thole92} have derived sum-rules relating the integrated intensity of the XMCD signal directly to the expectation value of $L_{z}$. This makes the use of XMCD spectroscopy rather easy and widely used. In chapter \ref{ChapterSumrules} we will discuss more about these sum-rules. The XMCD effect within CoO will be explained in more detail in chapter \ref{ChapterCoO}, where we will discuss in quite detail the interplay of spin, orbitals and the lattice in CoO. 

In conclusion; XAS is a very powerful type of spectroscopy. XAS spectra are about sensitive to any local operator one wants to measure. They give a huge amount of information about the ground-state of the system. XAS spectra consist of excitons and can not be compared to some kind of band structure, since $2p$ core hole - $3d$ electron attractions are important. XAS spectra are however so strongly excitonic that a small cluster (TMO$_6$) generally will be sufficient to calculate the spectra. From these cluster calculations information about operator values of the ground-state can be deduced.

\chapter{Cluster calculations}
\label{ChapterCluster}

In this chapter we will discuss some concepts of a cluster calculation. The main cluster we will be dealing with is a transition metal with 6 oxygens around, embedded in an electrical field due to the charges of the other atoms and electrons in the crystal. The electron hopping between the oxygen and the transition metal within the cluster will be dealt with within a tight-binding approach. The influence of spin and charge of the surrounding atoms will be treated within mean-field theory. For the transition metal we will only consider the $3d$ shell and for the oxygens only the $2p$ shell. Many good books have been been written on how to do cluster or ligand field calculations \cite{Ballhausen62, Sugano70, Atkins98Chapter14}, and the theory is quite old. With the calculation speed of the modern desktop PC-s and because of good source codes available \cite{Tanaka94, Thole97}, the calculation of a transition metal oxygen cluster is no problem any more. Within this thesis we used the program XTLS 8.0 written by A. Tanaka. \cite{Tanaka94}. In appendix \ref{ApendixXTLSManual} we present a short manual for his program.

Within cluster calculations we define a Hamiltonian matrix on a many-electron basis. The Hamiltonian will be diagonalized using a Lancsoz routine in order to find the ground state energy and wave function. For spectroscopy calculations we also define a final state basis and calculate the final state Hamiltonian. The spectra can now be calculated by diagonalizing the entire final state Hamiltonian and calculate the transition probability for each final state from the given initial state, or with much less computational burden by using a Greens-function method. In the following paragraphs we will introduce the idea of a many-electron basis and the full anti-symmetrized wave functions within that basis. Then we will discuss how to construct the Hamiltonian, in the end we will discuss how to calculate spectra.

 \section{Constructing the basis}

The basis we use for our cluster calculations is a configuration interaction basis. For a Cu $3d^{9}$ system interacting with one oxygen $2p^{6}$ shell we have two possible configurations, namely; Cu$3d^{9}$--O2$p^{6}$ and Cu$3d^{10}$--O2$p^{5}$. The configuration Cu$3d^9$--O$2p^6$ will have 10 wave functions, namely the five d orbitals with $m_{l}=-2$ to $m_{l}=2$, each with spin up or down. The configuration Cu$3d^{10}$--O$2p^{5}$ will have 6 wave functions, which are, the three $p$ orbitals with $m_{l}$=-1 to $m_{l}$=1, each with spin up or down. In total there will be 16 wave-functions building up the basis for a Cu--O cluster with one hole. In other words, we have for a Cu--O cluster a 16 by 16 Hamiltonian that we first have to fill in and then diagonallize, in order to find the ground-state energy and wave function. For a Ni $3d^8$ system with one oxygen, however, the number of basis states already drastically increases. There are 3 configurations which should be take into account. These configurations are; Ni$3d^8$--O$2p^6$, Ni$3d^9$--O$2p^5$, and Ni$3d^{10}$--O$2p^4$. The Ni$3d^8$--O$2p^6$ configuration has 45 wave-functions. The first hole on the Ni atom can be in 10 orbitals. For the second hole there are still 9 orbitals it can go to. This will give 90 possible functions, but one should realize that electrons are indistinguishable and that if the first electron goes in orbital one and the second in orbital two this will give the same wave function as when the first electron goes in orbital two and the second in orbital one. 45 wave functions thus for the Ni$3d^8$O$2p^6$ configuration. 60 wave-functions for the Ni$3d^9$--O$2p^5$ configuration and 15 ((6*5)/2) for the Ni$3d^{10}$--O$2p^4$ configuration. A total basis of 120 wave-functions. The largest basis one could encounter in a TmO$^6$ cluster is the basis of a $3d^0$ problem with 6 oxygen $2p^6$ shells around. If one would include all configurations, like $3d^02p^62p^62p^62p^62p^62p^6$, $3d^12p^52p^62p^62p^62p^62p^6$, ...., up to $3d^{10}2p^62p^62p^62p^62p^22p^0$, one would find a basis of $\frac{46!}{36! 10!}=4 076 350 421$ wave functions. Modern computers (2005) can not store one wave function of this size, so diagonalizing such a matrix is not possible at this time, but within ten to twenty years this might change. More important, configurations with 6 holes on one oxygen and 10 electrons at the $d$ site will have energies so high that there is no need to consider them. If one does not take into account the configurations which have more than 2 electrons at the $d$ site, there are only 28 configurations left with a total basis of 28711 wave-functions, which gives no computing problem.

 \section{Many-electron wave-functions}

For many-electron wave functions (2 electrons or more) it is important to anti-symmetrize the wave function with respect to the electron coordinates. If we have two electrons in orbital a and b, the total wave function would be $\sqrt{\frac{1}{2}}(\Psi_a(r_1)\Psi_b(r_2)-\Psi_a(r_2)\Psi_b(r_1))$. For more than two electrons, the full anti-symmetrized wave function can be written down as a Slater determinant, the determinant of a matrix in which the rows contain all different orbitals present in the wave function and the collums contain all different coordinates necessary to describe the positions of all electrons in the system. When we write a wave function we will implicitly assume that anti-symmetrizing has been taken care of. The wave function written as $\Psi_a\Psi_b$ should be read as $\sqrt{\frac{1}{2}}(\Psi_a(r_1)\Psi_b(r_2)-\Psi_a(r_2)\Psi_b(r_1))$. When calculating matrix elements proper anti-symmetrization has been taken into account.

 \section{Filling in the Hamilton matrix}
 
When filling in the Hamilton matrix it is important to realize that there are 2 very different types of Hamilton operators. The one-electron operator and the two-electron operator. Any one-electron operator can be written as $O^{\mu\nu}c_{\mu}^{\dagger}c_{\nu}$ where $c^{\dag}$ and $c$ are creation and annihilation operators. Whereas a two-electron operator is written as $O^{\mu\nu\eta\tau}c_{\mu}^{\dagger}c_{\nu}^{\dagger}c_{\eta}c_{\tau}$. All one-electron operators can be taken into account within a one-electron theory and it makes sense to draw a density of states for eigensystems of a one-electron Hamiltonian. For a two electron operator, however, one-electron theory fails drastically and a density of states is not defined for a Hamilton operator containing two-electron operators. Luckily most Hamilton operators are one-electron operators. Only electron-electron repulsion is a two-electron operator. 

We will first consider one-electron Hamilton operators. These operators will only have a contribution between two wave functions that differ no more then one electron from each other. We will start with the crystal-field and we will follow the line of thought of the first pages in S. Sugano, Y. Tanabe and H. Kamimura \cite{Sugano70}. With one difference. We will not assume a cubic point charge distribution responsible for the crystal field, but take an arbitrary charge distribution. After that we will discuss covalency. Covalency is responsible for the interaction terms between different configurations and shall be treated in a similar way as done by Slater and Koster within tight-binding theory \cite{Slater54}, but within a cluster not k dependent.

  \section{One-electron Hamilton Operators}
  
A one-electron Hamilton Operator has a non zero interaction between two wave functions only if the two functions do not differ by more than one electron. In that case the Hamiltonian between these two wave-functions is equal to the Hamiltonian between the two different one-electron basis functions, multiplied by -1 to the power of number of permutations needed to put the two basis functions in the same order. The factor -1 originates from the anti-symmetrization of the wave function. Realize that a fully anti-symmetrized wave function has all possible orders of one-electron wave functions with respect to the electron coordinates, but with different signs. To make things clearer lets have a look at an example. Let H$_1$ be a one-electron Hamiltonian, then: $\langle d_{xy}d_{yz}|H_1|d_{xy}d_{xz}\rangle = \langle d_{yz}|H_1|d_{xz} \rangle$ and $\langle d_{xy}d_{yz}|H_1|d_{xz}d_{xy} \rangle = -\langle d_{yz}|H_1|d_{xz}\rangle$. For diagonal elements one has to take the sum of all one-electron contributions: $\langle d_{xy}d_{yz}|H_1|d_{xy}d_{yz} \rangle = \langle d_{xy}|H_1|d_{xy} \rangle + \langle d_{yz}|H_1|d_{yz} \rangle$. In the next sections we will only describe the value of a Hamilton operator acting on a one-electron state. The Matrix elements for the many-electron wave-functions can be derived from this by using the above argumentation.

   \subsection{Crystal Field}
   
In the following section we want to discuss the effect of the charges, that surround the cluster we calculate, on the level splitting of that cluster. To do so we will use a mean-field approximation. The potential of all other charges on the site of one atom is known as the Madelung potential. It is custom to mention only the local value of the Madelung potential, but we will also use the derivatives of the Madelung potential in order to describe the local crystal field. These derivatives of the Madelung potential at the site of an atom will result in a level splitting similar to the Stark effect, but a few orders of magnitude stronger.

Let us define the Madelung potential at the site of the atom we consider as V($r,\theta,\phi$). Since our basis functions are expanded on spherical harmonics, $\Psi(r,\theta,\phi)=R_{n,l}(r)Y_l^m(\theta,\phi)$, we will also expand our Madelung potential on spherical harmonics:
\begin{equation}
V(r,\theta,\phi)=\sum_{k=0}^{\infty}\sum_{m=-k}^k A_{k,m} r^k C_k^m(\theta,\phi)
\end{equation}
Where $C_k^m(\theta,\phi)=\sqrt{\frac{4 \pi}{2k+1}}Y_k^m(\theta,\phi)$. The values of $A_{k,m}$ are Tailor expansions in r and are therefore related to the derivatives of V($r,\theta,\phi$). For $k\leq6$ we found the relation:
\begin{equation}
A_{k,m}=\frac{1}{\sqrt{(k-m)!}}\frac{1}{\sqrt{(k+m)!}}\partial_z^{k-|m|}(-\textmd{Sign}[m]\partial_x+\imath\partial_y)^{|m|}V(r,\theta,\phi)|_{r=0}
\end{equation}
If one knows the values of A$_{k,m}$ the matrix elements of the Hamiltonian can be calculated straightforward:
\begin{equation}\begin{split}
H_{i,j}=&
\langle R_{n_i}^{l_i}(r)Y_{l_i}^{m_i}(\theta,\phi)\mid\sum_{k=0}^{\infty}\sum_{m=-k}^k A_{k,m} r^k C_k^{m}(\theta,\phi)\mid R_{n_j}^{l_j}(r)Y_{l_j}^{m_j}(\theta,\phi)\rangle\\
=&\sum_{k=0}^{\infty}\sum_{m=-k}^k A_{k,m} 
\langle Y_{l_i}^{m_i}(\theta,\phi)\mid C_k^{m}(\theta,\phi)\mid Y_{l_j}^{m_j}(\theta,\phi)\rangle 
\langle R_{n_i}^{l_i}\mid r^k \mid R_{n_j}^{l_j}\rangle
\end{split}\end{equation}
The integrals over the radial part of the wave-function, $\langle R_{n_i}^{l_i}\mid r^k \mid R_{n_j}^{l_j}\rangle $, have been calculated in the Hartree-Fock approximation on a free atom with the use of Cowans's code \cite{Cowan81} and are tabulated in appendix \ref{ApendixSlaterIntegrals}. The integrals over the angular part can be calculated analytically and are normally expressed in Glebs-Gordan coefficients or 3J symbols. The 3J symbols are closed expressions and are explained in Cowans book, chapter 5 \cite{Cowan81}, for example:
\begin{equation}\begin{split}
\langle Y_{l_i}^{m_i}(\theta,\phi)\mid C_k^{m}(\theta,\phi)\mid Y_{l_j}^{m_j}(\theta,\phi)\rangle=&\\
(-1)^{m_i} \sqrt{(2l_i+1) (2l_j+1)}&\left(%
\begin{array}{ccc}
  l_i & k & l_j \\
  0 & 0 & 0 \\
\end{array}%
\right)
\left(%
\begin{array}{ccc}
  l_i & k & l_j \\
  -m_i & m & m_j \\
\end{array}%
\right)
\end{split}\end{equation}
The expansion of the Madelung potential on spherical harmonics does not have to be taken up to arbitrary high values of k in order to be exact. We know from the triangular equation on the 3J symbols that the crystal field only contributes to terms in the Hamiltonian for k$\leq l_i+l_j$. We can reduce the number of expansion coefficients, $A_{k,m}$, even further by realizing that k+l$_i$+l$_j$ has to be even. Furthermore, since the Hamiltonian has to be hermitian we find that: 
\begin{equation}\begin{split}
H_{ij}=H_{ji}^*\qquad\qquad&\\
<R_{n_i}^{l_i}(r)Y_{l_i}^{m_i}(\theta,\phi)\mid\sum_{k,m} r^k A_{k,m}C_{k}^{m}(\theta,\phi)&\mid Y_{l_j}^{m_j}(\theta,\phi)R_{n_j}^{l_j}(r)>=\qquad\qquad\\
\qquad\qquad<R_{n_j}^{l_j}(r)Y_{l_j}^{m_j}(\theta,\phi)\mid\sum_{k,m} r^k A_{k,m}^*&C_{k}^{-m}(\theta,\phi)\mid Y_{l_i}^{m_i}(\theta,\phi)R_{n_i}^{l_i}(r)>\\
A_{k,m}=&(-1)^mA_{k,-m}^*
\end{split}\end{equation}
A program to calculate the Madelung potential of an arbitrary crystal structure, using an Ewald summation over point charges and expanding this potential on spherical harmonics up to the 6$^{th}$ order, can be found in appendix \ref{ApendixAkm}.

In this thesis, however, we did not try to calculate the crystal field by the use of an Ewald summation. We fitted our crystal fields to our measured spectra. This is a more general approach that does not rely on an assumed point charge ordering within the crystal. The number of parameters with which the crystal field can be described can be greatly reduced if symmetries are present. If the system does not have a high symmetry, we used the help of LDA(+U) calculations in order to determine the crystal field.

In order to make fitting possible it is necessary to reduce the number of free parameters as far as possible. For a crystal field splitting within a $d$ shell we have l$_i$=l$_j$=2, so only values of k=0,2 or 4 contribute to the crystal field. This leaves us with 15 independent parameters A$_{k,m}$. The parameters A$_{k,m}$ always show up in the Hamiltonian multiplied by $\langle R_{n_i}^{l_i}\mid r^k \mid R_{n_j}^{l_j}\rangle$. Since the radial wave functions might differ from the atomic Hartree Fock radial wave functions within a solid, it seems logic to include the expectation value of the radial wave functions within the fitting parameters. Therefore we define the parameter B$_{k,m}$;
\begin{equation}
B_{k,m}=A_{k,m} \langle R_{n_i}^{l_i}\mid r^k \mid R_{n_j}^{l_j}\rangle
\end{equation}

Now we have to fit these 15 parameters B$_{k,m}$ in order to reproduce the measured spectra. 15 parameters is quite a lot, but if some symmetry is present many parameters will be related or have to be strictly zero. For a system with $O_h$ symmetry we will show that the number of parameters reduces to one, so it is worth to have a look into some symmetry arguments and group theory. A good starting-point would be chapter 3 and 4 of Ballhausen \cite{Ballhausen62} for example. We will go very fast through the symmetry arguments and quickly write down the final results for the crystal fields in higher symmetry.

  \begin{figure}[h]
   \begin{center}
    \includegraphics[width=120mm]{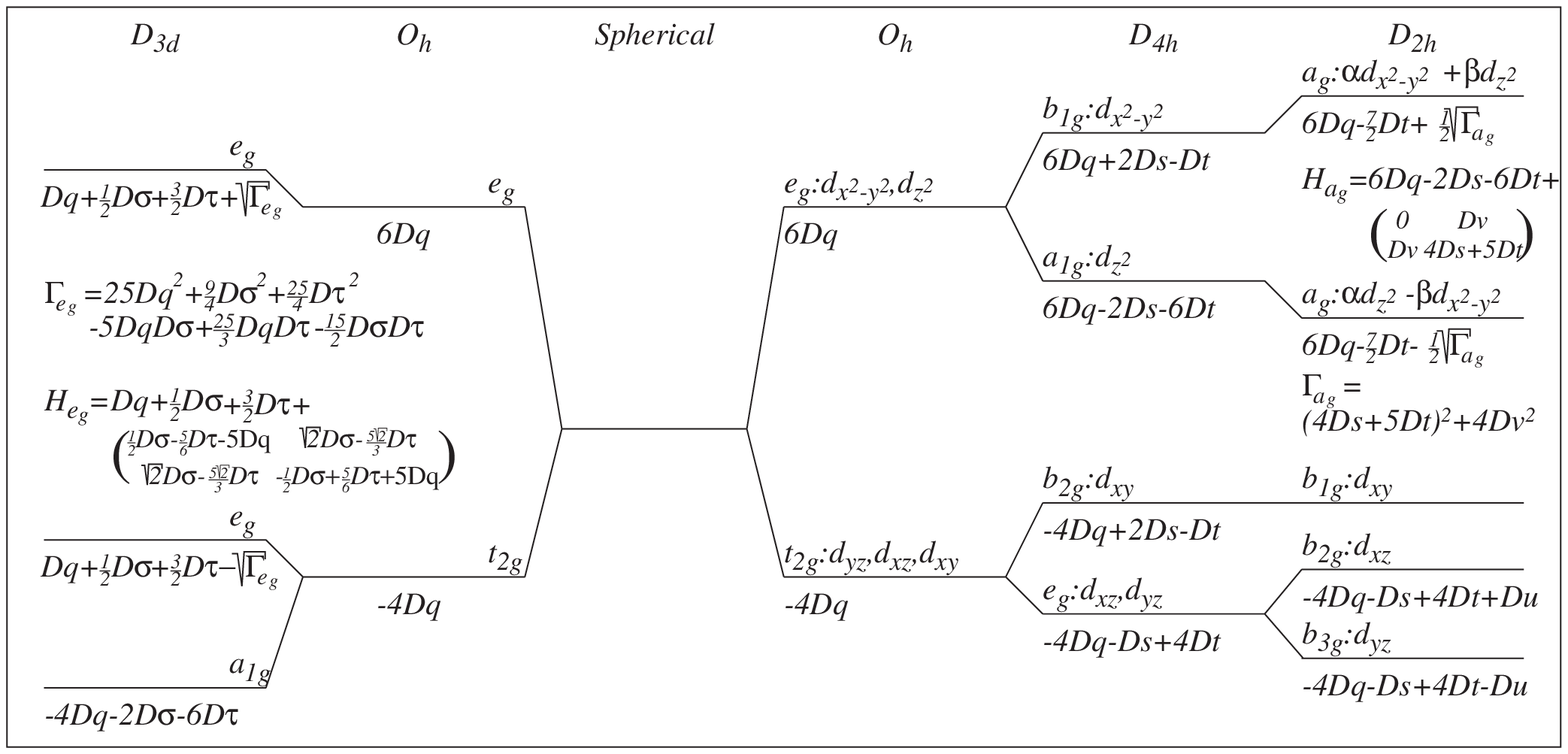}
    \caption{Energy level diagram for a $d$ shell in Trigonal ($D_{3d}$), Cubic ($O_h$), Tetragonal $D_{4h}$ and orthorhombic ($D_{2h}$) symmetry. \newline}
    \label{introductionEnergyDiagramSymmetry}
   \end{center}
  \end{figure} 

The Madelung potential at the position of the atom for which we want to calculate the crystal field has been written as $\int V(r,\theta,\phi) \delta r=V(\theta,\phi)=\sum_{k=0}^{\infty}\sum_{m=-k}^k B_{k,m} C_k^m(\theta,\phi)$. If we now know, for example, that the crystal goes back into itself after a rotation of $C_z^2$ which is a 180$^\circ$ rotation around the z axe, we can conclude that:
\begin{equation}
C_z^2 V(\theta,\phi)=V(\theta,\phi)
\end{equation}
A rotation of 180$^\circ$ around the z axe will transform $\phi$ into $\phi+\pi$. Spherical harmonics scale with $\phi$ as $e^{im\phi}$. From which one can see that:
\begin{equation}
C_z^2 C_k^m(\theta,\phi)=-1^m C_k^m(\theta,\phi)
\end{equation}
Combining the last two equations it is clear that if a crystal has a 2 fold symmetry around the z axe, all $B_{k,m}$ values with $m$ odd should be zero. A similar argumentation yields that for an $i$ fold symmetry around the z axe only $B_{k,m}$ values with $m=in$ are nonzero. Where $n \in \mathbb{N}$. Making use of all symmetry operations present we find that for an atom at an $O_h$ point-group site only one parameter remains to describe the crystal field splitting within a $d$-shell. This parameter is generally known as $10Dq$, the energy difference between the $t_{2g}$ and $e_{g}$ sub-shell. For an atom at a $D_{4h}$ point-group site there are 3 parameters. The $e_g$ levels are split by an additional parameter $\Delta_{e_g}$ and the $t_{2g}$ levels are split in a doublet ($d_{xz}$ and $d_{yz}$) and a singlet ($d_{xy}$) with an energy difference of $\Delta_{t_{2g}}$. Ballhausen \cite{Ballhausen62} uses the parameters $Ds$ and $Dt$ for this splitting, where $\Delta_{e_{g}}=4Ds+5Dt$ and $\Delta_{t_{2g}}=3Ds-5Dt$. Within $D_{2h}$ symmetry two more parameters are introduced. We will refer to these two parameters as $Du$ and $Dv$. $2Du$ will be the splitting between the $d_{xz}$ and $d_{yz}$ orbital. The parameter $Dv$ does not split the levels further, but introduces a mixing between the $d_{z^2}$ and $d_{x^2-y^2}$ orbital. The complete form of the crystal field coupling to $d$ orbitals within $O_{h}$, $D_{4h}$, $D_{2h}$ and $D_{3d}$ symmetry can now be written as a function of the parameters $10Dq$, $Ds$, $Dt$, $Du$, $Dv$, $D\sigma$, and $D\tau$.
\begin{equation}\begin{split}\nonumber
V_{O_{h\phantom{4}}}=&21Dq C_4^0(\theta,\phi) +21\sqrt{\frac{5}{14}}Dq (C_4^{4}(\theta,\phi) +C_4^{-4}(\theta,\phi))\\
V_{D_{4h}}=&-7Ds C_2^0(\theta,\phi) +21(Dq-Dt) C_4^0(\theta,\phi) \\
        &+21\sqrt{\frac{5}{14}}Dq (C_4^{4}(\theta,\phi) +C_4^{-4}(\theta,\phi))\\
V_{D_{2h}}=&-7Ds C_2^0(\theta,\phi) +(\frac{1}{2}\sqrt{6}Du-2\sqrt{2}Dv)(C_2^2(\theta,\phi)+C_2^{-2}(\theta,\phi)) \\         &+21(Dq-Dt) C_4^0(\theta,\phi)
            +3(\sqrt{\frac{2}{5}}Du+\sqrt{\frac{6}{5}}Dv)(C_4^2(\theta,\phi)+C_4^{-2}(\theta,\phi))\\
           &+21\sqrt{\frac{5}{14}}Dq (C_4^{4}(\theta,\phi) +C_4^{-4}(\theta,\phi)) \\
\end{split}\end{equation}
\begin{equation}\begin{split}
V_{D_{3d}}=&-7D\sigma C_2^0(\theta,\phi) -14(Dq+\frac{3}{2}D\tau) C_4^{0}(\theta,\phi) \\
           &-14\sqrt{\frac{10}{7}}Dq(C_4^{3}(\theta,\phi) +C_4^{-3}(\theta,\phi))
\end{split}\end{equation}

A complete energy level diagram can be seen in figure \ref{introductionEnergyDiagramSymmetry}. For the splitting within trigonal symmetry we will need three parameters. Ballhausen \cite{Ballhausen62} used the parameters $10Dq$, $D\sigma$ and $D\tau$. $10Dq$ is again the splitting between the $t_{2g}$ and $e_{g}$ orbitals. Within $D_{3d}$ symmetry, the $e_g$ levels stay degenerate and only the $t_{2g}$ levels split further with an energy of $\Delta_{t_{2g}}=-D\sigma+6\frac{2}{3}D\tau$. In figure \ref{introductionEnergyDiagramSymmetry} we also present the energy level splitting within $D_{3d}$ symmetry.

For high-symmetry systems we found that one can reduce the crystal field to a few understandable parameters. Part of this thesis, however, deals with systems that have very low symmetry at the metal site. This leaves us with 15 parameters $B_{k,m}$ which might have non-zero values. There is no intuitive meaning within these parameters $B_{k,m}$. We normally think in real-space orbitals. On a basis of real-space orbitals the 15 crystal field parameters are rather intuitively expressed as the 15 parameters of a 5 by 5 Hamiltonian coupling the real-space orbitals with each other. There are 5 parameters which define the on-site energies of the real-space orbitals, $V_{xy}, V_{yz}, V_{xz}, V_{x^2-y^2}$, and $V_{z^2}$. There are also ten parameters that couple the real-space orbitals with each other, $V_{xy,yz}, V_{xy,xz}, V_{xy,x^2-y^2}$, etc. These parameters are understandable. The relation between $B_{k,m}$ and $V_{\tau,\tau'}$ is as follows:
\begin{equation}\begin{split}\nonumber
B_{0,0}=&\frac{1}{5}(V_{x^2-y^2}+V_{xy}+V_{xz}+V_{yz}+V_{z^2})\\
B_{2,0}=&\frac{1}{2}(-2V_{x^2-y^2}-2V_{xy}+V_{xz}+V_{yz}+2V_{z^2})\\
B_{2,1}=&\frac{1}{2}({\rm{i}}\sqrt{6}V_{xy,xz} -\sqrt{6}V_{xy,yz} -\sqrt{6}V_{xz,x^2-y^2} \\
        &-\sqrt{2}V_{xz,z^2} -{\rm{i}}\sqrt{6}V_{yz,x^2-y^2} +{\rm{i}}\sqrt{2}V_{yz,z^2})\\
B_{2,2}=&\frac{1}{4}(-4\sqrt{2}V_{x^2-y^2,z^2} +4{\rm{i}}\sqrt{2}V_{xy,z^2} +\sqrt{6}V_{xz} -\sqrt{6}V_{yz} -2{\rm{i}}\sqrt{6}V_{yz,xz})\\
B_{4,0}=&\frac{3}{10}(V_{x^2-y^2}+V_{xy}-4V_{xz}-4V_{yz}+6V_{z^2})\\
B_{4,1}=&\frac{3}{10}(-{\rm{i}}\sqrt{5}V_{xy,xz} +\sqrt{5}V_{xy,yz} +\sqrt{5}V_{xz,x^2-y^2} \\
        &-2\sqrt{15}V_{xz,z^2} +{\rm{i}}\sqrt{5}V_{yz,x^2-y^2} +2{\rm{i}}\sqrt{15}V_{yz,z^2})\\
B_{4,2}=&\frac{3}{10}(\sqrt{30}V_{x^2-y^2,z^2} -{\rm{i}}\sqrt{30}V_{xy,z^2} +\sqrt{10}V_{xz} -\sqrt{10}V_{yz} -2{\rm{i}}\sqrt{10}V_{yz,xz})\\
\end{split}\end{equation}
\begin{equation}\begin{split}
B_{4,3}=&\frac{3\sqrt{35}}{10}({\rm{i}}V_{xy,xz} +V_{xy,yz} -V_{xz,x^2-y^2} +{\rm{i}}V_{yz,x^2-y^2})\\
B_{4,4}=&\frac{3\sqrt{70}}{20}(V_{x^2-y^2}-V_{xy}-2 {\rm{i}} V_{xy,x^2-y^2})\\
\end{split}\end{equation}

   \subsection{Covalency}
   
Covalency arises from the hopping of electrons between the oxygen and the transition metal site. Within our single cluster calculations it is the only term that mixes two different configurations, $d^np^6$ and $d^{n+1}p^5$. The size of the hopping can be expressed in the parameters $pd\sigma$ and $pd\pi$. In cubic symmetry the $t_{2g}$ electrons hop with the parameter $pd\pi$ and the $e_{g}$ electrons with the parameter $pd\sigma$. The angular dependence can be found in table 1 of the paper by Slater and Koster \cite{Slater54}. This paper is, although rather old, still an excellent introduction into tight-binding theory and cluster calculations.

  \begin{SCfigure}[][!h]
    \includegraphics[width=60mm]{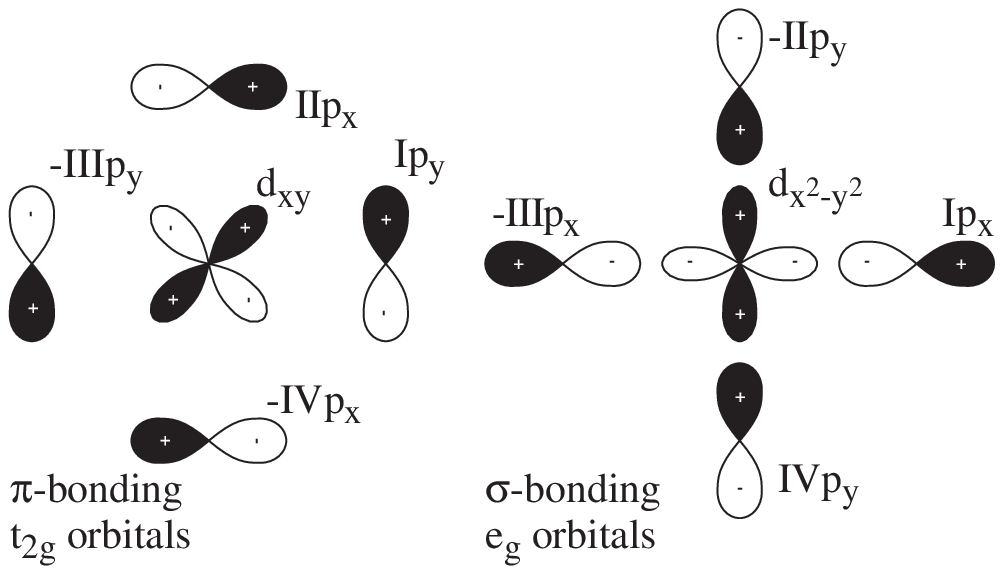}
    \caption{Bonding Oxygen orbitals in a Transition metal - Oxygen six cluster, within $O_{h}$ symmetry. Shown are the $d_{xy}$ and the $d_{x^2-y^2}$ orbital with the bonding linear combination of Oxygen $p$ orbitals.}
    \label{introductionBondingOrbitals}
  \end{SCfigure} 

We mostly calculate a cluster consisting of one transition metal and 6 oxygen atoms. One can use symmetry within the orbitals to reduce the basis. In figure \ref{introductionBondingOrbitals}, we show a $t_{2g}$ and an $e_{g}$ orbital together with the oxygen orbitals with which a bonding state can be made. By making a linear combination of the oxygen orbitals, bonding and non-bonding orbitals can be separated. In figure \ref{introductionBondingOrbitals} we see that for the $t_{2g}$ orbital the $IIp_x-IVp_x$ combination is bonding while the $IIp_x+IVp_x$ (not shown) combination is non-bonding. The $IIp_x+IVp_x$ has a positive overlap for oxygen II, but exactly the same amount of negative overlap for oxygen IV. Each $d$ orbital has exactly one linear combination of $p$ orbitals with which an overlap exists. The other orbitals are non-bonding. We show the bonding oxygen orbitals for a $d_{xy}$ and an $d_{x^2-y^2}$ orbital in figure \ref{introductionBondingOrbitals}.

Using a basis set of only bonding orbitals greatly reduces the size of the basis set. For a cluster consisting of one transition metal and 6 oxygens there are $10+6\times6=46$ orbitals. Of these 46 orbitals only 20 are important. Which are the 10 bonding and/or anti-bonding oxygen and transition metal orbitals. This reduction in basis set enormously reduces the computation time.

  \section[Two electron Hamiltonian Operators, electron-electron interaction]{Two electron Hamiltonian Operators, elec-tron-electron interaction}
\sectionmark{Two electron Hamiltonian Operators, electron-electron interaction} 

Electron-electron repulsion is the hardest Hamilton operator we have to deal with. This has everything to do with the fact that it is a two electron operator. It does not only couple wave-functions which are different in one electron, but the wave-functions may also be different in two electrons. In other words, if one electron moves from one orbital to another the rest of the electrons will react. The result is that one can not work with independent electrons anymore. An excellent short introduction into electron-electron repulsion can be found in chapter two of Ballhausen \cite{Ballhausen62}. A full, complete and lengthy account on electron-electron repulsion can be read in Cowan \cite{Cowan81}.

In second quantization we can write the electron-electron Hamilton as:
\begin{equation}\begin{split}
H_{e-e}=&\sum_{\langle m m' m'' m'''\sigma\sigma'\rangle}  U_{mm'm''m'''\sigma\sigma'}l_{m\sigma}^{\dagger}{l'}^{\dagger}_{m'\sigma'}{l_{m''\sigma}''}{l_{m'''\sigma'}'''}\\
U_{mm'm''m'''\sigma\sigma'}=&\delta(m+m',m''+m''') \\
                                 &\sum_{k=0}^{\infty} \langle lm|C^k_{m-m''}|l''m''\rangle
                                  \langle l'''m'''|C^k_{m'''-m'}|l'm'\rangle R^k(ll'l''l''')\\
R^k(ll'l''l''')=&e^2\int_{0}^{\infty}\int_{0}^{\infty}\frac{r_{<}^{k}}{r_{>}^{k+1}}R_{l}^{n}(r_1)R_{l'}^{n'}(r_2)R_{l''}^{n''}(r_1)R_{l'''}^{n'''}(r_2) r_1^2 r_2^2 \delta r_1 \delta r_2
\end{split}\end{equation}
With $r_{<}=\textmd{Min}(r_1,r_2)$ and $r_{>}=\textmd{Max}(r_1,r_2)$.

The integral over the radial wave functions has been calculated in the Hartree-Fock approximation on a free ion with the use of Cowans code \cite{Cowan81} and for $d$--$d$ and $p$--$d$ interactions they are tabulated in appendix \ref{ApendixSlaterIntegrals}.

One should be careful when one makes simplifications for electron-electron repulsion. In the introduction we compare the eigen-states of the full electron-electron repulsion Hamiltonian with different approximated Hamiltonians. The differences are very large.

  \section{Spectroscopy}

For calculating a spectrum from a cluster we start from the dipole approximation \cite{Schleich01, Slater60}:
\begin{equation}
\mu_{q}(h\nu)=\sum_{\Phi_f} |\langle \Phi_f|r C_q^{(1)}| \Phi_i \rangle|^2 \delta(E_i-E_f+h\nu)
\end{equation}
$\Phi_i$ is the initial state wave function that is occupied, $\Phi_f$ are all possible final state wave functions and q is the polarization of the light. The integral $|\langle \Phi_f|r C_q^{(1)}| \Phi_i \rangle|$ gives the excitation probability for a given final state and $\delta(E_i-E_f+h\nu)$ takes care of the energy conservation. If one wants to calculate a spectrum one has to find the lowest eigenstate of the initial state Hamiltonian and diagonalize the entire final state Hamiltonian. For each final state one has to calculate the excitation probability, which will give the spectrum as a set of delta-peaks. After broadening this can be compared with experiment. For larger clusters this becomes very time consuming and the spectra are calculated with less computational burden with the use of Greens functions.

The Greens function is defined as $\frac{1}{H_f +E_i-h\nu-\frac{1}{2} i \Gamma}$. The imaginary part of this function consists of Lorentzians with width $\Gamma$ that peak at all eigenvalues of the final-state Hamiltonian. The dipolar operator will be written in second quantization as $rC_q^{(1)}= r \sum_{m\sigma m' \sigma'} c_{m\sigma}l_{m'\sigma'}^{\dagger} \langle l_{m\sigma}|C_q^{(1)}|c_{m'\sigma'}\rangle$, which defines the non cubic transition matrix T which gives the transition probability from a given initial state basis function to a given final state basis function. The spectral intensity can now be written as:
\begin{equation}
\mu_q(h\nu)=\textmd{Im} \left\langle \Phi_i|T^{\dagger}GT|\Phi_i \right\rangle
\end{equation}
Instead of calculating all eigenvalues of the final state Hamiltonian one has to calculate once the inverse of this final state Hamiltonian.

\chapter{Sum-rules}
\label{ChapterSumrules}

Comparing cluster calculations with measured $2p$-XAS spectra gives an enormous amount of information about the system. This is not always an advantage, because it also means that all interactions that influence the ground-state and the spectrum have to be incorporated correctly. Sometimes it becomes quite hard to get a good fit of the spectra within a cluster calculation. In general, fitting becomes harder when the system has a lower symmetry and therefore more parameters to model the crystal field need to be incorporated. There is another method to derive information from XAS spectra. The integrated intensity of polarized spectra is related to a number of operator values of the initial state. These relations, first pointed out by Thole \textit{et al.} \cite{Thole92}, are extremely powerful since they are simple to use. Intuitively, one can understand that there should be something like a sum-rule. We know that the spectral intensity for a certain final state is proportional to
\begin{equation}
\langle\Phi_i|O|\Phi_f \rangle^2=
\langle\Phi_i|O|\Phi_f \rangle\langle\Phi_f|O|\Phi_i \rangle
\end{equation}
If one sums over all final states we get
\begin{equation}
\sum_{\Phi_f} \langle\Phi_i|O|\Phi_f \rangle\langle\Phi_f|O|\Phi_i \rangle=
\langle\Phi_i|O||O|\Phi_i \rangle
\end{equation}
Thereby it is used that, $\sum_{\Phi_f}\Phi_f \rangle\langle\Phi_f=1$. The only question is what is the operator $O||O$.

The original sum-rules have been derived for correlated wave-functions \cite{Thole92, Carra93a, Carra93b}. Later they have been re-derived within a one-electron approximation \cite{Ankudinov95}. It has been shown that the sum-rules are a result of symmetries already present in the 3J symbols \cite{Jo93}. In other words, the sum-rules are an intrinsic property of the dipole operator. They do not depend on the assumptions made on the local symmetry and they are independent of the framework in which one is working, namely, many-electron theory or one-electron theory. In this chapter we will show the derivation of the sum-rules in second quantization, along the lines of M. Altarelli \cite{Altarelli93}. The derivation of the sum-rules will start with the dipole approximation. The light interacting with the material will be written as $rC_q^{(1)}$, where $q$ is the polarization, $-1$ for left, $1$ for right circular polarized light and 0 for z linear polarized light. In the end one wants to have a sum-rule in which one does not need to integrate over the energy range from 0 eV to $\infty$ eV, but only over one edge. In order to achieve this we will rewrite the dipole operator in second quantization, active to one edge only:
\begin{equation}
rC_q^{(1)}=r \sum_{m\sigma m'\sigma'} c_{m\sigma}l_{m'\sigma'}^{\dagger} \langle l_{m\sigma}|C_q^{(1)}|c_{m'\sigma'}\rangle
\end{equation}
Now it is easy to define the dipole operator to be active into one edge only. To do so, one should not sum over all $l$ and $c$, but set $l$ and $c$ to a fixed value, depending on the edge one is looking at. For $2p$ XAS we have $c=1$ and $l=2$.

The dipole operator is expanded on spherical harmonics, $C$ and the same will be done with the wave functions. This can be done without loss of generality. One has to make the assumption that one can write the initial state and the final state locally as a product of the angular part times the radial part of the wave-function, $\Phi=\Psi(\theta,\phi) R(r)$. The spectral intensity for final state $\Psi_f R_f$ can now be written as:
 \begin{equation}\begin{split}
 |\langle R_{i}|r|R_{f}\rangle |^2
  \sum_{m,...,\sigma'''} 
  \langle \Psi_i|l_{m\sigma}c^{\dagger}_{m'\sigma'}|\Psi_f\rangle
  \langle \Psi_f|c_{m'''\sigma'''}l^{\dagger}_{m''\sigma''}|\Psi_i\rangle\\
  \langle l_{m\sigma}|C_q^{(1)}|c_{m'\sigma'}\rangle \langle l_{m''\sigma''}|C_q^{(1)}|c_{m'''\sigma'''}\rangle
 \end{split}\end{equation}
A summation over the final state wave-functions $\Psi_f$ and the use of the closure relation gives:
 \begin{equation}\begin{split}
  I_{q}=&|\langle R_{i}|r|R_{f}\rangle |^2\\
  &\sum_{m,...,\sigma'''} 
  \langle \Psi_i|c^{\dagger}_{m'\sigma'}c_{m'''\sigma'''}l_{m\sigma}l^{\dagger}_{m''\sigma''}|\Psi_i\rangle \\
  &\qquad\qquad\langle l_{m\sigma}|C_q^{(1)}|c_{m'\sigma'}\rangle \langle l_{m''\sigma''}|C_q^{(1)}|c_{m'''\sigma'''}\rangle
 \end{split}\end{equation}
The wave function $\Psi_i$ has a complete filled c shell. Therefore, the operator $c^{\dagger}_{m'\sigma'}c_{m'''\sigma'''}$ reduces to $\delta_{m',m'''}\delta_{\sigma',\sigma'''}$. One can drop the wave function $\Psi_i$ as it is known to take expectation values of operators for the initial state. Furthermore, one should note that $\sigma=\sigma'=\sigma''$ because otherwise the integral $<l_{m\sigma}|C_q^{(1)}|c_{m'\sigma'}>$ is zero:
 \begin{equation}
  I_{q}=|\langle R_{i}|r|R_{f}\rangle |^2
  \sum_{m,m',m'',\sigma} 
  l_{m\sigma}l^{\dagger}_{m''\sigma}
  \langle l_{m\sigma}|C_q^{(1)}|c_{m'\sigma}\rangle \langle l_{m''\sigma}|C_q^{(1)}|c_{m'\sigma}\rangle
 \end{equation}
The integrals $<l_{m\sigma}|C_q^{(1)}|c_{m'\sigma}>$ are integrals over 3 spherical harmonics and analytical expressions of $l,c,q,m$,and $m'$. A good way to see symmetries in these integrals is by writing them as 3J symbols. This can be done with the use of the following formula.
 \begin{equation}
  \langle l_{m}|C_q^{(k)}|l'_{m'}\rangle=-1^m\sqrt{(2l+1)(2l'+1)}  \left(
   \begin{array}{ccc}
     l & k & l' \\
     0 & 0 & 0 \\
   \end{array}
  \right)
  \left(
   \begin{array}{ccc}
     l & k & l' \\
     -m & q & m' \\
   \end{array}
  \right)
 \end{equation} 
Writing the integrals with the use of 3J symbols results in:
 \begin{equation}\begin{split}
  I_{q}=&|\langle R_{i}|r|R_{f}\rangle|^2\\
  &\sum_{m,m',m'',\sigma} 
  l_{m\sigma}l^{\dagger}_{m''\sigma}(-1)^{(m+m'')}(2l+1)(2c+1)\\
& \qquad\qquad\left(
   \begin{array}{ccc}
     l & 1 & c \\
     0 & 0 & 0 \\
   \end{array}
  \right)
  \left(
   \begin{array}{ccc}
     l & 1 & c \\
     -m & q & m' \\
   \end{array}
  \right)   
  \left(
   \begin{array}{ccc}
     l & 1 & c \\
     0 & 0 & 0 \\
   \end{array}
  \right)
  \left(
   \begin{array}{ccc}
     l & 1 & c \\
     -m'' & q & m' \\
   \end{array}
  \right)   
 \end{split}\end{equation} 
Absorbing the $c,l$ and $r$ dependence in one constant $P_{cl}^2$ this simplifies to: 
 \begin{equation}\begin{split}
  I_{q}=P_{cl}^2
  \sum_{m,m',m'',\sigma} &
  l_{m\sigma}l^{\dagger}_{m''\sigma}(-1)^{(m+m'')}\\
&  \left(
   \begin{array}{ccc}
     l & 1 & c \\
     -m & q & m' \\
   \end{array}
  \right)
  \left(
   \begin{array}{ccc}
     l & 1 & c \\
     -m'' & q & m' \\
   \end{array}
  \right)   
 \end{split}\end{equation}  
Using the triangular equations and using conservation of magnetic angular momentum for the 3J symbol one sees that $m''=q+m'=m$ and that $c=l\pm1$. As the excitation with $c=l-1$ is physical more relevant we will take $c=l-1$:
 \begin{equation}
  I_{q}=P_{cl}^2
  \sum_{m,\sigma} 
  \underline{n}_{m\sigma}
  \left(
   \begin{array}{ccc}
     l & 1 & l-1 \\
     -m & q & -q+m \\
   \end{array}
  \right)^2
 \end{equation}  
A 3J symbol with only one $j$ and one $m$ value unknown reduces to a simple polynomial expression. Since $q\in(-1,0,1)$ one can write three closed expressions:
 \begin{equation}\begin{split}
  I_{-1}=&P_{cl}^2 \sum_{m,\sigma} \underline{n}_{m\sigma} \frac{(l-m)(l-m-1)}{l(2l-1)(2l+1)}\\
  I_{0}=&P_{cl}^2 \sum_{m,\sigma} \underline{n}_{m\sigma} \frac{(l-m)(l+m)}{l(2l-1)(2l+1)}\\
  I_{1}=&P_{cl}^2 \sum_{m,\sigma} \underline{n}_{m\sigma} \frac{(l+m)(l+m-1)}{l(2l-1)(2l+1)}
 \end{split}\end{equation} 
These spectra can be recombined to give nice expectation values. The isotropic spectrum is $I_{iso}=I_{-1}+I_{0}+I_{1}$ and the $L_z$ spectrum is $I_{Lz}=I_{-1}-I_{1}$.
 \begin{equation}\begin{split}
  I_{Iso}=&P_{cl}^2 \sum_{m,\sigma} \underline{n}_{m\sigma} \frac{1}{(2l+1)}= P_{cl}^2\frac{\underline{n}}{2l+1}\\
  I_{L_z}=&P_{cl}^2 \sum_{m,\sigma} \underline{n}_{m\sigma} \frac{-m}{l(2l+1)}= P_{cl}^2\frac{\mathbf{L_z}}{l(2l+1)}
 \end{split}\end{equation}  
Because often only relative intensities can be measured, it is useful to define the normalized intensity $\mathcal{I}_q$ to be $\frac{I_q}{I_{-1}+I_0+I_1}$:
 \begin{equation}
  \mathcal{I}_{-1}-\mathcal{I}_{1}=\frac{1}{\underline{n}} \sum_{m,\sigma} \underline{n}_{m\sigma} \frac{-m}{l} =\frac{\mathbf{L_z}}{l\underline{n}}
 \end{equation}
This is the well known result, found by Thole \textit{et al.}\cite{Thole92}.

For linear polarized light one wants to compare x, y and z polarized light. The derivation of the sum-rules for x and y polarized light is very similar to the derivation for left, right and z polarized light. The dipole operator for x polarized light is $\sqrt{\frac{1}{2}}(C_{-1}^{(1)}-C_{1}^{(1)})$ and for y polarized light ${\rm{i}}\sqrt{\frac{1}{2}}(C_{-1}^{(1)}+C_{1}^{(1)})$. Using the formulas derived before one can write
 \begin{equation}\begin{split}
  I_{x,y}=&|\langle R_{i}|r|R_{f}\rangle |^2
  \sum_{m,m',m'',\sigma} 
  l_{m\sigma}l^{\dagger}_{m''\sigma}\\
&  \langle l_{m\sigma}|\sqrt{\frac{1}{2}}(C_{-1}^{(1)}\mp C_{1}^{(1)})|c_{m'\sigma}\rangle \langle l_{m''\sigma}|\sqrt{\frac{1}{2}}(C_{-1}^{(1)}\mp C_{1}^{(1)})|c_{m'\sigma}\rangle
 \end{split}\end{equation}
Writing again the integrals over 3 spherical harmonics as 3J symbols and absorbing the $l,c$, and $r$ dependence in a constant $P_{cl}^2$ one finds:
 \begin{equation}\begin{split}
  I_{x,y}=&\frac{1}{2}(I_{-1}+I_{1})\mp P_{cl}^2 \sum_{m,m',m'',\sigma} (-1)^{(m+m'')} l_{m\sigma}l^{\dagger}_{m''\sigma} \\
&  \frac{1}{2}  
  \left(
  \left(
   \begin{array}{ccc}
     l & 1 & c \\
     -m & 1 & m' \\
   \end{array}
  \right)
  \left(
   \begin{array}{ccc}
     l & 1 & c \\
     -m'' & -1 & m' \\
   \end{array}
  \right)+  \right.\\
&\left. \qquad \left(
   \begin{array}{ccc}
     l & 1 & c \\
     -m & -1 & m' \\
   \end{array}
  \right)
  \left(
   \begin{array}{ccc}
     l & 1 & c \\
     -m'' & 1 & m' \\
   \end{array}
  \right)
  \right)
 \end{split}\end{equation} 
Using the triangular equations and using conservation of magnetic orbital momentum one finds for the first term of two 3J symbols $m'=m-1, m''=m'-1=m-2$ and for the second term of two 3J symbols $m'=m+1, m''=m'+1=m+2$. Furthermore we will take $c=l-1$:
 \begin{equation}\begin{split}
  I_{x,y}=&\frac{1}{2}(I_{-1}+I_{1})\mp P_{cl}^2 \sum_{m,\sigma}\\
&  \frac{1}{2}  
  \left(
  \left(
   \begin{array}{ccc}
     l & 1 & l-1 \\
     -m & 1 & m-1 \\
   \end{array}
  \right)
  \left(
   \begin{array}{ccc}
     l & 1 & l-1 \\
     -m+2 & -1 & m-1 \\
   \end{array}
  \right)l_{m\sigma}l^{\dagger}_{m-2\sigma}+  \right.\\
&  \left.\qquad\left(
   \begin{array}{ccc}
     l & 1 & l-1 \\
     -m & -1 & m+1 \\
   \end{array}
  \right)
  \left(
   \begin{array}{ccc}
     l & 1 & l-1 \\
     -m-2 & 1 & m+1 \\
   \end{array}
  \right)l_{m\sigma}l^{\dagger}_{m+2\sigma}
  \right)
 \end{split}\end{equation}
One sees that the 3J symbols only depend on two variables ($l$ and $m$) and therefore can be expressed as simple polynomial functions:
 \begin{equation}\begin{split}
  I_{x,y}=&\frac{1}{2}(I_{-1}+I_{1})\mp P_{cl}^2 \sum_{m,\sigma}  \\
 & \frac{1}{2} \left( \frac{\sqrt{1+l-m}\sqrt{2+l-m}\sqrt{l+m-1}\sqrt{l+m}}{2l(2l+1)(2l-1)} l_{m\sigma}l^{\dagger}_{m-2\sigma}+ \right.\\
 & \left. \qquad\frac{\sqrt{l-m-1}\sqrt{l-m}\sqrt{l+m+1}\sqrt{l+m+2}}{2l(2l+1)(2l-1)} l_{m\sigma}l^{\dagger}_{m+2\sigma} \right)
 \end{split}\end{equation}
Changing $m$ to $m+2$ in the first term gives:
 \begin{equation}\begin{split}
  I_{x,y}=P_{cl}^2 &\sum_{m,\sigma} \left(\frac{l^2-l+m^2}{2l(2l-1)(2l+1)}l_{m\sigma}l_{m\sigma}^{\dagger}\mp\right.\\
 & \left.\frac{1}{2} \frac{\sqrt{l-m-1}\sqrt{l-m}\sqrt{l+m+1}\sqrt{l+m+2}}{2l(2l-1)(2l+1)}
(l_{m+2\sigma}l_{m\sigma}^{\dagger}+l_{m\sigma}l_{m+2\sigma}^{\dagger})\right)
 \end{split}\end{equation}
These equations are not very intuitive. This becomes better when one changes to real wave functions. For $l=1$ the relation between spherical harmonics and real wave-functions is $p_z=l_0$, $p_x=\sqrt{\frac{1}{2}}(l_{-1}-l_1)$, and $p_y=\sqrt{\frac{1}{2}}{\rm{i}}(l_{-1}+l_{1})$. We will write $\underline{n}_x$ for $p_xp_x^{\dagger}$. Now realizing that,
 \begin{equation}
  n_x=\frac{1}{2}(n_{-1}+n_{1})-\frac{1}{2}(l_{-1}l_{1}^{\dagger}+l_{1}l_{-1}^{\dagger})
 \end{equation}
we end up with:
 \begin{equation}\begin{split}
  \mathcal{I}_{x}=\frac{\underline{n}_{x}}{\underline{n}}\\
  \mathcal{I}_{y}=\frac{\underline{n}_{y}}{\underline{n}}\\
  \mathcal{I}_{z}=\frac{\underline{n}_{z}}{\underline{n}}
 \end{split}\end{equation}
We used again that $\mathcal{I}_q=\frac{I_{q}}{I_x+I_y+I_z}$. The same thing can be done for $d$ electrons. One should remember that the relations between the real wave functions and the spherical harmonics are: $d_{yz}=\sqrt{\frac{1}{2}}i(d_{-1}+d_{1})$, $d_{xz}=\sqrt{\frac{1}{2}}(d_{-1}-d_{1})$, $d_{xy}=\sqrt{\frac{1}{2}}i(d_{-2}-d_{2})$, $d_{x^2-y^2}=\sqrt{\frac{1}{2}}(d_{-2}+d_{2})$, and $d_{z^2}=d_{0}$. Knowing these relations one can relatively light simplify the sum-rules:
 \begin{equation}\begin{split}
  \mathcal{I}_{x}=&\frac{1}{\underline{n}}(\frac{1}{2}\underline{n}_{xy}+ \frac{1}{2}\underline{n}_{xz}+ \frac{1}{6}\underline{n}_{z^2}+ \frac{1}{2}\underline{n}_{x^2-y^2}- \sqrt{\frac{1}{12}}(d_{z^2}d_{x^2+y^2}^{\dagger}+d_{z^2}^{\dagger}d_{x^2+y^2}))\\
  \mathcal{I}_{y}=&\frac{1}{\underline{n}}(\frac{1}{2}\underline{n}_{xy}+ \frac{1}{2}\underline{n}_{yz}+\frac{1}{6}\underline{n}_{z^2}+ \frac{1}{2}\underline{n}_{x^2-y^2}+ \sqrt{\frac{1}{12}}(d_{z^2}d_{x^2+y^2}^{\dagger}+d_{z^2}^{\dagger}d_{x^2+y^2}))\\
  \mathcal{I}_{z}=&\frac{1}{\underline{n}}(\frac{1}{2}\underline{n}_{xz}+ \frac{1}{2}\underline{n}_{yz}+ \frac{2}{3}\underline{n}_{z^2})
 \end{split}\end{equation}
This can be written down more symmetric when realizing that $\underline{n}_{x^2}=\frac{1}{4}\underline{n}_{z^2}+ \frac{3}{4}\underline{n}_{x^2-y^2}- \sqrt{\frac{3}{16}}(d_{z^2}d_{x^2+y^2}^{\dagger}+d_{z^2}^{\dagger}d_{x^2+y^2}))$ and that $\underline{n}_{y^2}=\frac{1}{4}\underline{n}_{z^2}+ \frac{3}{4}\underline{n}_{x^2-y^2}+ \sqrt{\frac{3}{16}}(d_{z^2}d_{x^2+y^2}^{\dagger}+d_{z^2}^{\dagger}d_{x^2+y^2}))$.
 \begin{equation}\begin{split}
  \mathcal{I}_{x}=\frac{1}{\underline{n}}(\frac{1}{2}\underline{n}_{xy}+ \frac{1}{2}\underline{n}_{xz}+ \frac{2}{3}\underline{n}_{x^2})\\
  \mathcal{I}_{y}=\frac{1}{\underline{n}}(\frac{1}{2}\underline{n}_{xy}+ \frac{1}{2}\underline{n}_{yz}+ \frac{2}{3}\underline{n}_{y^2})\\
  \mathcal{I}_{z}=\frac{1}{\underline{n}}(\frac{1}{2}\underline{n}_{xz}+ \frac{1}{2}\underline{n}_{yz}+ \frac{2}{3}\underline{n}_{z^2})
 \end{split}\end{equation}

Not long after the publication of the first paper by Thole \textit{et al.} \cite{Thole92}, a second paper was published in which the normalized integrated intensity of the $L_3$ ($M_5$) edge minus the normalized integrated intensity of the $L_2$ ($M_4$) edge is related to the expectation value of $\frac{2}{3\underline{n}}\mathbf{S_z}+\frac{2l+3}{3l\underline{n}}\mathbf{T_z}$\cite{Carra93a}. In order to obtain this result one can use a similar derivation as shown above. For the $S_z$ sum-rule one has to assume, however, that $j_{c}$ is a good quantum number. Although this might sound reasonable, there are a large class of systems whereby $j$ of the core-hole is not a good quantum number. For the rare earths errors up to 230\% can exist \cite{Teramura96}. As shown before, the XAS for $L$ and $M$ edges consists of strong excitons. In other words, there is a strong interaction between the core hole and the valence electrons. This interaction mixes states with different $j$ of the core hole. It has been shown by Y. Teramura, A. Tanaka, B. T. Thole, and T. Jo\cite{Teramura96} that the shell specific sum-rules can not be used for $3d$ and $4f$ systems with less than half filled shells, since extremely large, low energy parameter dependent errors do occur.

In order to show the derivation of the sum-rules where integration over the spin-orbit spit edges is taken separately we start again by writing the dipole operator in second quantization. Now however, we do not use $m$ and $\sigma$ as quantum numbers for the $c$ shell, but $j$ and $m_j$, in the same way as done by P. Carra \textit{et al.} \cite{Carra93a}:
\begin{equation}
rC_q^{(1)}=r \sum_{m_jm_l\sigma} c_{jm_j}l_{m_l\sigma}^{\dagger} \langle jm_j|C_q^{(1)}|m_l\sigma\rangle
\end{equation}
The spectral intensity for a single final state $\Psi_f^j$ can now be written as:
 \begin{equation}\begin{split}
  I_{q}^j=&|\langle R_{i}|r|R_{f}\rangle |^2\\
&  \sum_{m_l,..\sigma'} 
  \langle\Psi_i|l_{m_l\sigma}c^{\dagger}_{jm_j}|\Psi_f^j\rangle
  \langle\Psi_f^j|c_{jm_j'}l^{\dagger}_{m_l'\sigma'}|\Psi_i\rangle\\
& \qquad\qquad  \langle l_{m_l\sigma}|C_q^{(1)}|c_{jm_j}\rangle\langle l_{m_l'\sigma'}|C_q^{(1)}|c_{jm_j'}\rangle
 \end{split}\end{equation}
Next, one has to sum over the final-state wave functions $\Psi_f^j$. Using the closure relation will remove the final-states in the expression for the integrated intensity. After that the expectation value of $c_{jm_j'}^{\dagger}c_{jm_j'''}$ has to be taken over the initial state. Since the initial state has a filled $c$ shell, $c_{jm_j'}^{\dagger}c_{jm_j'''}$ reduces to $\delta_{m_j',m_j'''}$. Doing all this at once, since it is merely a repetition of the steps done previously for integrating over both edges, one finds:
 \begin{equation}\begin{split}
  I_{q}^j=&|\langle R_{i}|r|R_{f}\rangle |^2\\
&  \sum_{m_l,m_l',\sigma,\sigma',m_j} 
  l_{m_l\sigma}l^{\dagger}_{m_l'\sigma'}
  \langle l_{m_l\sigma}|C_q^{(1)}|c_{jm_j}\rangle\langle l_{m_l'\sigma'}|C_q^{(1)}|c_{jm_j}\rangle
 \end{split}\end{equation}
In order to continue, one needs to express the wave functions $c_{jm_j}$ as linear combinations of $c_{m_c\sigma}$. To do so the following equation exists:
 \begin{equation}\begin{split}
  c_{jm_j}=\sum_{m_l\sigma}(-1)^{c-\frac{1}{2}+m_j}\sqrt{(2j+1)}\\
  \left(
   \begin{array}{ccc}
     c & \frac{1}{2} & j \\
     m_c & \sigma & m_j \\
   \end{array}
  \right)c_{m_c\sigma}
 \end{split}\end{equation}
Changing the basis of the $c$ shell from $jm_j$ orbitals to $m_c\sigma$ orbitals, writing the integrals over 3 spherical harmonics as 3J symbols, and absorbing the $c$, $l$ and $r$ dependence in a constant $P_{cl}^2$ gives:
 \begin{equation}\begin{split}
 I_{q}^j=P_{cl}^2\sum_{m_c,m_l,\sigma,m_c',m_l',\sigma',m_j}&(2j+1)l_{m_l\sigma}l_{m_l'\sigma'}^{\dagger}\\
 & \left(
   \begin{array}{ccc}
     c & \frac{1}{2} & j \\
     m_c & \sigma & -m_j \\
   \end{array}
  \right)
  \left(
   \begin{array}{ccc}
     c & 1 & l \\
     -m_c & q & m_l \\
   \end{array}
  \right)\\
& \left(
   \begin{array}{ccc}
     c & \frac{1}{2} & j \\
     m_c' & \sigma' & -m_j \\
   \end{array}
  \right)
    \left(
   \begin{array}{ccc}
     c & 1 & l \\
     -m_c' & q & m_l' \\
   \end{array}
  \right)
 \end{split}\end{equation}
From the triangular equations and from the conservation of magnetic angular momentum one finds that $m_j=m_c+\sigma$, $m_j=m_c'+\sigma'$, $m_c=q+m_l$, and $m_c'=q+m_l'$. Which means that $m_l+\sigma=m_l'+\sigma'$. Replacing $m_j$, $m_c$, and $m_c'$ with the use of these previous functions one ends up with:
 \begin{equation}\begin{split}
 I_{q}^j=P_{cl}^2\sum_{m_l,\sigma,m_l',\sigma'}&(2j+1)l_{m_l\sigma}l_{m_l'\sigma'}^{\dagger}\\
 & \left(
   \begin{array}{ccc}
     c & \frac{1}{2} & j \\
     q+m_l & \sigma & -(q+m_l+\sigma) \\
   \end{array}
  \right)
    \left(
   \begin{array}{ccc}
     c & 1 & l \\
     -(q+m_l) & q & m_l \\
   \end{array}
  \right)\\
&   \left(
   \begin{array}{ccc}
     c & \frac{1}{2} & j \\
     q+m_l' & \sigma' & -(q+m_l'+\sigma') \\
   \end{array}
  \right)
    \left(
   \begin{array}{ccc}
     c & 1 & l \\
     -(q+m_l') & q & m_l' \\
   \end{array}
  \right)
 \end{split}\end{equation}
Taking $c=l-1$ and $j=c\pm1$, one has for every $q$, $\sigma$ and $\sigma'$ a 3J symbol with no more than 2 unknowns. These are simple polynomial expressions. With some algebra the different expressions for $\sigma$, $\sigma'$, and $j$ can be combined to give the following result:
\begin{equation}\begin{split}
I_{q=-1}^j=P_{cl}^2\left(\vphantom{\sum_i}\right.&\sum_{m\sigma}\frac{(l+m-1)(l+m)(\frac{1}{2}+j+4(1+j-l)(m-1)\sigma)}{2(1-2l)^2l(1+2l)}l_{m\sigma}l_{m\sigma}^{\dagger}+\\
&\sum_m\left(\vphantom{\frac{l}{L}}\right.(-1)^{j-l-\frac{1}{2}}\frac{(l+m-1)(l+m)}{2(1-2l)^2l(1+2l)}\\
&\qquad\qquad\sqrt{l+m+1}\sqrt{l-m}(l_{m\frac{1}{2}}l_{m+1-\frac{1}{2}}^{\dagger}+l_{m+1-\frac{1}{2}}l_{m\frac{1}{2}}^{\dagger})\left.\vphantom{\frac{l}{l}}\right)\left.\vphantom{\sum_i}\right)\\
I_{q=0}^j=P_{cl}^2\left(\vphantom{\sum_i}\right.&\sum_{m\sigma}\frac{(l+m)(l-m)(\frac{1}{2}+j+4(1+j-l)(m)\sigma)}{(1-2l)^2l(1+2l)}l_{m\sigma}l_{m\sigma}^{\dagger}+\\
&\sum_m\left(\vphantom{\frac{l}{L}}\right.(-1)^{j-l-\frac{1}{2}}\frac{(l-m-1)(l+m)}{(1-2l)^2l(1+2l)}\\
&\qquad\qquad\sqrt{l+m+1}\sqrt{l-m}(l_{m\frac{1}{2}}l_{m+1-\frac{1}{2}}^{\dagger}+l_{m+1-\frac{1}{2}}l_{m\frac{1}{2}}^{\dagger})\left.\vphantom{\frac{l}{l}}\right)\left.\vphantom{\sum_i}\right)\\
I_{q=1}^j=P_{cl}^2\left(\vphantom{\sum_i}\right.&\sum_{m\sigma}\frac{(l-m)(l-m-1)(\frac{1}{2}+j+4(1+j-l)(m+1)\sigma)}{2(1-2l)^2l(1+2l)}l_{m\sigma}l_{m\sigma}^{\dagger}+\\
&\sum_m\left(\vphantom{\frac{l}{l}}\right.(-1)^{j-l-\frac{1}{2}}\frac{(l-m-1)(l-m-2)}{2(1-2l)^2l(1+2l)}\\
&\qquad\qquad\sqrt{l+m+1}\sqrt{l-m}(l_{m\frac{1}{2}}l_{m+1-\frac{1}{2}}^{\dagger}+l_{m+1-\frac{1}{2}}l_{m\frac{1}{2}}^{\dagger})\left.\vphantom{\frac{l}{l}}\right)\left.\vphantom{\sum_i}\right)
\end{split}\end{equation}
We dropped the $l$ as subscript from the $m$, because it is clear that we are talking about the magnetic orbital momentum. In order to end up with some simple expectation values of known operators one has to use difference spectra. P. Carra \textit{et al.} \cite{Carra93a, Carra93b} showed that it is useful to take the same linear combinations as for the $L_{z}$ sum-rule, but now one has to take $I_q^{c+\frac{1}{2}}-\frac{l}{l-1}I_q^{c-\frac{1}{2}}$ for each polarization.

For the isotropic branching ratio one finds:
\begin{equation}\begin{split}
(\mathcal{I}_{-1}^{c+\frac{1}{2}}&+\mathcal{I}_{0}^{c+\frac{1}{2}}+\mathcal{I}_{1}^{c+\frac{1}{2}})- \frac{l}{l-1}(\mathcal{I}_{-1}^{c-\frac{1}{2}}+\mathcal{I}_{0}^{c-\frac{1}{2}}+\mathcal{I}_{1}^{c-\frac{1}{2}})=\\
\frac{1}{l\underline{n}}&\left(\sum_{m\sigma}2m\sigma l_{m\sigma}l_{m\sigma}^{\dagger}- \sum_m\sqrt{l+m+1}\sqrt{l-m}(l_{m\frac{1}{2}}l_{m+1-\frac{1}{2}}^{\dagger}+l_{m+1-\frac{1}{2}}l_{m\frac{1}{2}}^{\dagger})\right)
\end{split}\end{equation}
If one now realizes that $\sqrt{l+m+1}\sqrt{l-m}(l_{m\frac{1}{2}}l_{m+1-\frac{1}{2}}^{\dagger}+l_{m+1-\frac{1}{2}}l_{m\frac{1}{2}}^{\dagger})$ can be written as $l^+s^-+l^-s^+$ and that $m\sigma l_{m\sigma}l_{m\sigma}^{\dagger}$ is equal to $-l_zs_z$ one arrives at:
\begin{equation}\begin{split}
(\mathcal{I}_{-1}^{c+\frac{1}{2}}+\mathcal{I}_{0}^{c+\frac{1}{2}}+\mathcal{I}_{1}^{c+\frac{1}{2}})- \frac{l}{l-1}(\mathcal{I}_{-1}^{c-\frac{1}{2}}+\mathcal{I}_{0}^{c-\frac{1}{2}}+\mathcal{I}_{1}^{c-\frac{1}{2}})=
-2\sum_{i}\frac{\mathbf{l}_i\centerdot \mathbf{s}_i}{l\underline{n}}
\end{split}\end{equation}
We made the operator $\mathbf{l}$ bold and the quantum number $l$ normal font in order to distinguish between quantum numbers and operators. The sum over i runs over all electrons. This is equal to a rather old result of Thole and van der Laan \cite{Thole88a}.

For circular polarized light one finds:
\begin{equation}\begin{split}
(\mathcal{I}_{-1}^{c+\frac{1}{2}}-\mathcal{I}_{1}^{c+\frac{1}{2}})- \frac{l}{l-1}&(\mathcal{I}_{-1}^{c-\frac{1}{2}}-\mathcal{I}_{1}^{c-\frac{1}{2}})=\\
&\sum_{m\sigma}\frac{2(l-2m^2)}{l(2l-1)\underline{n}}\sigma l_{m\sigma}l_{m\sigma}^{\dagger}+\\
\sum_{m}\frac{2m+1}{l(2l-1)\underline{n}}&\sqrt{l+m+1}\sqrt{l-m}(l_{m\frac{1}{2}}l_{m+1-\frac{1}{2}}^{\dagger}+l_{m+1-\frac{1}{2}}l_{m\frac{1}{2}}^{\dagger})
\end{split}\end{equation}
In order to simplify this, Carra \textit{et al.} \cite{Carey93a} used the magnetic dipole operator $\mathbf{T}$. $\mathbf{T}$ is defined as $\sum_{i}\mathbf{s}_i-3\hat{r}_i(\hat{r}_i\centerdot \mathbf{s}_i)$. One only needs the value of $\mathbf{T_z}$. One needs to write $\mathbf{T_z}$ in second quantization, in order to compare $\mathbf{T_z}$ with the calculations sofar. To do so we will first write $\mathbf{T_z}$ in spherical harmonics. $\mathbf{T_z}=\sum_i (1-3\hat{z}^2)\mathbf{s}_z^i-3\hat{z}\hat{x}\mathbf{s}_x^i-3\hat{z}\hat{y}\mathbf{s}_y^i= \sum_i -2C_0^{(2)} \mathbf{s}_z^i-\frac{1}{2}\sqrt{6}(-C_1^{(2)}\mathbf{s}^-_i+C_{-1}^{(2)}\mathbf{s}^+_i)$. By taking the expectation values over the spherical harmonics, writing them as 3J symbols and simplifying these to polynomial expressions, one can write $\mathbf{T}_z$ as $\sum_{m,\sigma} \frac{2(l^2+l-3m^2)}{(2l+3)(2l-1)}\sigma l_{m\sigma}l^{\dagger}_{m\sigma}+\sum_m \frac{3}{2}\frac{(1+2m)\sqrt{l-m}\sqrt{l+m+1}}{(2l+3)(2l-1)}(l_{m\frac{1}{2}}l^{\dagger}_{m+1-\frac{1}{2}}+l_{m+1-\frac{1}{2}}l^{\dagger}_{m\frac{1}{2}})$. Substituting the expression for $\mathbf{T}_z$ into the sum-rule for circular polarized light gives:
\begin{equation}\begin{split}
(\mathcal{I}_{-1}^{c+\frac{1}{2}}-\mathcal{I}_{1}^{c+\frac{1}{2}})- \frac{l}{l-1}(\mathcal{I}_{-1}^{c-\frac{1}{2}}-\mathcal{I}_{1}^{c-\frac{1}{2}})=
\frac{2}{3\underline{n}}\mathbf{S}_z+\frac{2(2l+3)}{3l\underline{n}}\mathbf{T}_z
\end{split}\end{equation}
Which is equal to the result found in the original publication of Carra, Thole, Altarelli, and Wang\cite{Carra93a}.

\section{Summery}

In summery, B. T. Thole, P. Carra, G. van der Laan \textit{et al.} showed that there are sum-rules for the isotropic spectrum \cite{Thole88a}, for linear polarized light \cite{Carra93b} and for circular polarized light \cite{Thole92, Carra93a}. One can integrate over both edges or compare the ratio of the $L_2$($M_4$) and $L_3$($M_5$) edge. Since normally only relative intensities are measured all spectra are divided by the isotropic intensity. For circular polarized light there are the two following sum-rules:
\begin{equation}\begin{split}
\mathcal{I}_{-1}-\mathcal{I}_1=&\frac{\mathbf{L_z}}{l\underline{n}}\\
(\mathcal{I}_{-1}^{c+\frac{1}{2}}-\mathcal{I}_{1}^{c+\frac{1}{2}})- \frac{l}{l-1}(\mathcal{I}_{-1}^{c-\frac{1}{2}}-\mathcal{I}_{1}^{c-\frac{1}{2}})=&
\frac{2}{3\underline{n}}\mathbf{S}_z+\frac{2(2l+3)}{3l\underline{n}}\mathbf{T}_z
\end{split}\end{equation}
For the isotropic branching ratio there exists:
\begin{equation}\begin{split}
(\mathcal{I}_{-1}^{c+\frac{1}{2}}+\mathcal{I}_{0}^{c+\frac{1}{2}}+\mathcal{I}_{1}^{c+\frac{1}{2}})- \frac{l}{l-1}(\mathcal{I}_{-1}^{c-\frac{1}{2}}+\mathcal{I}_{0}^{c-\frac{1}{2}}+\mathcal{I}_{1}^{c-\frac{1}{2}})=
-2\sum_{i}\frac{\mathbf{l}_i\centerdot \mathbf{s}_i}{l\underline{n}}\\
\end{split}\end{equation}
For linear polarized light one can write the sum rules in real orbitals and write different formulas for $l=p$ or $d$.
\begin{equation}\begin{split}
l=p:  &\mathcal{I}_{x}=\frac{\underline{n}_{x}}{\underline{n}}\\
      &\mathcal{I}_{y}=\frac{\underline{n}_{y}}{\underline{n}}\\
      &\mathcal{I}_{z}=\frac{\underline{n}_{z}}{\underline{n}}\\
l=d:  &\mathcal{I}_{x}=\frac{1}{\underline{n}}(\frac{1}{2}\underline{n}_{xy}+ \frac{1}{2}\underline{n}_{xz}+ \frac{2}{3}\underline{n}_{x^2})\\
      &\mathcal{I}_{y}=\frac{1}{\underline{n}}(\frac{1}{2}\underline{n}_{xy}+ \frac{1}{2}\underline{n}_{yz}+ \frac{2}{3}\underline{n}_{y^2})\\
      &\mathcal{I}_{z}=\frac{1}{\underline{n}}(\frac{1}{2}\underline{n}_{xz}+ \frac{1}{2}\underline{n}_{yz}+ \frac{2}{3}\underline{n}_{z^2})
\end{split}\end{equation}
We would like to warn again that for the derivation of the sum-rules, where one integrates separately over edges with different $j_c$ it is assumed that $j_c$ is a good quantum number. This is in general not the case due to excitonic effects! These problems do not occur when using the sum-rules where the integral is taken over both edges at once.

\chapter{Magnetic versus crystal field linear dichroism in NiO thin films \\ \textmd{Phys. Rev. B} \textbf{69}\textmd{, 020408 (2004)} }
\chaptermark{Magnetic versus crystal field linear dichroism in NiO thin films}

\label{ChapterNiO}

\begin{center}
\begin{minipage}{0.8\textwidth}
We have detected strong dichroism in the Ni $L_{2,3}$ x-ray absorption spectra
of monolayer NiO films. The dichroic signal appears to be very similar to the
magnetic linear dichroism observed for thicker antiferromagnetic NiO films. A
detailed experimental and theoretical analysis reveals, however, that the
dichroism is caused by crystal field effects in the monolayer films, which is a
non trivial effect because the high spin Ni $3d^{8}$ ground state is not split
by low symmetry crystal fields. We present a practical experimental method for
identifying the independent magnetic and crystal field contributions to the
linear dichroic signal in spectra of NiO films with arbitrary thicknesses and
lattice strains. Our findings are also directly relevant for high spin $3d^{5}$
and $3d^{3}$ systems such as LaFeO$_{3}$, Fe$_{2}$O$_{3}$, VO, LaCrO$_{3}$,
Cr$_{2}$O$_{3}$, and Mn$^{4+}$ manganate thin films.
\end{minipage}
\end{center}

Magnetic linear dichroism (MLD) in soft-x-ray absorption spectroscopy (XAS) has
recently developed into one of the most powerful tools to study the magnetic
properties of antiferromagnetic thin films
\cite{Thole85b,Sinkovic90,Kuiper93,Alders95,Alders98}. The contrast that one can
obtain as a result of differences in the magnitude and orientation of local
moments is essential to determine the spin anisotropy and important parameters
like the N\'{e}el temperature $(T_{N})$, as well as to map out spatially the
different magnetic domains that are present in antiferromagnetic films
\cite{Stohr98a,Spanke98,Stohr99,Scholl00,Nolting00,Ohldag01a,Ohldag01b,Zhu01,Hille01}.
Such information is extremely valuable for the research and application of
magnetic devices that make use of exchange-bias.

  \begin{SCfigure}[][h]
    \includegraphics[width=60mm]{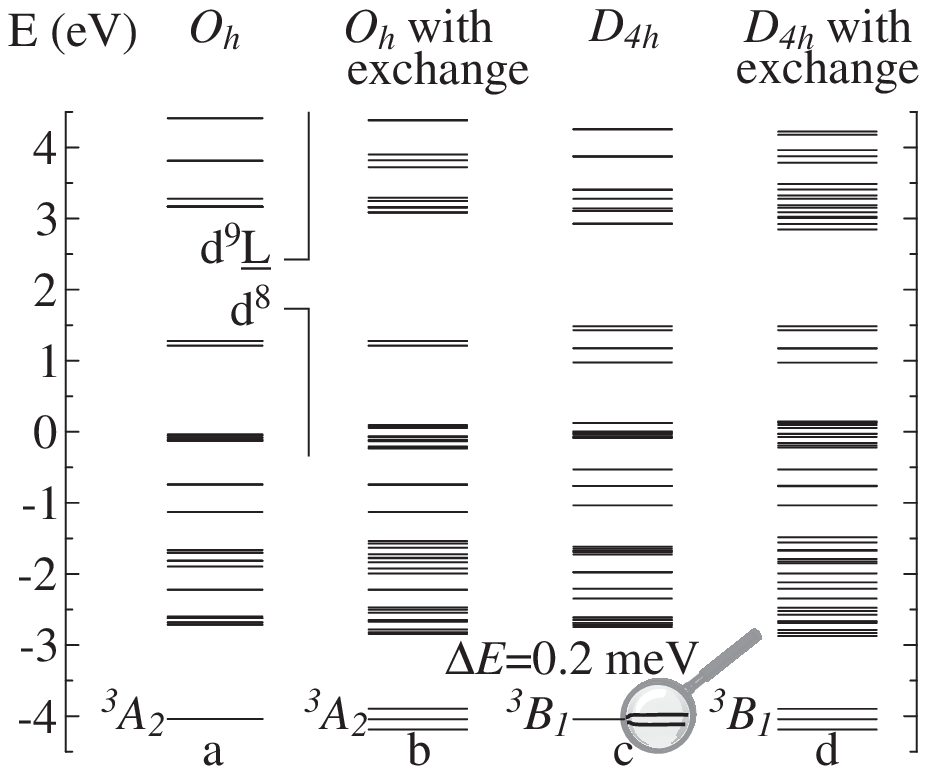}
    \caption{Energy level diagram for Ni$^{2+}$ ($3d^{8}$) in:
            (a) $O_{h}$ symmetry with $pd\sigma$=-1.29 and $10Dq$=0.85 eV;
            (b) $O_{h}$ symmetry with additional exchange field of 0.16 eV;
            (c) $D_{4h}$ symmetry with $pd\sigma$=-1.29, $10Dq$=0.85, and $Ds$=0.12 eV;
            (d) $D_{4h}$ symmetry with additional exchange.
            The $3d$ spin-orbit interaction is included, but the states are labelled
            as if the spin-orbit interaction was not present \newline
            }\label{NiOfig1}
  \end{SCfigure}

Much of the modern MLD work has been focussed on NiO and LaFeO$_{3}$ thin
films, and the observed dichroism has been attributed entirely to magnetic
effects
\cite{Stohr98a,Spanke98,Stohr99,Scholl00,Nolting00,Ohldag01a,Ohldag01b,Zhu01,Hille01}.
Other sources that could contribute to linear dichroism, however, such as
crystal fields of lower than octahedral symmetry, have been neglected or not
considered. Indeed, one would expect that such low symmetry crystal fields are
negligible for bulk-like NiO and LaFeO$_{3}$ films, and, more fundamentally,
that such crystal fields will not split the high-spin Ni $3d^{8}$ or Fe
$3d^{5}$ ground state. We have illustrated this insensitivity in Fig. 1 for the
Ni$^{2+}$ case, where the energy level diagram in an $O_{h}$ environment is
compared to that in a $D_{4h}$ point group symmetry \cite{SOremark}. In
contrast, an exchange field will split the Ni$^{2+}$ ground state into three
levels with $M_{S}$=-1,0,1 with an energy separation given by the exchange
coupling $J$, see Fig. \ref{NiOfig1}. The basis for obtaining strong dichroism in the Ni
$L_{2,3}$ ($2p$$\rightarrow$$3d$) absorption spectra is that dipole selection
rules dictate which of the quite different final states can be reach and with
what probability for each of the initial states. The isotropic spectrum of each
of these three states will be the same, but each state with a different
$|M_{S}|$ value will have a different polarization dependence
\cite{Thole85b,Sinkovic90,Kuiper93,Alders95,Alders98}. A completely analogous
argumentation can be given for the orbitally highly symmetric high spin
$3d^{5}$ and $3d^{3}$ cases, e.g. Mn$^{2+}$, Fe$^{3+}$, V$^{2+}$, Cr$^{3+}$,
Mn$^{4+}$.

In this paper we report on XAS measuments on single monolayer (ML) NiO films
which are grown on a Ag(100) substrate and capped by a 10 ML MgO(100) film. We
have observed strong linear dichroism in the Ni $L_{2,3}$ spectra, very similar
to that measured for thicker NiO films. From a detailed theoretical and
experimental analysis, however, we discovered that the dichroism can not be
attributed to the presence of some form of magnetic order, but entirely to
crystal field effects. The analysis provides us also with a practical guide of
how to disentangle quantitatively the individual contributions to the linear
dichroic signal, i.e. the contribution from magnetic interactions versus that
from low symmetry crystal fields. This is important for a reliable
determination of, for instance, the spin moment orientation in NiO as well as
LaFeO$_{3}$, Fe$_{2}$O$_{3}$, VO, LaCrO$_{3}$, Cr$_{2}$O$_{3}$, and Mn$^{4+}$
manganate ultra thin films, surfaces and strained films, where the low symmetry
crystal field splittings may not be negligible as compared to the exchange
field energies.

The polarization dependent XAS measurements were performed at the Dra\-gon
beamline of the National Synchrotron Radiation Research Center in Taiwan. The
spectra were recorded using the total electron yield method in an XAS chamber
with a base pressure of $3\times10^{-10}$ mbar. The photon energy resolution at the
Ni $L_{2,3}$ edges ($h\nu \approx 850-880$ eV) was set at 0.3 eV, and the
degree of linear polarization was $\approx 98 \%$. A NiO single crystal is
measured \textit{simultaneously} in a separate chamber upstream of the XAS
chamber in order to obtain a relative energy reference with an accuracy of
better than 0.02 eV. The 1 ML NiO film on Ag(100) was prepared in Groningen, by
using NO$_{2}$ assisted molecular beam epitaxy. Immediately after the NiO
growth, the sample was capped \textit{in-situ} with an epitaxial 10 ML MgO(100)
film. Reflection high energy electron diffraction (RHEED) intensity
oscillations recorded during growth of thicker films demonstrated the
layer-by-layer growth mode and provided an accurate thickness calibration
\cite{Altierithesis,Altieri99}.

  \begin{SCfigure}[][h]
    \includegraphics[width=60mm]{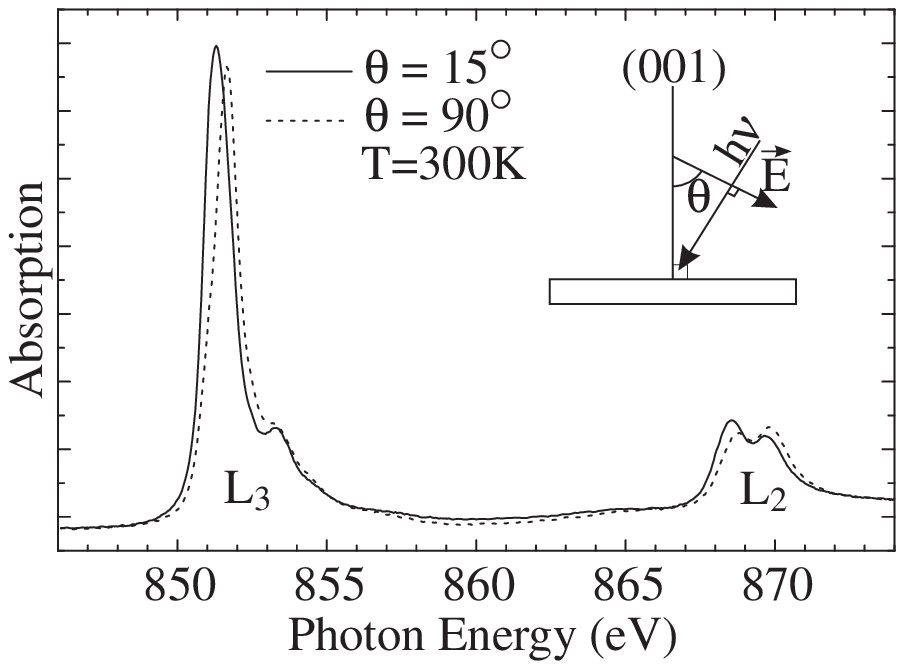}
    \caption{Experimental polarization dependent Ni $L_{2,3}$ XAS of 1 ML
             NiO(100) on Ag(100) covered with MgO(100). $\theta$ is the angle
             between the light polarization vector and the (001) surface normal
             ($\theta$=$90^{\circ}$ means normal light incidence). \newline\newline
    }\label{NiOfig2}
  \end{SCfigure}

Fig. \ref{NiOfig2} shows the polarization dependent Ni $L_{2,3}$ XAS spectra of the 1 ML
NiO film, taken at room temperature. The angle between the light polarization
vector and the (001) surface normal is given by $\theta$ ($\theta = 90^{\circ}$
means normal light incidence). The general lineshape of the spectra is very
similar to that of thicker NiO films and bulk NiO \cite{Alders98}.

  \begin{SCfigure}[][h]
    \includegraphics[width=60mm]{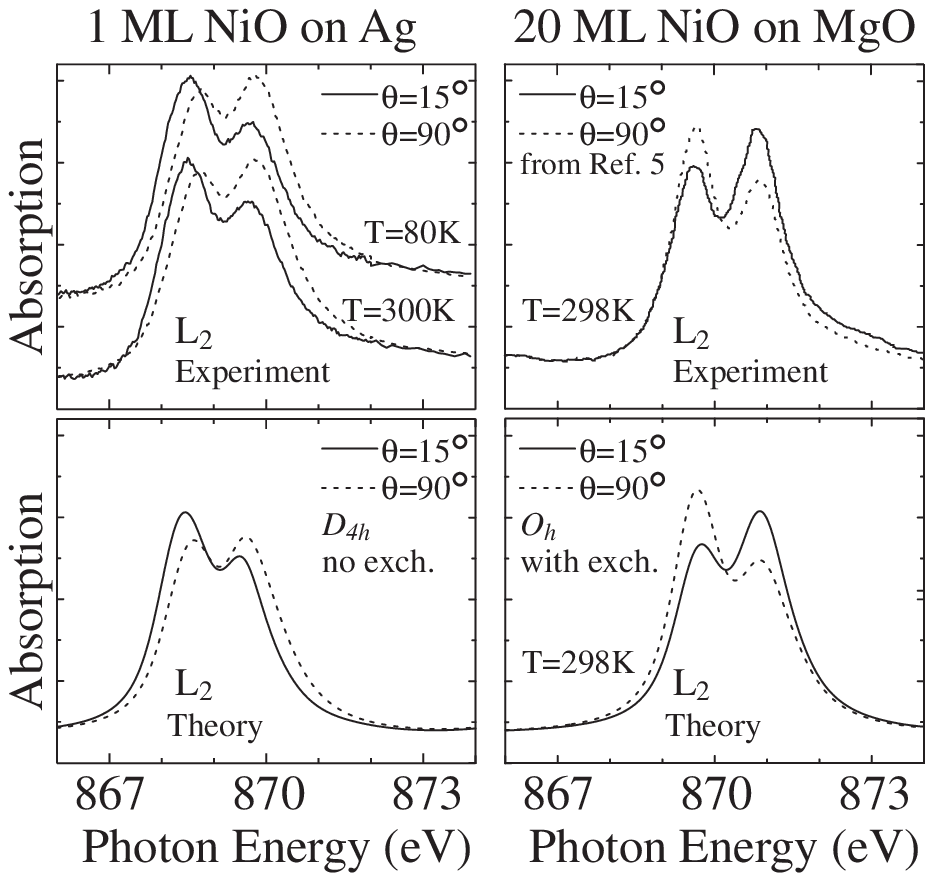}
    \label{NiOfig3}
    \caption{Polarization dependence of the Ni $L_{2}$ XAS of 1 ML NiO on
    Ag(100) covered by MgO(100). The 20 ML NiO on MgO spectra are taken from Ref. \cite{Alders98}.
    The theoretical spectra for the 1 ML NiO
    are calculated in $D_{4h}$ symmetry without
    exchange, and for the 20 ML in $O_{h}$ with exchange. \newline }
  \end{SCfigure}

Fig. \ref{NiOfig3} presents a close-up of the $L_{2}$ edge, the region most often used to
measure the magnitude of the magnetic linear dichroic effect in
antiferromagnetic NiO films
\cite{Stohr98a,Spanke98,Stohr99,Ohldag01a,Ohldag01b,Zhu01,Hille01}. The spectra
of the 1 ML NiO film show a very clear polarization dependence. This linear
dichroic effect is as strong as that for a 20 ML NiO film grown on MgO(100)
(taken from Alders \textit{et al.} \cite{Alders98}), albeit with an opposite
sign, as can be seen from Fig. \ref{NiOfig3}. Relying on the analysis by Alders \textit{et
al.} \cite{Alders98} for the antiferromagnetic 20 ML film, one may be tempted
to conclude directly that the spin orientation in the 1 ML film is quite
different to that of the 20 ML film, i.e. that the spins for the 1 ML would be
lying more parallel to the interface while those of the thicker films are
pointing more along the interface normal. However, Alders \textit{et al.}
\cite{Alders98} have also shown that the magnetic ordering temperature of NiO
films decreases strongly if the film is made thinner. In fact, for a 5 ML NiO
film on MgO(100), it was found that $T_{N}$ is around or below room
temperature, i.e. that no linear dichroism can be observed at room temperature.
A simple extrapolation will therefore suggest that 1 ML NiO will not be
magnetically ordered at room temperature. This is in fact supported by the 80 K
data of the 1 ML NiO on Ag(100) as shown in Figs. 2 and 3: the spectra and the
dichroism therein are identical to those at 300 K, indicating that $T_{N}$ must
be at least lower than 80 K.

  \begin{SCfigure}[][h]
    \includegraphics[width=60mm]{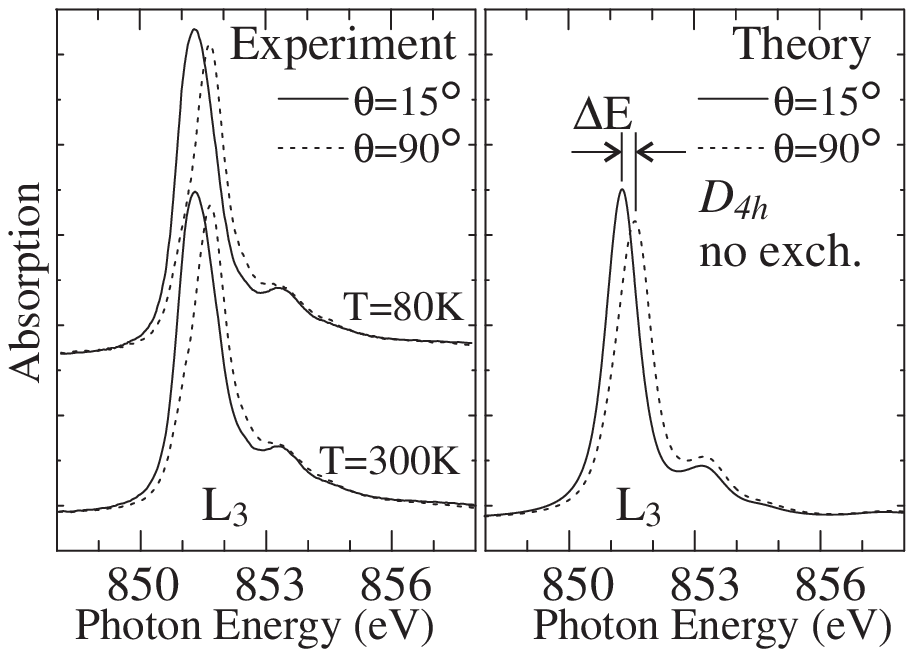}
    \caption{Theoretical and experimental polarization dependence of the Ni
    $L_{3}$-XAS of a 1 ML NiO(100) on Ag(100) covered with MgO(100). The
    theoretical spectra are calculated in $D_{4h}$ symmetry without
    exchange. \newline
    }\label{NiOfig4}
  \end{SCfigure}

In order to resolve the origin of the linear dichroism in the 1 ML NiO system,
we now resort to the Ni $L_{3}$ part of the spectrum. A close-up of this region
is given in Fig. \ref{NiOfig4}. We can easily observe that the strong polarization
dependence of the spectra is accompanied by an energy shift $\Delta E$ of 0.35
eV in the main peak of the $L_{3}$ white line. This shift seems small compared
to the 852 eV photon energy being used, but it is very reproducible and well
detectable since the photon energy calibration is done with an accuracy of
better than 0.02 eV thanks to the simultaneous measurement of a NiO single
crystal reference. We now take this energy shift as an indicator for the
presence and strength of local crystal fields with a symmetry lower than
$O_{h}$, i.e. crystal fields that do not split the ground state but do alter
the energies of the XAS final states, and, via second order processes, also
causes spectral weight to be transferred between the various peaks as we will
show below.

To understand the Ni $L_{2,3}$ spectra quantitatively, we perform calculations
for the atomic $2p^{6}3d^{8} \rightarrow 2p^{5}3d^{9}$ transitions using the
same method as described earlier by Alders \textit{et al.} \cite{Alders98}, but
now in a $D_{4h}$ point group symmetry. The method uses the full atomic
multiplet theory and includes the effects of the solid. It accounts for the
intra-atomic $3d$-$3d$ and $2p$-$3d$ Coulomb and exchange interactions, the
atomic $2p$ and $3d$ spin-orbit couplings, the O $2p$ - Ni $3d$ hybridization
with $pd\sigma$ = -1.29 eV, and an $O_{h}$ crystal field splitting of $10Dq$ =
0.85 eV. The local symmetry for the Ni ion sandwiched between the Ag(100)
substrate and the MgO(100) film is in principle $C_{4v}$, but for $d$ electrons
one can ignore the odd part of the crystal field, so that effectively one can
use the tetragonal $D_{4h}$ point group symmetry. As we will explain below, the
$D_{4h}$ parameters $Ds$ and $Dt$ \cite{Ballhausen62} are set to 0.12 and 0.00
eV, respectively, and the exchange field (the molecular field acting on the
spins) to zero. The calculations have been carried out using the XTLS 8.0
programm\cite{Tanaka94}.

The right panel of Fig. \ref{NiOfig4} shows the calculated $L_{3}$ spectrum for the light
polarization vector perpendicular and parallel to the $C_{4}$ axis ($\theta =
90^{\circ}$ and $\theta = 0^{\circ}$, respectively). One can clearly see that
the major experimental features are well reproduced, including the 0.35 eV
energy shift between the two polarizations. This shift can be understood in a
single electron picture. The ground state has the
$\underline{3d}_{x^{2}-y^{2}}\underline{3d}_{z^{2}}$ configuration, where the
underline denotes a hole. The final state has a
$2p^{5}\underline{3d}_{x^{2}-y^{2}}$ or $2p^{5}\underline{3d}_{z^{2}}$
configuration. For $z$ polarized light the $3d_{z^{2}}$ state can be reached,
but the $3d_{x^{2}-y^{2}}$ can not, and the final state will be of the form
$2p^{5}\underline{3d}_{x^{2}-y^{2}}$. For $x$ polarized light the final state
will be of the form $2p^{5}\underline{3d}_{z^{2}-y^{2}}$$=
$$\sqrt{3/4}(2p^{5}\underline{3d}_{z^{2}})$$+
$$\sqrt{1/4}(2p^{5}\underline{3d}_{x^{2}-y^{2}})$. In a pure ionic picture, the
$2p^{5}\underline{3d}_{x^{2}-y^{2}}$ state will be 4$Ds$+5$Dt$ lower in energy
than the $2p^{5}\underline{3d}_{z^{2}}$ state. In the presence of the O $2p$ -
Ni $3d$ hybridization, we find that $Ds$=0.12 and $Dt$=0.00 eV reproduce the
observed 0.35 eV shift.

Going back to the $L_{2}$ edge, we can see in Fig. 3 that the calculations can
also reproduce very well the observed linear dichroism in the 1 ML NiO spectra.
In fact, one now could also see the same 0.35 eV shift at this edge, although
it is not as clear as in the $L_{3}$ edge. We would like to stress here that
the good agreement has been achieved without the inclusion of an exchange
splitting, i.e. the dichroism is solely due to the low symmetry crystal field
splitting. It is a final state effect and the change in the ratio between the
two peaks of the $L_{2}$ edge as a function of polarization can be understood
as follows. In $O_{h}$ symmetry the first peak is due to two final states, one
of $T_{2}'$ and one of $E_{1}'$ symmetry. The second peak is due to a final
state of $T_{1}'$ symmetry. All three states have the
$2p^{5}\underline{3d}_{{e}_{g}}$ configuration. If one reduces the crystal
field to $D_{4h}$ symmetry the peaks will split. The $T_{2}'$ state will split
into two states of $B_{2}'$ and $E_{1}'$ symmetry, the $E_{1}'$ into $A_{1}'$
and $B_{1}'$, and the $T_{1}'$ into $A_{2}'$ and $E_{1}'$. The energy splitting
can be measured, but this is much easier done using the $L_{3}$ edge. We note
that each of the two peaks in the L$_{2}$ edge will have a state of $E_{1}'$
symmetry, so that these two will mix and transfer spectral weight. This can be
seen with isotropic light, but will show up more pronounced as a linear
dichroic effect if polarized light is used.

In contrast to the 1 ML NiO case, the good agreement between theory and
experiment for the polarization dependent spectra of a 20 ML NiO film
\cite{Alders98} have been achieved by assuming the presence of an
antiferromagnetic order with an exchange field of about 0.16 eV in a pure local
$O_h$ symmetry. It is surprising and also disturbing that a low symmetry
crystal field could induce a spectral weight transfer between the two peaks of
the $L_{2}$ white line such that the resulting linear dichroism appears to be
very similar as a dichroism of magnetic origin. It is obvious that the ratio
between the two peaks can not be taken as a direct measure of the spin
orientation or magnitude of the exchange field in NiO films \cite{Alders98} if
one has not first established what the crystal field contribution could be.

  \begin{SCfigure}
    \includegraphics[width=60mm]{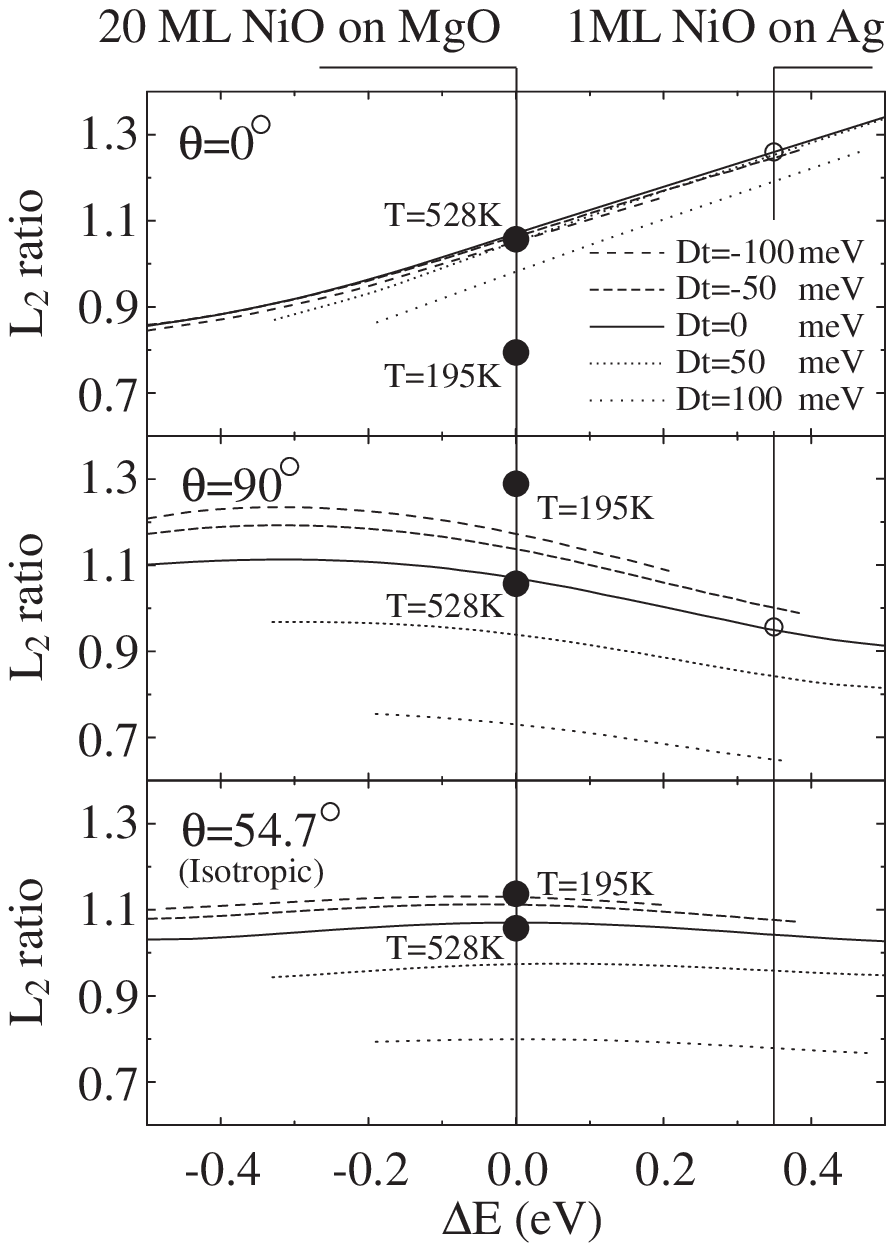}
    \caption{Calculated ratio of the two peaks in the $L_{2}$ edge as a
     function of $\Delta E$, which is the shift in energy of the $L_{3}$
     main peak in going from normal ($\theta = 90^{\circ}$) to grazing
     ($\theta = 0^{\circ}$) incidence of the linearly polarized light.
     The $L_{2}$ ratio is calculated for $\theta = 90^{\circ}$,
     $\theta = 0^{\circ}$, and for the isotropic spectrum. \newline
     }\label{NiOfig5}
  \end{SCfigure}

We now can identify two strategies for finding out which part of the linear
dichroism is due to low symmetry crystal field effects. The first one is to
study the temperature dependence as we have done above. Here we have made use
of the fact that those crystal fields do not split the high spin ground state,
so that there are no additional states to be occupied with different
temperatures other than those already created by the presence of exchange
fields. Thus there should not be any temperature dependence in the crystal
field dichroism. The linear dichroism due to magnetism, however, is temperature
dependent and scales with $<$$M^2$$>$. By going to temperatures high enough
such that there is no longer any temperature dependence in the linear
dichroism, i.e. when all magnetic ordering has been destroyed, one will find
the pure crystal field induced dichroism.

The second strategy to determine the low symmetry crystal field contribution is
to measure carefully the energy shift $\Delta E$ in the main peak of the Ni
$L_{3}$ white line for $\theta$=$0^{\circ}$ vs. $\theta$=$90^{\circ}$. We now
calculate the ratio between the two peaks of the Ni $L_{2}$ edge as a function
of $\Delta E$, and the results are plotted in Fig. \ref{NiOfig5} for $\theta =90^{\circ}$,
$\theta =0^{\circ}$, and $\theta =54.7^{\circ}$ (isotropic spectrum). Since
$\Delta E$ is a function of $Ds$ and $Dt$ combined, we have carried out the
calculations with $Ds$ as a running variable for several fixed values of $Dt$,
and plotted the resulting $L_{2}$ ratios vs. $\Delta E$. We now can use Fig. 5
as a road map to determine how much of the linear dichroism in the Ni $L_{2}$
edge is due to crystal field effects and how much due to magnetism. We can see
directly that the 1 ML data lie on curves with the same $Dt$, meaning that the
measured $L_{2}$ ratios are entirely due to crystal fields. The same can also
be said for the 20 ML NiO on MgO at 528 K, which is not surprising since this
temperature is above $T_{N}$. However, for the 20 ML NiO at 195 K, one can see
that the data points do not lie on one of the $Dt$ curves (one may look for
larger $Dt$ curves, but this results in lineshapes very different from
experiment) or let alone on curves with the same $Dt$, indicating that one need
magnetism to explain the $L_{2}$ ratios. In other words, knowing the $L_{2}$
ratio and $\Delta E$ together allows us to determine the magnitude of the
exchange interaction and the orientation of the spin moments. It is best to use
the $\theta = 0^{\circ}$ spectra since here the $L_{2}$ ratio is determined
almost by $\Delta E$ alone and is not too sensitive to the individual values of
$Ds$ and $Dt$.

To conclude, we have observed strong linear dichroism in the 1 ML NiO on Ag,
very similar to the well known magnetic linear dichroism found for bulk like
antiferromagnetic NiO films.  The dichroism in the 1 ML, however, can not be
attributed to the presence of some form of magnetic order, but entirely to
crystal field effects. We provide a detailed analysis and a practical guide of
how to disentangle quantitatively the magnetic from the crystal field
contributions to the dichroic signal. This is important for a reliable
determination of, for instance, the spin moment orientation in NiO as well as
LaFeO$_{3}$, Fe$_{2}$O$_{3}$, VO, LaCrO$_{3}$, Cr$_{2}$O$_{3}$, and Mn$^{4+}$
manganate ultra thin films, surfaces and strained films, where the low symmetry
crystal field splittings may not be negligible as compared to the exchange
field energies.

We acknowledge the NSRRC staff for providing us with an extremely stable beam.
We would like to thank Lucie Hamdan for her skillful technical and
organizational assistance in preparing the experiment. The research in Cologne
is supported by the Deutsche Forschungsgemeinschaft through SFB 608.

\chapter{CoO}
\label{ChapterCoO}

CoO is a rock-salt antiferromagnetic insulator with a N\'eel temperature of about 291 K \cite{Singer56}. At first sight it may look like a simple material, but it is actually quite an amazing material with many unusual properties. Above the N\'eel temperature, CoO orders in a simple rock-salt crystal structure with a small tetragonal elongation of about 0.25 \% \cite{Rechtin71a}. Below the N\'eel temperature, there exists a relatively large tetragonal contraction, more or less proportional to the magnetic ordering, saturating at about 1.2\% at 0 K \cite{Roth58a, Jauch01}. Unlike MnO and NiO, however, the spins do not orientate in a $\{112\}$ direction. The exact direction of the spin has given rise to a lot of debate in the literature. This debate continues up to date \cite{Li55, Nagamiya58, Roth58a, Roth58b, Uchida64, Nagamiya65, Laar65a, Laar65b, Laar66, Khan70, Jauch01, Neubeck01, Tomiyasu04}. CoO is one of the few materials that has a large magnetic orbital momentum and therefore a strong coupling between magnetic and crystal structure \cite{Uchida64, Neubeck01,Jauch02,Ghiringhelli02}. It is most likely that spin-orbit coupling will play an important role in understanding the fascinating magnetic properties of the class of cobaltates, which we will discuss in chapter \ref{ChapterCobaltates}. The effects of spin-orbit coupling in those materials are not widely known and CoO is a good reference system. As we will show below, spin-orbit coupling results in a few amazing properties of CoO, which should be true in general for systems where spin-orbit coupling is important.

From a technological point of view CoO is quite interesting. About 50 years ago the exchange-bias phenomenon has been discovered in surface-oxidized cobalt particles \cite{Meiklejohn56, Meiklejohn57}. This discovery marks the beginning of a new research field in magnetism. Since then several combinations of antiferromagnetic and ferromagnetic thin-film materials have been fabricated and investigated \cite{Nogues99, Berkowitz99}, motivated by the potential for applications in information-storage technology, such as magnetoresistive devices involving spin valves. Therefore it is not surprising that there is a lot of research on how exchange-bias functions. Thin films of CoO are widely used as antiferromagnets within the field of exchange-bias \cite{Carey93a, Carey93b, Takano97, Borchers98, Ijiri98a, Ijiri98b,  Zaag00a, Zaag00b, Ohldag01a, Ohldag03, Radu03, Scholl04}.

A good understanding of the magnetic and electronic properties of CoO is desirable. However, antiferromagnetic thin films are not so easy to investigate. The total volume is very small and standard techniques like magnetic susceptibility measurements or neutron scattering would require the use of multiple repetition of these films in the form of multi-layers. Moreover, susceptibility measurements do not work if the antiferromagnetic CoO is grown adjoint to a ferromagnetic layer. Fortunately, soft x-ray absorption spectroscopy has made a large development in the last decades and has become a good option for the investigation of antiferromagnetic thin films. It can be combined with microscopy in order to resolve different domains \cite{Stohr00, Groot05, Alders98}. For NiO this has been used with great success \cite{Alders98, Stohr99, Ohldag01a, Ohldag01b, Zhu01, Oppeneer03, Altieri03, Weber03, Finazzi04, Scholl04}. For CoO, however, no polarization-dependent data have been reported so far from which one could deduce the magnetic structure and the underlaying electronic structure.

In this chapter we will present detailed measurements on; bulk CoO, a poly-crystalline CoO thin film and single-crystal CoO thin films epitaxial grown on different substrates. After an introduction about the general properties of CoO and our measurement setup, we will discuss the temperature dependence of the isotropic CoO spectra. From the temperature dependence of the isotropic spectra the importance of spin-orbit coupling can be deduced. Next we will show the x-ray magnetic circular dichroism (XMCD) in CoO, induced by a high magnetic field. From the XMCD spectra, information about the orbital momentum and the spin momentum are obtained. We also obtain a value for the exchange coupling in the paramagnetic phase. The last part of this chapter will be about linear dichroism in thin CoO films. We will give a detailed understanding of the magnetic and crystal field induced linear dichroism within these thin films. Using CoO grown on different substrates (Ag and MnO), we impose different kinds of stress in our thin films. CoO on Ag is elongated in plane and CoO on MnO is compressed in plane. We will show that these different kinds of in-plane stress will lead to different crystal-fields, a reorientation of orbital momentum and a reorientation of spin direction. We will present a quantitative model to explain this spin reorientation.

Our analysis is based on a detailed fit of spectra, calculated within a CoO$_6$ cluster, to the measured spectra, as well as the use of sum-rules. Sum-rules are powerful due to there simplicity and give a nice double check for the results of the cluster fit. The parameters used, for the cubic non-distorted and non-magnetic spectra, within our cluster fit are, in agreement with the literature \cite{paramCoO, Tanaka92, Tanaka94}. Below the N\'eel temperature we added an effective exchange field. The exchange field at 0 K has been taken to agree with neutron measurements \cite{Rechtin72, Sakurai68}. The temperature dependence has been taken to be according to a $J=\frac{3}{2}$ Brillouin function. We scaled the temperature axis in order to get the correct N\'eel temperature. The spin-orbit coupling constant has been taken to be the Hartree-Fock value. The tetragonal crystal field distortions and the spin direction for our thin films have been fitted to reproduce the correct spectra.

When we refer to CoO, we always refer to type I CoO \cite{Ok68}. A face centered cubic structure with no vacancies on the Co or O site. CoO can have vacancies at the Co and O site in a similar way as VO always has \cite{Rata04}. CoO with vacancies could be denoted as Co$_{1-y}$O$_{1-y}$ or type II CoO, as introduced by Hang Nam Ok \textit{et al.} \cite{Ok68}. Our samples are stable when heated up to 400 K and cooled down again over several measurement cycles, show sharp x-ray diffraction peaks, are good insulators and have sharp peaks in the x-ray absorption data. Within Co$_{1-y}$O$_{1-y}$ the Co ions have a different local symmetry, resulting in locally different crystal fields for each ion. As we will show below, Co 2p x-ray absorption spectra are very sensitive to the local crystal field. If many different crystal fields would be present, the x-ray absorption spectra would also become very broad. This broadening can be seen very nicely in VO \cite{Rata04}, but is completely absent in CoO. In contrast, the first peak of the XAS spectrum is so sharp that the width is limited by the resolution of the beam-line. We therefore conclude that the amount of double vacancies is rather low.

\section{Crystal and magnetic structure of cobalt oxide.}

In antiferromagnetic CoO the spin direction has given rise to a lot of debate. Different angles between the spin and the tetragonal axis have been reported \cite{Shull51, Nagamiya58, Roth58a, Jauch01}. A tilt of the spin with respect to the tetragonal axis is not compatible with a tetragonal space-group. In order to resolve this discrepancy, a multi-axis spin structure has been proposed \cite{Roth58b, Laar65b, Laar66, Khan70, Laar65a}. The multi-axis spin structure is, however, inconsistent with the findings that the magnetic anisotropy does not have its minimum in the direction of the tetragonal axis \cite{Uchida64, Laar65a, Nagamiya65}. Thereby an additional rhombohedral elongation of about $5\times10^{-4}$ lowering the total symmetry to monoclinic was found \cite{Nagamiya65, Jauch01}. These experiments, together with diffraction experiments of CoO under uniaxial stress \cite{Herrmann-Ronzaud78}, resolved this discrepancy in favor of a collinear alignment of the spins. Recently a new spin structure has been proposed, where a type I antiferromagnet has been mixed into a type II antiferromagnet, in order to explain all neutron reflections \cite{Tomiyasu04}. It has been suggested that the discrepancies between these measurements might be due to the neglect of orbital momentum of the Co ions \cite{Nagamiya65}. Neutron measurements give an exchange constant $J_2$ between next nearest neighbors of about 1.4--1.5 meV \cite{Rechtin72, Sakurai68}. This results in a temperature dependent exchange field of $H_{ex}=6 J \langle S \rangle$, or 12.6 meV when the Co spin is fully aligned. The orbital momentum of CoO is quite large. The presence of strong $K$-edge resonant x-ray magnetic scattering indicates a large orbital contribution to the ordered momentum \cite{Neubeck99}. A $\gamma$-ray diffraction study found an electron distribution at 10 K that is consistent with an ordered orbital momentum of 1.6 $\mu_B$, where they assumed the spin and orbital momentum to be parallel \cite{Jauch02}. The total ordered momentum at low temperatures lies between $3.36 \mu_B$ and $3.98 \mu_B$ \cite{Roth58a, Laar65b, Khan70, Herrmann-Ronzaud78, Jauch01}. The angle between the orbital momentum and the spin momentum is not known.

\section{Experimental setup}

\begin{SCfigure}[][h]
     \includegraphics[width=60mm]{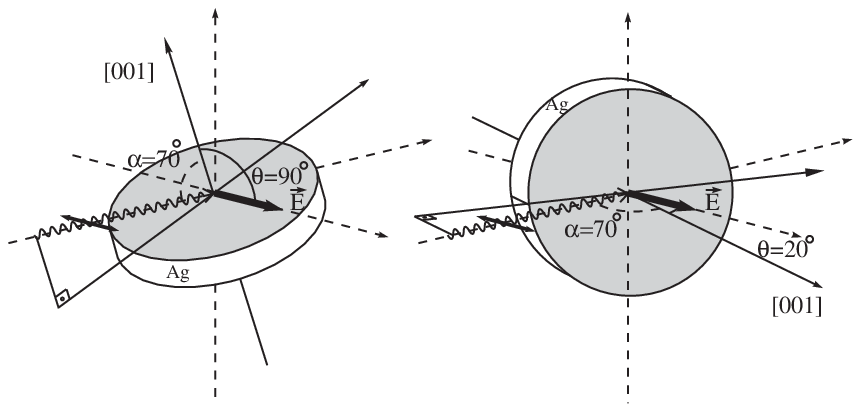}
     \caption{Experimental XAS geometry for the linear dichroism experiments, with polarization
     of the light in the horizontal plane. $\theta$ gives the angle
     between the electric field vector $\vec{E}$ and the [001] surface
     normal and can be varied between $20^{\circ}$ and
     $90^{\circ}$. The tilt angle $\alpha$ between the Poynting vector and the
     surface normal is kept at $70^{\circ}$.}
     \label{Geometry}
\end{SCfigure}

Three different samples have been measured. In order to gain information about the isotropic spectra and their temperature dependence a poly-crystalline CoO thin film on Ag has been measured. In order to gain information about the local spin momentum, orbital momentum and the exchange coupling, circular-dichroism spectra have been taken on a bulk CoO sample. The linear dichroism experiments have been preformed on a thin film of CoO, epitaxially grown on different substrates.

All spectra have been recorded using the total electron yield method in a chamber with a base pressure of $3\times10^{-10}$ mbar. The absolute-energy calibration of the monochromator has been done by setting the energy scale of a bulk NiO spectrum to the absolute-energy scale in the NiO electron energy loss spectrum, as measured by Reinert \textit{et al.} \cite{Reinert95}. The relative energy calibration between spectra taken at different scans has been done by measuring, simultaneously with the measured sample, a bulk CoO crystal in a separate upstream chamber. This way an energy calibration of better than 20 meV per scan has been achieved.

The isotropic spectra have been measured at the Dragon beamline of the NSRRC in Taiwan. The photon energy resolution at the Co $L_{2,3}$ edges ($h\nu \approx 770$--800 eV) was set to 0.3 eV. The sample has been grown by evaporating elemental Co from alumina crucibles in a pure oxygen atmosphere of $10^{-7}$ to $10^{-8}$ mbar onto sandpapered poly-crystalline Ag. The base pressure of the chamber is in the low $10^{-10}$ mbar range. The sample thickness is about 90 \AA. The sample has been measured \textit{in situ}. The spectra have been measured at normal incidence. By growing a thin poly-crystalline film, we measured an isotropic spectrum, since our beam-spot of 1 by 1 mm$^2$ is larger than the size of each ordered crystallite.

The magnetic circular dichroic spectra have been measured at the Dragon beamline ID08 of the ESRF in Grenoble. The photon energy resolution at the Co $L_{2,3}$ edges ($h\nu \approx 770$--800 eV) was set to 0.17 eV, and the degree of circular polarization was $\approx 100\%$. The magnetic field used to create the circular dichroic effect was 6 Tesla. The XMCD spectra have been broadened with a Gaussian of FWHM of 0.12 eV, resulting in a total resolution of 0.21 eV. The broadening was done in order to gain better statistics, needed for the energy calibration. A good energy calibration was necessary to measure small differences between different scans. After averaging over 36 scans a final energy calibration of better than 2 meV between different polarizations has been achieved. The sample is a pressed pellet of CoO powder mixed with 10\% Au to decrease the resistivity. The pallet has been sintered at 600K before and after pressing in order to reduce the Co$^{3+}$ content. The sample has been checked with x-ray diffraction patterns for the presence of undesired Co$_{2}$O$_{3}$, which content was found to be negligible small. The pellet has been cleaved \textit{in situ} to create a clean surface.

The linear dichroic spectra have been measured at the Dragon beamline of the NSRRC in Taiwan. The photon energy resolution at the Co $L_{2,3}$ edges ($h\nu \approx 770-800$ eV) was set to 0.3 eV, and the degree of linear polarization was $\approx 98 \%$. In order to assure an equivalent light path for the different polarizations measured we recorded the spectra with an angle of $\alpha=0^{\circ}$ between the $[001]$ surface normal and the Poynting vector of the light. To change the polarization, the sample was rotated around the Poynting vector as depicted in figure \ref{Geometry}. The angle $\theta$, between the electric field vector $\vec{E}$ and the $[001]$ surface normal, can be varied between $20^{\circ}$ and $90^{\circ}$. This measurement geometry allows for an optical path of the incoming beam which is independent of $\theta$, guaranteeing a reliable and accurate comparison of the spectral line shapes as a function of $\theta$.

We measured linear dichroism on two different samples, (14\AA)MnO/(10\AA) CoO/(100\AA)MnO/Ag(001) and (90\AA)CoO/Ag(001). The two samples were grown by molecular beam epitaxy (MBE), evaporating elemental Mn and Co from alumina crucibles in a pure oxygen atmosphere of $10^{-7}$ to $10^{-6}$ mbar. The base pressure of the MBE system is in the low $10^{-10}$ mbar range. The thickness and epitaxial quality of the films are monitored by reflection high energy electron diffraction measurements. With the lattice constant of bulk Ag (4.09 \AA) being smaller than that of bulk CoO (4.26 \AA) and MnO (4.444 \AA), we find from x-ray diffraction that CoO on Ag is slightly compressed in-plane ($a_{\parallel}\approx4.235$ \AA, $a_{\perp}\approx4.285$ \AA), and from reflection high energy electron diffraction (RHEED) that CoO sandwiched by MnO is about 4\% expanded in-plane ($a_{\parallel}\approx 4.424$ \AA). The sandwich structure was used to maximize the CoO thickness with full in-plane strain. Details about the growth will be published elsewhere \cite{Csiszar05b}.

\begin{SCfigure}[][h]
     \includegraphics[width=60mm]{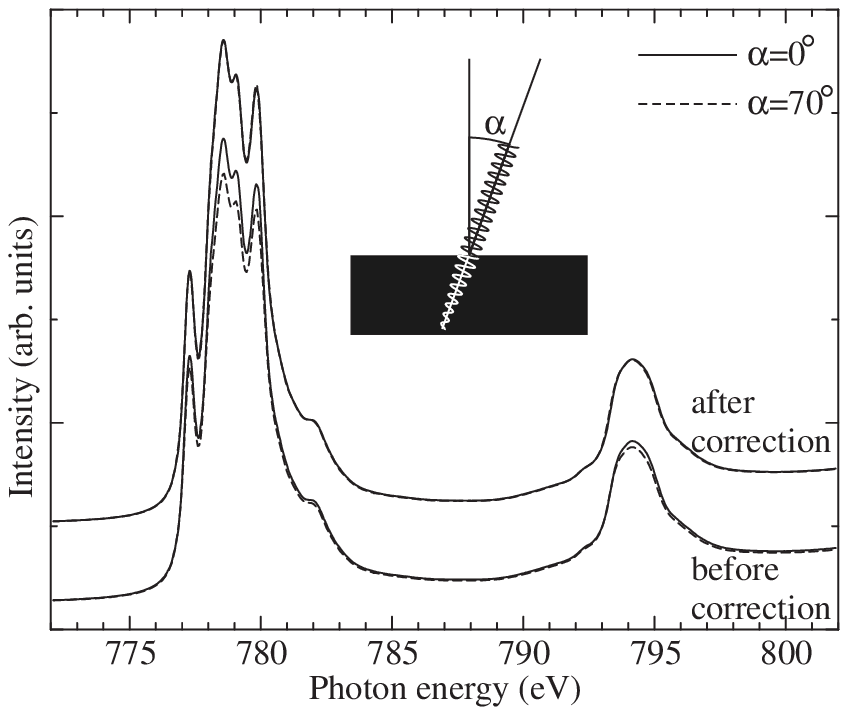}
     \caption{CoO $L_{2,3}$ spectra taken at normal incidence ($\alpha=0^{\circ}$) and taken at grazing incidence ( $\alpha=70^{\circ}$), before correction for electron-yield saturation effects (bottom) and after (top). After correction the two spectra fall perfectly on top of each other.\newline}
     \label{Saturation}
\end{SCfigure}

The spectra have been corrected for electron-yield saturation effects. Saturation effects are very small for spectra taken with light of normal incidence. For light that comes in with an angle of $70^{\circ}$ the saturation effect is about $10\%$ at the maximum of the $L_3$ edge. The corrections have been made using the fact that light with grazing incidence has larger saturation effects than light with normal incidence. We know that the measured electron-yield, $Y_e$ is proportional to the absorption coefficient $\mu$ with the following formula \cite{Nakajima99}:
\begin{equation}
Y_e=Y_0 \frac{G_e}{G_0} \frac{1}{1+\frac{\lambda_e \mu}{ \cos[\alpha]}} \left(1-e^{-\frac{t}{\lambda_e}(1+\frac{\lambda_e \mu}{ \cos[\alpha]})}\right) \frac{\lambda_e \mu}{\cos[\alpha]} 
\end{equation}
Where $Y_e$ is the measured sample current. $Y_0$ is the current of a gold mesh, mounted upstream in the beam-spot used for normalization of the incident photon flux. $G_e$ is the electron gain factor of the sample, $G_0$ is the electron gain factor of the gold reference mesh, $\alpha$ is the angle between the Poynting vector of the light and the surface normal, $t$ is the sample thickness, $\lambda_e$ is the electron escape depth and $\mu$ is the absorption coefficient we want to know. For bulk samples $t=\infty$ and $\mu \lambda_e$ is expressed easily as a function of $\frac{Y_e}{Y_0}$. There remains one unknown in this equation, namely $\frac{G_e}{G_0}$. This constant has been fitted such that the spectra $\lambda_e \mu$ taken at different incidence angles, $\alpha$ overlap over the entire spectra range. The fitted constant $\frac{G_e}{G_0}$ determines for a given beam-line and material the effect of saturation uniquely. For thin films one has to make a Taylor expansion of $e^{-\frac{t}{\lambda_e}(1+\frac{\lambda_e \mu}{ \cos[\alpha]})}$ in order to create $\mu \lambda_e$ as a function of $\frac{Y_e}{Y_0}$. This function has two fitting parameters, $\frac{G_e}{G_0}$ and $\frac{t}{\lambda_e}$, that can be determined uniquely by comparing spectra taken at different incidence angles, $\alpha$. In figure \ref{Saturation} we show the spectra of a $90$ {\AA} thick CoO film on Ag, taken with normal incidence and at an angle of $\alpha=70^{\circ}$. The polarization of both spectra is such that $\theta=90^{\circ}$, the electric vector $\vec{E}$ is in the plane of the thin film. One can clearly see that the spectra taken with $\alpha=70^{\circ}$ show a larger saturation effect at the top of the $L_3$ edge than the spectra taken with normal incidence. We also show the spectra after correction. We would like to note that the corrections made are quite small. The orbital momentum, as derived from the $L_z$ sum-rule, does not change more then 0.05 $\mu_B$ due to these corrections. The branching ratio changes by about 5\%.

\section{Isotropic line-shape and temperature dependence.}

\begin{SCfigure}[][h]
     \includegraphics[width=0.50\textwidth]{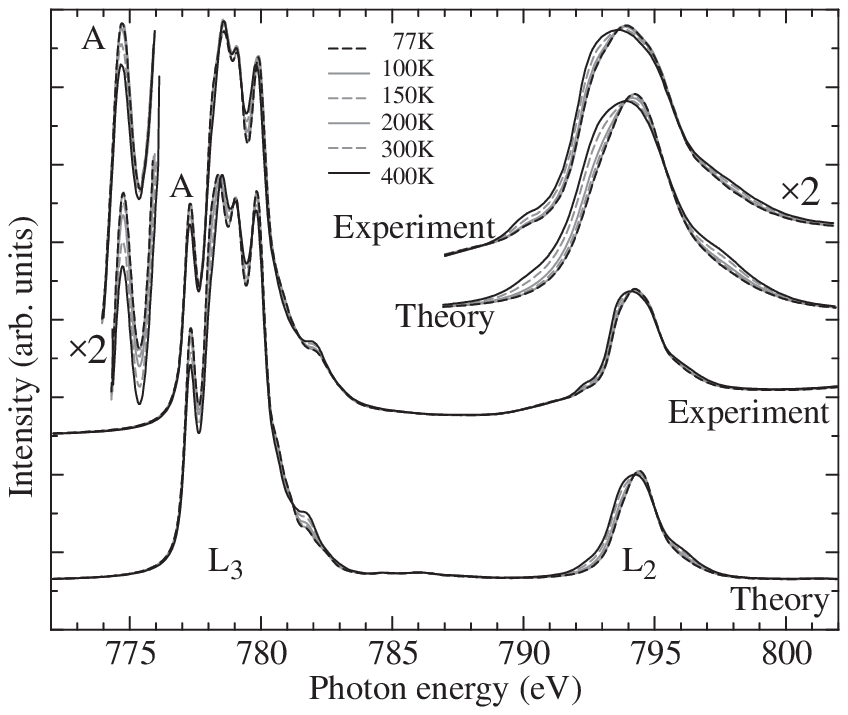}
     \caption{Temperature dependence of the experimental and theoretical isotropic CoO $L_{2,3}$ spectra.\newline}
     \label{Isotropic}
\end{SCfigure}

In order to get a better understanding of the CoO spectra and its many properties, we start with the temperature dependence in the isotropic spectra. In figure \ref{Isotropic} we show the temperature dependence of the isotropic CoO spectra. Within the $L_{3}$ edge one sees that the first sharp peak at 777 eV becomes less in intensity as the temperature goes up. This peak has been enlarged by a factor of two on the left in the same figure. One might think that this reduction in intensity is due to phononic broadening. However, the onset of the peak does not go up with temperature, showing that the peak does not become broader but really looses intensity as the temperature is increased. At the $L_{2}$ edge one can see, beyond any doubt, that something happens with the spectrum when the temperature changes which can not be related to phonons. The intensity of the $L_{2}$ edge shifts to the left when the temperature is increased. Thereby the total intensity of the $L_{2}$ edge increases with increasing temperature. 

Let's figure out where this temperature dependence comes from. One could assume that it has to do with magnetic interactions, in the same way as found for NiO \cite{Alders98}. When CoO is cooled down below the N\'eel temperature the system starts to order antiferromagnetically. This ordering will result in an exchange field of the form $\sum_i J \langle S_i \rangle$, which is the driving force for the spin-ordering. This exchange field will not only order the magnetic moments in the ground-state, but can also influence the excitonic final states. These changes in final state can modify the spectra as has been shown for NiO \cite{Alders98}. The exchange field for CoO, is much smaller than the exchange field for NiO, therefore the measured changes of the isotropic spectra, due to the increase of an exchange field, should be quite small in CoO. Thereby the change in spectra in the temperature range from 300 K to 400 K is about as large as the change in spectra in the temperature range from 0 K to 300 K. If the change in the spectra was due to an exchange field, one would not expect any, or very small changes above the N\'eel temperature. We have to conclude that the changes in isotropic spectra of CoO with temperature are not due to magnetic interactions. We have to look for another mechanism.

\begin{figure}[h]
     \includegraphics[width=120mm]{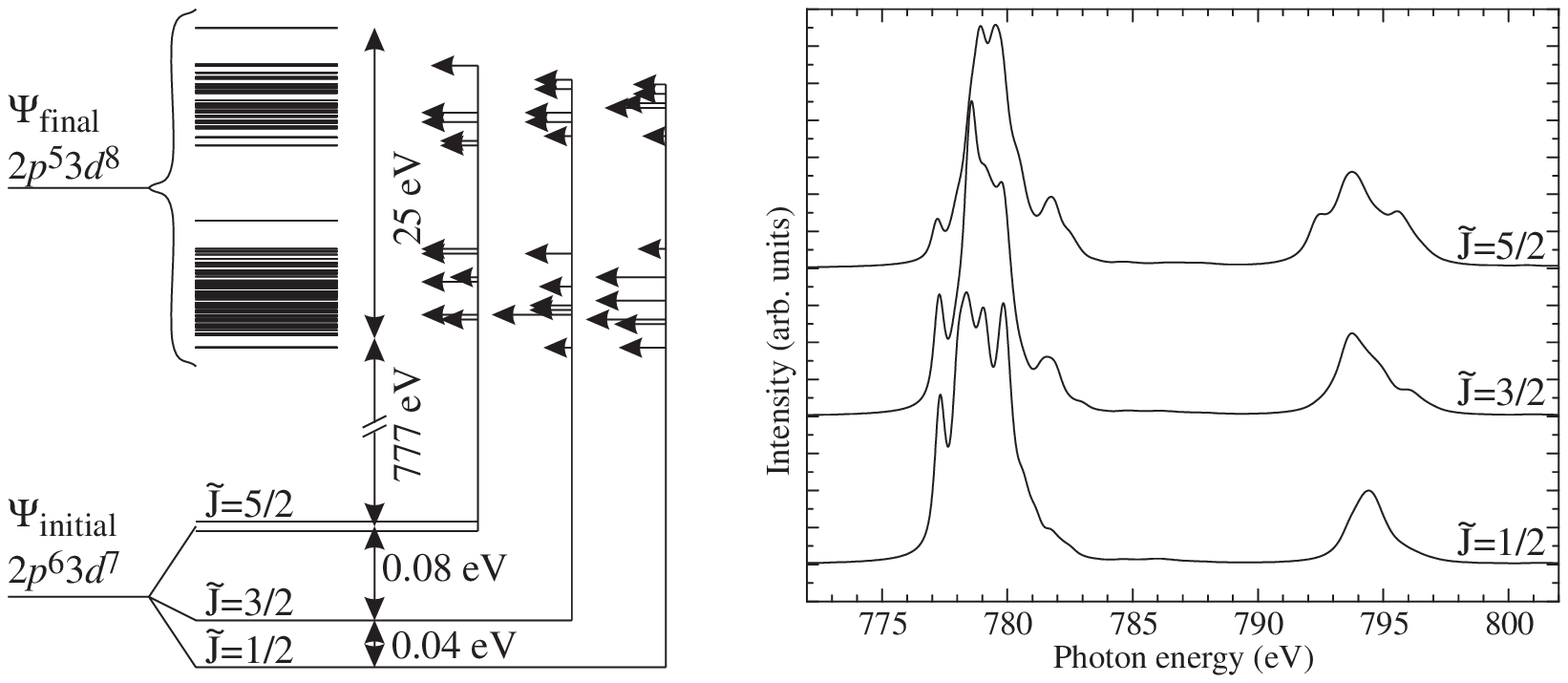}
     \caption{Left bottom: Energy level diagram of the 12 lowest states of the Co$^{2+}$ ion split by spin-orbit coupling. Left top: Energy level diagram of the many final states possible for a Co$^{2+}$ ion with a 2p core hole. Middle: The lines and arrows indicate the few possible excitations that are allowed due to the strict selection rules. Each initial state can reach a different set of final states. Right: 2p XAS spectra for the three different states with \~J=$\frac{1}{2}$, \~J=$\frac{3}{2}$ and \~J=$\frac{5}{2}$.}
     \label{IsotropicExplanation}
\end{figure}

In CoO the cobalt ions are $2+$ and have 7 electrons in the $3d$ shell. In cubic symmetry the $3d$ shell is split into a lower-lying $t_{2g}$ shell and a higher-lying $e_{g}$ shell. The $t_{2g}$ shell is 3-fold orbital degenerate and the $e_{g}$ shell is 2-fold orbital degenerate. The first five electrons of the cobalt ion populate the $t_{2g}$ and $e_{g}$ orbitals with spin up. The last two electrons occupy the $t_{2g}$ orbitals with spin down. Two electrons in three orbitals means that there is a 3-fold orbital degeneracy present at the Co site. In other words, there is one hole in the $t_{2g}$ shell, which can be in the $d_{xy}$, $d_{yz}$, or $d_{xz}$ orbital. One is used to chose a real basis set for the $t_{2g}$ orbitals. However, with a linear combination of the $d_{yz}$ and $d_{xz}$ orbitals one can create a $d_{1}=\sqrt{\frac{1}{2}}(-d_{xz}-{\rm{i}}d_{yz})$ or a $d_{-1}=\sqrt{\frac{1}{2}}(d_{xz}-{\rm{i}}d_{yz})$ orbital, containing a magnetic orbital momentum of $\pm1 \mu_B$ in the z direction. Since the $t_{2g}$ shell has three orbitals with a magnetic orbital momentum of $-1$,0, and 1 $\mu_B$, it is natural to assign a pseudo orbital momentum of \~l $=1$ to the $t_{2g}$ shell. This pseudo orbital momentum couples with the local spin of $\frac{3}{2}$ and forms three separate states. These states are separated due to the $3d$ spin-orbit coupling and have their spin and pseudo orbital momentum coupled to a pseudo total momentum of \~J $=\frac{1}{2}$, \~J $=\frac{3}{2}$ and \~J $=\frac{5}{2}$. The ground state is twofold degenerate and has \~J $=\frac{1}{2}$. The first excited state has \~J $=\frac{3}{2}$ and is fourfold degenerate. The second excited state is sixfold degenerate and has \~J $=\frac{5}{2}$. The Hartree-Fock value of the spin-orbit coupling constant for Co$^{2+}$ is about 66 meV. However, the effective splitting is reduced due to covalency, so we find for a CoO$_{6}^{10-}$ cluster with a coupling constant of 66 meV a splitting between ground state and first excited state of about 40 meV.

At 0 K, only the state with \~J $=\frac{1}{2}$ is populated, but when the temperature is increased, the first and second excited state become populated. At 400 K, the state with \~J $=\frac{3}{2}$ is populated appreciable. Since we are talking about many-electron wave-functions one can calculate the occupation with the use of Boltzmann statistics. Knowing that the ground-state of CoO is split by spin-orbit coupling and that for higher temperatures excited states become populated, it is quite easy to understand why the x-ray absorption spectra are temperature dependent. The $2p$ to $3d$ transition is dipole allowed and strong dipole selection rules are present. One of the dipole selection rules is $\Delta J=0,\pm1$. For each state with a different \~J, a different set of final states can be reached. On the left side of figure \ref{IsotropicExplanation} we show this graphically. On the bottom we show the energy level diagram of the lowest 12 states of the $d^7$ configuration split by spin-orbit coupling. On the top we show the many states present in the final $2p^53d^8$ configuration. The arrows indicate different excitation probabilities from a given initial state with a specific \~J value to the different final states. On the right of figure \ref{IsotropicExplanation} the 3 spectra for the different initial states with different values of \~J are shown. As one can see they are quite different. 

Now we have all ingredients in order to calculate the temperature dependence of the isotropic spectrum. We have calculated the spectrum for each of the initial states with different values of \~J. We have calculated the splitting between the different initial states. Next we use Boltzmann statistics to calculate the different populations of each of the initial states at different temperatures. Finally we take a weighted sum of the spectra according to the population calculated with the use of Boltzmann statistics. In figure \ref{Isotropic} we compare the temperature-dependence in the theoretical spectrum for a cluster calculation with the experimental spectrum. All features are represented and the temperature dependent change of spectra can be related to the population of excited states with different \~J. Our calculations are in good agrement with previous predictions made about the temperature changes in the CoO spectra on the basis of the same cluster calculations \cite{Jo98, Tanaka92, Groot93a}. The parameters used are in agreement with parameters noted in the literature \cite{paramCoO, Tanaka92, Tanaka94}.

\begin{SCfigure}[][h]
     \includegraphics[width=60mm]{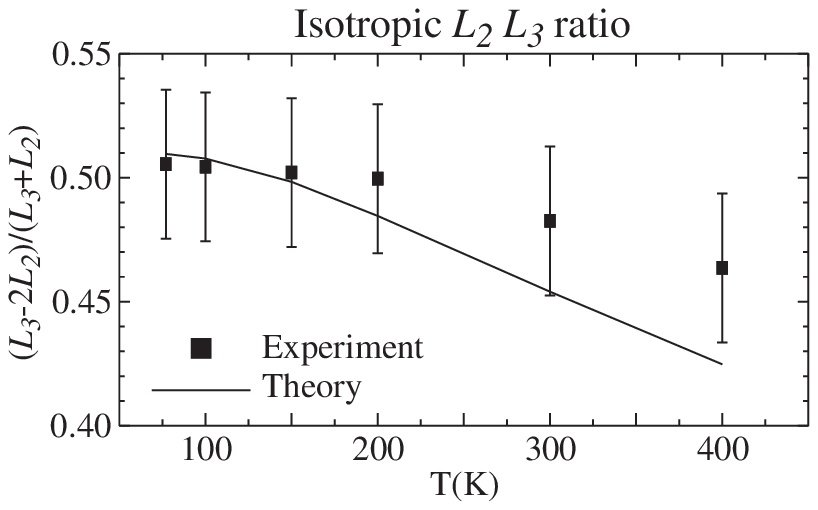}
     \caption{Isotropic branching ratio as a function of temperature.\newline}
     \label{IsotropicBranchingratio}
\end{SCfigure}

There is another way to check the influence of $3d$ spin orbit coupling in the initial state. Thole and Van der Laan have studied the reason for the nonstatistical branching ratio and found that the branching ratio scales linear with the expectation value of $\sum_i\langle l_i\cdot s_i\rangle$ \cite{Thole88a, Thole88b}. They already suggested that by doing temperature-dependent x-ray absorption measurements one could deduce, if excited states with different spin-orbit coupling become populated by looking at the branching ratio. The branching ratio is related to an exact number for the spin-orbit coupling operator and for $p$ to $d$ transitions the relation they found is:
\begin{equation}
\frac{\int_{L_3}-2 \int_{L_2}}{\int_{L_3}+\int_{L_2}}=\frac{\sum_i\langle l_i\cdot s_i \rangle}{\underline{n}}
\end{equation}
Where $\underline{n}$ is the number of holes in the 3d shell. $\sum_i\langle l_i\cdot s_i\rangle$ is different for each state with a different \~J. In our case $\sum_i\langle l_i\cdot s_i \rangle=-1.31$ for states with \~J=$\frac{1}{2}$, $\sum_i\langle l_i\cdot s_i\rangle=-0.72$ for states with \~J=$\frac{3}{2}$ and $\sum_i\langle l_i\cdot s_i\rangle=0.35$ for the average over states with \~J=$\frac{5}{2}$. In figure \ref{IsotropicBranchingratio} we show the measured branching ratio as well as the branching ratio of the cluster calculations. A clear degreasing trend can be seen, as one should expect from the sum rule. We find for T=77K an expectation value of $\sum_i<l_i\cdot s_i>=-1.30$ and for T=400 K an expectation value of $\sum_i\langle l_i\cdot s_i\rangle=-1.00$. These values are derived from our cluster fit and not from the sum rule. The values for $\sum_i\langle l_i\cdot s_i\rangle$ are in perfect agreement with circular polarized spin resolved photo-emission measurements that measured the expectation value of $\sum_i\langle l_i\cdot s_i\rangle=-0.99$ at 390 K. If we now compare these values with the values obtained from the sum rule we find that the sum rule gives values for $\sum_i\langle l_i\cdot s_i\rangle$ that are about 10\% larger than the values found from our cluster fit. However our fit does get the correct $L_3$--$L_2$ ratio. The deviation of the sum rule is due to an approximation made in the derivation of sum rule. In the derivation of the sum rule it is assumed that the $L_2$ and the $L_3$ edge are well separated. In other words, that $j_{2p}$ is a good quantum number. This is not the case. The $2p$--$3d$ electron electron repulsion mixes both edges and the prerequisites for the sum rule are not fulfilled. A 10\% deviation might not seem too bad, but this deviation depends on other parameters within the system. In figure six of reference \cite{Thole88b} Thole and Van der Laan showed the theoretical branching ratio for different initial state terms as a function of $2p$--$3d$ electron-electron interaction. If the $2p$--$3d$ electron-electron interaction is made zero, the statistical branching ratio is found and the $l\cdot s$ sum-rule works perfectly. If the $2p$--$3d$ electron-electron interaction is increased, deviations from the statistical branching ratio occur. These deviations are different for each multiplet term of the $d^7$ ion. This can be seen in figure seven of reference \cite{Thole88b}. The $^4F$ term has a larger deviation from the statistical branching ratio than the $^4P$ term. In CoO, the Co ions are in a high-spin $d^7$ state. In spherical symmetry there are two terms that have $S=\frac{3}{2}$, a $P$ (L = 2) and a $F$ (L = 3) term. One should realize that the orbital momentum is not a good quantum number in $O_h$ symmetry. This means that the $F$ and $P$ term mix in order to build the ground-state. The amount of $F$ and $P$ depends on the splitting between the $t_{2g}$ and $e_{g}$ orbitals. If this splitting is zero, we are back in spherical symmetry and find that the ground-state belongs to the $F$ term. If this splitting is very large, we have a ground-state that is built up of about 20\% $P$ and 80\% $F$. Since the branching ratio for the $F$ and $P$ terms is different, we will also find a change in branching ratio if we change the splitting between the $t_{2g}$ and $e_{g}$ orbitals. Therefore the decreasing trend found in the branching ratio when temperature is increased, as shown in figure \ref{IsotropicBranchingratio}, is very nice, but without the analysis done before, based on cluster calculations, we would not have been able to determine wether the change in branching ratio is due to population of excited states with a different value of \~J or due to an increase of the splitting between the $t_{2g}$ and $e_{g}$ orbitals. From our cluster fits there is no doubt. Within CoO we can quantify a localized ground-state and excited states that are split due to spin-orbit coupling and become thermally populated when temperature is increased.

\sectionmark{Circular dichroism; magnetic spin momentum, magnetic...}
\section{Circular dichroism; magnetic spin momentum, magnetic orbital momentum, and exchange coupling.}
\sectionmark{Circular dichroism; magnetic spin momentum, magnetic...}

In order to get a better understanding of the exchange field and magnetic properties of CoO and its influence on the CoO spectra, we set out an experiment to do x-ray magnetic circular dichroism (XMCD) on bulk CoO. The XMCD signal is proportional to the magnetization of the Co ion. CoO has the largest susceptibility at its N\'eel temperature of 291 K. However, the susceptibility at that temperature is still rather small. To get a well measurable signal, we needed a strong magnetic field. We used a magnetic field of 6 Tesla, present at ID08 of the ESRF.

\begin{SCfigure}[][h]
     \includegraphics[width=0.50\textwidth]{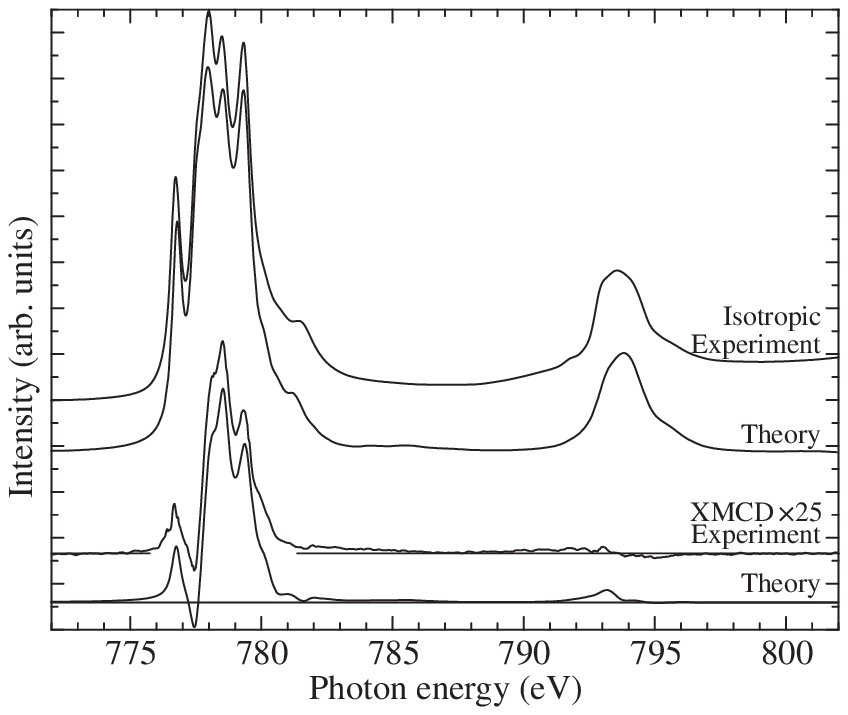}
     \caption{Magnetic circular dichroism in CoO at the N\'eel temperature of 291 K. Magnetic ordering is induced by a magnetic field of 6 Tesla.\newline}
     \label{XMCD}
\end{SCfigure}

In figure \ref{XMCD} we show the isotropic CoO spectrum at 291 K as well as the XMCD signal. The isotropic spectrum is defined as the average of the four different spectra, taken with positive or negative helicity and positive or negative magnetic field. The XMCD signal is defined as the difference between the anti-parallel (e.g. positive helicity and negative magnetic field) and parallel spectra. The XMCD signal is about 3\% of the isotropic spectrum and is highly reproducible for different scans and positions of the sample. When within a cluster calculation only a magnetic field of 6 Tesla is introduced to calculate the XMCD signal, we find that the theoretical XMCD spectra does have the correct line shape, but is about 2.5 times too large. This should be expected, as a magnetic field will order the spins and therefore induce an exchange field. This exchange field acts in mean-field theory as a magnetic field, which only operates on the spins. This exchange field is anti-parallel to the applied magnetic field, therefore it reduces the size of the XMCD effect. We fitted the size of the exchange field to be such that the total size of the XMCD signal is correctly reproduced. We find that at 291 K we need to include an exchange field of 0.58 meV. The magnetic moments then become; $L_z=0.022$ and $S_z=0.016$. If we assume that the magnetic susceptibility is equal to $\frac{M}{B}$ we find that $\chi=\frac{(2S_z+L_z) \mu_B N_A}{B 10}=5.1\times10^{-3}$ [emu mol$^{-1}$]. This is in reasonable agrement with the magnetically measured value of $5.3\times10^{-3}$ [emu mol$^{-1}$] \cite{Singer56}. If we compare our value of the exchange field with the value expected from neutron measurements we do not find an agreement. 
We know that $H_{ex}=\sum_{n.n.} J_1 \langle S_{n.n.} \rangle + \sum_{n.n.n.} J_2 \langle S_{n.n.n.} \rangle$, where $n.n.$ stands for the nearest-neighbors and $n.n.n.$ for the next-nearest-neighbors. The nearest-neighbors can be found in the $\{011\}$ directions, the next-nearest-neighbors in the $\{001\}$ directions. Within the paramagnetic phase all sites are equal. If we now realize that there are 12 nearest-neighbors and 6 next-nearest-neighbors we find that  $H_{ex}=(12 J_{1} + 6 J_{2})\langle S \rangle$. From our measurements we have $2 J_1 + J_2=\frac{0.58}{0.016\cdot6} = 6$ meV. This value is 3 to 4 times larger than found from neutron measurements \cite{Sakurai68, Rechtin72}. Sakurai \textit{et al.} \cite{Sakurai68} deduced from inelastic neutron scattering and paramagnetic susceptibility measurements a value of $J_1=0.1$ meV and $J_2=1.5$ meV. The spin waves have been measured at 110 K. Rechtin \textit{et al.} \cite{Rechtin72} measured the diffuse scattered neutron intensity in the paramagnetic phase at temperatures close to the N\'eel temperature. In the same temperature regime as our XMCD measurements have been done. Rechtin \textit{et al.} found $J_1=0.3$ meV and $J_2=1.4$ meV. The discrepancies between neutron measurements and our XMCD results are outside the error-bars of our measurement and should be taken seriously. 
It is not clear what the reason for this discrepancy is, but a few considerations can be given. An exchange-field is introduced in our cluster calculations in order to reduce the calculated magnetic moments in the paramagnetic phase and make them consistent with the measured moments. This exchange-field is taken into account in a mean-field way, which is for antiferromagnets not so good. It would be better to do spin-wave theory, whereby it is known that within spin-wave theory the ordered moments are reduced. However, for a spin-only system this can never explain the factor of 3 to 4 with which the exchange field has to be reduced.

Spin-wave theory for CoO is not so easy. CoO is a cubic antiferromagnet with orbital degree of freedom. Above the N\'eel temperature the tetragonal distortion is small and the $t_{2g}$ levels can be taken to be degenerate. Spin-orbit coupling is important and with different temperatures different orbitals are occupied. One should expect different Co--O--Co hopping for each different orbital, resulting in a temperature dependent exchange coupling constant. Thereby direct orbital-orbital interactions might play an important role in describing the size of the exchange coupling constants and might be an essential ingredient for understanding the magnetic interactions. Should CoO be described in terms of an orbital liquid in the presence of strong spin-orbit coupling as proposed for LaTiO$_{3}$ \cite{Khaliullin00}? Within an orbital liquid orbital fluctuations are important for describing the magnetic interactions and are responsible for reducing the magnetic moments.

\section[Linear dichroism; orbital occupation and spin direction.]{Linear dichroism; orbital occupation and \- spin direction.}

Next we turn our attention to the linear dichroism within the CoO spectra. Linear dichroism can give us information about spin-orientation and exchange fields, as well about orbital occupation and crystal fields. It is important to have a good understanding of the linear dichroism in CoO, since with the use of XAS and the linear dichroism therein, an enormous amount of information can be obtained about the properties of CoO.

We compare our theoretical calculations with two different CoO thin films, one under tensile and another under compressive in-plane stress. This allows us to present a detailed understanding of CoO and the XAS spectra of CoO with a tetragonal distortion. We do not present data on perfect cubic CoO, as these samples simply do not exist. Bulk CoO becomes tetragonally distorted when cooled down below the N\'eel temperature. We decided to take thin films of CoO grown on well defined substrates with a well-defined lattice mismatch, in order to fully control the crystal fields in the CoO thin film. We took CoO thin films epitaxially grown on MnO(100) and on Ag(100), as model systems for CoO under either tensile or compressive in-plane stress. We will show how XAS can be used to establish how the magnetic anisotropy, as well as the spin and orbital contributions to the magnetic moments, depend on the lowering of the local crystal-field symmetry by epitaxial strain.

The actual compositions of the CoO/MnO(100) and CoO/Ag(100) systems are (14\AA)MnO/(10\AA)CoO/ (100\AA)MnO/Ag(001) and (90\AA)CoO/Ag(001), respectively. With the lattice constant of bulk Ag (4.09 \AA) being smaller than that of bulk CoO (4.26 \AA) and MnO (4.444 \AA), we find from x-ray diffraction that CoO on Ag is slightly compressed in-plane ($a_{\parallel}\approx4.235$ \AA, $a_{\perp}\approx4.285$ \AA), and from reflection high energy electron diffraction (RHEED) that CoO sandwiched by MnO is about 4\% expanded in-plane ($a_{\parallel}\approx 4.424$ \AA). The sandwich structure was used to maximize the CoO thickness with full in-plane strain.

\begin{figure}[h]
     \includegraphics[width=120mm]{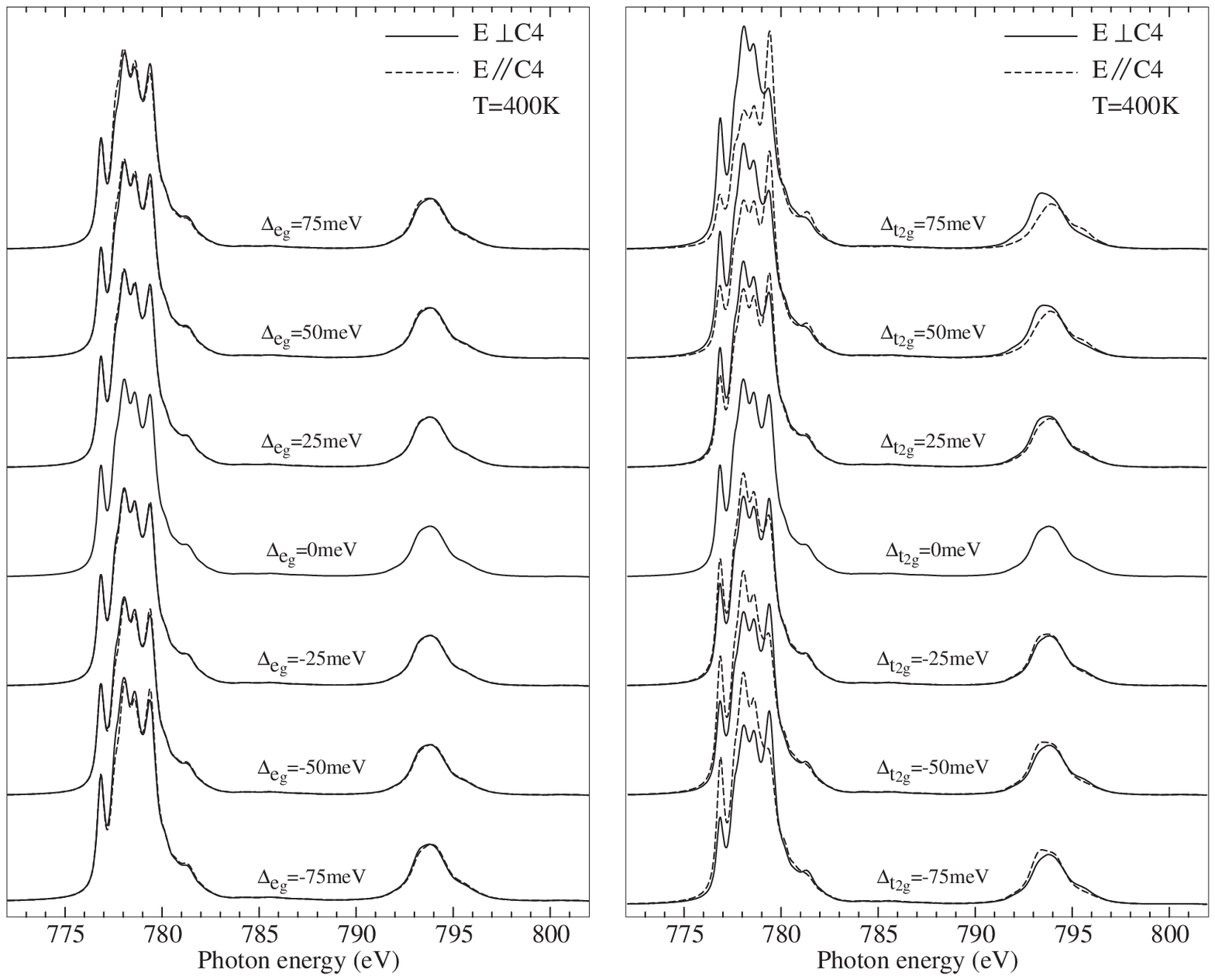}
     \caption{Theoretical CoO spectra as a function of tetragonal distortion parameter $\Delta_{e_g}$ (left) and $\Delta_{t_{2g}}$ (right).}
     \label{LDcrystalfield}
\end{figure}

A tetragonal distortion can be modelled by two parameters, $\Delta_{t_{2g}}$ and $\Delta_{e_g}$. Here $\Delta_{t_{2g}}$ is defined as the energy difference between the $d_{xy}$ orbital, and the $d_{xz}$ and $d_{yz}$ orbitals. $\Delta_{e_{g}}$ is defined as the energy difference between the $d_{x^2-y^2}$ and $d_{z^2}$ orbital. The parameters $\Delta_{t_{2g}}$ and $\Delta_{e_g}$ are related to the parameters $Ds$ and $Dt$, used in many textbooks, via $\Delta_{t_{2g}}=3Ds-5Dt$ and $\Delta_{e_{g}}=4Ds+5Dt$ \cite{Ballhausen62}. In figure \ref{LDcrystalfield} we show the theoretical linear dichroism of the $L_{2,3}$ edges of CoO for different values of the tetragonal crystal field parameter $\Delta_{e_{g}}$ on the left hand side and $\Delta_{t_{2g}}$ on the right hand side. Since CoO has an open $t_{2g}$ shell, the orbitals will align when $\Delta_{t_{2g}}$ is set nonzero. The spectra are therefore rather sensitive to changes in the parameter $\Delta_{t_{2g}}$. In a first approximation, the $e_{g}$ shell is half filled with two electrons, one electron in the $d_{z^2}$ and one electron in the $d_{x^2-y^2}$ orbital. Therefore the $e_g$ shell does not show any anisotropy when the parameter $\Delta_{e_g}$ is varied. One might expect that the spectra then do not change at all when $\Delta_{e_g}$ is varied. This, however, is not completely true. The changes in spectral line-shape are much smaller when one varies $\Delta_{e_{g}}$, but still clearly visible. There is a simple reason for that. The final-state does change when one changes $\Delta_{e_{g}}$, therefore one finds changes in the spectra. The mechanism is very similar as has been shown to be important for NiO \cite{Haverkort04}.

\begin{figure}[h]
     \includegraphics[width=120mm]{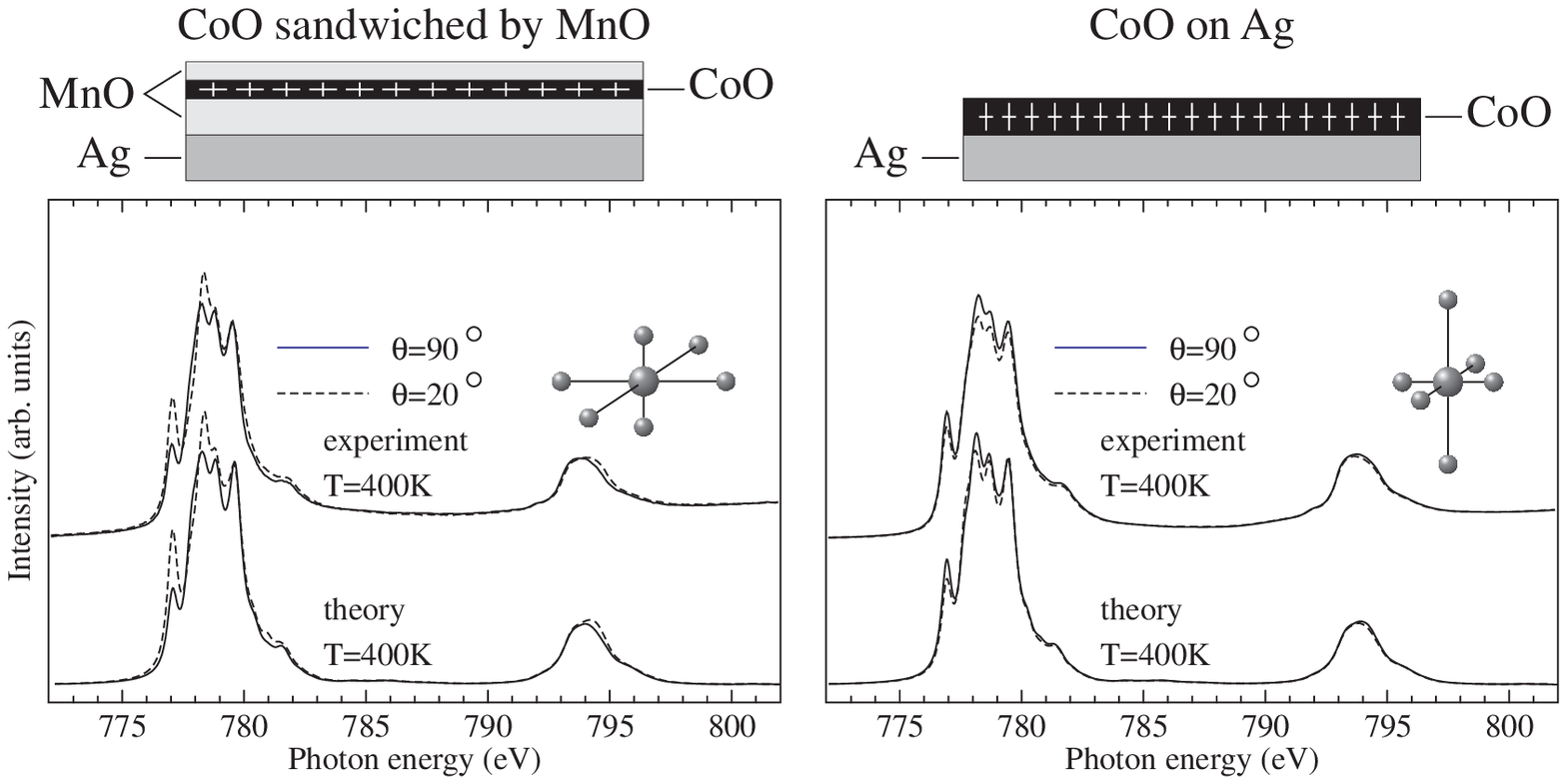}
     \caption{Experimental spectra of CoO for CoO sandwiched by MnO on the left and CoO on Ag on the right taken at 400 K, well above the N\'eel temperature. The theoretical fits have been made by choosing the best value of $\Delta_{t_{2g}}$ and by assuming that $\Delta_{e_{g}}=4\Delta_{t_{2g}}$.\newline}
     \label{LDmeasuredCF}
\end{figure}

In figure \ref{LDmeasuredCF} we show the experimental spectra of CoO sandwiched by MnO on the left and CoO on Ag on the right. These spectra are taken at 400 K, well above the N\'eel temperature. At these temperatures there should not be any linear dichroism due to magnetic ordering and all linear dichroism should be due to the crystal field. This is confirmed by our cluster calculations. The theoretical spectra and the linear dichroism therein can be found below the experimental spectra in figure \ref{LDmeasuredCF}. We took the value of $\Delta_{eg}$ to be $4 \Delta_{t_{2g}}$ for all our simulations. We find for CoO sandwiched by MnO a crystal field of $\Delta_{t_{2g}}=-56$ meV ($\Delta_{e_{g}}=-224$ meV) and for CoO on Ag a crystal field of $\Delta_{t_{2g}}=18$ meV ($\Delta_{e_{g}}=72$ meV). The opposite sign of the crystal field reflects perfectly the opposite strain opposed in these thin films. For CoO sandwiched by MnO we find that the $d_{xy}$ orbital is lower in energy then the $d_{xz}$ and $d_{yz}$ orbitals, which also should be so, since the oxygen atoms in the xy-plane are further away then the oxygen atoms in the z direction. For CoO on Ag the energies of the orbitals is opposite, as one should expect from the crystal structure. The size of the crystal fields is also fully consistent with our structural data that shows that the CoO sandwiched by MnO(001) experiences a large in-plane expansion and therefore has a large negative crystal field ($-56$ meV) while the CoO in CoO on Ag(001) is only slightly contracted in-plane and therefore has a small positive crystal field ($+18$ meV).

If we look at figure \ref{LDcrystalfield} or figure \ref{LDmeasuredCF} we first notice that the largest contrast can be found at the first peak of the spectrum around 777 eV. That is nice, since this is also the only part of the spectrum that can be understood, more or less quantitatively, in a one-electron picture. The lowest exciton one can make with one hole in the $2p$ shell and 8 electrons in the $3d$ shell is a state with all $t_{2g}$ electrons occupied and 2 electrons in the $e_g$ shell. This means that the first peak originates from an excitation from the $2p$ core level to the $t_{2g}$ shell. CoO has one hole in the $t_{2g}$ shell and now the selection rules for the dipole operator become important. The intensity for an excitation is proportional to the square of $\langle p_i|q|d_j\rangle$, where i can be x,y, or z depending on the p orbital excited, q can be x,y, or z depending on the polarization and j can be xy, xz, or yz, since we excite to the $t_{2g}$ shell. This integral is integrated over all space and therefore 0 if the integrand is odd. For z polarized light the only way to create an even integrand and therefore to excite an electron into the $t_{2g}$ shell is by exciting the $p_x$ orbital into the $d_{xz}$ orbital ($\langle p_x|z|d_{xz}\rangle$), or the $p_y$ orbital into the $d_{yz}$ orbital ($\langle p_y|z|d_{yz}\rangle$). The integrand of the integral $\langle p_i|z|d_{xy}\rangle$ with i=x,y or z is always odd. Therefore the $d_{xy}$ orbital can not be reached with z polarized light. In the same way the selection rules for x and y polarized light can be deduced. We find that for every polarization there is one $t_{2g}$ orbital that can not be reached. With z polarized light one can not reach the $d_{xy}$ orbital, with x polarized light one can not reach the $d_{yz}$ orbital and with y polarized light one can not reach the $d_{xz}$ orbital. CoO has only one hole in the $t_{2g}$ shell. Therefore if this hole is in the $d_{xy}$ orbital one will not find any intensity with z polarized light for the first peak of the Co$^{2+}$ XAS spectra. If we now take a crystal field with positive $\Delta_{t_{2g}}$ then the $d_{xy}$ orbital is raised in energy with respect to the $d_{xz}$ and $d_{yz}$ orbitals, it becomes unpopulated. The intensity of the first peak goes down when we look with z polarized light ($E\parallel C_4$). This can also be seen in the right collum of figure \ref{LDcrystalfield} where for the topmost spectrum, $\Delta_{t_{2g}}=75$ meV the intensity of the first peak at 777 eV is almost gone for z polarized ($E\parallel C_4$) light. What one should notice is that the intensity contrast for the first peak does not change abrupt, but slowly as the crystal field is varied. This is due to the spin-orbit coupling that favors a spherical charge distribution. Without spin-orbit coupling, one has zero or full intensity for the first peak already for any crystal field larger than the temperature of the system. Spin-orbit coupling turns this into a slow change when $\Delta_{t_{2g}}$ is varied from $+75$ meV to $-75$ meV. 

\begin{SCfigure}[][h]
     \includegraphics[width=60mm]{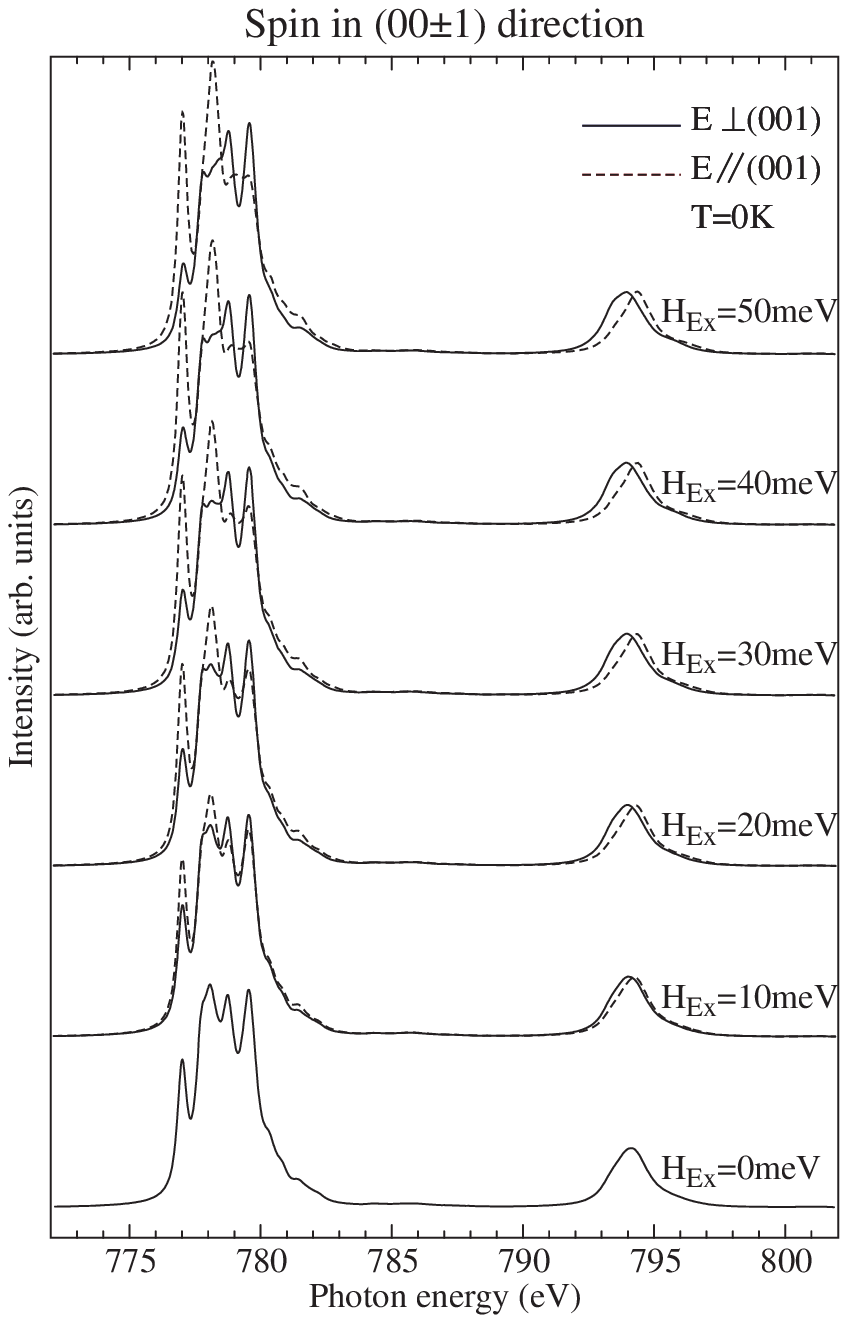}
     \caption{Polarization-dependent theoretical CoO 2p XAS spectra for different values of the exchange field $H_{ex}=6J_2 \langle S \rangle$. \newline \newline}
     \label{LDexchangestrenght}
\end{SCfigure}

Ferro-orbital ordering and crystal fields are not the only reason for linear dichroism. Magnetic ordering and exchange fields can give linear dichroism as well \cite{Laan91, Alders98}. In figure \ref{LDexchangestrenght} we show the magnetic linear dichroism of cubic CoO as a function of the exchange field at 0 K. Up to an exchange field of 30 meV one sees that the dichroic effect scales with the size of the exchange field. Above 30 meV the changes become roughly independent of the exchange field. This is radically different from the magnetic linear dichroism found in NiO for example. Within NiO there is hardly any dependence of the magnetic linear dichroism on the size of the exchange field. For NiO at T=0 K the linear dichroism is already present at its full extend for infinitesimally small exchange fields. This can be understood if one takes into account that within CoO strong spin-orbit coupling and a free orbital momentum is present. Spin-orbit coupling favors a spherical state. Therefore one needs a finite exchange field to align the spins. For NiO this is not the case and the spins align already with an infinitesimal exchange field. If we have a look at the spectra as shown in figure \ref{LDexchangestrenght}, one sees that there is a large contrast between different polarizations for the first peak at 777 eV. This is the same peak where we saw the large contrast in the linear dichroism that was induced due to crystal fields. The reason why there is linear dichroism due to an exchange field in the first peak is actually the same as for a crystal field. An exchange field induces a preferred orbital occupation. If the spins of CoO are aligned in the z direction the spin-orbit coupling will also align the orbital momentum  in the z direction. Now one needs to realize that the $d_{xz}$ and the $d_{yz}$ orbital can be recombined to give a $d_{1}=\sqrt{\frac{1}{2}}(-d_{xz}-{\rm{i}}d_{yz})$ orbital. This orbital has an orbital momentum in the z direction. If we align the spin in the z direction, then the orbital momentum will also be aligned in the z direction. In other words there will be half a hole in the $d_{xz}$ orbital and half a hole in the $d_{yz}$ orbital, but no holes in the $d_{xy}$ orbital. For z polarized light (E$\parallel$ 001) we find a large intensity of the first peak and for x or y polarized light (E$\bot$001) a small intensity. If one regards figure \ref{LDexchangestrenght}, this is exactly what shows in the theoretical spectra.

An exchange field is doing more then changing the initial state orbital occupation. If we look at the $L_{2}$ edge we see an effect that can not be described by the orbital occupation, but measures directly the spin alignment. The mean-energy position of the $L_{2}$ edge depends on the polarization direction with respect to the spin. For light polarized parallel to the spin direction, the mean-energy position of the $L_{2}$ edge is lower than for light polarized perpendicular to the spin direction. This effect is also present as one varies the tetragonal crystal field, but much weaker. Using both properties of the linear dichroism of CoO, one is able to determine the spin and orbital momentum direction. These do not have to be parallel for a Co ion in a symmetry lower then $D_{2}$.

\begin{figure}[h]
     \includegraphics[width=120mm]{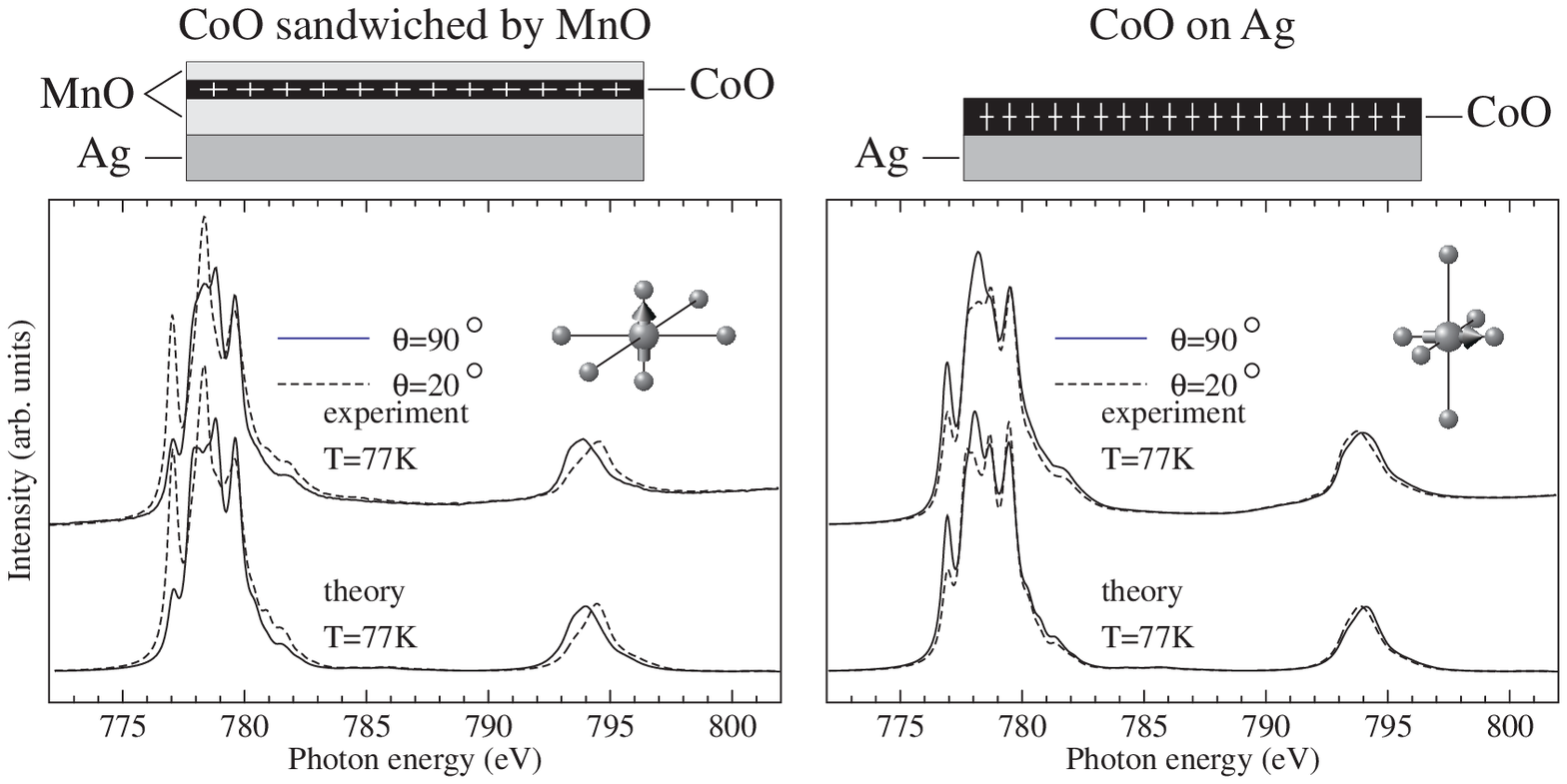}
     \caption{Experimental spectra of CoO for CoO sandwiched by MnO on the left and CoO on Ag on the right taken at 77 K, below the N\'eel temperature. The theoretical fits have been made by including an exchange field of 12.6 meV as found from neutron measurements \cite{Sakurai68, Rechtin72}. The direction of the exchange field has been fitted to the best agreement between theory and experiment. For CoO sandwiched by MnO we find the exchange field to be perpendicular to the thin film surface and for CoO on Ag we find the spin to be in the plane of the thin-film surface.\newline}
     \label{LDmeasuredHex}
\end{figure}

In figure \ref{LDmeasuredHex} we show the low temperature spectra of the CoO thin films. On the left we show the spectra of CoO sandwiched by MnO and on the right the spectra of CoO on Ag. The spectra are taken at a temperature of T=77 K, well below the N\'eel temperature of about 300 K for these thin films as we will show below. These temperatures guarantee an almost complete alignment of the spins. The theoretical fits have been made by including an exchange field of 12.6 meV, as found from neutron measurements \cite{Sakurai68, Rechtin72}. The direction of the exchange field has been fitted to the best agreement between theory and experiment. For CoO sandwiched by MnO, we find the exchange field to be perpendicular to the thin-film surface and for CoO on Ag we find the spin to be in the plane of the thin-film surface. Theoretically, one would expect that dipolar interactions would tilt the spin towards the {112} direction. Our findings hint, that these tilts should be rather small. A tilt of the spin, away from the surface normal for CoO sandwiched by MnO or away from the film plane for CoO on Ag, would reduce the measured linear dichroism. It has, exactly the correct size if one assumes an exchange field of 12.6 meV and spins perfectly in or out off the film plane. The canting of the spins could not be measured directly since different domains are present. The average in-plane spin ordering over these domains is isotropic. In this sense it would be extremely nice to combine x-ray spectroscopy with microscopy (PEEM). By doing so one should be able to see different domains and determine the angle with which the spins are canted. For single-domain measurements information about the spin direction as well as information about the orbital momentum direction can be obtained separately from the XAS spectra.

\begin{figure}[h]
     \includegraphics[width=120mm]{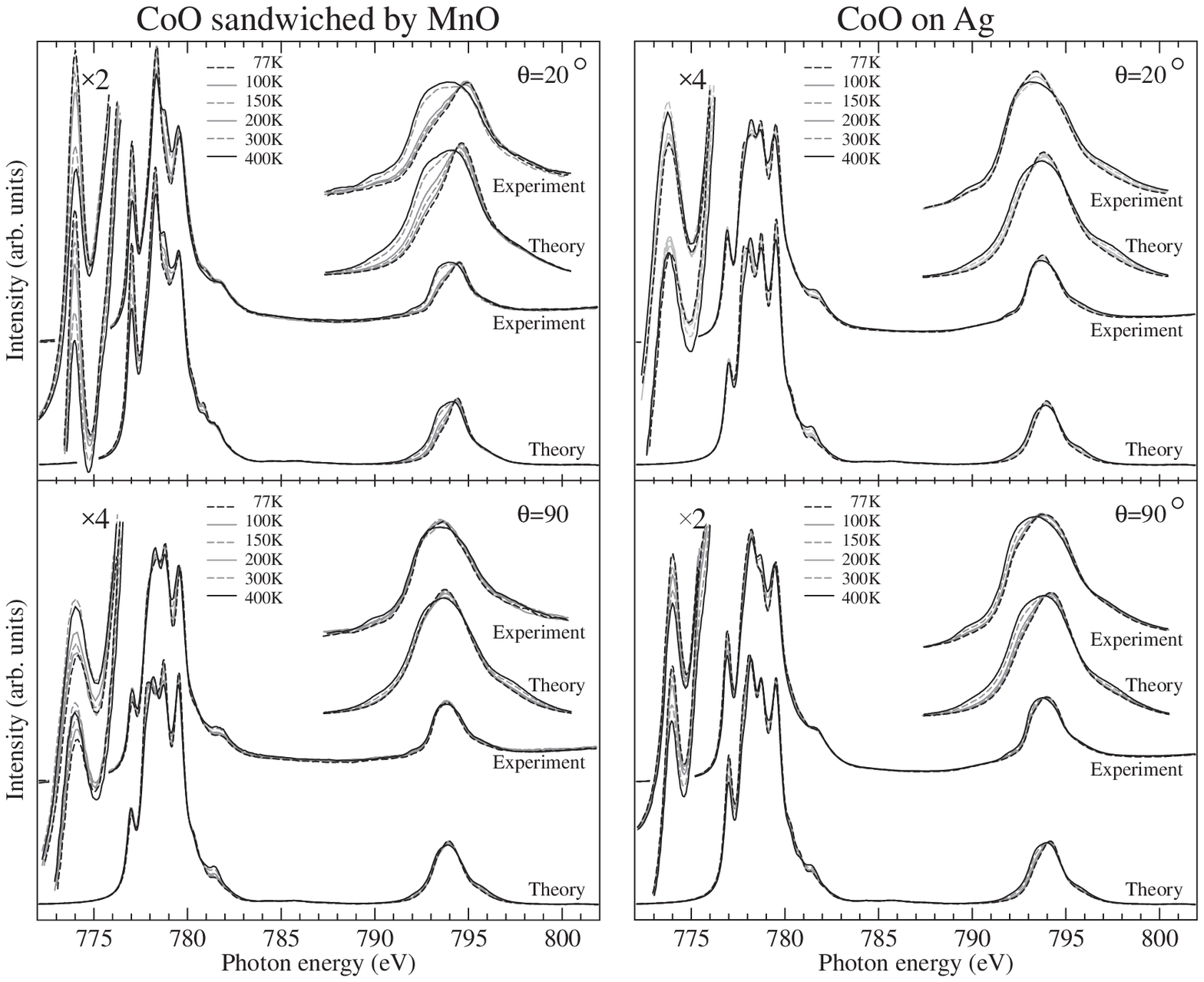}
     \caption{Experimental and calculated Co $L_{2,3}$ XAS spectra
     and their temperature dependence of: left panel) CoO in (14\AA)MnO/(10\AA)CoO/(100\AA)MnO/Ag(001) at
     $\theta=20^{\circ}$ (top panel) and $\theta=90^{\circ}$ (bottom panel);
     right panel) the same for CoO in (90\AA)CoO/Ag(001)\newline}
     \label{LDtemperature}
\end{figure}

We have shown that assuming an exchange field of 12.6 meV as measured by neutrons \cite{Rechtin72, Sakurai68}, we find good fits for our spectra if we take the spins for CoO on Ag in the film plane and for CoO sandwiched by MnO parallel to the surface normal. In order to calculate spectra at a temperature below the N\'eel temperature, but not so low that the Co spins are fully ordered, we assumed that the exchange field scales like a Brillouin function. In figure \ref{LDtemperature} we show the linear polarized spectra for a temperature range of 77 K to 400 K.

If we compare the temperature dependence of the linear polarized spectra in figure \ref{LDtemperature} with the temperature dependence of the isotropic spectrum shown in figure \ref{Isotropic} we find some remarkable similarities. The intensity of the first peak, found at 777 K, goes down when the temperature is increased for the isotropic spectrum. This also happens for the spectra taken with $\theta=20^{\circ}$ on the CoO between MnO sample and for the spectra taken with $\theta=90^{\circ}$ on the CoO on Ag sample. If one looks more carefully one will notice that the effect is even larger for these thin films. The intensity of the first peak goes up with increase of temperature between 77 K and 300 K for the CoO sample sandwiched by MnO taken at $\theta=90^{\circ}$ and for the CoO on Ag sample taken at $\theta=20^{\circ}$. Surprisingly, above 300 K this upward trend is broken and also for the CoO on Ag spectra taken at $\theta=20^{\circ}$ and the CoO sandwiched by MnO spectra taken at $\theta=90^{\circ}$ the intensity of the first peak goes down when the temperature is increased above 300 K. All these changes can be understood by using the argumentation given in the previous paragraphs. At low temperature only the state with \~J = $\frac{1}{2}$ is occupied. At higher temperatures also the state with \~J = $\frac{3}{2}$ becomes occupied. As can be seen in figure \ref{IsotropicExplanation} the state with \~J = $\frac{3}{2}$ has less intensity at the first peak than the state with \~J = $\frac{1}{2}$. When an exchange field is included, the states with different values of \~J start to mix in order to create a non-negative spin momentum. However, there still will be many (12) excited states that, with increase of temperature, start to get populated. The general trend for the polarized temperature-dependent intensity of the spectra is equal to the general trend of the temperature-dependent isotropic spectra. As we will show later this is a direct consequence of the fact that the exchange field (12.6 meV) is much smaller than the spin-orbit coupling constant (66 meV). There are a few places in the spectra where the temperature dependence of the polarized spectra is different from the temperature dependence of the isotropic spectra. First of all the intensity of the first peak goes down faster with temperature if the polarization is parallel to the spin direction, compared to the isotropic spectra. Second the intensity measured with the polarization of the first peak perpendicular to the spin direction goes up with increase of temperature up to the N\'eel temperature and than starts to fall again. In order to understand this temperature behavior we have to look at the effect of an exchange field on the orbital ordering. If the spins become orientated due to an exchange field, the orbital momentum will follow, in order to maximize the spin-orbit coupling. This means that the hole in the $t_{2g}$ orbital will not be spread out evenly. The orbital occupation depends on the exchange field, as shown in figure \ref{LDexchangestrenght}. If one now increases the temperature from 77 K to 300 K the exchange-field decreases. This will also decrease the ordered orbital momentum and the orbital occupation. Therefore the intensity of the first peak, taken with the polarization perpendicular to the spin direction, will increase when the temperature is increased from 77 to 300 K. For temperatures larger than 300K the intensity will decrease again, as states with \~J = $\frac{3}{2}$ start to become thermally populated.

\begin{SCfigure}[][h]
     \includegraphics[width=60mm]{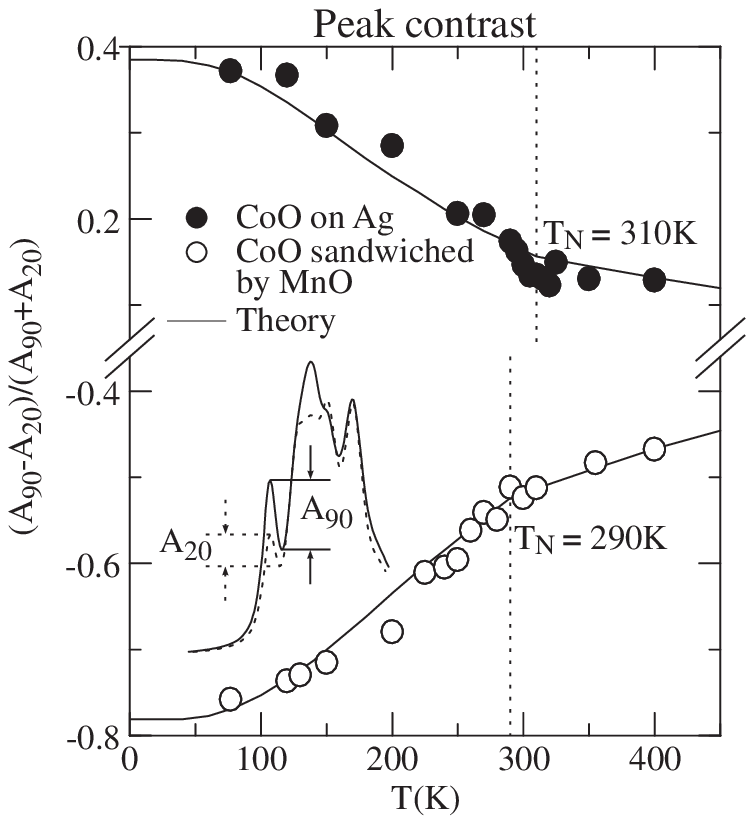}
     \caption{Temperature dependence of the polarization peak contrast, defined as the difference divided by the sum
     of the height of the first peak at $h\nu$ = 775 eV, taken with
     $\theta=20^{\circ}$ and $\theta=90^{\circ}$ polarizations. Filled and empty
     circles are the experimental data points. The solid lines are the
     theoretical simulations.\newline}
     \label{LDpeakAratio}
\end{SCfigure}

This we can make more quantitative by comparing the polarization-dependent peak contrast measured at the first peak at 777 eV, which we will call peak A, as a function of temperature. In figure \ref{LDpeakAratio} we show this contrast for temperatures between 77 K and 400 K. We see that there is a kink in the line around 300 K, which can be related to the N\'eel temperature. Below the N\'eel temperature the exchange field increases, which enlarges the absolute value of the polarization-dependent peak contrast in peak A. Furthermore we see that the polarization contrast does not go to zero above the N\'eel temperature, which is a direct consequence of the tetragonal distortions present in these films due to the strain induced by the substrate. We finally also note that the polarization contrast above the N\'eel temperature is still temperature dependent. This does not mean that the crystal fields are changing, but is a consequence of the population of higher excited states. These states exist due to the spin-orbit coupling and due to the crystal field. The interaction between spin-orbit coupling and crystal fields will be discussed more in the next section.

\begin{SCfigure}[][h]
     \includegraphics[width=60mm]{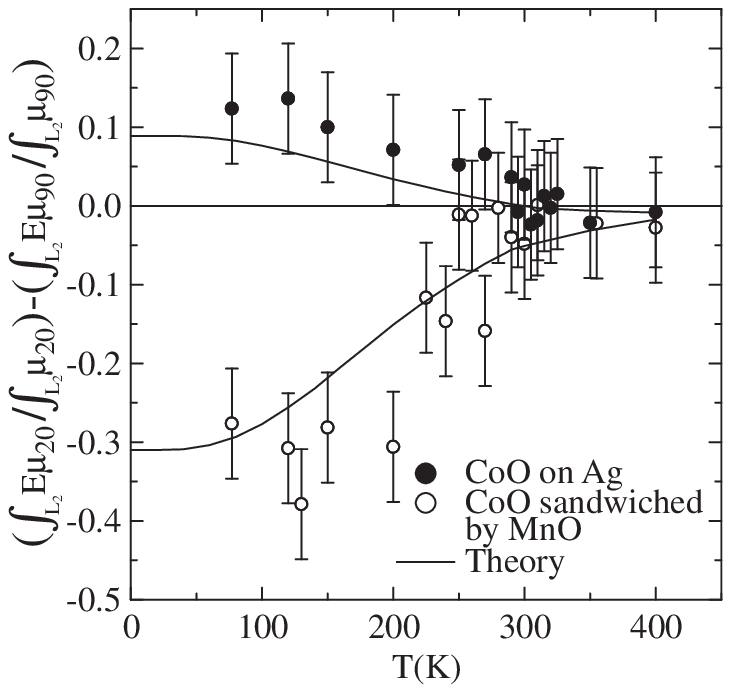}
     \caption{Temperature dependence of the polarization-dependent energy shift of the $L_2$ edge as defined as $\frac{\int_{L_2} E \mu^{(\theta=20^{\circ})}}{\int_{L_2} \mu^{(\theta=20^{\circ})}} - \frac{\int_{L_2} E \mu^{(\theta=90^{\circ})}}{\int_{L_2}\mu^{(\theta=90^{\circ})}}$. Filled and empty
     circles are the experimental data points. The solid lines are the
     theoretical simulations.\newline}
     \label{LDL2energy}
\end{SCfigure}

We can also look at the polarization-dependent energy position of the $L_2$ edge. We therefore plotted in figure \ref{LDL2energy} the integral $\frac{\int_{L_2} E \mu^{(\theta=20^{\circ})}}{\int_{L_2} \mu^{(\theta=20^{\circ})}} - \frac{\int_{L_2} E \mu^{(\theta=90^{\circ})}}{\int_{L_2}\mu^{(\theta=90^{\circ})}}$. This integral is a measure for the polarization-dependent energy shift of the $L_2$ edge and does not depend on the absolute energy calibration or resolution of the monochromator. We see a similar behavior for the polarization-dependent energy position of the $L_2$ edge as found for the peak contrast of peak A. With one difference, the contrast in peak A is mainly given due to the orbital occupation. The contrast of peak A is largely sensitive to crystal field and for exchange fields picks up the induced orbital momentum that starts to order. The polarization-dependent energy position of the $L_{2}$ edge is mainly sensitive to the spin ordering. This opens up great opportunities for combined measurements of core-level spectroscopy and microscopy (PEEM). With this technique one can look at different domains without averaging over these domains. Therefore accurate determination between the spin, orbital momentum and the crystal axis of each domain can be obtained.

\begin{SCfigure}
     \includegraphics[width=60mm]{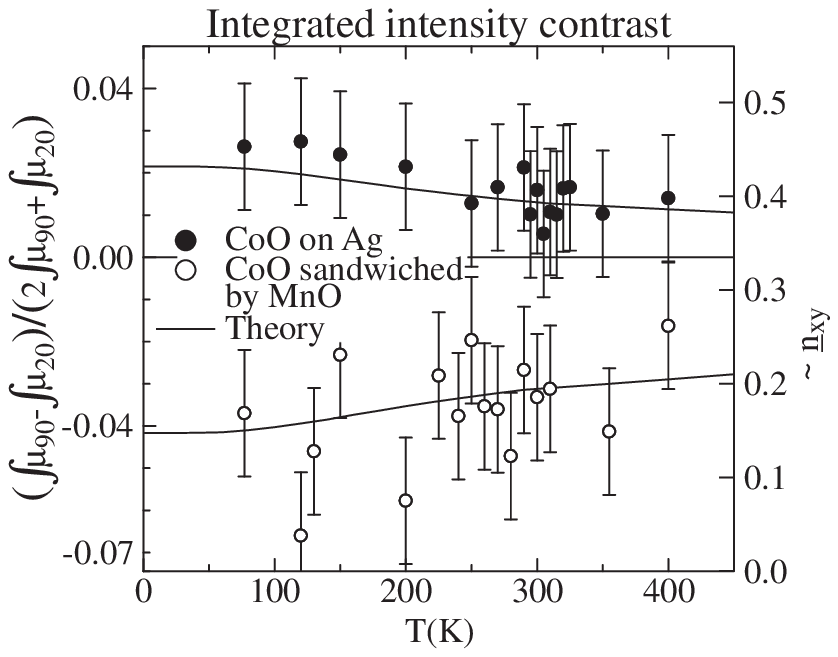}
     \caption{Temperature dependence of the polarization integrated intensity contrast, defined as the difference
     divided by the sum of the intensity, integrated over the entire
     $L_{2,3}$ range, taken with the two polarizations. Filled and empty
     circles are the experimental data points. The solid lines are the
     theoretical simulations.\newline}
     \label{LDsumrule}
\end{SCfigure}

For linear polarized light there have been sum-rules derived, relating the total integrated intensity contrast over the $L_{2,3}$ edges to the operator value of the quadrupole moment, $Q_{zz}$. \cite{Carra93b} These sum-rules are extremely powerful while they are so simple in use and generate a quantitative number for the quadrupole moment. For the cuprates, $d^9$ systems, it has been shown that the quadropole moment is directly related to the question if the hole is in the $d_{z^2}$ or the $d_{x^2-y^2}$ orbital \cite{Regueiro95}. With the following argumentation we will show that this is true more general and that the linear polarized integrated intensity contrast over the $L_{2,3}$ edges can be related to the occupied orbitals.

The simplest example is an excitation from an $s$ orbital to a $p$ orbital. The excitation probability is proportional to the square of the integral $\langle s|q|p_i\rangle$ where $q$ can be x,y, or z depending on the polarization and $i$ can be x,y, or z depending on the $p$ orbital to which the $s$ orbital is excited to. The integrand has to be even. An $s$ orbital is always even. So for x-polarized light, one can excite into $p_x$ holes, for y-polarized light one can excite into $p_y$ holes and for z-polarized light one can excite into $p_z$ holes. We know that the total integrated intensity for x plus y plus z polarized light is proportional to the total number of holes in the $p$ shell. If we now express the integral over the intensity taken with one polarization divided by the integral over the spectra taken with all polarizations we find for an $s$ to $p$ excitation.
\begin{eqnarray}
\frac{\int_{K} \mu^x}{\int_{K} \mu^x + \mu^y + \mu^z}=\frac{\underline{n}_x}{\underline{n}}\nonumber\\
\frac{\int_{K} \mu^y}{\int_{K} \mu^x + \mu^y + \mu^z}=\frac{\underline{n}_y}{\underline{n}}\nonumber\\
\frac{\int_{K} \mu^z}{\int_{K} \mu^x + \mu^y + \mu^z}=\frac{\underline{n}_z}{\underline{n}}
\end{eqnarray}
Where $\underline{n}$ denotes the total number of holes present and $\underline{n}_i$ denotes the number of holes in orbital i. The prefactors, in this case 1, are chosen such that the sum of the three normalized integrals, for x, y and z polarization is 1. This has to be the case as it is given by our normalization to the isotropic spectrum.

We can do the same thing for a $p$ to $d$ excitation. For $2p$ to $3d$ excitations the electron can be excited from three different $2p$ orbitals to five different $3d$ orbitals. We have to find all combinations of $p$ orbital, polarization and $d$ orbital for which the integrand of the integral $\langle p_i|q|d_j\rangle$ is even. For $p$ to $t_{2g}$ excitations we already showed that this means that for z polarized light one can excite a $p_x$ orbital to the $d_{xz}$ orbital and the $p_y$ orbital to the $d_{yz}$ orbital and that the $d_{xy}$ orbital can not be reached. For the $e_{g}$ shell one can excite the $p_{z}$ orbital into the $d_{z^2}$ orbital with z polarized light but can not excite into the $d_{x^2-y^2}$ orbital. The integrand of the integral $\langle p_z|z|d_{x^2-y^2}\rangle$ is even in x, y and z, but changes sign when x and y are interchanged. For x and y polarization, one can use cyclic permutation of the orbitals in order to obtain to which orbital one can excite. Combining these argumentations, we arrive at the following formulas.
 \begin{eqnarray}
\frac{\int_{L_{2,3}} \mu^x}{\int_{L_{2,3}} \mu^x + \mu^y + \mu^z}=\frac{\frac{1}{2}\underline{n}_{xy}+ \frac{1}{2}\underline{n}_{xz}+ \frac{2}{3}\underline{n}_{x^2}}{\underline{n}}\nonumber\\
\frac{\int_{L_{2,3}} \mu^y}{\int_{L_{2,3}} \mu^x + \mu^y + \mu^z}=\frac{\frac{1}{2}\underline{n}_{xy}+ \frac{1}{2}\underline{n}_{yz}+ \frac{2}{3}\underline{n}_{y^2}}{\underline{n}}\nonumber\\
\frac{\int_{L_{2,3}} \mu^z}{\int_{L_{2,3}} \mu^x + \mu^y + \mu^z}=\frac{\frac{1}{2}\underline{n}_{xz}+ \frac{1}{2}\underline{n}_{yz}+ \frac{2}{3}\underline{n}_{z^2}}{\underline{n}}
\label{CoOsumrules}
 \end{eqnarray}
Where the prefactors $\frac{1}{2}$ and $\frac{2}{3}$ are chosen such that the sum of these three integrals, $\frac{\frac{1}{2}\underline{n}_{xy}+ \frac{1}{2}\underline{n}_{xz}+ \frac{2}{3}\underline{n}_{x^2}}{\underline{n}}+ \frac{\frac{1}{2}\underline{n}_{xy}+ \frac{1}{2}\underline{n}_{yz}+ \frac{2}{3}\underline{n}_{y^2}}{\underline{n}}+ \frac{\frac{1}{2}\underline{n}_{xz}+ \frac{1}{2}\underline{n}_{yz}+ \frac{2}{3}\underline{n}_{z^2}}{\underline{n}}=1$, which it has to be, since we normalized our integrals over the spectra in that way. One has to consider that the $d_{x^2}$ orbital is not orthogonal to the $d_{z^2}$ orbital and in tetragonal symmetry it is better to rewrite the $d_{x^2}$ orbital as a linear combination of the $d_{z^2}$ and the $d_{x^2-y^2}$ orbital. The number operator $\underline{n}_i$ can be written as a creation and annihilation operator, $\underline{n}_i=d_id_i^{\dagger}$, resulting in the total expression:
\begin{eqnarray}
\underline{n}_{x^2}=\frac{1}{4}\underline{n}_{z^2}+ \frac{3}{4}\underline{n}_{x^2-y^2}- \sqrt{\frac{3}{16}}(d_{z^2}d_{x^2+y^2}^{\dagger}+c.c.))\nonumber\\ \underline{n}_{y^2}=\frac{1}{4}\underline{n}_{z^2}+ \frac{3}{4}\underline{n}_{x^2-y^2}+ \sqrt{\frac{3}{16}}(d_{z^2}d_{x^2+y^2}^{\dagger}+c.c.))
\end{eqnarray}
In tetragonal symmetry, the expectation values of the operator $d_{z^2}d_{x^2+y^2}^{\dagger}$ is exactly zero. In lower symmetry, even in $D_{2}$ where the x,y and z axe are still orthogonal to each other, the operator $d_{z^2}d_{x^2+y^2}^{\dagger}$ becomes important. This operator counts the mixing between the $d_{z^2}$ and the $d_{x^2-y^2}$ orbital.

These sum-rules are nice, since we now can exactly quantify where the holes are. This is important, since for CoO it will directly relate to the direction and size of orbital momentum present. The sum-rules relate to the holes in the $t_{2g}$ shell as well as to the holes in the $e_{g}$ shell. We are only interested in the hole of the $t_{2g}$ shell. Let us therefore make a few assumptions for the ground-state of CoO. First of all we assume that the Co$^{2+}$ ion has 5 electrons in the $t_{2g}$ shell and 2 electrons in the $e_{g}$ shell. Since the Co$^{2+}$ ion is in the high-spin state we assume that the holes are equally spaced over the $e_{g}$ orbitals, even when a tetragonal distortion is induced, therefore we have that $\underline{n}_{z^2}=\underline{n}_{x^2-y^2}=\underline{n}_{x^2}=\underline{n}_{y^2}=1$. Using these approximations we can simplify equation \ref{CoOsumrules} to relate to the occupation of only one $t_{2g}$ orbital, where we will choose the $d_{xy}$ orbital. In tetragonal symmetry we know that $\underline{n}_{xz}=\underline{n}_{yz}$ and since we only have one hole in the $t_{2g}$ orbital we have $\underline{n}_{xz}+\underline{n}_{yz}=1-\underline{n}_{xy}$. Furthermore we know from our approximation that $\underline{n}=3$. Plugging all these numbers in equation \ref{CoOsumrules} we get:
\begin{equation}
\frac{\int \frac{1}{2}(\mu^{x}+\mu^{y})-\int \mu^{z}}{\int \mu^{x} + \mu^y + \mu^{z}} =\frac{1}{4}\underline{n}_{xy}-\frac{1}{12}
\end{equation}

In figure \ref{LDsumrule} we show the integrated intensity contrast for our two different films as a function of temperature. We also show the integrated intensity contrast as obtained from our calculated spectra. On the right hand side we now also show a scale where one can read of the number of holes in the $d_{xy}$ orbital obtained from the sum-rule. Above the N\'eel temperature, one sees that for CoO on Ag there are more then $\frac{1}{3}$ holes in the $d_{xy}$ orbital. This is directly related to the in-plane contraction of the CoO on Ag crystal structure. For the CoO sandwiched by MnO there are less than $\frac{1}{3}$ holes in the $d_{xy}$ orbital. This is what one would expect while in CoO sandwiched by MnO there exists an in plane elongation. One can also see that with increase of temperature the number of holes in the $d_{xy}$ orbital becomes closer to $\frac{1}{3}$. This is due to population of higher excited states that are present because of the crystal fields and the spin-orbit coupling. If the temperature would have been raised high enough to populate all states equally the number of holes in the $d_{xy}$ orbital would reduce again to $\frac{1}{3}$. One can also see that there is a \textit{kink} in the theoretical value of the number of holes in the $d_{xy}$ orbital at the N\'eel temperature. An exchange field will introduce spin ordering. Spin ordering will order the orbital momentum due to spin orbit coupling and a state with orbital momentum does not have all the $t_{2g}$ orbitals equally occupied. What should be noted is that for a system with a full magnetic orbital momentum of $1\mu_B$ in the z direction due to the $t_{2g}$ orbitals one would expect that the hole is in the $d_{1}=\sqrt{\frac{1}{2}}(-d_{xz}-{\rm{i}}d_{yz})$ orbital. In other words we would expect no holes in the $d_{xy}$ orbital anymore. For the CoO film sandwiched by MnO the number of holes in the $d_{xy}$ orbital goes down to 0.18 at low temperatures and for the CoO on Ag the number of holes in the $d_{xy}$ orbital goes up to 0.44 at low temperatures. Clearly indicating that spin-orbit coupling is larger, or approximately of the same order, than the crystal field distortions we have made.

\sectionmark{Spin direction in thin films of CoO under tensile or...}
\section[Spin direction in thin films of CoO under tensile or compressive strain]{Spin direction in thin films of CoO under tensile or compressive strain \cite{Csiszar05a}.}
\sectionmark{Spin direction in thin films of CoO under tensile or...}

By fitting the magnetic linear dichroism we have observed that CoO films with an in-plane compression (CoO on Ag) have their spin in the plane of the film whereas CoO films with an in-plane elongation (CoO on MnO) have there spins perpendicular to the plane of the thin film. This can be understood very nicely within one-electron theory as we will show in this section. Within the next section we will discuss the properties of CoO within the framework of many-electron theory. The advantage of many-electron theory above one-electron theory is that it allows us to derive quantitative numbers for the ordered magnetic orbital moment, ordered magnetic spin moment and the single ion anisotropy.

The spin orientation of CoO sandwiched by MnO is the easiest to understand. The CoO octahedral is tetragonal distorted and the in-plane Co--O distance is larger than the out-off-plane Co--O distance. This results in a crystal field where the $d_{xy}$ orbital is 56 meV lower in energy then the $d_{xz}$ and $d_{yz}$ orbital. There is one hole in the $t_{2g}$ orbitals. The cheapest orbital to place this hole in will be the $d_{xz}$ or $d_{yz}$ orbital. One is normally used to talk about real orbitals, but these orbitals can be recombined to spherical harmonics, which have an orbital momentum. The $d_1$ orbital can be written as $d_1=\sqrt{\frac{1}{2}}(-d_{xz}-{\rm{i}} d_{yz})$ and the $d_{-1}$ orbital can be written as $d_{-1}=\sqrt{\frac{1}{2}}(d_{xz}-{\rm{i}} d_{yz})$. With the use of the $d_{xz}$ and the $d_{yz}$ orbital only an orbital momentum in the z direction can be created and not in the x or y direction. The spin (of each electron) wants to order anti-parallel to the orbital momentum (of that same electron). The two $t_{2g}$ electrons of Co ion with spin down will therefore be in the $d_{xy}$ orbital and the $d_{1}$ orbital. This results in an orbital momentum of $1 \mu_B$ in the z direction and a spin in the z direction.

For CoO on Ag the $d_{xy}$ orbital is raised by 18 meV with respect to the $d_{xz}$ and $d_{yz}$ orbital. On first sight one would think that this would result in double occupation of the $d_{xz}$ and $d_{yz}$ orbital and a hole in the $d_{xy}$ orbital. There would be no orbital momentum and the spin is free to orientate to any direction it likes. This is not fully true. The crystal field is only 18 meV whereas the spin-orbit coupling constant is equal to 66 meV. Spin orbit coupling is more important than the crystal field and we should have a look at the spherical harmonics as a basis. If we would place the hole in the $d_{-1}$ orbital, spin-orbit coupling is \textit{happy}, but the crystal field is fully \textit{unhappy}. The $d_{xy}$ orbital is fully occupied, and raised in energy by the crystal field. So orbital momentum and thus spin in the z direction is not a good choice. We can make things better by placing the spin and orbital momentum in the x direction. For that it would be better to quantize also around the x axis. Thereby the tetragonal $C_4$ axis stays the z axis. This is more easy to realize than one might think. A cyclic permutation of the coordinates is enough. Replace all x'es by y's all y's by z's and all z's by x'es, but leave the $d_{xy}$ orbital the orbital in which the holes want to be. Then we will find that the $d_{-1}^x$ orbital, an orbital with a magnetic orbital momentum of $-1 \mu_B$ in the x direction, is build up from $d_{-1}^x=\sqrt{\frac{1}{2}}(d_{yx}-{\rm{i}} d_{zx})$. If we place the hole in this orbital we have half a hole in the $d_{xy}$ orbital and made the spin-orbit interaction completely \textit{happy}.

CoO under tensile in-plane strain will have the spin out-of-plane and CoO under compressive in-plane strain will have the spin in the plane of the thin film. These findings are extreme important for understanding the mechanism(s) responsible for exchange-biasing. It is well agreed that the interface between ferromagnet and antiferromagnet is important for the functionality of exchange-biassing. It is not well studied what happens at this interface and strain might reorientate the spins of the antiferromagnet at the interface completely as we have shown. These findings, however, also give great opportunities. By applying strain to the antiferromagnet one can control the spin direction and choose whether one wants to have a system with the spins parallel to the surface of the film plane or perpendicular to the surface of the film plane.

\section{Electronic structure of CoO}

We have done polarization-dependent $2p$ XAS on CoO bulk samples, and thin films. Using cluster fits we have been able to derive the important parameters for CoO. Knowing these parameters, we now can derive other useful information about CoO and its electronic structure. CoO is a strongly correlated insulator with a band-gap of about 2.6--3.0 eV\cite{Lee91, Parmigiani99}. There is an energy difference of about 1.1 eV between the $t_{2g}$ and $e_{g}$ electrons. The system is not very covalent as only 7.12 electrons reside per Co atom in a CoO$_6$ cluster calculation.

\begin{SCfigure}[][h]
     \includegraphics[width=60mm]{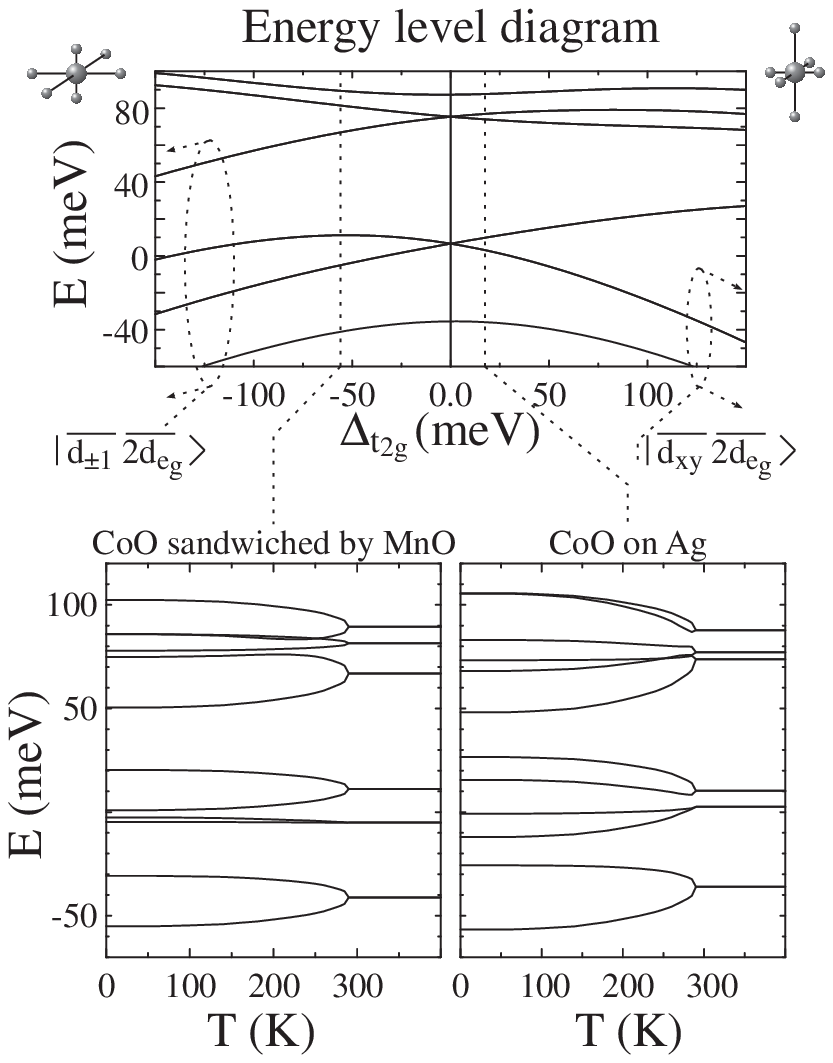}
     \caption{Energy level diagram of the twelve lowest states of CoO. Top panel: Energy level diagram as a function of tetragonal distortion. Bottom left panel: Energy level diagram for crystal fields of CoO sandwiched by MnO as a function of temperature, where below the N\'eel temperature the exchange field is changed with temperature. Bottom right panel: same but for crystal fields belonging to CoO on Ag.\newline}
     \label{EnergyLevelDiagram}
\end{SCfigure}

The Co ions have a formal valence of $d^7$, which means that in a simple picture there are 5 electrons with spin up and 2 electrons with spin down in the $t_{2g}$ shell. In other words we have one hole in the $t_{2g}$ shell and therefore a threefold orbital degeneracy. We have locally a spin $\frac{3}{2}$ system and thus a fourfold spin degeneracy. Combining both gives a twelvefold totally degenerated ground-state. This degeneracy is lifted due to spin-orbit coupling, crystal fields of lower than cubic symmetry and exchange fields. The hardest thing about CoO is that none of these interactions is so much larger than the others that one can introduce the largest one first and treat the rest within perturbation theory. Spin-orbit coupling is the largest of the three interactions and we will start with its influence. 

In figure \ref{EnergyLevelDiagram} we show the energy level diagram of the 12 lowest eigen-states of a CoO cluster. We start with the top graph. Here we show the energy level diagram as a function of the $t_{2g}$ crystal-field splitting. We start in the middle where $\Delta_{t_{2g}}=0$. Here we find the energy level diagram of cubic CoO. We see that the 12 states are split due to spin orbit coupling into a ground-state with \~J = $\frac{1}{2}$, a first excited state with \~J = $\frac{3}{2}$ and a highest excited state with \~J = $\frac{5}{2}$. There is a small splitting in the state with \~J = $\frac{5}{2}$ into a lower lying quartet and higher lying doublet. This should be there, already on the basis of group theory, but we do not want to discuss its presence in this chapter.

If we now look what happens when we introduce a tetragonal distortion, it is good to first have a look at the extreme cases. If we make an in-plane elongation then $\Delta_{t_{2g}}$ becomes negative and the $d_{xy}$ orbital is lowered in energy. The hole in the $t_{2g}$ shell will therefore be in the $d_{xz}$ or $d_{yz}$ orbital. Due to spin orbit coupling, these will recombine to a $d_{1}$ and $d_{-1}$ orbital. The magnetic orbital momentum of these states can be parallel or anti-parallel to the magnetic total spin momentum of $\pm\frac{3}{2}$ or $\pm\frac{1}{2}$. In total this will give four twofold degenerate states that go down with decrease of $\Delta_{t_{2g}}$. The ground-state in the extreme case of $-\Delta_{t_{2g}}\gg\zeta$ would consist of a wave-function with a hole in the $d_{-1}$ orbital and a total magnetic spin momentum of $\frac{3}{2}$ or a hole in the $d_{1}$ orbital and a total magnetic spin momentum of $-\frac{3}{2}$. These two states form together a Kramers doublet. The four twofold degenerate states coming down with a tetragonal in-plane elongation can be seen as four lines going down on the left in the top graph of figure \ref{EnergyLevelDiagram}. On the right side of the top graph in figure \ref{EnergyLevelDiagram} we show what happens when one makes an in-plane contraction. For an in-plane contraction $\Delta_{t_{2g}}$ becomes positive and the energy of the $d_{xy}$ orbital goes up. Therefore in the extreme case that $\Delta_{t_{2g}}\gg\zeta$ the $t_{2g}$ hole will be in the $d_{xy}$ orbital and the orbital momentum is totaly quenched.

In the top graph of figure \ref{EnergyLevelDiagram} we show how the different energies of the 12 eigen-states evolve when $\Delta_{t_{2g}}$ is changed from $-150$ to plus 150 meV. The actual parameters found for our thin films of CoO are $\Delta_{t_{2g}}=-56$ meV for CoO sandwiched by MnO which is about 4\% expanded in-plane ($a_{\parallel}\approx 4.424$\AA). For CoO on Ag which is slightly compressed in-plane ($a_{\parallel}\approx4.235$ \AA, $a_{\perp}\approx4.285$ \AA) we find $\Delta_{t_{2g}}=18$ meV. These values for the crystal field are somewhat smaller than the spin-orbit coupling energy scale. For the $\Delta_{t_{2g}}$ values of these distortions, marked by the dotted line in the top graph of figure \ref{EnergyLevelDiagram}, one can still recognize the doublet, quartet and sextet energy levels set up due to spin-orbit coupling. For the CoO sandwiched by MnO with a crystal field of $-56$ meV, one can also already recognize a lower lying eightfold degenerate state with the hole in the $d_1$ or $d_{-1}$ orbital (four separate lines) and a higher lying quartet with the hole in the $d_{xy}$ orbital (two separate lines).

The effect of an exchange field on the energy level diagram can be seen in the bottom part of figure \ref{EnergyLevelDiagram}. We did not plot the energy level diagram as a function of exchange field, but as a function of temperature. The exchange field changes with temperature according to a J = $\frac{3}{2}$ Brillioun function. We set the exchange field at 0 K to be 12.6 meV and scaled the temperature scale in order to reproduce the experimental N\'eel temperatures of 290 K for the CoO sandwiched by MnO and 310 for CoO on Ag. On the bottom left of figure \ref{EnergyLevelDiagram} we show the energy level diagram as a function of temperature for CoO sandwiched by MnO and on the bottom right we show the energy level diagram for CoO on Ag. The crystal field is -56 meV and 18 meV respectively. An exchange field splits the Kramers doublets into a state with spin up and spin down. However, not all states are split in the same way. Thereby an exchange field also mixes the states and should not be treated as a perturbation on only the \~J = $\frac{1}{2}$ states.

\begin{center}
\begin{table}
\label{TableCoOvalues}
\begin{center}
\begin{tabular}{|c|c|c|c|}
  \hline
    & CoO sandwiched & CoO on Ag & bulk CoO \\
    &  by MnO        &           &          \\
  \hline
  $\zeta$              &  66 meV &  66 meV & 66 meV\\
  $\Delta_{t_{2g}}$    & $-56$ meV &  18 meV & $\approx-25$ meV\\
  $H_{ex}$             & 12.6 meV& 12.6 meV& 12.6 meV\\
  \hline
  $M_L$                & 1.36 $\mu_B$& 1.00 $\mu_B$&*\\
  $M_S$                & 2.46 $\mu_B$& 2.14 $\mu_B$&*\\
  $M_T$                & 0.04 $\mu_B$& 0.02 $\mu_B$&*\\
  $\sum_i\langle l_i.s_i\rangle$    &$-1.25$ $\mu_B^2$&$-1.28$ $\mu_B^2$& $-1.28$ $\mu_B^2$\\
  $\underline{n}_{xy}$ & 0.18 & 0.44 & *\\
  $E_{H_{ex}^{z}}-E_{H_{ex}^{(x,y)}}$&-4.8 meV&1.6 meV & -2.3 meV\\
  \hline
\end{tabular}
\end{center}
\caption{Crystal-field, Exchange field and spin-orbit coupling parameter for CoO sandwiched by MnO, CoO on Ag and bulk CoO at 0K. As well as the magnetic orbital and spin momentum, the expectation value of the magnetic dipole operator, the expectation value of the spin-orbit operator, the number of holes in the $d_{xy}$ orbital and the single-ion anisotropy $E_{H_{ex}^{z}}-E_{H_{ex}^{(x,y)}}$. \newline
         $^*$These values depend crucially on the direction of the spin.
}
\end{table}
\end{center}

In table \ref{TableCoOvalues} we list the most important operator values for the ground-state at 0 K of our thin films, found from our cluster calculations. The number of holes in the $d_{xy}$ orbital as calculated from our cluster calculation is fully consistent with the values found from the sum-rules. The magnetic orbital momentum is surprisingly large. Even larger than the maximum of $1\mu_B$ that one can have for a single $t_{2g}$ electron. The reason for this lies in the presence of electron-electron repulsion. We assumed that the Co ion has two holes in the $e_g$ shell and one hole in the $t_{2g}$ shell. This is not completely true. Electron-electron repulsion is minimized when there are about 1.2 holes in the $t_{2g}$ shell and 1.8 holes in the $e_g$ shell. If one increases the splitting between the $t_{2g}$ and $e_{g}$ shell, electrons move from the $e_{g}$ shell to the $t_{2g}$ shell in order to minimize the crystal field energy, but thereby paying energy due to electron-electron repulsion. More exactly, in spherical symmetry there are two terms with $S=\frac{3}{2}$. One term has $L=3 (F)$ and is the ground-state in spherical symmetry, the other term has $L=2 (P)$. When one introduces a cubic crystal field, the $^4F$ term is split into a lowest $T_1$ level, a higher lying $T_2$ level and an even higher lying $A_1$ level. The $4P$ term does not split when a cubic crystal field is introduced, but its symmetry becomes $T_1$, the same as the ground-state. The $A_1$ level has three holes in the $t_{2g}$ shell. The $T_2$ level has two holes in the $t_{2g}$ shell and one hole in the $e_{g}$ shell. The $T_1$ level split from the $^4F$ term, has for an infinitesimal small crystal field, about 1.2 holes in the $t_{2g}$ shell and 1.8 holes in the $e_{g}$ shell. The $T_1$ level that originates from the $^4P$ term, has for an infinitesimal small crystal field, about 1.8 holes in the $t_{2g}$ shell and 1.2 holes in the $e_{g}$ shell. If the splitting between the $t_{2g}$ shell and the $e_{g}$ shell is made large, with respect to the electron-electron repulsion energy difference between the $^4F$ and $^4P$ terms (about 2 eV), the $T_1$ level originating from the $^4P$ term will mix in for about 20\% to the ground-state, resulting in an orbital occupation of one hole in the $t_{2g}$ shell and two holes in the $e_g$ shell. Covalency makes things a bit more complicated, since when covalency is included the total number of electrons is not 7, but becomes 7.13. If we now restrict ourselves to the fitted parameters, we find that from those 7.13 electrons, 4.93 electrons are in the $t_{2g}$ shell and 2.20 are in the $e_g$ shell. The $^4P$ term has mixed in for about 10\% into the ground-state. In order to explain the large orbital momentum, we now can combine the 0.1 extra hole in the $t_{2g}$ shell with the extra electrons in the $e_{g}$ shell to give additional orbital momentum. Within the $t_{2g}$ shell we could combine the $d_{xz}$ and the $d_{yz}$ orbital to a $d_{1}$ and $d_{-1}$ orbital. When we combine the $d_{xy}$ orbital with the $d_{x^2-y^2}$ orbital we can create a $d_{2}$ or $d_{-2}$ state with magnetic orbital momentum of 2 and $-2$ respectively. We find that, for CoO sandwiched by MnO 0.72 of the total orbital momentum comes from a $d_{1}$ orbital and that 0.65 of the orbital momentum comes from a $d_{2}$ orbital. CoO is one of the few materials where mixing of the $t_{2g}$ and $e_{g}$ shell due to electron-electron repulsion is important. (Based on group theory this only happens for $d^2$ and $d^7$ configurations.) In order to get the orbital momentum correct one has to include this mixing. That this mixing is also present in the real material and is not just an artifact of a cluster calculation can be seen very nicely at the XMCD spectra. Here we find that about $\frac{2}{5}$ of the magnetic susceptibility is due to magnetic orbital momentum; an abnormally large portion. This is found from the $L_z$ sum-rule, as well as from cluster calculations.

We also show the single ion anisotropy energy in table \ref{TableCoOvalues}. The single ion anisotropy energy depends on the size of the exchange field, spin-orbit coupling and low symmetry crystal field distortions \cite{Alders01}. These parameters are deduced from our fits of the spectra and therefore we are able to calculate the single ion anisotropy energies. The single ion anisotropy has been calculated by comparing the total energy of a CoO$_6$ cluster for different assumed directions of the exchange field.

Knowing all cluster parameters for CoO thin films from our fits to experiment, we now can have a look at the properties of bulk CoO. At 0 K there is a tetragonal distortion of 1.2\%, which means that $\Delta_{t_{2g}}$ is about $-25$ meV. The parameter values and expectation values of operators can be found in table \ref{TableCoOvalues}. We first have a look at the single ion anisotropy. We find that the energy for the exchange-field in the direction of the tetragonal contraction is lower. This can be understood quite easily. Spin orientation induces an orbital momentum. A state with orbital momentum does not have all orbitals equally occupied. For an orbital momentum in the z direction the hole will be in a $d_1$ orbital $(-d_{xz}-{\rm{i}}d_{yz})$ and therefore not in the $d_{xy}$ orbital. If we have to fit this electron density cloud in a crystal structure that has one short and two large axis, the energy will be lowest if we place the filled $d_{xy}$ orbital in the direction of the two larger axis. One probably should see the tetragonal distortion in CoO the other way around. Since below the N\'eel temperature the spins order, orbital momentum is induced. This turns the spherical electron cloud into something cylindrical, that is larger in the direction perpendicular to the orbital momentum. This will introduce a tetragonal crystal deformation.

For bulk CoO neutron measurements tell us that the spins are not perfectly aligned with the tetragonal axis, but have an angle of 27.4$^{\circ}$ with this axis. \cite{Jauch01} In order to calculate this scenario we placed our exchange field in the $[\overline{0.325}\:\overline{0.325}\:0.888]$ direction. The total energy found is 0.44 meV higher than the energy for the spins orientated parallel to the tetragonal axis. One should notice that these calculations are not self-consistent. If we place an exchange field at an angle of 27.4$^{\circ}$ with the tetragonal axis we do not find as a result that the spins are also orientated with an angle of 27.4$^{\circ}$ to the tetragonal axis. We find that the angle between the spin and the tetragonal axis is 20.3$^{\circ}$ and that the angle of the orbital momentum with the tetragonal axis is 16.5$^{\circ}$. The large angles found between orbital momentum and spin and between spin and exchange field can be understood again if one realizes that the electron cloud is not spherical while there is some orbital momentum induced. The orbital momentum wants to be in the tetragonal axis direction due to the crystal field and the shape of a $d^7$ system with orbital momentum (the $d_{xy}$ orbital is double occupied and the hole is half in the $d_{xz}$ and half in the $d_{yz}$ orbital). The spin wants to be parallel to the orbital momentum due to spin-orbit coupling and the spin wants to be in the direction of the exchange field. Since non of these interactions is larger than the others, one finds a system where there is a relative large angle between the easy axis and the orbital momentum, between the orbital momentum and the spin and between the spin and the exchange field. Hereby one should realize that the exchange field is by definition parallel to the spins on the next-nearest neighboring sites. 

\section{Conclusion}

We have shown that with the use of polarized $2p$-core level spectroscopy one can deduce an enormous amount of information for CoO systems. We showed that there is a temperature dependence in the isotropic spectra due to the presence of spin-orbit coupling in the initial state. We showed how one can deduce the size of the crystal field in thin films by going above the N\'eel temperature and look at the linear dichroism that still is present at these elevated temperatures. We calculated the spectra for different values of $\Delta_{t_{2g}}$, so that other systems can be fit to these spectra. We showed how one can find out, which orbitals are occupied by using sum-rules relating the linear polarized integrated intensity contrast to the number of holes in the different orbitals. We showed that by using the linear polarized intensity contrast of the first peak at 777 eV, one can get approximate information about where the holes in the $t_{2g}$ shell are. For magnetically ordered systems this also relates to the orbital momentum. We also showed how one can relate the linear polarization dependent energy position shift of the $L_2$ edge to the approximate direction of spin momentum. We have based our cluster calculations on parameters fitted to two systems, one time CoO on Ag, with an in-plane contraction, and one time CoO sandwiched by MnO with an in-plane elongation. Our cluster fits are satisfactory and give confidence in the findings and experimental method. The orbital momentum found for CoO is surprisingly large, this can be understood quite well in a cluster approach. Thereby more confidence in this large orbital momentum is obtained by direct XMCD measurement of the orbital momentum of bulk CoO in the paramagnetic phase at 300 K. These XMCD measurements show an orbital momentum which is also extremely large and well reproduced by our cluster calculations.

It would be extremely nice to use PEEM to look at the films studied here and see if one can find different domains present in these films. One then could measure the angle between the surface normal and the spin in a more accurate way and also find the angle between spin and orbital momentum. When one finds a way to solve the problem of charging of bulk CoO at low temperatures it would be good to look at bulk CoO and determine the spin and orbital momentum direction of bulk CoO with the use of XAS.

\chapter[Aligning spins in antiferromagnetic films using antiferromagnets]{Aligning spins in antiferromagnetic films using antiferromagnets \cite{Csiszar05c}}
\label{ChapterMnO}

\begin{center}
\begin{minipage}{0.8\textwidth}
We have explored the possibility to orient spins in
antiferromagnetic thin films with low magnetocrystalline
anisotropy via the exchange coupling to adjacent antiferromagnetic
films with high magnetocrystalline anisotropy. We have used MnO
as a prototype for a system with negligible single-ion
anisotropy. We were able to control its spin direction very
effectively by growing it as a film on antiferromagnetic CoO
films with different predetermined spin orientations. This result
may pave the way for tailoring antiferromagnets with low
magnetocrystalline anistropy for applications in exchange-bias
systems. Very detailed information concerning the
exchange-coupling and strain effects was obtained from the Mn
$L_{2,3}$ soft x-ray absorption spectroscopy.
\end{minipage}
\end{center}

The study of the exchange-bias phenomena in multilayered magnetic
systems is a very active research field in magnetism, not the
least motivated by the high potential for applications in
information technology. Various combinations of antiferromagnetic
(AFM) and ferromagnetic (FM) thin film materials have been
fabricated and intensively investigated
\cite{Nogues99,Berkowitz99}. There seems to be an agreement among
the experimental and theoretical studies that the largest
exchange-bias effects can be found in systems containing AFMs
with a high magnetocrystalline anisotropy, such as CoO. The
simple underlying idea is that the anisotropy helps to fix the
spin orientation in the AFM while switching the magnetization in
the FM.

Our objective is to explore the possibilities to control and to
pin the spin direction in AFM oxides having low
magnetocrystalline anisotropy, e.g. transition metal oxides with
the $3d^{3}$, $3d^{5}$, or $3d^{8}$ ionic configurations. If
successful, this would help to extend the materials basis for the
AFMs used in exchange-bias systems. One could then consider thin
films of not only NiO, but also LaCrO$_{3}$, LaFeO$_{3}$,
$\gamma$-Fe$_{2}$O$_{3}$, and R$_{3}$Fe$_{5}$O$_{12}$
\cite{Goodenough63}. At first sight, our chances may seem bleak
since a recent study on ultra thin NiO films reveal that the
magnetic anisotropy results from a detailed balance between the
influence of strain and thickness on the already very weak
dipolar interactions in the AFM \cite{Altieri03,Finazzi03}.
On the other hand, the few studies available in the literature on
combinations of AFM/AFM films revealed that the interlayer
exchange coupling can be very strong
\cite{Ramos90,Carrico92,Wang92,Lederman93,Borchers93,Carey93a, Carey93b,Abarra96}.
We took these findings as starting point of our work.

We have used MnO as an ideal model for an antiferromagnetic system
with negligible single-ion anisotropy. We have grown the MnO as a
thin film epitaxially on two different types of CoO single crystal
films. In one CoO film the spin direction is oriented
perpendicular to the surface, and in the other parallel to the
surface \cite{Csiszar05a}. Using soft x-ray absorption
spectroscopy at the Mn $L_{2,3}$ edges, we observed that the spin
direction of the MnO film strongly depends on the type of CoO
film the MnO is grown on, and that it is dictated by the spin
orientation of the CoO film and not by the strain or dipolar
interactions in the MnO film. Interlayer exhange coupling is thus
a very effective manner to control spin directions and may be
used for tailoring AFMs with low magnetocrystalline anistropy for
exchange-bias applications.

The actual MnO/CoO systems studied are
(14\AA) MnO/ (10\AA) CoO/ (100\AA) MnO/ Ag(001) and
(22\AA) MnO/( 90\AA) CoO/ Ag(001). The two samples were grown on a
Ag(001) single crystal by molecular beam epitaxy (MBE),
evaporating elemental Mn and Co from alumina crucibles in a pure
oxygen atmosphere of $10^{-7}$ to $10^{-6}$ mbar. The base
pressure of the MBE system is in the low $10^{-10}$ mbar range.
The thickness and epitaxial quality of the films are monitored by
reflection high energy electron diffraction (RHEED) measurements.
With the lattice constant of bulk Ag, CoO and MnO being 4.09\AA,
4.26\AA, and 4.444\AA, respectively, we find from x-ray
diffraction (XRD) and RHEED that the in-plane lattice constants
in each film are essentially given by the thickest layer 
which is almost bulk like. Compared to the bulk, the 10\AA~CoO
sandwiched by MnO is about 4\% expanded in-plane
($a_{\parallel}\approx 4.424$\AA), while the 90\AA~CoO directly
on Ag is slightly compressed in-plane
($a_{\parallel}\approx4.235$\AA, $a_{\perp}\approx4.285$\AA). The
MnO is in both samples compressed, but much more so for the one
on the 90\AA~CoO film. Details about the growth will be published
elsewhere \cite{Csiszar05b}. We have shown recently that the spin
direction is oriented perpendicular to the surface in the CoO
film under tensile in-plane stress , and that it is parallel to
the surface in the film with the slightly compressive in-plane
stress \cite{Csiszar05a}.

\begin{figure}[h]
     \includegraphics[width=120mm]{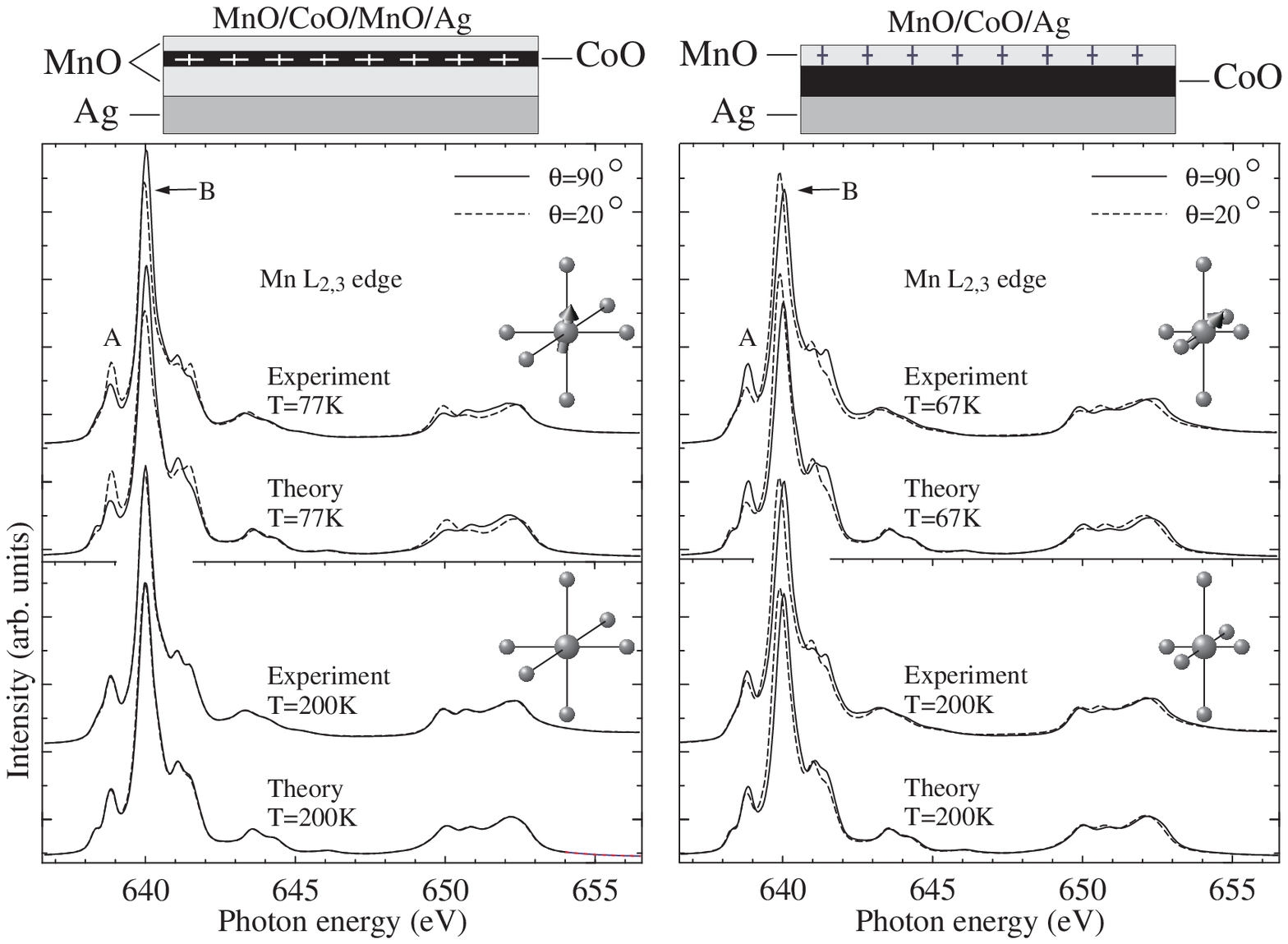}
     \caption{Experimental and calculated Mn $L_{2,3}$ XAS spectra
     of: left panel) MnO in (14\AA)MnO/(10\AA)CoO/(100\AA)MnO/Ag(001) for
     $\theta=20^{\circ}$ and $\theta=90^{\circ}$, below (top panel, T=77K) and
     above (bottom panel, T=200K) the $T_{N}$ of the MnO thin
     film; right panel) the same for MnO in (22\AA)MnO/(90\AA)CoO/Ag(001)}
    \label{MnOspectra}
\end{figure}

The XAS measurements were performed at the Dragon beamline of the
NSRRC in Taiwan using \textit{in-situ} MBE grown samples. The
spectra were recorded using the total electron yield method in a
chamber with a base pressure of $3\times10^{-10}$ mbar. The photon
energy resolution at the Mn $L_{2,3}$ edges ($h\nu \approx
635-655$ eV) was set at 0.3 eV, and the degree of linear
polarization was $\approx 98 \%$. The sample was tilted with
respect to the incoming beam, so that the Poynting vector of the
light makes an angle of $\alpha=70^{\circ}$ with respect to the
[001] surface normal. To change the polarization, the sample was
rotated around the Poynting vector axis, so that $\theta$, the
angle between the electric field vector $\vec{E}$ and the [001]
surface normal, can be varied between $20^{\circ}$ and
$90^{\circ}$ \cite{Csiszar05a}. This measurement geometry allows
for an optical path of the incoming beam which is independent of
$\theta$, guaranteeing a reliable comparison of the spectral line
shapes as a function of $\theta$. A MnO single crystal is measured
\textit{simultaneously} in a separate chamber to obtain a
relative energy reference with an accuracy of better than 0.02 eV.

Fig. \ref{MnOspectra} shows the polarization dependent Mn $L_{2,3}$ XAS spectra
of the MnO/CoO samples with CoO spin-orientation perpendicular
(left panels) and parallel (right panels) to the surface, taken at
temperatures far below (top panels) and far above (bottom panels)
the N\'{e}el temperature ($T_{N}$) of the MnO thin film, which is
about 130 K as we will show below. The spectra have been
corrected for electron yield saturation effects
\cite{Nakajima99}. The general line shape of the spectra shows
the characteristic features of bulk MnO \cite{Groot94},
ensuring the good quality of our MnO films. Very striking in the
spectra is the clear polarization dependence, which is the
strongest at low temperatures. Important is that below $T_{N}$
the dichroism, i.e. the polarization dependence, of the two
samples are opposite: for instance, the intensity of the first
peak at $h\nu$ = 639 eV is higher for $\theta=20^{\circ}$ than
for $\theta=90^{\circ}$ in MnO/CoO where the spin orientation of
the CoO is out-of-plane, while it is smaller in the other sample.
Above $T_{N}$, the dichroism almost vanishes. Nevertheless, small
but clear and reproducible shifts in the spectra as a function of
polarization can be seen: the main peak at 640 eV has a shift of
about 30 meV for the MnO sandwiching the 10\AA~CoO film, and 150
meV for the MnO overlaying the 90\AA~CoO film.

\begin{SCfigure}[][h]
\includegraphics[width=60mm]{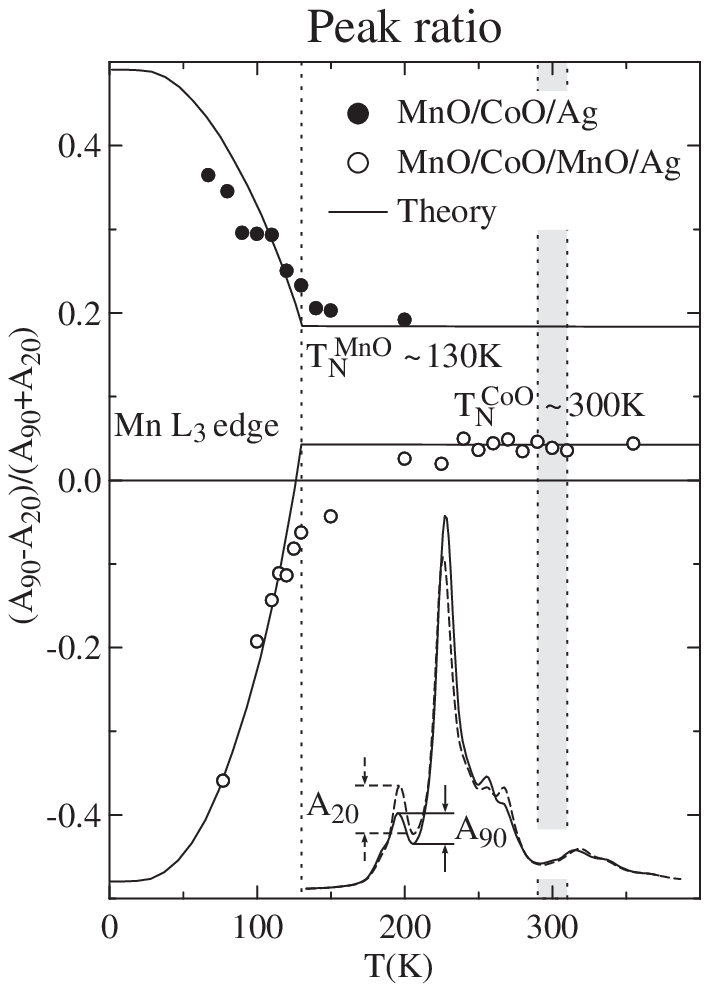}
     \caption{Temperature dependence of the polarization contrast
     in the Mn $L_{2,3}$ spectra, defined as the difference divided by the sum
     of the height of the first peak at $h\nu$ = 639 eV, taken with
     $\theta=20^{\circ}$ and $\theta=90^{\circ}$ polarizations. Filled and
     empty circles are the experimental data. The solid lines are the
     theoretical simulations. The shaded area represents the $T_{N}$
     of the CoO layers under the MnO film.\newline}
     \label{MnObranching}
   \end{SCfigure}

In order to resolve the origin of the dichroism in the spectra, we
have investigated the temperature dependence in more detail. Fig.
\ref{MnObranching} depicts the polarization contrast of the peak at $h\nu$ = 639
eV, defined as the difference divided by the sum of the peak
height in the spectra taken with the $\theta=20^{\circ}$ and
$\theta=90^{\circ}$ polarizations. In going from low to high
temperatures, one can see a significant temperature dependence
for both samples (with opposite signs), which flattens off at
about 130K, indicating the $T_{N}$ of these MnO thin films. We
therefore infer that at low temperatures the strong dichroic
signal is caused by the presence of magnetic ordering. Important
is to note that the opposite sign in the dichroism for the two
samples implies that the orientation of the magnetic moments is
quite different.

To analyze the Mn $L_{2,3}$ spectra quantitatively, we perform
calculations for the atomic-like $2p^{6}3d^{5} \rightarrow
2p^{5}3d^{6}$ transitions using a similar method as described by
Kuiper \textit{et al.} \cite{Kuiper93} and Alders \textit{et al.}
\cite{Alders98}, but now in a $D_{4h}$ point group symmetry and
including covalency. The method uses a MnO$_6$ cluster which
includes the full atomic multiplet theory and the local effects
of the solid \cite{Groot94, Tanaka94}. It accounts for the
intra-atomic $3d$-$3d$ and $2p$-$3d$ Coulomb and exchange
interactions, the atomic $2p$ and $3d$ spin-orbit couplings, the
O $2p$ - Mn $3d$ hybridization, local crystal field parameters
$10Dq$, $Ds$ and $Dt$, and a Brillouin type temperature dependent
exchange field which acts on spins only and which vanishes at
$T_N$. The calculations have been carried out using the XTLS 8.0
programm\cite{Tanaka94}.

\begin{SCfigure}[][h!]
\includegraphics[width=60mm]{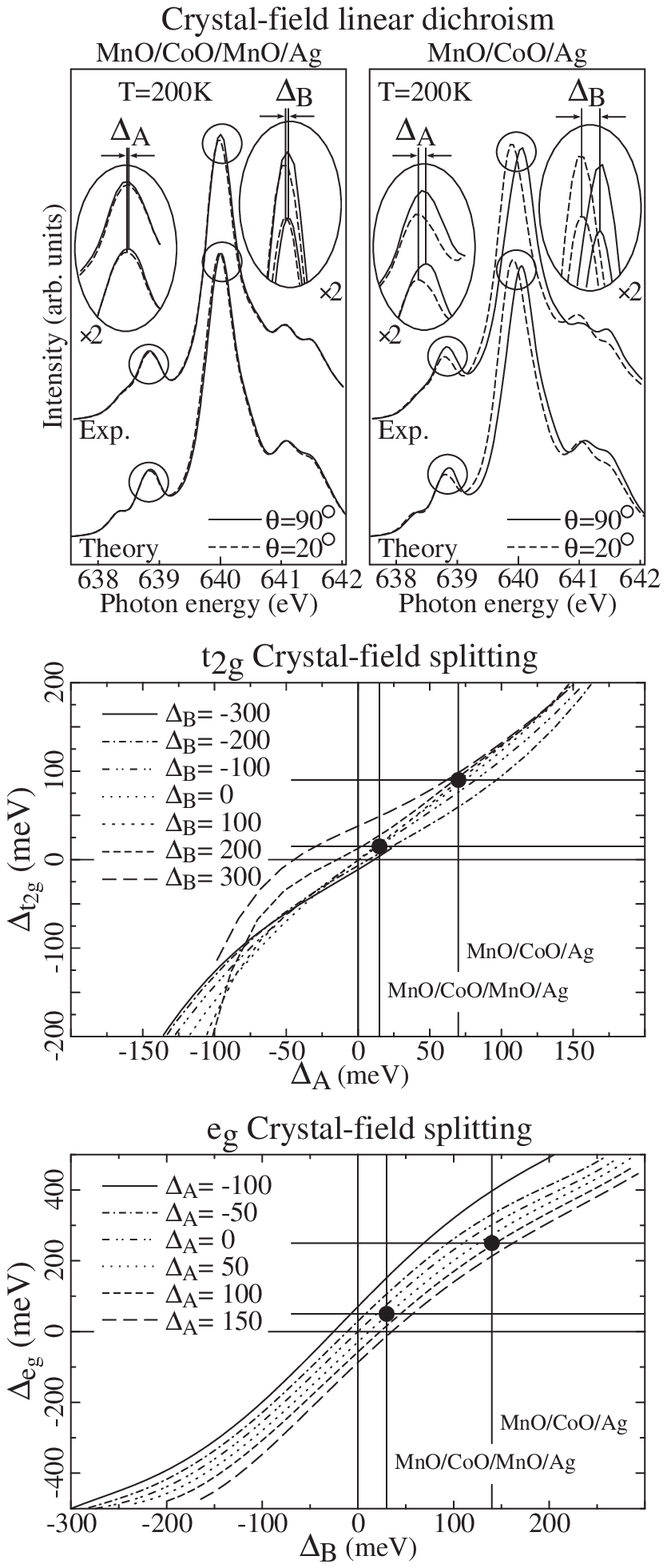}
 \caption{top panel) Close-up of the experimental and
     calculated Mn $L_{3}$ XAS spectra at 200K, i.e. above
     $T_{N}$ of (left) (14\AA) MnO/ (10\AA) CoO/ (100\AA) MnO/ Ag(001)
     and (right) (22\AA) MnO/ (90\AA) CoO/ Ag(001); $\Delta_A$ and
     $\Delta_B$ are the polarization dependent shifts in the peaks
     at 639 and 640 eV, respectively; (middle and bottom panels)
     Relationship between \{$\Delta_A$, $\Delta_B$\} and the
     tetragonal \{$\Delta_{e_g}$, $\Delta_{t_{2g}}$\} splittings.\newline}
 \label{MnOcrystal}
\end{SCfigure}

The results of the calculations are shown in Fig. \ref{MnOspectra}. We have used
the parameters already known for bulk MnO
\cite{Tanaka94,paramMnO}, and have to tune only the parameters
for $Ds$, $Dt$ and the direction of the exchange field. For the
MnO sandwiching the 10\AA~CoO we find an excellent simulation of
the experimental spectra for $Ds$ = 9.3 meV, $Dt$ = 2.6 meV and
an exchange field parallel to the [112] direction. For the MnO
overlaying the 90\AA~CoO we obtained the best fit for $Ds$ = 48.6
meV, $Dt$ = 11.1 meV and an exchange field along the [211]
direction. These two sets of parameters reproduce extremely well
the spectra at all temperatures. This is also demonstrated in
Fig. 2, showing the excellent agreement between the calculated and
measured temperature dependence of the dichroism in the first
peak. Most important is obviously the information concerning the
spin direction that can be extracted from these simulations. We
thus find that the magnetic moments in the MnO are oriented
towards the surface normal when it is grown on the CoO film which
has the spin direction perpendicular to the surface, and that it
is lying towards the surface when it is attached to the CoO film
which has the parallel alignment. In other words, it seems that
the MnO tries to follow the CoO magnetically.

In order to find out whether the spin direction in the MnO thin
films is determined by the exchange coupling with the CoO, or
whether it is given by strain and dipole interactions in the
films as found for NiO thin films on non-magnetic substrates
\cite{Altieri03,Finazzi03}, we now have to look more closely into
the tetragonal crystal fields in the MnO films. The values for
the tetragonal crystal field parameters $Ds$ and $Dt$, which we
have used to obtain the excellent simulations as plotted in Fig.
1, can actually be extracted almost directly from the high
temperature spectra, where the magnetic order has vanished and
does not contribute anymore to the polarization dependence.

The top panels of Fig. \ref{MnOcrystal} show a close-up of the spectra taken at
200K, i.e. above $T_{N}$. One can now observe the small but clear
and reproducible shifts in the spectra as a function of
polarization: the shift in the first peak at 639 eV is denoted by
$\Delta_A$ and in the main peak at 640 eV by $\Delta_B$. In order
to understand intuitively the origin of these shifts, we will
start to describe the energetics of the high spin Mn$^{2+}$
($3d^{5}$) ion in a one-electron like picture. In $O_{h}$ symmetry
the atomic $3d$ levels are split into 3 $t_{2g}$ and 2 $e_{g}$
orbitals, all containing a spin-up electron. The $L_3$ edge of 
Mn$^{2+}$ should then consist of two
peaks: in exciting an electron from the $2p$ core level to the
$3d$, one can add an extra spin-down electron either to the lower
lying $t_{2g}$ or the higher lying $e_g$ shell, producing the
peaks at 639 and 640 eV, respectively. In the presence of a
tetragonal distortion, both the $t_{2g}$ and $e_{g}$ levels will
be split. This will result in a polarization dependent energy
shifts $\Delta_{A}$ and $\Delta_{B}$, analogous as found for NiO
\cite{Haverkort04}.

Due to the intra-atomic $2p$-$3d$ and $3d$-$3d$ electron
correlation effects, the relationship between the shifts in the
spectra and the crystal field splittings become non-linear. Using
the cluster model we are able to calculate this relationship for
a Mn$^{2+}$ $3d^{5}$ system and the results are plotted in Fig. 3.
Using this map, we find for the MnO sandwiching the 10\AA~CoO
that $\Delta_{t_{2g}}$ = 15 meV and $\Delta_{e_g}$ = 50 meV, and
for the MnO overlaying the 90\AA~CoO that $\Delta_{t_{2g}}$ = 90
meV and $\Delta_{e_g}$ = 250 meV
($Ds$=($\Delta_{e_g}$+$\Delta_{t_{2g}}$)/7,
$Dt$=(3$\Delta_{e_g}$-4$\Delta_{t_{2g}}$)/35). The crystal field
splittings for the second sample are much larger than for the
first sample, fully consistent with our structural data in that
the MnO in the second sample experiences a much stronger
compressive in-plane strain. Important is now to recognize that
the crystal field splittings for the two samples have the same
sign, i.e. that both MnO films are compressed in-plane. This
implies that strain together with the dipolar interactions cannot
explain the quite different spin-orientations of the two MnO
systems. We conclude that the magnetic anisotropy mechanism
present in, for instance, NiO thin films on non-magnetic
substrates \cite{Altieri03, Finazzi03}, is overruled by the
stronger interlayer exchange coupling
\cite{Ramos90, Carrico92, Wang92, Lederman93, Borchers93, Carey93a, Carey93b, Abarra96}
between the CoO and MnO layers.

The $T_{N}$ for these thin MnO layers is found to be at about 130
K. It is surprising that it is not reduced as compared to the bulk
value of 121 K \cite{Shull51}, since generally one would expect
such to happen with decreasing thickness as was observed for NiO
on MgO \cite{Alders98}. The origin for this is not clear at this
moment. It is possible that the in-plane compressive stress gives
an increase of the Mn $3d$ - O $2p$ hybridization, which in turn
could produce an increase of the superexchange interaction
strength \cite{Goodenough63} and thus also of $T_{N}$. Another,
more exciting, possibility emerges from the recent experimental
and theoretical work on AFM/AFM multilayers such as
FeF$_{2}$/CoF$_{2}$ and CoO/NiO
\cite{Ramos90, Carrico92, Wang92, Lederman93, Borchers93, Carey93a, Carey93b, Abarra96}.
Experiments have revealed that multilayers could even have a
single magnetic ordering transition temperature lying in between
the two $T_{N}$s of the constituent materials. The phenomenon has
been ascribed to the very strong interlayer exchange coupling.

In conclusion we have shown that it is possible to control the
spin direction in MnO very effectively by growing them as thin
films on antiferromagnetic CoO films with different predetermined
spin orientations. Using detailed Mn $L_{2,3}$ soft x-ray
absorption spectroscopy, we are also able to show that it is not
strain but interlayer exchange coupling which plays a decisive
role herein. This result may pave the way for tailoring
antiferromagnets with low magnetocrystalline anistropy for
applications in exchange-bias.

We acknowledge the NSRRC staff for providing us with an extremely
stable beam. We would like to thank Lucie Hamdan and Henk
Bruinenberg for their skillful technical and organizational
assistance in preparing the experiment. The research in Cologne
is supported by the Deutsche Forschungsgemeinschaft through SFB
608.

\chapter{The spin-state puzzle in the cobaltates}
\label{ChapterCobaltates}

The cobaltates are one of the most fascinating and richest class of materials. Effects like metal-insulator transitions, large magneto resistance, non-magnetic to magnetic transitions, superconductivity and different kinds of magnetic ordering are found in the cobaltates. The large diversity, is closely related to the many possibilities in which the Co ion can be stabilized. The Co ion can be 2+, 3+ or 4+. Not only differences in valence are important, but also the variations in the spin-state, which make the cobaltates unique.

  \begin{SCfigure}[][h]
    \includegraphics[width=60mm]{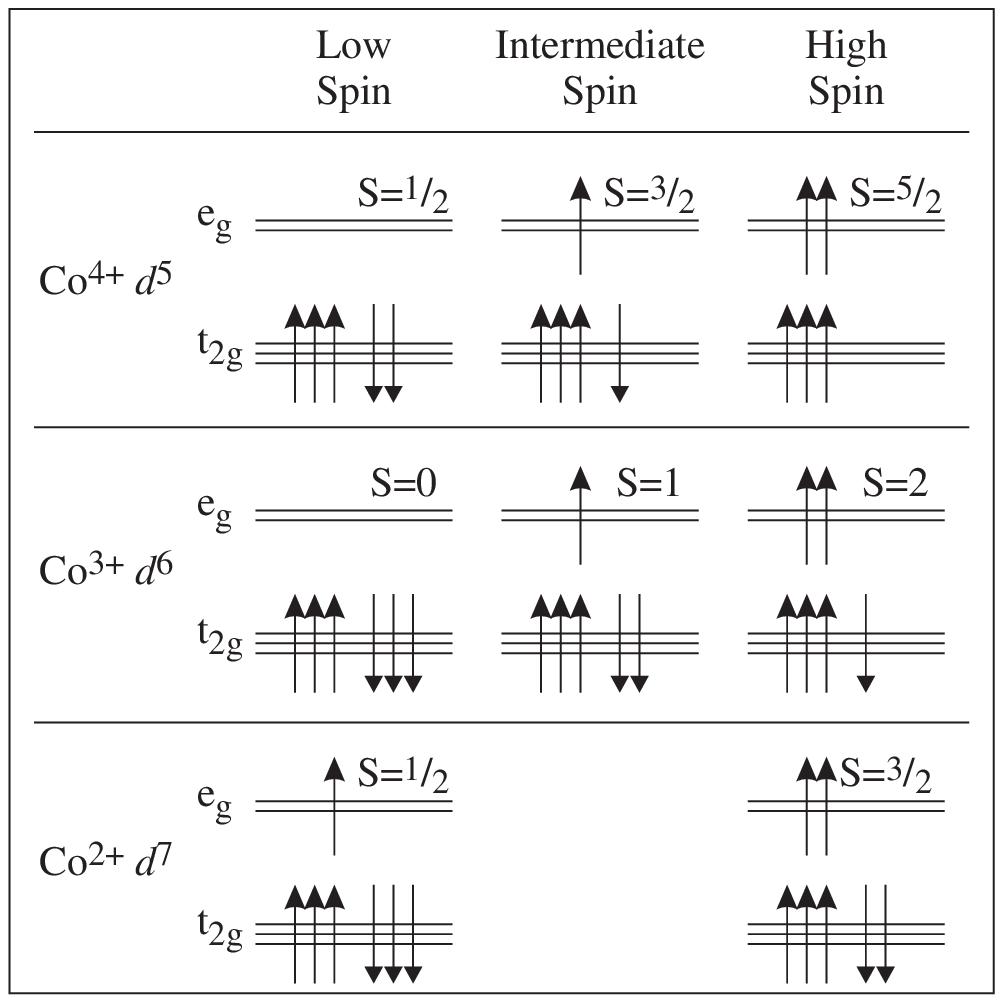}
    \caption{Possible valences and spin states of the Co ion in cubic symmetry. The $d$ shell is split in cubic symmetry into a $t_{2g}$ and $e_{g}$ sub-shell.}
    \label{spinfigPosibilities}
  \end{SCfigure}
  
Within cubic symmetry, the $d$ orbitals are split into three energetically lower-lying $t_{2g}$ orbitals and two higher-lying $e_{g}$ orbitals. The crystal field favors the occupation of the $t_{2g}$ orbitals. Hund's rule exchange tries to align the spins to be parallel. This competition leads to the variation in spin-state. In figure \ref{spinfigPosibilities} we show all possible spin-states of Co 2+, 3+ and 4+.

The crystal-field splitting between the $t_{2g}$ and $e_{g}$ orbitals is generally called $10Dq$. This means that every electron in the $t_{2g}$ shell has an energy of $-6Dq$ and every electron in the $e_{g}$ shell an energy of $4Dq$. Within the simple scheme as explained in the introduction, one can calculate the electron-electron repulsion energy easily. The electron-electron repulsion energy is $U$ times the number of electron pairs found minus $J_H$ times the number of pairs of electrons with parallel spin. A full explanation of the simple scheme can be found within the introduction. If we compare the energy of different spin-states for a $d^6$ system, one finds that all $d^6$ spin states have a contribution of $15 U$ and are only different in $J_H$ and $10Dq$. Ignoring $U$ as it does not change the relative energies of the different spin-states within a single-ion approximation, we find for the total energies of the different spin-states: $E_{\rm{HS}}=-10J_H-4Dq$, $E_{\rm{IS}}=-7J_H-14Dq$ and $E_{\rm{LS}}=-6J_H-24Dq$. 

  \begin{figure}[h]
    \includegraphics[width=120mm]{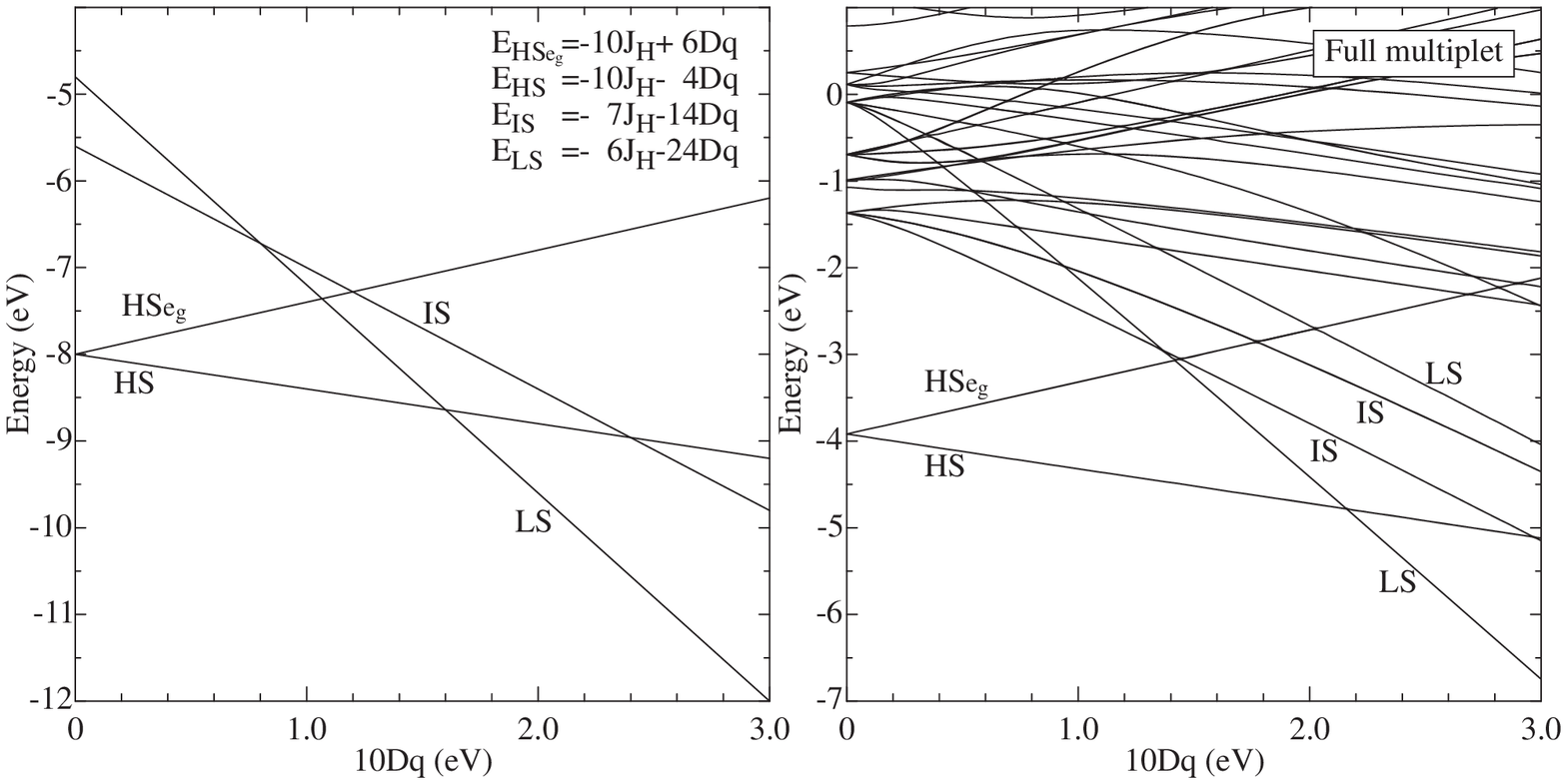}
    \caption{Energy-level diagram of the Co$^{3+}$ multiplet as a function of crystal-field parameter $10Dq$. Left: electron-electron interaction is approximated, only Hund's rule exchange energy is included. Right: the full multiplet structure is included.}
    \label{spinEnlevDq}
  \end{figure}

In figure \ref{spinEnlevDq}, we show the energy of the low-, intermediate-, and high-spin state of a $d^6$ configuration for different values of the crystal-field splitting $10Dq$. In the left graph, we approximated the electron-electron repulsion in the simple scheme. Each pair of parallel spins lowers the energy by an amount of $J_{H}\approx0.8$. $10Dq$, the energy between the $e_{g}$ and the $t_{2g}$ orbitals, is varied from 0 to 3 eV. One can see that for small values of $10Dq$ the high-spin state is the ground-state whereas for larger values of $10Dq$ Hund's rule is broken and the low-spin state becomes the ground-state. On the right of figure \ref{spinEnlevDq} we did the same calculation, but now included the full electron-electron interaction Hamiltonian. There are many more states in the full multiplet calculation and, most important for our discussions later, there are two different intermediate-spin states. For the simplified calculation we find a low-spin to high-spin transition at $10Dq=1.6$ eV. For the full multiplet calculation we find this transition at a crystal field of about $10Dq=2.2$ eV. Within a single-ion calculation the intermediate-spin state never becomes the ground-state. The large differences between the energy-level diagram as calculated within the simple scheme and with the use of the full electron-electron repulsion Hamiltonian make clear that it is absolutely necessary to use the full electron-electron repulsion Hamiltonian when theoretically comparing spin-states. Here we would like to stress again, that the multiplet splitting is not screened within solids. See for example \cite{Antonides77} or the references in the introduction of this thesis.

The question that arises is: Are these states realized in nature? We first look at some examples. For Co$^{2+}$, we know that almost all systems are in the high-spin state. In some molecular magnets the low-spin state can be stabilized. See reference \cite{Brooker02} for examples. In these systems the low-spin is stabilized by short Co-N bonds of about 1.92--1.98 \AA$ $ within the plane of the molecule. For Co$^{2+}$ we can conclude that both spin states are realized. Next we will have a look at the Co$^{3+}$ spin-states. Many examples of Co$^{3+}$ with a low-spin configuration are known. The double perovskite La$_{4}$LiCoO$_{8}$ is a clear example \cite{Hu02}. Also all Lanthanide perovskites LnCoO$_{3}$ have a low-spin ground-state. Although, LaCoO$_{3}$ has a first excited magnetic state very close in energy to the ground-state, it is generally excepted that the ground-state at 0 K is a low-spin state. There are also undisputed examples of Co$^{3+}$ ions being in the high-spin state. Cs$_{3}$CoF$_{6}$, Rb$_{3}$CoF$_{6}$ and CoF$_{3}$ are well-known fluorides with a Co$^{3+}$ high-spin state \cite{Wollan58}. Up to now, however, no $d^6$ system has been found that is undisputed in the intermediate-spin state. Many claims for an intermediate $d^6$ Co ion have been made, but for each of these systems also contradicting claims have put forward. Next we have a look at Co$^{4+}$. For Co$^{4+}$, we have Ba$_{2}$CoO$_{4}$ as an example for a high-spin system \cite{Candela73}. The intermediate-spin state has been suggested to be realized in SrCoO$_3$ \cite{Potze95}. A low-spin state can be found in BaCoO$_{3}$ \cite{Candela73, Cacheiro03, Pardo04}. In La$_{1.8}$Sr$_{0.2}$Co$_{0.5}$Li$_{0.5}$O$_{4}$ the 20\% hole doping creates Co$^{4+}$ ions which are in the low-spin state \cite{Hu02}. If the number of doped holes is increased, so that all Co ions are Co$^{4+}$, we get La$_{1.5}$Sr$_{0.5}$Co$_{0.5}$Li$_{0.5}$O$_{4}$ which has all its Co ions in the low-spin state \cite{Hu02}. In conclusion we can say that for all valences low-spin and high-spin states are known to us, but that the existence of an intermediate-spin state is still debated. There are some candidates, but the discussion is still going on.

Care has to be taken about the spin-state assignments made in the literature. Often different papers disagree with each other, especially for some newly synthesized materials. Take TbBaCo$_{2}$O$_{5.5}$ for example. This material belongs to the group of layered cobaltates where CoO$_{2}$ planes are separated by BaO and TbO$_{0.5}$ planes. Within TbBaCo$_{2}$O$_{5.5}$, half of the Co sites are surrounded by 6 O atoms in a more or less octahedral environment and the other Co sites are surrounded by only 5 O atoms with a pyramidal arrangement. All Co atoms are 3+, i.e. all Co atoms have a $d^6$ configuration. This system has a metal-insulator (or perhaps an insulator-insulator?) transition around 340K. The proposed spin states, are however, contradictory \cite{Moritomo00, Kusuya01, Soda03}. For the high-temperature phase a low-spin, an intermediate-spin and a high-spin state have been proposed for the octahedral site, and for the low-temperature phase a low-spin and an intermediate-spin state. For the low-temperature phase there is an agreement that the pyramidal sites are in the intermediate-spin state. For the high-temperature phase the Co ion is thought to be in the high-spin or intermediate-spin state.

The class of layered cobaltates is extremely confusing \cite{Martin97, Maignan99, Yamaura99a, Yamaura99b, Loureiro01, Vogt00, Suard00, Fauth01, Burley03, Mitchell03, Moritomo00, Respaud01, Kusuya01, Frontera02, Fauth02, Taskin03, Soda03, Loureiro00, Knee03, Wu00, Kwon00, Wang01, Wu01, Wu02, Wu03}, but not the only class where problems with spin-state assignments arise. For Ca$_{2}$Co$_{3}$O$_{6}$ \cite{Fjellvag96, Aasland97, Sampathkumaran04, Vidya03, Whangbo03} or LaCoO$_{3}$ \cite{Raccah67, Bhide72, Abbate93, Itoh95, Takahashi97, Saitoh97, Zhuang98, Asai98, Radwanski99, Kobayashi00, Xu01, Zobel02, Noguchi02, Maris03, Ropka03}, there are also different experimental assignments in the literature.

The problem of the spin states in the cobaltates becomes even more confusing when one realizes that even within theory there are conflicting spin-state assignments. Take YBaCo$_{2}$O$_{5}$ for example. YBaCo$_{2}$O$_{5}$ also belongs to the class of layered cobaltates. It has all Co atoms surrounded by 5 O atoms in a pyramidal environment. Half of the Co atoms are 2+ and the other half 3+. One LDA+U study found all Co atoms to be in the high-spin state \cite{Wu00}. Another LDA+U study found the Co$^{2+}$ ions to be in the high-spin state and the Co$^{3+}$ ions to be in the intermediate-spin state \cite{Kwon00}. Hartree-Fock calculations found both ions to be in the high-spin state again \cite{Wang01}.

In this chapter we will concentrate on the spin state of Co$^{3+}$ and especially on the option of having the Co ion in an intermediate-spin state. The intermediate spin state is an unusual state and it is \textit{a-priori} not clear how it can be stabilized to become the ground-state. As we have explained above, in the limit that the crystal field is much more important than the Hund's exchange energy, the system will be in a low spin state, whereas when the crystal field is smaller than the Hund's exchange energy, the system will be in the high-spin state. In this simple picture the intermediate-spin state never will be the ground-state. We will start with some theory and try to explain why the intermediate-spin state can become the ground-state. We will start our theory within a single CoO$_{6}$ cluster calculation, which explicitly includes the O-$2p$ Co-$3d$ hybridization. We will show that, for Co$^{3+}$ in cubic symmetry, the intermediate-spin state can not be stabilized within such a single-cluster calculation. For Co$^{4+}$ it can be stabilized. Next we will look at the effects of band-formation and distortions. In the end we turn to experiment and try to find systems that are in the intermediate-spin state.

\sectionmark{Stabilization of the intermediate-spin state within...}
\section{Stabilization of the intermediate-spin state within cluster calculations}
\sectionmark{Stabilization of the intermediate-spin state within...}

Within a CoO$_{6}$ cluster there are two reasons for the energy difference between the $t_{2g}$ and $e_{g}$ orbitals. One is the ionic crystal field, an electric field due to the charges of the other atoms and their electrons. This energy difference is called $10Dq$. The other reason is the covalency between Co and O atoms. The covalency is larger for $e_{g}$ orbitals then for $t_{2g}$ orbitals, resulting in an effective splitting between the $e_{g}$ and $t_{2g}$ orbitals.

  \begin{figure}[h!]
   \begin{center}
    \includegraphics[width=90mm]{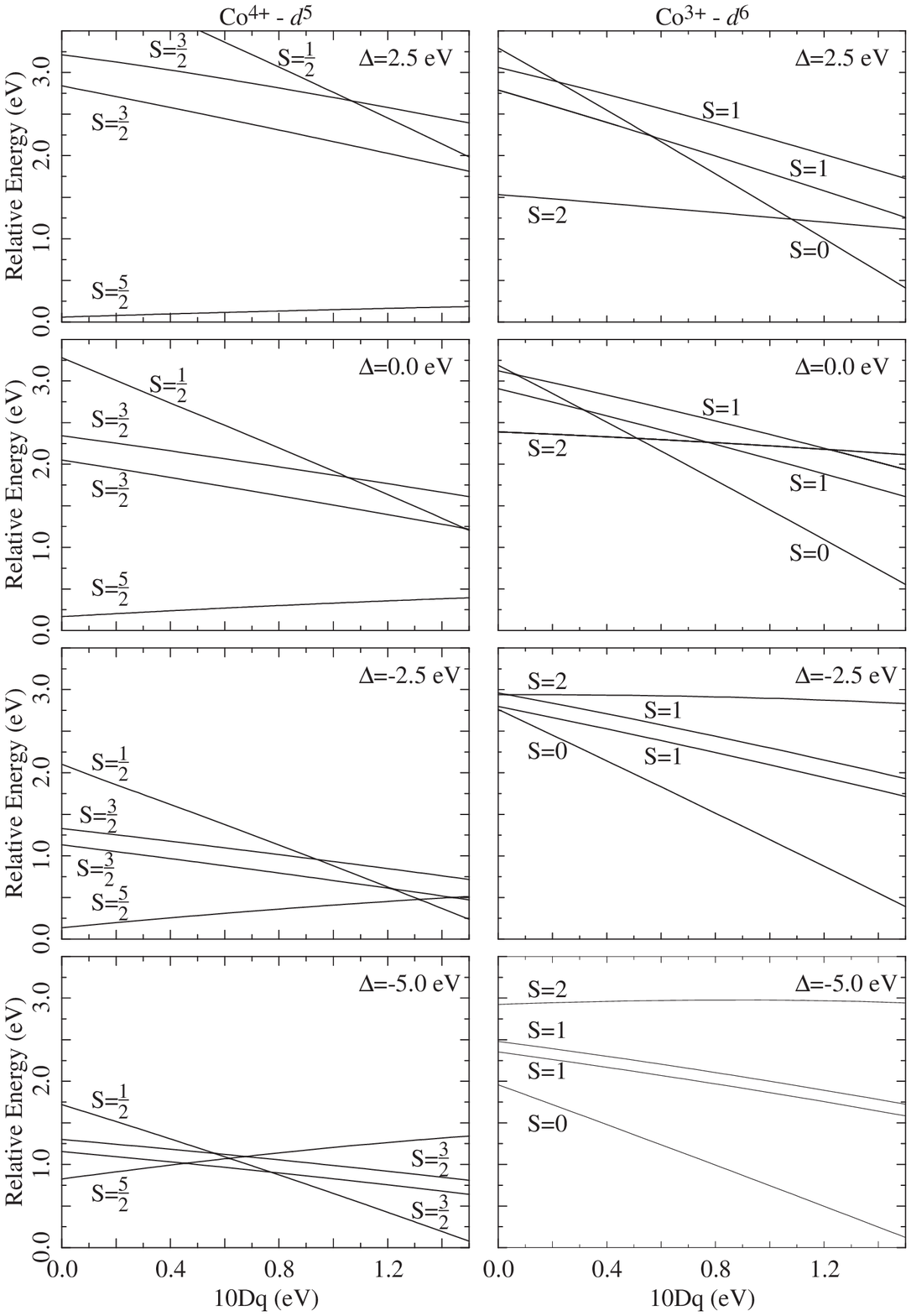}
    \caption{Energy level diagram of a CoO$_{6}$ cluster, for Co$^{4+}$, $d^5$ in the left side collum and for Co$^{3+}$, $d^6$ in the right collum. The energy level diagram is calculated as a function of $10Dq$ the ionic contribution to the crystal-field splitting between the $e_{g}$ and the $t_{2g}$ orbitals. The calculations have been done for 4 different values of $\Delta$, i.e. the energy it costs to hop with one electron from the O atom to the Co atom.}
    \label{spinfigEnlevDelta}
   \end{center}
  \end{figure}

In figure \ref{spinfigEnlevDelta} we show the energy-level diagram of Co$^{4+}$ and Co$^{3+}$ calculated within a CoO$_6$ cluster. We vary the size of $10Dq$, the crystal-field splitting, in order to see which states can become the ground-state. We have calculated these graphs for 4 different values of $\Delta$, the charge transfer energy, which is the energy it costs to hop with one electron from an oxygen atom to the Co atom. $\Delta$ has been defined with respect to the multiplet center. In the top graph, we show the energy-level diagram for $\Delta=2.5$ eV. This diagram is more or less equal to the ionic energy-level diagram as shown in figure \ref{spinEnlevDq}. The biggest difference is that the point of zero crystal field has been shifted. In first approximation, covalency gives an equal effect as an ionic crystal field.

For $\Delta=2.5$, plotted in the top graphs in figure \ref{spinfigEnlevDelta}, we find that a $d^5$ ion is rather stable in the high-spin state. Only for crystal fields larger than $10Dq=3.3$ eV (not shown in figure \ref{spinfigEnlevDelta}) the low-spin state can be stabilized. A $d^6$ ion can, for $\Delta=2.5$, be either in the low- or high-spin state, depending on the size of the crystal field. If we now increase the covalency by reducing the value of $\Delta$ we see that the intermediate-spin state is reduced in energy, but the low-spin state is reduced even more in energy. This can be understood when one realizes how many orbitals there are to which the oxygen electrons can hop. Thereby one should take into account that the hopping from the oxygens to the $e_{g}$ electrons is much larger than the hopping from the oxygens to the $t_{2g}$ electrons. In other words, covalent mixing for the $e_g$ shell is larger than for the $t_{2g}$ shell. A low-spin state has four holes in the $e_{g}$ shell and is therefore very covalent. The intermediate-spin state has three holes in the $e_{g}$ shell and is therefore less covalent than the low-spin state. The high-spin state has only two holes in the $e_{g}$ shell and is therefore also the least covalent. For a $d^5$ system we find that, if we make $\Delta=-5$ eV, there is a range of values of $10Dq$ for which the intermediate-spin state is the ground-state. This is the same result as previously found by Potze \textit{et al.} \cite{Potze95}. If we decrease $\Delta$ even further the low-spin state becomes the ground-state for all values of $10Dq$. For a $d^6$ system, we find that in cubic symmetry the intermediate-spin state can never be the ground-state. Within a CoO$_6$ cluster it is always the high-spin or low-spin state that is the ground-state. We do find, however, that for $\Delta \approx 0$ the energy difference between the low-spin - high-spin crossing and the intermediate-spin state is not so big, $\approx 0.25$ eV. If one now realizes that a solid is more than one cobalt ion with six oxygens around, there might be possibilities to stabilize the intermediate-spin state.
  
\section[Stabilizing the intermediate-spin state by band formation]{Stabilizing the intermediate-spin state by \- band formation}

Korotin \textit{et al.} \cite{Korotin96} have done band-structure calculations for LaCoO$_{3}$ on the basis of the LDA+U approximation. They found that the intermediate-spin state is the ground-state for a certain range of lattice-constant parameters. They also found that in LaCoO$_3$ the high-spin state energy is about 0.5 eV per formula unit higher than the low- and intermediate-spin state.

  \begin{SCfigure}[][h]
    \includegraphics[width=60mm]{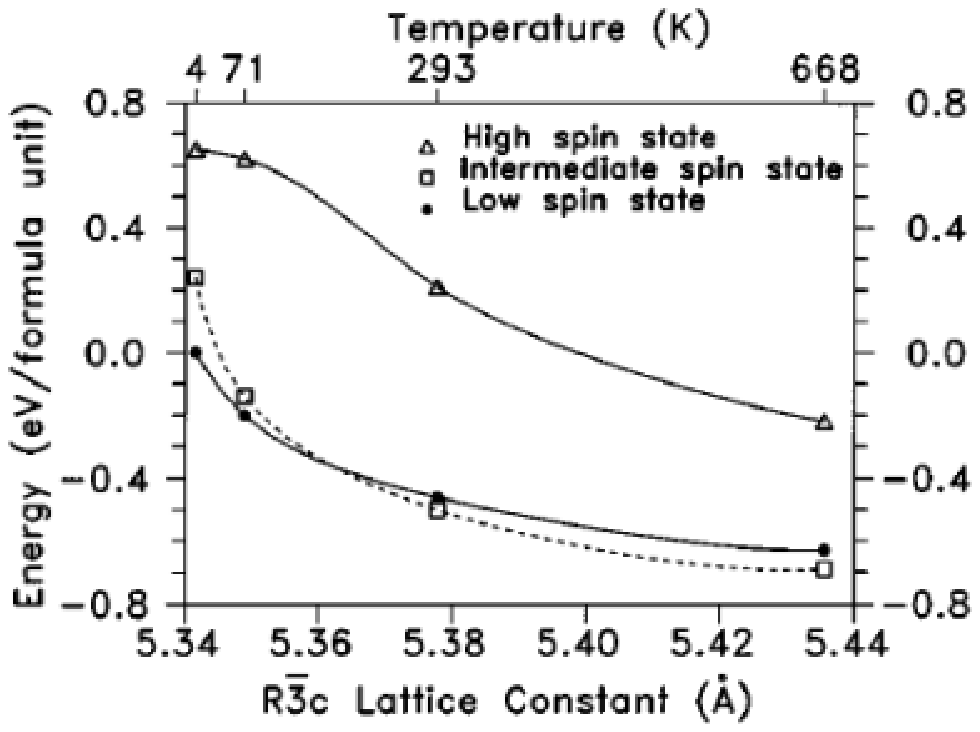}
    \caption{Total-energy level diagram as calculated within the LDA+U approximation, as a function of crystal-lattice size. The graph has been copied from Korotin \textit{et al.} \cite{Korotin96}. \newline \newline  }
    \label{spinfigEnlevLDApU}
  \end{SCfigure}
  
In figure \ref{spinfigEnlevLDApU} we show the results of the calculation by Korotin \textit{et al.} \cite{Korotin96}. To simulate the temperature effect, when heating up the sample, the lattice parameters are varied. One can see that for low temperatures the low-spin state is the ground-state and for higher temperatures the intermediate-spin state is the ground-state. This is a very nice result, as it explains the temperature behavior of the susceptibility in LaCoO$_3$. At low temperatures the system is non-magnetic and at higher temperatures the system becomes magnetic, with a maximum of the susceptibility at about 100 K. Following the paper of Korotin \textit{et al.} \cite{Korotin96} many experimental efforts have been made to confirm that the low-temperature spin-state transition in LaCoO$_3$ is from a low-spin to an intermediate-spin state \cite{Kobayashi00, Xu01, Zobel02, Maris03}, whereas in the 'old' picture the transition was from the low-spin to the high-spin state \cite{Raccah67, Bhide72, Abbate93, Itoh95}.

The reason why band-formation lowers the intermediate-spin state relative to the low- and the high-spin state is not so obvious. It has often been noted that the intermediate-spin state is stabilized due to covalency. The intermediate-spin state is more covalent than the high-spin state because the intermediate-spin state has 3 holes in the $e_{g}$ shell whereas the high-spin state has only two holes in the $e_{g}$ shell. This is true, but the low-spin state is even more covalent. Covalency within a single cluster does not stabilize the intermediate-spin state as we have shown in the previous section. But then, why do Korotin \textit{et al.} \cite{Korotin96} find the intermediate-spin state to be so low, and why do they find the high-spin state so high in energy?

In order to answer this question, we have to look into the details of their calculation. Korotin \textit{et al.} assumed a ferromagnetic ordering and a $U$ so large that the high-spin state is an insulator. With this knowledge their calculation can be understood when considering the super-exchange paths present between two cobalt sites within a double cluster. The hopping between oxygen and cobalt is much larger for $e_{g}$ orbitals than for $t_{2g}$ orbitals, so we neglect the $t_{2g}$ shell. There are two super-exchange paths present, coupling two cobalt atoms joined over a $180^{\circ}$ oxygen bond. The most important path for LaCoO$_{3}$ is the path where first one of the oxygen electrons hops from the oxygen $p$ orbital pointing to the cobalt atom, into the $e_{g}$ orbitals of that cobalt atom. Next, one of the $e_{g}$ electrons of the cobalt atom on the opposite site of the oxygen atom hops into the hole at the oxygen atom. This intermediate states cost an energy $U_{dd}$. In order to return to the original state, one has to hop back the same way as the electrons came. If possible this hopping will reduce the energy of the state in a double cluster with respect to the energy found in a single cluster.

For the ferromagnetic intermediate-spin state this super-exchange path is very well possible. For the low-spin state this super-exchange path is not possible. There are no $e_{g}$ electrons present at the cobalt site that can hop to the hole left behind when an electron hoped from the oxygen atom to an other cobalt $e_{g}$ orbital. For the ferromagnetic high-spin state this hopping path is also not present. There are holes in the cobalt $e_{g}$ shell, but only holes with spin down and none with spin up. Only electrons with spin down can hop from the oxygen to the cobalt site. There are no $e_{g}$ electrons with spin down that can hop to the hole left behind on the oxygen atom. For a ferromagnetic spin ordering only the intermediate-spin state can gain super-exchange energy and is therefore lowered with respect to the low-spin and high-spin state. This is the result found by Korotin \textit{et al.} \cite{Korotin96}.

If one wants to calculate which of the spin-states has the lowest energy, one should consider all possible magnetic orderings. The intermediate-spin state is ferromagnetic and gains energy with respect to the low- and high-spin state when a ferromagnetic calculation is done. The high-spin state can not gain super-exchange energy when the spins are aligned ferromagnetically, but it can gain super-exchange energy when the spins are aligned antiferromagnetically. For an antiferromagnetic alignment of the cobalt spins the super-exchange path discussed before becomes possible for the high-spin state. The oxygen electron with spin up can hop  into the $e_{g}$ shell of a neighboring cobalt atom. This means that the cobalt that accepts the electron had its spin down. The cobalt atom on the other site of the oxygen atom will have therefore its spin up and one of its $e_{g}$ electrons can hop into the spin up hole of the oxygen atom.

  \begin{SCfigure}[][h]
    \includegraphics[width=60mm]{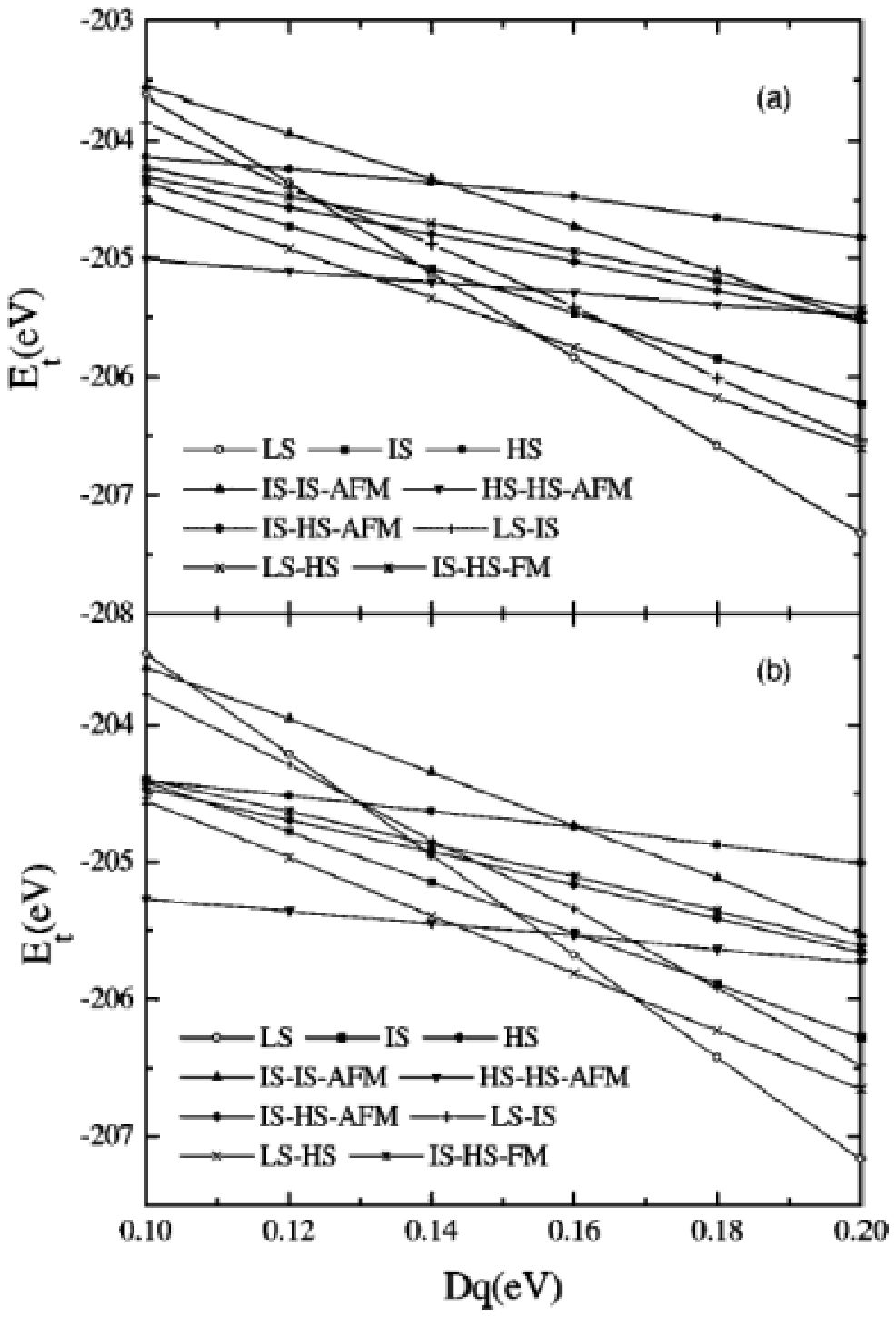}
    \caption{Energy-level diagram of the different spin-states in LaCoO$_{3}$ assuming different magnetic or spin-state ordering, calculated within the Hartree-Fock approximation. The graph has been copied from the paper by Min Zhuang, \textit{et al.} \cite{Zhuang98}.\newline \newline}
    \label{spinfigEnlevHF}
  \end{SCfigure}

The effect of different magnetic ordering on the energy of the different spin-states has been calculated within the Hartree-Fock approximation by Min Zhuang, Weiyi Zhang and Naiben Ming \cite{Zhang03}. In figure \ref{spinfigEnlevHF} we show their results. They calculated the energy for different spin-states and different magnetic orderings within LaCoO$_{3}$ as a function of the crystal field $10Dq$. From this calculation, one can clearly see that the intermediate-spin state never becomes the lowest state. The high-spin antiferromagnetic state or the low-spin state are always lower in energy than the intermediate-spin state. One can also see a third possibility. If there is a low-spin - high-spin ordering, the energy is even lower. Hua Wu is doing calculations in the LDA+U approximation \cite{Wuprivate} along these lines at the moment. More work has to be done to find out which state has the lowest energy within the LDA+U approximation.

Another issue is the question how well LDA or LDA+U can handle the electron-electron interaction for the different spin states. The electron-electron interaction does not only depend on the local electron density. In operator form we could write the electron-electron repulsion Hamiltonian as $U_{m\sigma m'\sigma' m''\sigma'' m'''\sigma'''}$ $d_{m\sigma}^{\dagger} d_{m'\sigma'}^{\dagger} d_{m''\sigma''} d_{m'''\sigma'''}$. This operator can not be written in density operators. Thereby one should realize that $\langle d_{m\sigma}^{\dagger} d_{m'\sigma'}^{\dagger} d_{m''\sigma''} d_{m'''\sigma'''} \rangle\neq$ $\langle d_{m\sigma}^{\dagger} d_{m''\sigma''} \rangle$ $\langle d_{m'\sigma'}^{\dagger} d_{m'''\sigma'''} \rangle$. In the introduction we showed how severe these approximations can be. We compared the multiplet calculated in spherical symmetry as found in the Kanamori scheme with a mean-field approximation for the Kanamori scheme. The differences are large. Within the introduction we also compared full multiplet calculations with the Kanamori scheme. For the Hund's rule ground-state, i.e. the high-spin state, the energy is reproduced quite reasonable. The other multiplets as well as the degeneracies are completely different. This can also be seen in figure \ref{spinEnlevDq} where we compare the energy level diagram of a $d^6$ state as a function of $10Dq$. On the left we use the simple scheme, on the right we use the full multiplet calculation. There are remarkable differences. It is not clear how the electron-electron interactions are exactly implemented within LDA+U. These kind of errors may very well be present within the LDA+U codes used, due to the approximations made in the calculation of the electron-electron repulsion energy. Maybe this is one of the reasons why different codes show different total energies for the same spin-state \cite{Wuprivate}.

\sectionmark{Stabilizing the Intermediate-spin state by Jahn...}
\section[Stabilizing the intermediate-spin state by Jahn-Teller distortions]{Stabilizing the intermediate-spin state by \- Jahn-Teller distortions}
\sectionmark{Stabilizing the Intermediate-spin state by Jahn...}

There is, however, yet another way to stabilize the intermediate-spin state. The intermediate-spin state has one electron in the $e_{g}$ shell and is therefore Jahn-Teller active. If the structure becomes distorted, the intermediate-spin state can gain energy, but the low-spin state can not. The high-spin state has one electron in the $t_{2g}$ spin down shell and should therefore be mildly Jahn-Teller active, since the Jahn-Teller effect for the $t_{2g}$ orbitals is much smaller than for the $e_{g}$ orbitals. This scenario for stabilizing the intermediate spin state due to Jahn-Teller distortions has been proposed by G. Maris \textit{et al.} \cite{Maris03} based on their single-crystal x-ray diffraction experiments.

  \begin{SCfigure}[][h]
    \includegraphics[width=60mm]{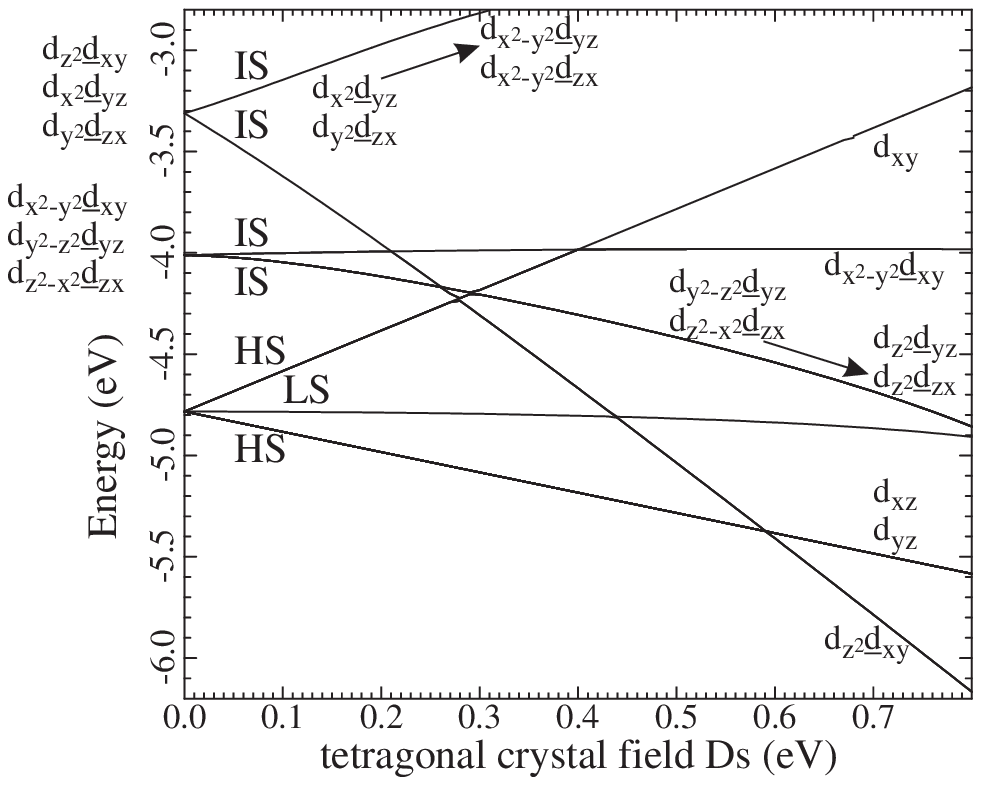}
    \caption{Ionic energy level diagram of a Co$^{3+}$ ion as a function of tetragonal distortion parameter $Ds$. The crystal field is chosen such that for zero distortion the high-spin and low-spin state are degenerate. \newline}
    \label{spinfigEnlevDs}
  \end{SCfigure}
  
\begin{figure}[h]
   \begin{center}
    \includegraphics[width=90mm]{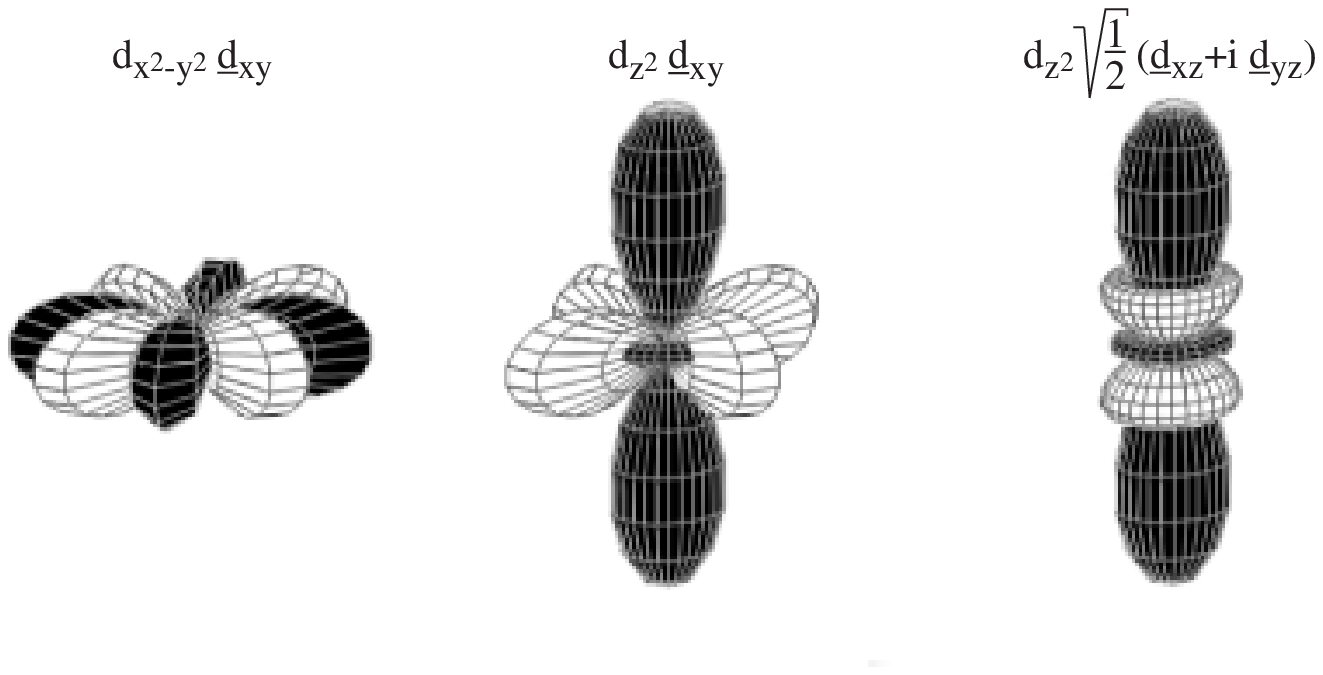}
    \caption{Possible orbital arrangement of the intermediate-spin state. The intermediate-spin state has one hole in the $t_{2g}$ shell, shown in white and one electron in the $e_{g}$ shell, drawn in black. Left: the lowest energy configuration, the $t_{2g}$ hole and the $e_{g}$ electron are close to each other. Middle: the highest energy configuration, the $t_{2g}$ hole and the $e_{g}$ electron are far away from each other. Right: Not an eigenstate of the electron-electron repulsion Hamiltonian, but if there is a tetragonal distortion (elongation) the $d_{z^2}$ orbital is lowered. For crystal-fields only the $d_{z^2}\underline{d}_{xy}$ configuration  would be best, the $d_{z^2}\sqrt{\frac{1}{2}}(\underline{d}_{xz}+{\rm{i}} \underline{d}_{yz})$ configuration is, however, better for electron-hole attraction.}
    \label{spinfigISorbitals}
   \end{center}
  \end{figure}

In figure \ref{spinfigEnlevDs}, we show the energy-level diagram of Co$^{3+}$ as a function of the tetragonal-distortion parameter $Ds$ \cite{Ballhausen62}. As one can see for $Ds>0.6$, the intermediate-spin state becomes the ground-state. One should notice that in cubic symmetry, i.e. $Ds=0$, there exist two different intermediate-spin states with different energies. The lowest intermediate-spin state is about 0.8 eV higher than the high-spin state. The highest intermediate-spin state is about 1.5 eV higher than the high-spin state. One should further notice that not the lowest intermediate-spin state, but the highest intermediate-spin state becomes the ground-state due to a tetragonal distortion. In order to understand this behavior, we have to understand the cause of the splitting in the intermediate-spin state. The intermediate-spin state has one hole in the $t_{2g}$ shell and one electron in the $e_{g}$ shell. In total, the intermediate-spin state is $2\cdot3=6$ times orbitally degenerate. This degeneracy is lifted since the $t_{2g}$ hole and the $e_{g}$ electron have a strong Coulomb interaction. In order to create the maximum $e_{g}$-electron-$t_{2g}$-hole density overlap and therefore the lowest energy, the hole will occupy the $d_{xy}$ orbital and the electron the $d_{x^2-y^2}$ orbital. This lowest intermediate-spin state is threefold degenerate and the other two wave-functions can be found by cyclic permutation of x, y and z. The highest intermediate-spin state can be found by choosing orbitals that are orthogonal to the lowest intermediate-spin state. In figure \ref{spinfigEnlevDs}, we have labelled the states accordingly. One should realize that the lowest intermediate-spin state does not have a $d_{z^2}$ occupation, but always a $d_{x^2-y^2}$-like occupation. This is also the occupation found by LDA+U for an intermediate-spin state \cite{Korotin96}. The fact that the $d_{x^2-y^2}$-electron is combined with a $d_{xy}$ hole has important consequences. In figure \ref{spinfigISorbitals}, we show these two orbitals with the hole in white and the electron in black. This is a very symmetric electron distribution and is thus not as Jahn-Teller active as one would expect for a half-filled $e_{g}$ shell, which is the reason why this state does not become the ground-state when tetragonal distortions are made. The second intermediate-spin state does have a $d_{z^2}$ orbital, combined with a $d_{xy}$ hole, and is very Jahn-Teller active.

We have calculated the energy-level diagram as a function of $Ds$ with $Dt$ equal to zero. This is not the most favorable way to lower the intermediate-spin state, but it makes very clear what happens. If $Dt$ is taken into account, the lowest and highest intermediate-spin state start to mix. This mixing already happens if only Ds is considered for the two states with $d_{y^2-z^2}\underline{d}_{yz}$ and $d_{z^2-x^2}\underline{d}_{xz}$ occupation. These occupations slowly change to $d_{z^2}\underline{d}_{yz}$ and $d_{z^2}\underline{d}_{xz}$. When $Dt$ is included, this mixing is increased and the $d_{z^2}\underline{d}_{xz}$ and $d_{z^2}\underline{d}_{yz}$ states can become lower in energy than the $d_{z^2}\underline{d}_{xy}$ state. One should realize that the double-degenerate intermediate-spin state has a magnetic orbital momentum. In figure \ref{spinfigISorbitals} we also drew the shape of the intermediate-spin orbital that can be stabilized when $Dt$ is included. One can see that the electron density of the $\sqrt{\frac{1}{2}}(-d_{xz}-{\rm{i}} d_{yz})$ orbital is quite different from the $d_{xz}$ or $d_{xz}$ orbital. This state, is neither an eigenfunction of the electron-electron repulsion Hamiltonian nor of the crystal-field Hamiltonian. If both are present, this state can be a good compromise and becomes the ground-state.

We have to conclude that distortions might stabilize the intermediate-spin state, but that important details concerning the multiplet structure need to be considered. Furthermore, we notice that distortions do not automatically reduce the magnetic orbital momentum. Even for a locally tetragonal elongation it is on forehand not so easy to tell whether the magnetic orbital momentum of the intermediate-spin state will be quenched or not.
  
\sectionmark{Different look at the spin-state of Co$^{3+}$ ions in a CoO$_5$...}
\section[Different look at the spin-state of Co$^{3+}$ ions in a CoO$_5$ pyramidal coordination.\\ \textmd{Phys. Rev. Lett.} \textbf{92}\textmd{, 207402 (2004)}]{Different look at the spin state of Co$^{3+}$ ions in a CoO$_5$ pyramidal coordination.}
\sectionmark{Different look at the spin-state of Co$^{3+}$ ions in a CoO$_5$...}

In order to find out if an intermediate-spin state is realized in nature, we set out an experimental search for systems with an intermediate-spin state. To maximize the chance of finding an intermediate-spin state, we took a system that has a large Jahn-Teller distortion. We also took a system where we find $180^{\circ}$ bonds between Co-O-Co atoms. This maximizes the hopping and band-formation. The material we choose is Sr$_2$CoO$_3$Cl which has CoO$_{5}$ pyramids that are connected by a corner-sharing oxygen. The results of our measurements can be seen in the publication below.

\begin{center}

\begin{Large}

\textmd{Phys. Rev. Lett.} \textbf{92}\textmd{, 207402 (2004)}
\\$\quad$
\\$\quad$
\end{Large}

\begin{minipage}{0.8\textwidth}
Using soft-x-ray absorption spectroscopy at the Co-$L_{2,3}$ and O-$K$ edges,
we demonstrate that the Co$^{3+}$ ions with the CoO$_{5}$ pyramidal
coordination in the layered Sr$_2$CoO$_3$Cl compound are unambiguously in the high-spin
state. Our result questions the reliability of the spin-state assignments made
so far for the recently synthesized layered cobalt perovskites, and calls for a
re-examination of the modeling for the complex and fascinating properties of
these new materials.
\end{minipage}
\end{center}

The class of cobalt-oxide based materials has attracted considerable interest
in the last decade because of expectations that spectacular properties may be
found similar to those in the manganites and cuprates. Indeed, giant magneto
resistance effects have been observed in the La$_{1-x}$$A$$_{x}$CoO$_{3}$
($A$=Ca,Sr,Ba) perovskites \cite{Briceno95} and $R$BaCo$_{2}$O$_{5+x}$
($R$=Eu,Gd) layered perovskites \cite{Martin97,Maignan99}. Very recently,
superconductivity has also been found in the Na$_{x}$CoO$_{2}\cdot y$H$_{2}$O material
\cite{Takada03}. In fact, numerous one-, two-, and three-dimensional cobalt-oxide materials have been synthesized or rediscovered in the last 5 years, with
properties that include metal-insulator and ferro-ferri-antiferro-magnetic
transitions with various forms of charge, orbital and spin ordering
\cite{Fjellvag96,Aasland97,Kageyama97,Yamaura99a,Loureiro01,
Vogt00,Suard00,Fauth01,Burley03,Mitchell03,Moritomo00,Respaud01,Kusuya01,
Frontera02,Fauth02,Taskin03,Soda03,Loureiro00,Knee03}.

A key aspect of cobalt oxides that distinguishes them clearly from the manganese
and copper materials, is the spin-state degree of freedom of the Co$^{3+/{\rm{III}}}$
ions: it can be low spin (LS, $S$=0), high spin (HS, $S$=2) and even
intermediate spin (IS, $S$=1) \cite{Sugano70}. This aspect comes on top of the
orbital, spin (up/down) and charge degrees of freedom that already make the
manganite and cuprate systems so exciting. It is, however, also precisely this
aspect that causes considerable debate in the literature. For the classic
LaCoO$_{3}$ compound, for instance, various early studies attributed the low-temperature spin-state change to be of LS-HS nature \cite{Goodenough65, Raccah67}, while
studies in the last decade put a lot of effort to propose a LS-IS scenario
instead \cite{Korotin96,Saitoh97}. More topical, confusion has arisen about the
Co spin state in the newly synthesized layered cobalt perovskites
\cite{Martin97,Maignan99,Yamaura99a,Yamaura99b,Loureiro01,Vogt00,Suard00,Fauth01,Burley03,
Mitchell03,Moritomo00,Respaud01,Kusuya01,Frontera02,Fauth02,Taskin03,Soda03,
Loureiro00,Knee03}. In fact, all possible spin states have been claimed for
each of the different Co sites present.  There is even no consensus in the
predictions from band structure calculations \cite{Wu00,Wu01,Wu02,Wu03,Kwon00,Wang01}.

In this letter we are questioning the reliability of the spin states as
deduced from magnetic, neutron and x-ray diffraction measurements for the newly
synthesized layered cobalt perovskites
\cite{Martin97,Maignan99,Yamaura99a,Yamaura99b,Loureiro01,
Vogt00,Suard00,Fauth01,Burley03,Mitchell03,Moritomo00,Respaud01,Kusuya01,
Frontera02,Fauth02,Taskin03,Soda03}.  We carried out a test experiment using a
relatively simple model compound, namely Sr$_2$CoO$_3$Cl, in which there are no spin
state transitions present and in which there is only one kind of Co$^{3+}$ ion
coordination \cite{Loureiro00,Knee03}. Important is that this coordination is
identical to the pyramidal CoO$_{5}$ present in the heavily debated layered
perovskites \cite{Martin97,Maignan99,Yamaura99a,Yamaura99b,Loureiro01,
Vogt00,Suard00,Fauth01,Burley03,Mitchell03,Moritomo00,Respaud01,Kusuya01,
Frontera02,Fauth02,Taskin03,Soda03}. Using a \textit{spectroscopic} tool, that
is soft x-ray absorption spectroscopy (XAS), we demonstrate that pyramidal
Co$^{3+}$ ions are not in the often claimed IS state but unambiguously in a HS
state. This outcome suggests that the spin states and their temperature
dependence in layered cobalt perovskites may be rather different in nature from
those proposed in the recent literature.

Bulk polycrystalline samples of Sr$_2$CoO$_3$Cl were prepared by a solid state reaction
route \cite{Loureiro00}. The magnetic susceptibility is measured to be very
similar to the one reported by Loureiro \textit{et al.} \cite{Loureiro00} and
Knee \textit{et al.} \cite{Knee03}. We find that up to 600 K the susceptibility
does not follow a Curie-Weiss behavior, making a simple determination of the
spin state impossible. Spectroscopic measurements were carried out using soft
x-rays in the vicinity of the Co-$L_{2,3}$ ($h\nu$ $\approx$ 780--800 eV) and
O-$K$ ($h\nu$ $\approx$ 528--535 eV) absorption edges. The experiments were
performed at the Dragon beamline at the NSRRC in Taiwan, with a photon energy
resolution of about 0.30 eV and 0.15 eV, respectively. Clean sample surfaces
were obtained by scraping \textit{in-situ} with a diamond file, in an
ultra-high vacuum chamber with a pressure in the low $10^{-9}$ mbar range. The
Co-$L_{2,3}$ XAS spectra were recorded in the total electron yield (TEY) mode
by measuring the sample drain current. The O-$K$ XAS spectra were collected by
both the TEY and the bulk-sensitive fluorescence yield (FY) mode
simultaneously. The close similarity of the spectra taken with these two modes
is used to verify that the TEY mode spectra are representative for the bulk
material. A single crystal of EuCoO$_{3}$ is included as an unambiguous
reference for a LS Co$^{\rm{III}}$ system \cite{Baier03}.

\begin{SCfigure}[][h]
\includegraphics[width=60mm]{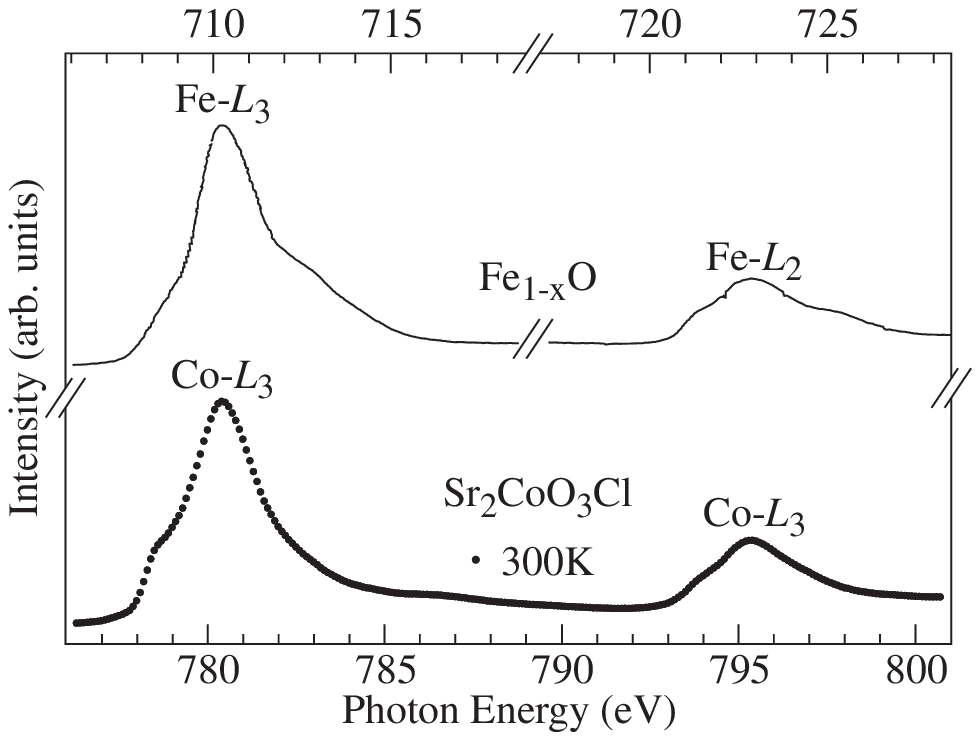}
\caption{Co-$L_{2,3}$ XAS spectrum of Sr$_2$CoO$_3$Cl measured at 300 K ($\bullet$) and
Fe-$L_{2,3}$ XAS spectrum of Fe$_{1-x}$O ($x$$\leq$$0.05$) reproduced from reference \cite{Pen97} (solid line). \newline}
\label{SSPfig1}
\end{SCfigure}

Figure \ref{SSPfig1} shows the Co-$L_{2,3}$ XAS spectrum of Sr$_2$CoO$_3$Cl taken at room temperature.
It is dominated by the Co $2p$ core-hole spin-orbit coupling which splits the
spectrum roughly in two parts, namely the $L_{3}$ ($h\nu \approx 780$ eV) and
$L_{2}$ ($h\nu \approx 796$ eV) white lines regions. The line shape of the
spectrum depends strongly on the multiplet structure given by the Co $3d$-$3d$
and $2p$-$3d$ Coulomb and exchange interactions, as well as by the local
crystal fields and the hybridization with the O $2p$ ligands. Unique to soft
x-ray absorption is that the dipole selection rules are very effective in
determining which of the $2p^{5}3d^{n+1}$ final states can be reached and with
what intensity, starting from a particular $2p^{6}3d^{n}$ initial state ($n$=6
for Co$^{3+}$) \cite{Groot94,Thole97}. This makes the technique extremely
sensitive to the symmetry of the initial state, i.e. the valence \cite{Chen90a},
orbital \cite{Chen92,Park00} and spin \cite{Laan88,Thole88b,Cartier92,Pen97}
state of the ion.

Utilizing this sensitivity, we compare the Co-$L_{2,3}$ XAS spectrum of Sr$_2$CoO$_3$\-Cl
to that of another $3d^{6}$ compound, namely Fe$_{1-x}$O ($x$$\leq$$0.05$),
reproduced from the thesis of J.-H. Park \cite{Park94}. This spectrum was
taken at room temperature. Except for the different photon energy scale and
the smaller $2p$ core-hole spin-orbit splitting, the Fe$_{1-x}$O spectrum as
shown in Figure \ref{SSPfig1} is essentially identical with that of Sr$_2$CoO$_3$Cl. From this we can
immediately conclude that the Co$^{3+}$ ions in Sr$_2$CoO$_3$Cl are in the HS state,
since the Fe$^{2+}$ ions are also unambiguously HS.

\begin{SCfigure}[][h]
\includegraphics[angle=0,width=60mm]{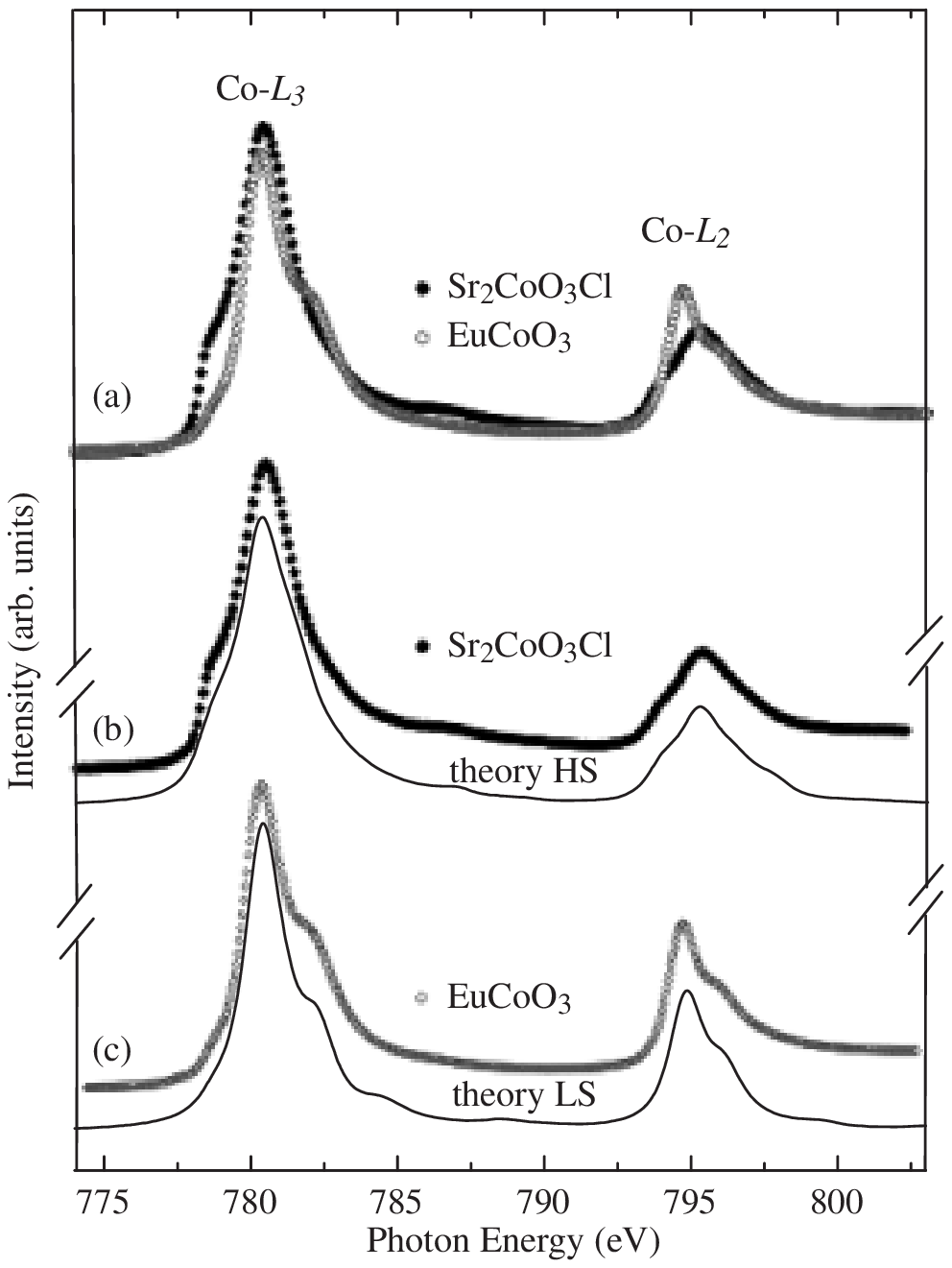}
\caption{(a) Co-$L_{2,3}$ XAS spectra of Sr$_2$CoO$_3$Cl ($\bullet$) and EuCoO$_{3}$
($\circ$); (b) Comparison between the Sr$_2$CoO$_3$Cl spectrum ($\bullet$) and a
theoretical simulation for a high-spin (HS) CoO$_{5}$ pyramidal cluster (solid
line); (c) Comparison between the EuCoO$_{3}$ spectrum ($\circ$) and a
theoretical simulation for a low-spin (LS) CoO$_{6}$ octahedral cluster (solid
line). \newline}
\label{SSPfig2}
\end{SCfigure}

To find further support for our conclusion, we also compare the Co-$L_{2,3}$
XAS spectrum of Sr$_2$CoO$_3$Cl with that of EuCoO$_{3}$, which is known to be a LS
system \cite{Baier03}. From Figure \ref{SSPfig2} one now can clearly see large discrepancies
between the spectra of the two compounds. Not only are the line shapes
different, but also the ratios of the integrated intensities of the $L_{3}$ and
$L_{2}$ regions: in comparison with Sr$_2$CoO$_3$Cl, the LS EuCoO$_{3}$ has relatively
less intensity at the $L_{3}$ and more at the $L_{2}$, characteristic for a
spin state difference \cite{Laan88,Thole88b,Cartier92,Pen97}. Figure \ref{SSPfig2} thus
demonstrates that Sr$_2$CoO$_3$Cl is definitely not a LS system.

It would have made our case even easier to prove, if we could have excluded
experimentally the IS scenario for Sr$_2$CoO$_3$Cl by comparing the spectrum to that of a
known Co$^{3+}$ IS reference system. However, there is to date no consensus for
such an oxide reference system. Nevertheless, the spin state can also be
deduced from theoretical simulations of the experimental spectra. To this end,
we use the successful configuration interaction cluster model that includes the
full atomic multiplet theory and the hybridization with the O $2p$ ligands
\cite{Groot94,Thole97,Tanaka94}. We have carried out the calculations for a
Co$^{3+}$ ion in the CoO$_{5}$ pyramidal cluster as present in Sr$_2$CoO$_3$Cl and for
the ion in the CoO$_{6}$ octahedral cluster found in EuCoO$_{3}$. We use
parameter values typical for a Co$^{3+}$ system \cite{Saitoh97}. The Co $3d$ to
O $2p$ transfer integrals are adapted for the various Co-O bond lengths
according to Harrison's prescription \cite{Harrison89,paramSrCoOCl}. This together with
the crystal field parameters determines whether the Co$^{3+}$ ion is in the HS
or LS state \cite{Sugano70}. The results are shown in Figure \ref{SSPfig2} and one can clearly
see that the calculated spectrum of the HS pyramidal CoO$_{5}$ cluster
reproduces very well the experimental Sr$_2$CoO$_3$Cl spectrum, and that the calculated
LS octahedral CoO$_{6}$ spectrum matches nicely the experimental EuCoO$_{3}$
spectrum. This demonstrates that our spectroscopic assignments are firmly
founded.

\begin{SCfigure}
\includegraphics[width=60mm]{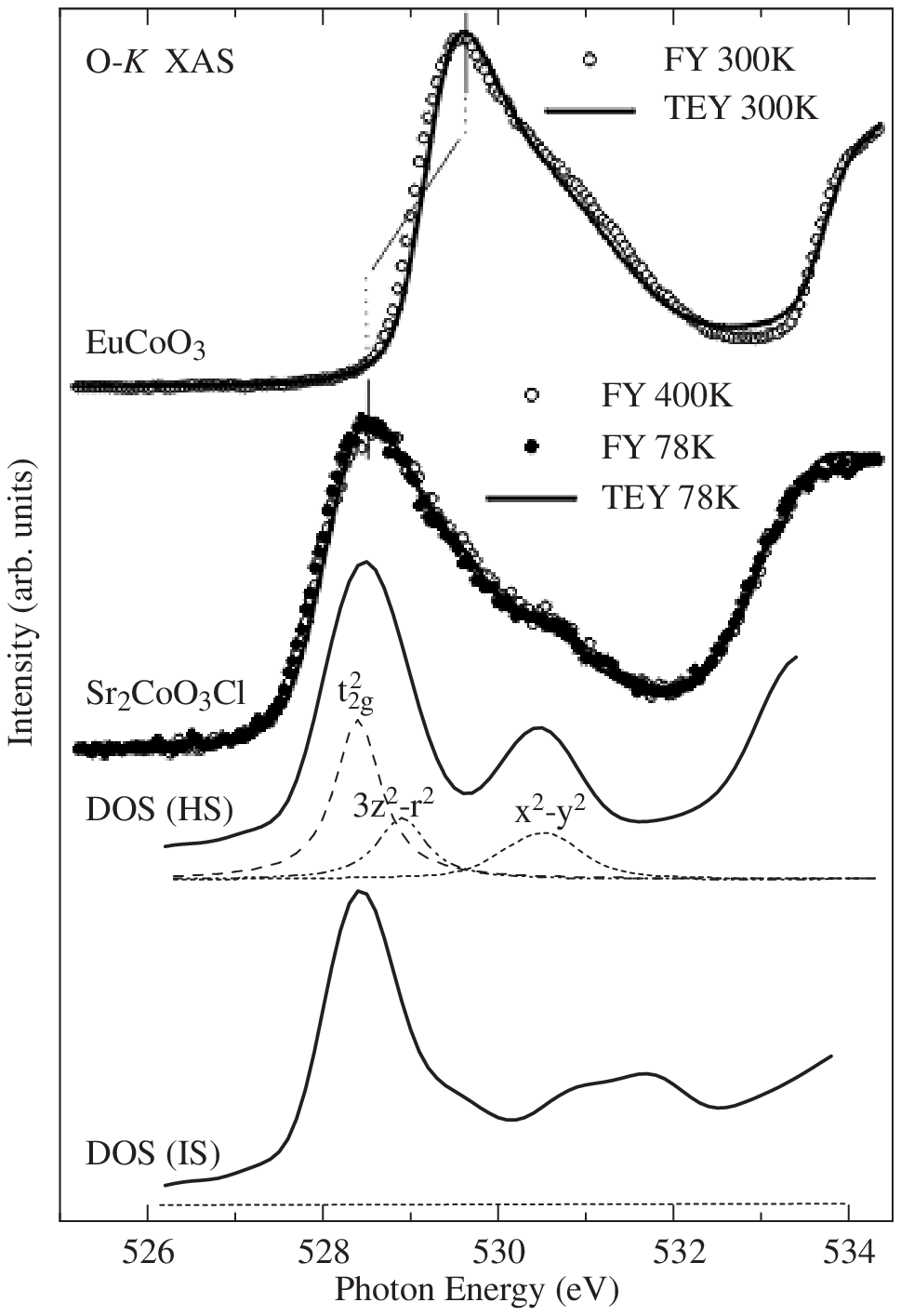}
\caption{O-$K$ XAS spectra of Sr$_2$CoO$_3$Cl and EuCoO$_{3}$. The solid lines below the
experimental curves depict the LDA+U calculated unoccupied O $2p$ partial DOS
for Sr$_2$CoO$_3$Cl in the real crystal structure with the HS state (upper) and in the
artificial structure with the IS state (lower). The dashed, dashed-dotted, and
dotted lines are the $t_{2g}^{2}$, $3z^{2}$-$r^{2}$, and $x^{2}$-$y^{2}$
projections, respectively. \newline}
\label{SSPfig3}
\end{SCfigure}

More spectroscopic evidence for the HS nature of the Co$^{3+}$ in the pyramidal
CoO$_{5}$ coordination can be found from the O-$K$ XAS spectrum as shown in
Figure \ref{SSPfig3}. The structures from 528 to 533 eV are due to transitions from the O
$1s$ core level to the O $2p$ orbitals that are mixed into the unoccupied Co
$3d$ $t_{2g}$ and $e_{g}$ states. The broad structures above 533 eV are due to
Sr $4d$, Co $4s$ and Cl $3p$ related bands. For comparison, Figure \ref{SSPfig3} also
includes the spectrum of the LS EuCoO$_{3}$, and clear differences can be seen
in the line shapes and energy positions of the Co $3d$--O $2p$ derived states.
This again is indicative that Sr$_2$CoO$_3$Cl is not a LS system. To interpret the
spectra, we also have carried out full-potential band structure calculations
\cite{Wu00, Wu01, Wu02, Wu03} for Sr$_2$CoO$_3$Cl in the local-density approximation with correction for
electron-correlation effects (LDA+U) \cite{Anisimov91}. We find the ground
state of the system to be an antiferromagnetic insulator with a band gap of 1.3
eV and a magnetic moment of 3.2 $\mu_B$. Although less than $4 \mu_B$, this
indicates that the Co is in the HS state since in an antiferromagnet the moment
is reduced due to covalency. The calculated unoccupied O $2p$ partial density
of states (DOS) are depicted in Figure \ref{SSPfig3}, and good agreement with the
experimental spectrum can be observed.

It is now interesting to look with more detail into the character of the states
relevant for the O $K$ XAS spectra. For the LS EuCoO$_{3}$ with the $3d$
$t_{2g}^{6}$ configuration, the lowest energy structure in the spectrum at
about 529.5 eV is due to transitions into the unoccupied Co $3d$ $e_{g}$
states. The fact that Sr$_2$CoO$_3$Cl has a lower energy structure thus indicates that
transitions to the lower lying $t_{2g}$ are allowed, i.e. that the $t_{2g}$
states are not fully occupied. In other words, Sr$_2$CoO$_3$Cl is in the HS
$t_{2g}^{4}e_{g}^{2}$ or IS $t_{2g}^{5}e_{g}^{1}$ state. At first sight, one
might then expect a much larger spectral weight for the higher lying $e_{g}$
level, since the hybridization with the O $2p$ is larger for the $e_{g}$ than
for the $t_{2g}$. However, our LDA+U calculations in which we find the HS
ground state, indicate that, because of the missing apical oxygen in the
CoO$_{5}$ coordination, the unoccupied $3z^{2}$-$r^{2}$ level is pulled down by
1.6 eV from the $x^{2}$-$y^{2}$, and comes close to the unoccupied
$t_{2g}^{2}$. Moreover, because of the large displacement (0.33 \AA) of the Co
ion out of the O$_{4}$ basal plane of the pyramid \cite{Loureiro00}, the
hybridization of the $x^{2}$-$y^{2}$ with the O $2p$ ligands is strongly
reduced. Therefore, the dominant lower energy structure at 528.3 eV consists of
the unoccupied minority $t_{2g}^{2}$ (dashed line in Figure \ref{SSPfig3}) and minority
$3z^{2}$-$r^{2}$ (dashed dotted line) levels, and the shoulder at 530.4 eV of
the minority $x^{2}$-$y^{2}$ (dotted line).

From the LDA+U calculations, we have found that the IS state \cite{Kwon00} is
unstable with respect to HS ground state for the real crystal structure of
Sr$_2$CoO$_3$Cl. We have also found nevertheless, that the IS state \textit{can} be
stabilized by \textit{artificially} moving the Co ion into the O$_{4}$ basal
plane of the CoO$_{5}$ pyramid. For the latter, however, the calculated
unoccupied O $2p$ partial DOS does not reproduce the experimental O-$K$ XAS
spectrum that well, as one can see from the discrepancies in the 531-532 eV
range in Fig. 3. What happens is that the $x^{2}$-$y^{2}$ level is pushed up by
the increased hybridization with the O $2p$ ligands, since the Co ion is within
the O$_{4}$ basal plane in this artificial crystal structure. Moreover, the
up-rising majority $x^{2}$-$y^{2}$ becomes unoccupied, resulting in the IS
state. Apparently, the actual large base corrugation of the CoO$_{5}$ pyramid
helps to stabilize the HS state \cite{Wu00, Wu01, Wu02, Wu03}, a trend that should not be
overlooked if one is to understand the real spin state of CoO$_{5}$ pyramids.
We find from our LDA+U calculations that the HS is more stable than the IS for
out-of-basal-plane Co displacements larger than a critical value of about 0.15
\AA.

Having established that the pyramidal coordinated Co$^{3+}$ ions in Sr$_2$CoO$_3$Cl are
in the HS state, we now turn our attention to other layered cobalt materials
that have the same structural units. Neutron diffraction experiments on
$R$BaCo$_{2}$O$_{5.0}$ ($R$ = rare earth) have revealed the existence of
alternating Co$^{3+}$ and Co$^{2+}$ ions, both in pyramidal CoO$_{5}$
coordination. The magnetic structure is $G$-type antiferro with moments of 2.7
and 4.2 $\mu_{B}$ \cite{Vogt00}, or 2.7 and 3.7 $\mu_{B}$, respectively
\cite{Fauth01}. For the $R$=Nd compound, charge ordering was not observed, but
an average moment of 3.5 $\mu_{B}$ was measured \cite{Burley03,Mitchell03}.
These studies suggested two possible scenarios for the Co$^{3+}$ ions, namely
either HS with spin-only moments or IS with orbital moment. Our findings based
on Sr$_2$CoO$_3$Cl on the other hand, strongly suggest the HS state of such pyramidal
Co$^{3+}$ ions. Here we keep in mind that the out-of-plane Co displacements of
the pyramids in $R$BaCo$_{2}$O$_{5.0}$ are larger than 0.35 \AA~
\cite{Vogt00,Fauth01,Burley03}, i.e. much larger than the above mentioned 0.15
\AA~ critical value. The first scenario is thus favored, with the remark that
neutron diffraction techniques tend to observe smaller magnetic moments due to
the Co-O covalency, which is responsible for the antiferromagnetic
superexchange interactions present in these materials.

The experimental situation for the $R$BaCo$_{2}$O$_{5.5}$ system is more
complicated. Neutron and x-ray diffraction measurements indicate the presence
of all Co$^{3+}$ ions in alternating pyramidal CoO$_{5}$ and octahedral
CoO$_{6}$ units \cite{Burley03,Mitchell03,Moritomo00,Respaud01,Kusuya01,
Frontera02,Fauth02,Soda03}. The magnetic structure is most likely not a simple
$G$-type \cite{Fauth02,Taskin03}, and depending on the model, values between
0.7 and 2.0 $\mu_{B}$ have been extracted for the pyramidal Co$^{3+}$
\cite{Fauth02,Soda03}. The IS state is thus proposed, and in fact most other
studies also assume this starting point
\cite{Burley03,Mitchell03,Moritomo00,Respaud01,Kusuya01, Frontera02,Taskin03}.
Nevertheless, structural data indicate that the CoO$_{5}$ pyramids in these
compounds have very similar Co-O bond lengths and angles as in Sr$_2$CoO$_3$Cl. The
out-of-plane Co$^{3+}$ displacements in the pyramids are larger than 0.3 \AA~
\cite{Burley03,Respaud01,Kusuya01,Frontera02}, and again, much larger than the
0.15 \AA~ critical value. We therefore infer that also in these compounds the
pyramidal Co$^{3+}$ must be HS, which is supported by the observation that the
effective magnetic moment as extracted from the high temperature Curie-Weiss
behavior indicates a HS state for all Co$^{3+}$
\cite{Martin97,Maignan99,Moritomo00}. In fact, the average Co-O bond length for
the CoO$_{5}$ pyramids even increases at lower temperatures
\cite{Kusuya01,Frontera02}, thereby stabilizing the HS state even more. The
fact that neutron diffraction detects lower moments may indicate a complex
magnetic structure as a result of a delicately balanced spin state of the
octahedral Co$^{3+}$ ions affecting the various exchange interactions in the
compounds in which the pyramidal Co$^{3+}$ remains HS.

Summarizing, we have found an overwhelming amount of evidence for the HS nature
of the pyramidal coordinated Co$^{3+}$ ions in Sr$_2$CoO$_3$Cl: (1) the Co $L_{2,3}$
spectrum has essentially an identical line shape as the Fe $L_{2,3}$ in
Fe$_{1-x}$O ($x$$\leq$$0.05$); (2) the Co $L_{2,3}$ spectrum can be reproduced
to a great detail by model calculations with the Co ion in the HS state; (3)
the O $K$ spectrum can be well explained by LDA+U calculations with the Co in
the HS state, but not with the Co in the IS state; and (4) LDA+U calculations
yield the HS ground state and no stable IS state for the real crystal
structure. With other newly synthesized layered cobalt oxides having very
similar pyramidal CoO$_{5}$ units, we infer that those Co$^{3+}$ ions must
also be in the HS state, contradicting the assignments made so far. It is
highly desirable to investigate the consequences for the modeling of the
properties of these new materials.

We would like to thank Lucie Hamdan for her skillful technical and
organizational assistance in preparing the experiment, and Daniel Khomskii for
stimulating discussions. The research in K\"oln is supported by the Deutsche
Forschungsgemeinschaft through SFB 608.

\chapter{Determination of the orbital moment and crystal field splitting in LaTiO$_{3}$\\ \textmd{Phys. Rev. Lett.} \textbf{94}\textmd{, 056401 (2005)}}
\label{ChapterLaTiO3}
\chaptermark{Determination of the orbital moment and crystal field...}

\begin{center}
\begin{minipage}{0.8\textwidth}
Utilizing a sum-rule in a spin-resolved photoelectron spectroscopic experiment
with circularly polarized light, we show that the orbital moment in LaTiO$_3$
is strongly reduced from its ionic value, both below and above the N\'{e}el
temperature. Using Ti $L_{2,3}$ x-ray absorption spectroscopy as a local probe,
we found that the crystal field splitting in the $t_{2g}$ subshell is about
0.12-0.30 eV. This large splitting does not facilitate the formation of an
orbital liquid.
\end{minipage}
\end{center}

LaTiO$_3$ is an antiferromagnetic insulator with a pseudocubic perovskite
crystal structure \cite{MacLean79,Eitel86,Cwik03}. The N\'{e}el temperature
varies between $T_N$ = 130 and 146 K, depending on the exact oxygen
stoichiometry \cite{Cwik03,Goral83,Meijer99}. A reduced total moment of about
0.45-0.57 $\mu_B$ in the ordered state has been observed
\cite{Cwik03,Goral83,Meijer99}, which could imply the presence of an orbital
angular momentum that is antiparallel to the spin momentum in the Ti$^{3+}$
$3d^1$ ion \cite{Meijer99,Mizokawa96}. In a recent Letter, however, Keimer {\it
et al.} \cite{Keimer00} have reported that the spin wave spectrum is nearly
isotropic with a very small gap, and concluded that therefore the orbital
moment must be quenched. To explain the reduced moment, they proposed the
presence of strong orbital fluctuations in the system. This seems to be
supported by the theoretical study of Khaliullin and Maekawa
\cite{Khaliullin00}, who suggested that LaTiO$_3$ is in an orbital liquid
state. If true, this would in fact constitute a completely novel state of
matter. By contrast, Cwik \textit{et al.} \cite{Cwik03}, Mochizuki and Imada
\cite{Mochizuki03}, as well as Pavarini \textit{et al.} \cite{Pavarini04}
estimated that small orthorhombic distortions present in LaTiO$_{3}$ would
produce a crystal field (CF) splitting strong enough to lift the Ti $3d$
$t_{2g}$ orbital degeneracy. However, one of the latest theoretical papers
finds a much smaller CF splitting, leaving open the possibility for an orbital
liquid state \cite{Solovyev04}.

In view of these controversies, it is highly desirable to have experimental
tests which would allow to uniquely choose between different possibilities. On
the experimental side, however, very little is known about the energetics of
the LaTiO$_{3}$ system. We have carried out spin-resolved photoemission (PES)
experiments using circularly polarized light, and by applying a sum-rule we
have determined unambiguously that the orbital moment is indeed strongly
reduced from its ionic value in a wide temperature range. We have also
performed temperature dependent Ti $L_{2,3}$ x-ray absorption (XAS)
measurements, and found from this local probe that the Ti $3d$ $t_{2g}$
orbitals are split by about 0.12-0.30 eV. Our results are consistent with the
conclusion of Keimer \textit{et al.} in that the orbital moment is very small.
However, the sizable CF splitting does not provide conditions favorable for the
realization of an orbital liquid.

Twinned single crystals of LaTiO$_{3}$ with $T_N$ = 146 K have been grown by
the traveling floating-zone method. The PES experiments were performed at the
ID08 beamline of the ESRF in Grenoble. The photon energy was set to 700 eV,
sufficiently high to ensure bulk sensitivity \cite{Weschke91,Sekiyama00}. The
degree of circular polarization was close to $100\%$ and the spin detector had
an efficiency (Sherman function) of 17\%. The combined energy resolution for
the measurements was 0.6 eV and the angle $\theta$ between the Poynting vector
of the light and the analyzer was 60$^\circ$. The XAS measurements were carried
out at the Dragon beamline of the NSRRC in Taiwan, with a photon energy
resolution set at 0.15 eV for the Ti $L_{2,3}$ edges ($h\nu \approx 450-470$
eV). The spectra were recorded using the total electron yield method. Clean
sample areas were obtained by cleaving the crystals inside the measuring
chambers with a pressure of low 10$^{-10}$ mbar.

  \begin{SCfigure}[][h!]
    \includegraphics[width=60mm]{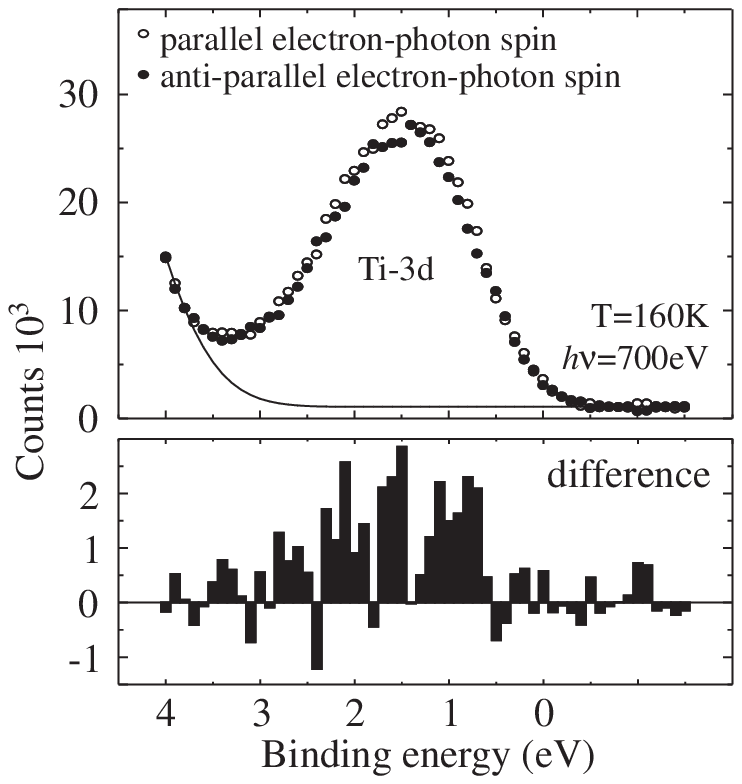}
    \caption{Spin-resolved photoemission spectra of twinned LaTiO$_{3}$
    single crystal taken with circularly polarized light. \newline \newline}
    \label{LaTiO3fig1}
  \end{SCfigure}

Fig. \ref{LaTiO3fig1} shows the spin-resolved photoemission spectra of the LaTiO$_3$ $3d$
states in the valence band, taken with circularly polarized light. The spectra
are corrected for the spin detector efficiency. One can observe a small but
reproducible difference between the spectra taken with the photon spin (given
by the helicity of the light) parallel or antiparallel to the electron spin.
The relevant quantity to be evaluated here is the integrated intensity of the
difference spectrum ($\int_{dif}$) relative to that of the integrated intensity
of the sum spectrum ($\int_{sum}$). This can be directly related to the
expectation value of the spin-orbit operator (\textbf{l.s}) applied to the
\textit{initial} state, thanks to the sum rule developed by van der Laan and
Thole \cite{Laan93}. For a randomly oriented sample \cite{Ghiringhelli02}, and
for a $3d$ system in which the final states are mainly of $f$ character due to
the high photon energies used \cite{LaTiO3pfinalstates, Tanaka94}, and we obtain:
\begin{equation}
\frac{\int_{dif}}{\int_{sum}} = -\frac{1}{U(\theta)}
 \frac{\langle \sum_{i} \textbf{l}_{i} \cdot \textbf{s}_{i} \rangle }
{3 \langle n \rangle} \label{Eq1}
\end{equation}
where $U(\theta)$=$(2$-$cos^{2}(\theta))$/$(3$-$4cos^{2}(\theta))$ is a
geometrical factor to account for the angle between the Poynting vector of the
light and the outgoing photoelectron, the index \textit{i} runs over the
electrons in the $3d$ shell and $\langle n \rangle$ is the number of $3d$
electrons contributing to the spectra.

\begin{SCfigure}[][h!]
    \includegraphics[width=60mm]{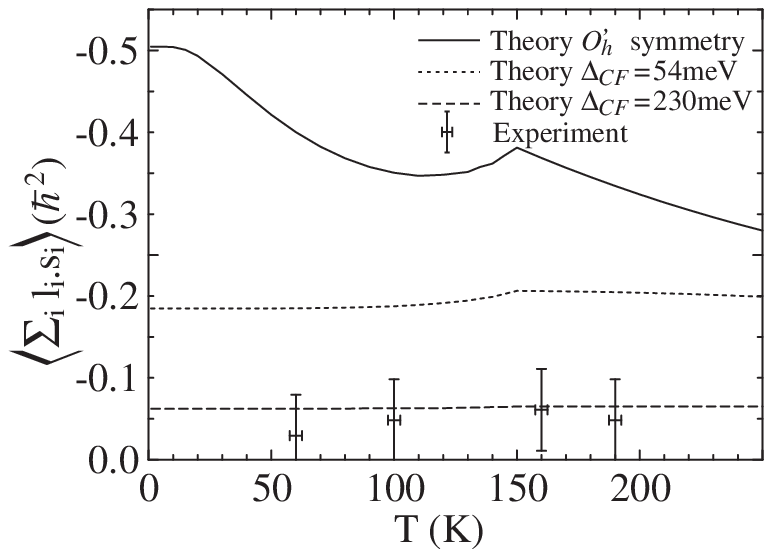}
    \caption{$\langle \sum_{i} \textbf{l}_{i} \cdot \textbf{s}_{i}
    \rangle$ values extracted from the spin-resolved circularly polarized
    photoemission data, together with theoretical predictions for various
    crystal field parameters. \newline}
    \label{LaTiO3fig2}
  \end{SCfigure}

With $\int_{dif}$/$\int_{sum}$$\approx$0.03, $\theta$=60$^\circ$, and $\langle
n \rangle$$\approx$0.8 from our cluster calculations \cite{Tanaka94}, we arrive
at $\langle \sum_{i} \textbf{l}_{i} \cdot \textbf{s}_{i} \rangle$$\approx$-0.06
(in units of $\hbar^2$), see Fig. \ref{LaTiO3fig2}. This is, in absolute value, an order of
magnitude smaller than the maximum possible value of -0.50 for a $3d^{1}$
$t_{2g}$ ion with $s_z$=1/2 and $l_z$=-1 (in units of $\hbar$). In fact, the
-0.06 value is so small, that we can directly conclude that for this $3d^{1}$
ion the orbital momentum is practically quenched. Fig. \ref{LaTiO3fig2} shows that this is the
case for a wide range of temperatures, both below and above $T_{N}$.

Having established that LaTiO$_3$ has a strongly reduced orbital moment, we now
focus on the issue whether this is caused by strong orbital fluctuations
\cite{Keimer00,Khaliullin00} or rather by strong local CF effects as
theoretically proposed \cite{Cwik03,Mochizuki03,Pavarini04}. To this end, we
carry out temperature dependent XAS measurements at the Ti $L_{2,3}$ ($2p
\rightarrow 3d$) edges. Here we make use of the fact that the $2p$ core hole
produced has a strong attractive Coulomb interaction with the $3d$ electrons.
This interaction is about 6 eV, and is more than one order of magnitude larger
than the band width of the $3d$ $t_{2g}$ states. The absorption process is
therefore strongly excitonic, making the technique an ideal and extremely
sensitive local probe \cite{Tanaka94,Groot94,Thole97}.

 \begin{figure}[h!]
 \begin{center}
    \includegraphics[width=115mm]{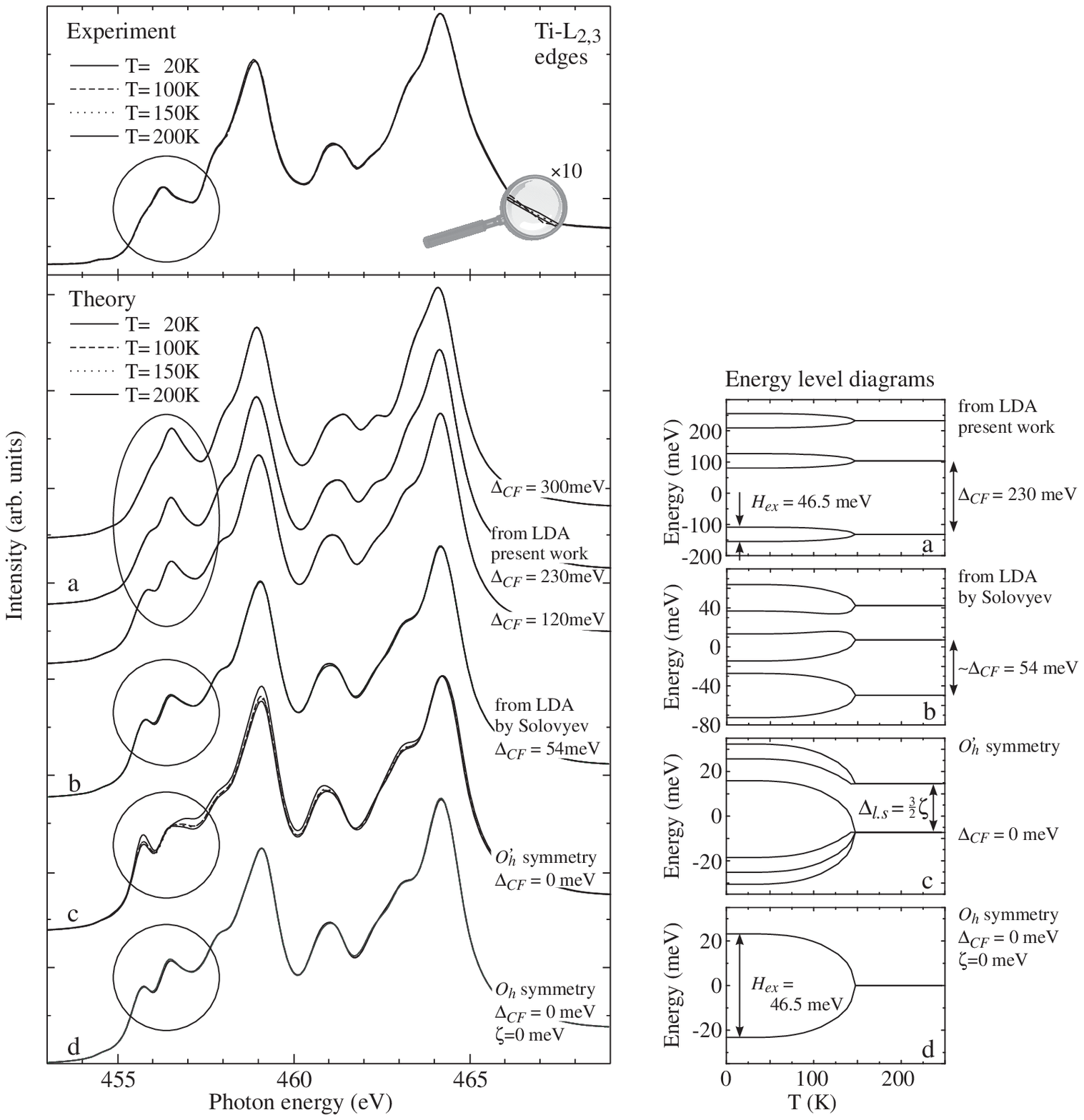}
    \caption{Top panel: experimental Ti $L_{2,3}$ XAS spectra taken from
    a twinned LaTiO$_{3}$ single crystal at 20, 100, 150 and 200 K.
    Left panel: simulated isotropic spectra calculated for a TiO$_{6}$ cluster
    at 20, 100, 150 and 200 K for several CF parameters. Right panel:
    corresponding energy level diagrams for the cluster
    in an exchange field of $H_{ex}$ = 46.5 meV (from Keimer \textit{et al.}
    \cite{Keimer00}) at $T$ = 0 K and vanishing at $T_N$ = 146 K.
    Four scenarios are presented: (a) non-cubic symmetry with
    $\Delta_{CF}$ = 230 meV from our LDA calculation \cite{Streltsov04}
    and with $\Delta_{CF}$ = 120 and 300 meV,
    (b) non-cubic symmetry with $\Delta_{CF}$ = 54 meV from
    Solovyev \cite{Solovyev04}, (c) $O_{h}^{'}$ and (d) $O_{h}$ symmetry.
    The spin-orbit constant $\zeta$ is 15.2 meV for (a), (b) and (c), and 0
    for (d). Note the very different energy scales. \newline}
    \label{LaTiO3fig3}
  \end{center}
  \end{figure}

The top panel of Fig. \ref{LaTiO3fig3} shows the experimental Ti $L_{2,3}$ XAS spectra for
several temperatures below and above $T_N$. One can clearly observe that the
spectra are temperature independent. In the subsequent sections we will discuss
two aspects of the spectra that are relevant for the determination of the
energetics and symmetry of the ground state and the lowest excited states of
LaTiO$_{3}$. The first is the detailed line shape of the spectra, and the
second is their temperature insensitivity.

To start with the first aspect, we have performed simulations in order to
obtain the best match with the experimental spectra, and by doing so, to
determine the magnitude of the CF splitting in the $t_{2g}$ levels. For this we
have used the well-proven configuration interaction cluster model that includes
the full atomic multiplet theory and the hybridization with the O $2p$ ligands
\cite{Tanaka94,Groot94,Thole97}. Curves (a) in left panel of Fig. \ref{LaTiO3fig3} are the
calculated isotropic spectra of a TiO$_{6}$ cluster with a non-cubic crystal
field splitting of $\Delta_{CF}$ = 230 meV, as obtained, using a Wannier
function projection procedure, from our LDA calculation \cite{Streltsov04} on
the refined orthorhombic crystal structure \cite{Cwik03}. One can see that the
experimental data are well reproduced. We have also carried out simulations
with other $\Delta_{CF}$ values, and found that $\Delta_{CF}$ should be in the
range of about 120 to 300 meV in order to maintain the good agreement. If we
chose, for example, $\Delta_{CF}$ = 54 meV as proposed from the LDA
calculations by Solovyev \cite{Solovyev04}, we find that the simulated line
shapes are less satisfactory: curves (b) show deviations from the experimental
spectra, especially in the encircled region. More important is that the
situation without CF splitting, i.e. in $O_{h}^{'}$ symmetry as shown by curves
(c), definitely does not agree with the experiment. Also the case as depicted
by curves (d), in which the spin-orbit interaction in $O_{h}$ symmetry is
artificially switched off as to obtain fully degenerate $t_{2g}$ levels, which
was the starting point of the treatment of Khaliullin and Maekawa
\cite{Khaliullin00}, does not agree with the measurement. From the line shape
analysis we can thus firmly conclude that the crystal field splitting in
LaTiO$_{3}$ is quite appreciable.

\begin{SCfigure}[][h!]
    \includegraphics[width=60mm]{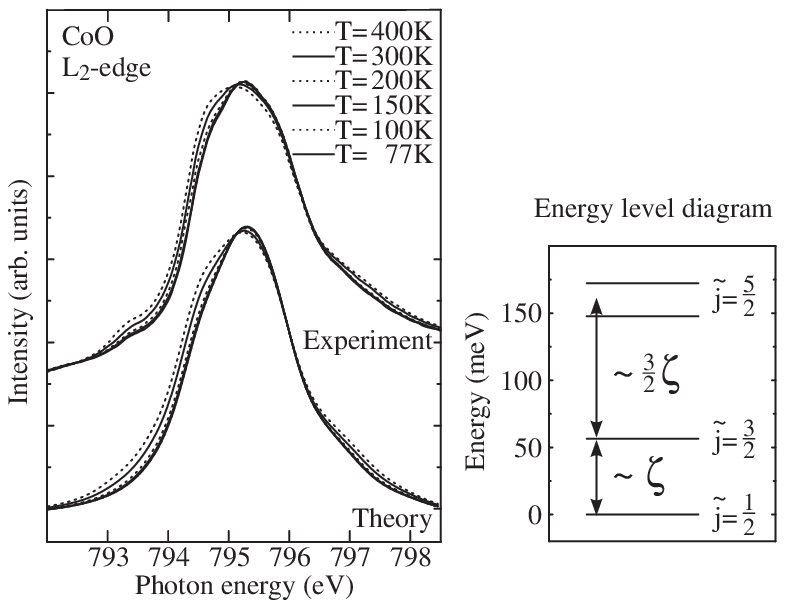}
    \caption{Experimental temperature dependent Co $L_2$ XAS spectra of
    polycrystalline CoO, together with the simulated isotropic spectra
    and corresponding energy level diagram. \newline}
    \label{LaTiO3fig4}
  \end{SCfigure}

The second aspect of the Ti $L_{2,3}$ XAS spectra is their temperature
insensitivity. This may look like a trivial observation, but actually it is
not. For a $3d$ system with an open $t_{2g}$ shell, one usually expects to see
an appreciable temperature dependence in the isotropic spectrum: for instance,
in Fig. 4 we depict the Co $L_{2}$ XAS spectra of polycrystalline CoO, and
indeed, we do see a strong temperature dependence. The reason for this behavior
is that for a system with an unquenched orbital moment like CoO, the ground
state and the lowest excited states are split in energy by the spin-orbit
interaction and are separated in energy by an amount of the order of the
spin-orbit coupling \cite{Groot93a}. Since the final states that can be
reached from the ground state and from the lowest excited states are very
different, the spectrum will change with temperature depending on how much each
of the initial states is thermally populated. In Fig. \ref{LaTiO3fig4} we have also simulated
the CoO spectra using a CoO$_{6}$ cluster model, and clearly the temperature
dependence is reproduced. In the left panel of Fig. \ref{LaTiO3fig3}, we have calculated the
LaTiO$_3$ spectra assuming a perfect $O_{h}^{'}$ local symmetry, and the
resulting curves (c) show indeed also a strong temperature dependence. However,
the fact that experimentally the LaTiO$_3$ spectra are temperature independent,
indicates directly that the spin-orbit interaction is inactive in LaTiO$_3$.
Indeed, simulations carried out for CF splittings much larger than the
spin-orbit interaction, e.g. curves (a) and (b) in Fig. \ref{LaTiO3fig3}, are not temperature
sensitive. The experimentally observed temperature insensitivity is therefore
fully consistent with the very small orbital moment found from the
spin-resolved photoemission measurements. We would like to note that the XAS
simulations were carried out including, for completeness, the presence of an
exchange field as depicted in the right panel of Fig. \ref{LaTiO3fig3}, although this had a
negligible influence on the isotropic spectra.

Returning to the spin-resolved photoemission data, we are able to reproduce the
very low $\langle \sum_{i} \textbf{l}_{i} \cdot \textbf{s}_{i} \rangle$ of
about -0.06 if we use $\Delta_{CF}$ values in the range of 120 and 300 meV. In
Fig. 2 we show the results calculated for the $\Delta_{CF}$ = 230 meV as found
from our LDA. The corresponding extracted orbital moment is $L_{z}$ = -0.06.
The $\Delta_{CF}$ = 54 meV value as proposed by Solovyev \cite{Solovyev04},
however, clearly gives a $\langle \sum_{i} \textbf{l}_{i} \cdot \textbf{s}_{i}
\rangle$ value that deviates substantially from the experimental one. The
orbital moment in this scenario is quite large: $L_{z}$ = -0.24. It is almost
superfluous to note that the calculation with $\Delta_{CF}$ = 0 meV, i.e. in
perfect $O_{h}^{'}$ symmetry, gives results that are in strong disagreement
with the experiment.

To conclude, we have observed that the orbital moment in LaTiO$_3$ is strongly
reduced from its ionic value, supporting the analysis from the neutron
experiment by Keimer {\it et al.} \cite{Keimer00}. Our experiments have also
revealed the presence of non-cubic crystal fields sufficiently strong to split
the Ti $t_{2g}$ levels by about 0.12-0.30 eV, confirming several of the
theoretical estimates \cite{Cwik03,Mochizuki03,Pavarini04,Streltsov04}. Such a
large crystal field splitting provides a strong tendency for the Ti $3d$
orbitals to be spatially locked, i.e. the quadrupole moment measured at 1.5 K
by NMR \cite{Kiyama03} should also persist at the more relevant higher
temperatures, making the formation of an orbital liquid in LaTiO$_{3}$ rather
unfavorable.

We acknowledge Lucie Hamdan for her skillful technical assistance. The research
in K\"oln is supported by the Deutsche Forschungsgemeinschaft through SFB 608
and the research in Ekaterinburg by grants RFFI 04-02-16096 and yp.01.01.059.

\chapter[Orbital-assisted metal-insulator transition in VO$_{2}$]{Orbital-assisted metal- insulator transition in VO$_{2}$ \cite{Haverkort05b}}
\label{ChapterVO2}

\begin{center}
\begin{minipage}{0.8\textwidth}
We have found direct experimental evidence for an orbital switching in the V $3d$
states across the metal-insulator transition in VO$_{2}$. We have used
soft-x-ray absorption spectroscopy at the V $L_{2,3}$ edges as a sensitive
local probe, and have determined quantitatively the orbital polarizations.
These results strongly suggest that, in going from the metallic to the
insulating state, the orbital occupation changes in a manner which makes the
system more 1-dimensional and more susceptible to a Peierls-like transition,
and that the required massive orbital switching can only be made if the system
is close to a Mott insulating regime.
\end{minipage}
\end{center}

The problem of metal-insulator transitions (MIT) in transition metal compounds attracts considerable attention already for a long time. Among the best studied of such systems are the V oxides, especially V$_2$O$_3$ and VO$_2$ \cite{Tsuda91,Imada98}. The long-standing problem in these systems is the relative role of electron-lattice interactions and corresponding structural distortions versus electron correlations. This problem is especially acute for the MIT in VO$_2$, which was described either as predominantly a Peierls transition \cite{Wentzcovitch94} or as a Mott-Hubbard transition \cite{Rice94}.

An intriguing aspect that has largely been neglected in the discussions about MIT in TM oxides is the possible role of magnetic correlations and especially the orbital structure of constituent TM ions \cite{Park00}. Very recently, a theoretical model for spinels such as MgTi$_2$O$_4$ and CuIr$_2$S$_4$ has been proposed in which a specific orbital occupation effectively leads to the formation of one-dimensional bands, making the systems, in a natural manner, susceptible to a Peierls-like MIT \cite{Khomskii05}.

\begin{SCfigure}[][h]
    \includegraphics[width=60mm]{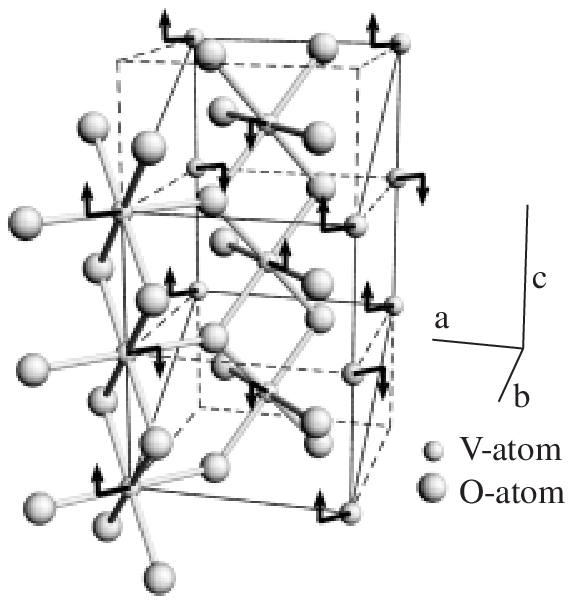}
    \caption{Crystal structure of VO$_{2}$ in the metallic rutile (R)
    phase and in the insulating monoclinic (M$_1$) phase. The arrows show
    the direction of the displacements of the V ions in the M$_1$-phase.
    The $a$, $b$ and $c$-axes are defined with respect to the rutile
    structure.\newline}
    \label{VO2fig1}
\end{SCfigure}

\begin{figure}[h]
    \includegraphics[width=120mm]{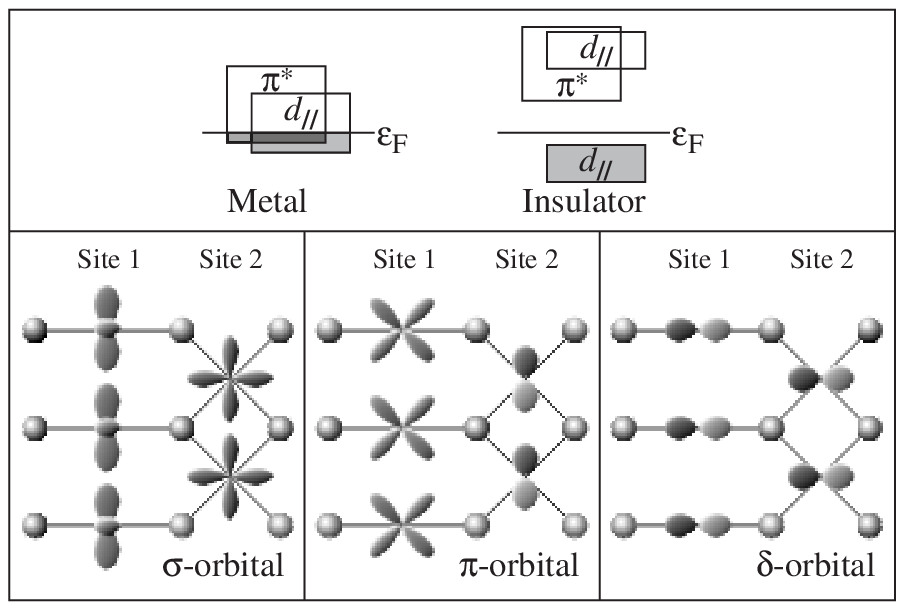}
    \caption{Top panel: schematic electronic structure of VO$_{2}$
    according to Goodenough \cite{Goodenough71}; bottom panel:
    definitions of the relevant V $3d$ $t_{2g}$ orbitals used in this work,
    drawn in the (110) plane spanned by the $a$, $b$ and
    $c$-axes of Fig. 1. Site 1 and 2 are related by a 90$^{\circ}$ rotation
    around the $c$-axis.}
    \label{VO2fig2}
\end{figure}

The crystal and electronic structure of VO$_{2}$ is in this respect more intricate. The MIT in VO$_2$ is a structural transition from the high-temperature rutile (R) structure to a monoclinic (M$_1$) structure, in which there appears simultaneous \textit{dimerization} in each V chain along the $c$-axis and a \textit{twisting} of V-V pairs due to an antiferroelectric shift of neighboring V atoms towards the apex oxygens, which lay at the axis perpendicular to the crystal $c$-axis, as shown in Fig. \ref{VO2fig1}. As argued already long ago by Goodenough \cite{Goodenough71}, one should discriminate between two types of orbitals and corresponding bands: $d_{\parallel}$-orbitals/bands, formed by the $t_{2g}$-orbitals with strong direct overlap with the neighboring V in the chains, and $\pi^{*}$-orbitals/bands, made of the two other $t_{2g}$-orbitals. In the R-phase, the $d_{\parallel}$ band overlaps with the $\pi^{*}$-band, resulting in a orbitally isotropic metallic state, see Fig. \ref{VO2fig2}. The twisting in the M$_1$-phase increases the V - apex O hybridization and moves the $\pi^{*}$ band up, so that only the $d_{\parallel}$ band is occupied. The later one then becomes split by the dimerization, leading to the insulating state.

Many theoretical \textit{ab-initio} studies were performed to test the Goodenough picture. LDA calculations indicated indeed that the $d_{\parallel}$ band becomes more occupied in the M$_1$-phase \cite{Wentzcovitch94, Eyert02}, but they in fact failed to reproduce the insulating state. The LDA+U approach predicts more dramatic changes in the orbital occupations, but unfortunately it does not give the metallic solution for the R-phase \cite{Korotin03, Liebsch05}. Similar changes were also the outcome of a study with the three-band Hubbard model by means of exact diagonalization of finite size clusters \cite{Tanaka04}. Very recently, various LDA+DMFT methods have been applied to explain the MIT, and also here the orbital occupations are an important issue \cite{Liebsch05, Laad03,  Laad04, Biermann05}. In view of the fact that the orbital occupation and changes thereof are central to the MIT theories for VO$_{2}$, it is quite surprising that an experimental proof of it is completely lacking.

In this paper we give a direct experimental evidence of this orbital redistribution at the MIT in VO$_{2}$. We present a polarization-dependent x-ray absorption spectroscopy (XAS) study on VO$_{2}$ single crystals at the V $L_{2,3}$ ($2p \rightarrow 3d$) edges. Here we make use of the fact that the Coulomb interaction of the $2p$ core hole with the $3d$ electrons is much larger than the $3d$ $t_{2g}$ band width, so that the absorption process is strongly excitonic and therefore can be well understood in terms of atomic-like transitions to multiplet split final states subject to dipole selection rules. This makes the technique an extremely sensitive local probe \cite{Tanaka94,Groot94,Thole97}, ideal to study the orbital character \cite{Park00,Chen92} of the ground or initial state. For our experiment on VO$_{2}$, we redefine the orbitals in terms of $\sigma$, $\pi$, or $\delta$ with respect to the V chain as shown in Fig. 2. The $\sigma$ orbital is then equivalent to the $d_{\parallel}$, and the $\pi$ and $\delta$ to the $\pi^{*}$. The transition probability will strongly depend on which of the $\sigma$, $\pi$, or $\delta$ orbitals are occupied and on how the polarization vector $\vec{E}$ of the light is oriented. Our measurememts reveal a dramatic switching of the orbital occupation across the MIT, even more than in V$_{2}$O$_{3}$ \cite{Park00}, indicating the crucial role of the orbitals and lattice in the correlated motion of the electrons.

Single crystals of VO$_{2}$ with $T_{MIT}$ = 67 $^{\circ}$C have been grown by the vapor transport method \cite{VObook}. The XAS measurements were performed at the Dragon beamline of the NSRRC in Taiwan. The spectra were recorded using the total electron yield method in a chamber with a base pressure of $3\times10^{-10}$ mbar. Clean sample areas were obtained by cleaving the crystals \textit{in-situ}. The photon energy resolution at the V $L_{2,3}$ edges ($h\nu \approx 510-530$ eV) was set at 0.15 eV, and the degree of linear polarization was $\approx 98 \%$. The VO$_{2}$ single crystal was mounted with the $c$-axis perpendicular to the Poynting vector of the light. By rotating the sample around this Poynting vector, the polarization of the electric field vector can be varied continuously from $\vec{E} \parallel c$ to $\vec{E} \perp c$. This measurement geometry allows for an optical path of the incoming beam which is independent of the polarization, guaranteeing a reliable comparison of the spectral line shapes as a function of polarization.

\begin{SCfigure}[][h]
    \includegraphics[width=60mm]{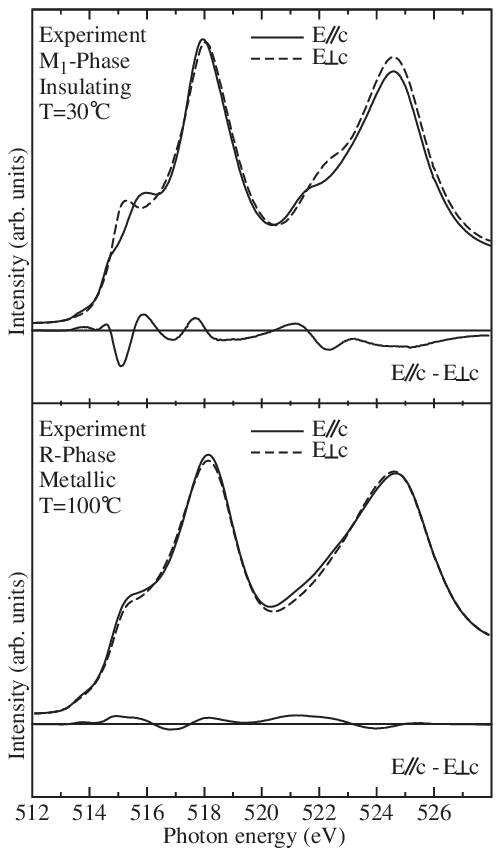}
    \caption{Experimental V $L_{2,3}$ XAS spectra of VO$_{2}$ in
    the insulating M$_1$-phase (top panel, T=30$^{\circ}$C) and
    metallic R-phase (bottom panel, T=100$^{\circ}$C),
    taken with the light polarization $\vec{E} \parallel c$
    (solid lines) and $\vec{E} \perp c$ (dashed lines). The metal-insulator
    transition temperature is 67 $^{\circ}$C.\newline}
    \label{VO2fig3}
\end{SCfigure}

Fig. \ref{VO2fig3} shows the V $L_{2,3}$ XAS spectra of VO$_{2}$ taken in the insulating M$_1$-phase (top panel), and in the metallic R-phase (bottom panel). The general lineshape of the spectra is similar to that in earlier works, each of which only reports spectra for one particular polarization \cite{Abbate91,Goering94}. In our work, we have measured for each phase the spectra with two different light polarizations, namely $\vec{E} \parallel c$ (solid lines) and $\vec{E} \perp c$ (dashed lines). We observe a clear polarization dependence for the insulating phase. By contrast, the polarization dependence is quite weak for the metallic phase. Fig. \ref{VO2fig3} shows for each phase also the dichroic spectrum, i.e. the difference between the spectra taken with the two polarizations. One now can see that the dichroic spectrum of the insulating phase has not only a larger amplitude, but also a very different lineshape then that of the metallic phase.

\begin{figure}[h]
     \includegraphics[width=120mm]{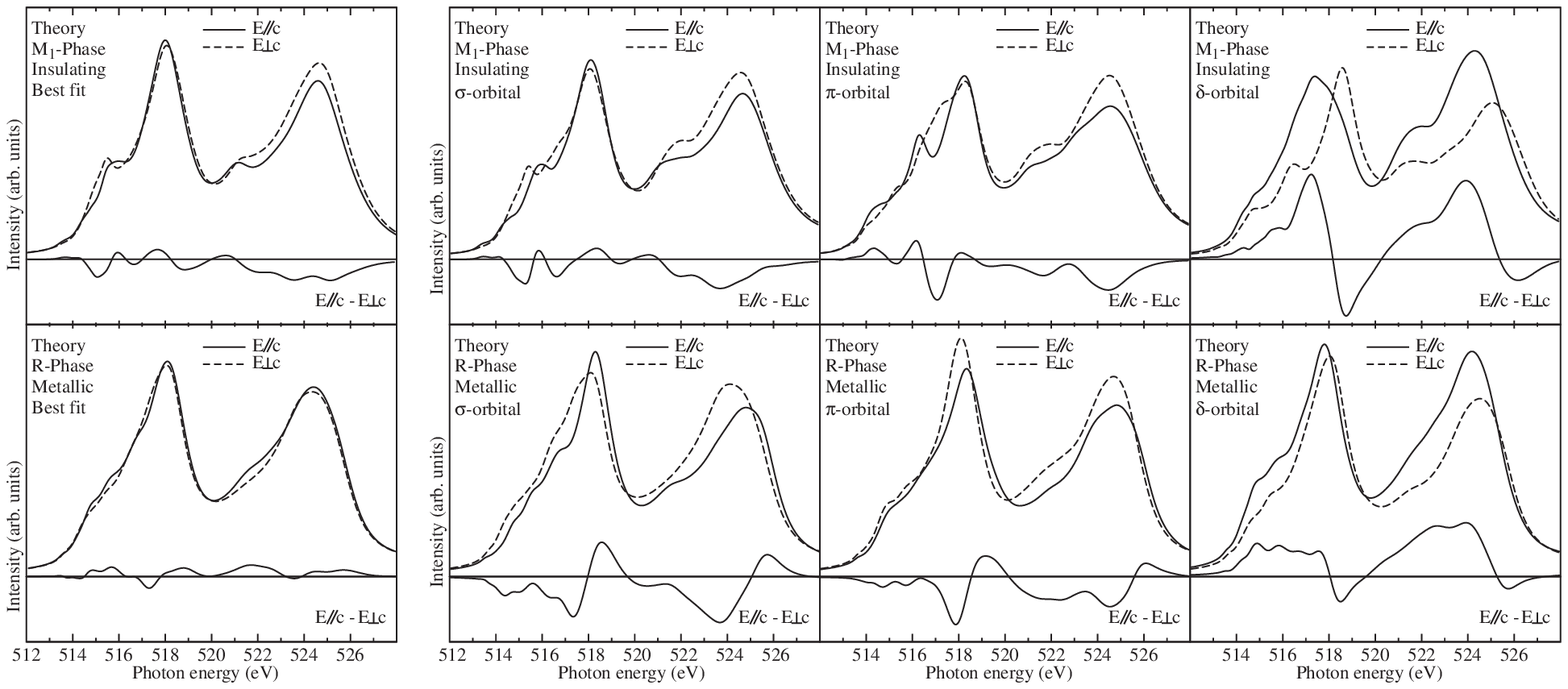}
     \caption{Theoretical simulations for the polarization dependent
     V $L_{2,3}$ XAS spectra.
     Left panel: best fit to the experimental spectra, using orbital
     occupations as indicated in Table I.
     Right panel: simulations for the V $3d^{1}$ ion assuming a pure
     $\sigma$, $\pi$, or $\delta$ orbital occupation.
     Top part of the panels: simulations for the insulating M$_1$-phase.
     Bottom part of the panels: \textit{idem} for the metallic R-phase.}
     \label{VO2fig4}
\end{figure}

In the subsequent section we will describe how one can determine the orbital occupation of the V $3d$ shell from the lineshape of the spectra and the dichroism in those spectra. First of all, it is important to note, that one may be tempted to think that the polarization dependence in the insulating phase, although clear, is still rather small since the amplitude of the dichroic spectrum is not larger than 10\% of the strongest peaks in the spectra, and that therefore  the orbital occupation will not be strongly polarized. This, however, would be an incorrect statement, since one should not see a strong polarization for a system which contain many holes in the $3d$ shell. In the particular case of VO$_{2}$: even if the $3d^{1}$ state would be fully orbitally polarized, there still will be 9 possible orbitals left to choose for the $2p$ core electron to make the $L_{2,3}$ XAS transition, so that the maximum dichroic signal cannot be expected to be very large.

To extract information concerning the orbital occupation from the V $L_{2,3}$ XAS spectra, we have performed simulations of the atomic-like $2p^{6}3d^{1} \rightarrow 2p^{5}3d^{2}$ transitions using the well-proven configuration interaction cluster model \cite{Tanaka94,Groot94,Thole97}. The method uses a VO$_6$ cluster which includes the full atomic multiplet theory and the local effects of the solid. It accounts for the intra-atomic $3d$-$3d$ and $2p$-$3d$ Coulomb and exchange interactions, the atomic $2p$ and $3d$ spin-orbit couplings, the O $2p$ - V $3d$ hybridization, and local crystal field parameters. Parameters for the multipole part of the Coulomb and exchange interactions were given by the Hartree-Fock values \cite{Tanaka94}, while the monopole parts ($U_{dd}$, $U_{pd}$) as well as the O $2p$ - V $3d$ charge transfer energy were determined from photoemission experiments \cite{Bocquet96}. The one-electron parameters such as the O $2p$ - V $3d$ and O $2p$ - O $2p$ transfer integrals as well as the local crystal fields were extracted from the LDA band structure results \cite{Korotin03,Streltsov05} for the crystal structure corresponding to each of the two phases of VO$_{2}$. The simulations have been carried out using the XTLS 8.0 programm \cite{Tanaka04, Tanaka94, paramVO2}.

Fig. \ref{VO2fig4} shows the results of our theoretical simulations of the spectra. In the top part of the right panel we have simulated the insulating M$_1$-phase spectra for the following three scenarios: the V $3d^{1}$ ion is set either in the pure $\sigma$, $\pi$, or $\delta$-orbital symmetry. One can clearly observe that the different orbital symmetries will lead to very different spectra with quite different polarization dependence. One can notice that the $\sigma$-orbital scenario resembles the experimental spectra the most, especially when one focuses on the most excitonic part of the spectrum, namely between 512 and 516 eV. In a simulation with the V ion in the true ground state symmetry, belonging to the proper local crystal fields of the M$_1$-phase, we find, as shown in the left panel, even a better fit to the experimental data. The corresponding orbital symmetry, as listed in the left column of Table \ref{VO2table}, has indeed overwhelmingly the $\sigma$ character (0.81), and only very little $\pi$ (0.10) and $\delta$ (0.09).

We have also simulated the spectra in the metallic R-phase, again for the three scenarios in which the V ion is set to have either the pure $\sigma$, $\pi$, or $\delta$-orbital symmetry. The bottom part of the right panel shows that each scenario results in quite different spectra and polarization dependence. We also note that each of the R-phase scenario gives spectra different from the corresponding M$_1$-phase, simply because of the differences in the local electronic structure, resulting from the different crystal structure. Important is now that none of the three scenarios of the R-phase give good agreement with the experimental spectra. Apparently, the V ion has an orbital symmetry which is very far from a pure $\sigma$, or $\pi$, or $\delta$. We now approximate the initial state symmetry of the V ion by a linear combination of those three symmetries, and optimize the relative weights to obtain the best fit to the experiment, with the emphasis on the excitonic part. The left panel of Fig. \ref{VO2fig4} shows that this state is built up of 0.33 $\sigma$, 0.16 $\pi$ and 0.51 $\delta$ symmetries, see also the 5th column of Table \ref{VO2table}. It seems thus that in the metallic phase the V orbital occupation is almost isotropic.

\begin{table}\label{occupationtable}
\begin{tabular}{|l|c|c|c|c||c|c|c|c|}
\hline
     & \multicolumn{4}{c||}{M$_1$-phase}& \multicolumn{4}{c|}{R-phase}\\
     & \multicolumn{2}{c|}{fit to exp.}&\multicolumn{2}{c||}{\textit{ab-initio}}
     & \multicolumn{2}{c|}{fit to exp.}&\multicolumn{2}{c|}{\textit{ab-initio}}\\
     & sym.& $3d$ occ. & LDA & LDA+U & sym.& $3d$ occ. & LDA & LDA+U\\
\hline
$\sigma$ & 0.81& 0.86& 0.64& 0.89& 0.33& 0.41& 0.43& 0.20 \\
$\pi$    & 0.10& 0.21& 0.39& 0.23& 0.16& 0.25& 0.35& 0.24 \\
$\delta$ & 0.09& 0.17& 0.41& 0.25& 0.51& 0.58& 0.67& 0.97 \\
$e_{g1}$ & 0.00& 0.27& 0.46& 0.40& 0.00& 0.27& 0.47& 0.42 \\
$e_{g2}$ & 0.00& 0.32& 0.53& 0.51& 0.00& 0.27& 0.48& 0.48 \\
\hline
  tot.   & 1.00& 1.83& 2.43& 2.48& 1.00& 1.78& 2.40& 2.31 \\
\hline
\end{tabular}
\caption{Symmetry and orbital occupation of the $3d$ shell of VO$_{2}$ in the
M$_1$- and R-phase.}
\label{VO2table}
\end{table}

In Table \ref{VO2table} we have also listed the $3d$ orbital occupation as found from the simulations of the experimental spectra. These numbers are not identical to the symmetry occupation numbers because of the covalency, i.e. the hybridization of the V $3d$ with the surrounding O $2p$ ligands. We now can compare our findings directly with the numbers from our LDA and LDA+U calculations \cite{Korotin03,Streltsov05}. We note that our LDA band structure is quite similar to the one published earlier \cite{Wentzcovitch94,Eyert02}, and that the occupation numbers of our LDA+U is in close agreement with the one published very recently \cite{Liebsch05}. For the insulating M$_1$-phase, we find that the orbital occupation, which is highly $(\sigma)$ polarized, is well reproduced by the LDA+U model but not so by the standard LDA, see Table I. On the other hand, for the metallic R-phase, we observe that the almost isotropic orbital occupation as experimentally determined is well reproduced by the LDA, but not so by the LDA+U. It seems that the LDA tends to underestimate the orbital polarization, which makes the method less suitable for the insulating phase. The LDA+U, on the other hand, tend to overestimate it, which puts this approach in disadvantage for the metallic phase. These problems are likely to be related to the fact that the LDA cannot reproduce the insulating state in the M$_1$-phase, while the LDA+U does not give the metallic state for the R-phase. Nevertheless, the general trend that the orbital occupation is more $\sigma$-polarized in the M$_1$-phase is predicted correctly in both approaches.

In comparing our experimental results with the DMFT calculations, we note that in one implementation of the standard LDA+DMFT method the change in orbital polarization is too small, which has been attributed to the fact that the insulating phase cannot be reproduced \cite{Liebsch05}. Very exciting is that an LDA+cluster/DMFT study \cite{Biermann05} has been very successful in reproducing the strong change in orbital polarization, indicating the importance of the $k$-dependence of the self-energy correction.

The significant outcome of our experiments is that the orbital occupation changes from almost isotropic in the metallic phase to the almost completely $\sigma$-polarized in the insulating phase, in close agreement with the two-site cluster model \cite{Tanaka04}. This very strong orbital polarization leads in fact to a change of the electronic structure of VO$_{2}$ from a 3-dimensional to effectively an one-dimensional system \cite{Khomskii05}. The V ions in the chain along the c-axis are then very susceptible to a Peierls transition. In this respect, the MIT in VO$_{2}$ can indeed be regarded as a Peierls transition \cite{Wentzcovitch94}. However, to achieve the required dramatic change of the orbital occupation one also need the condition that strong electron correlations bring this narrow band system close to the Mott regime \cite{Rice94}. The MIT in VO$_{2}$ may therefore be labelled as a "collaborative" Mott-Peierls transition.

To conclude, we have found direct experimental evidence for an orbital switching in the V $3d$ states across the metal-insulator transition in VO$_{2}$. We have used soft-x-ray absorption spectroscopy at the V $L_{2,3}$ edges as a sensitive local probe, and have determined quantitatively the orbital occupation on both sides of the transition. These results strongly suggest that, in going from the metallic to the insulating state, the orbital occupation changes in such a manner that the system becomes more 1-dimensional and more susceptible to a Peierls-like transition, and that this orbital change can only be so dramatic if the system is close to a Mott insulating regime.

We acknowledge the NSRRC staff for providing us with an extremely stable beam. We acknowledge Lucie Hamdan for her skillful technical assistance. The research in K\"oln is supported by the Deutsche Forschungsgemeinschaft through SFB 608 and the research in Ekaterinburg by grants RFFI 04-02-16096 and yp.01.01.059.

\appendix

\chapter{Slater integrals for $3d$ and $4d$ elements}
\label{ApendixSlaterIntegrals}

The Slater integrals used in this thesis are calculated within the Hartree-Fock approximation with the use of R. D. Cowans coded RCN36K. The basics of the functionality of this code is described in the book: "The theory of atomic structure and spectra" \cite{Cowan81}. Some calculations on transition metal ions with low valences converged with an extremely large ionic radius. When the value of $r^2$ ($r^4$) exceeded 10 \AA$^2$ (\AA$^4$) we replaced it by an $*$. For these configurations non of the Slater-integrals should be trusted to work within a solid.
\\
\\
The values of $r^2$ and $r^4$ are in \AA$^2$ and \AA$^4$ respectively. All other values are in eV.
\\
\\
\begin{fontsize}{8}{8}
\begin{huge}$3d$ elements\end{huge}
\\
\\
\\
\begin{tabular}{|l|r@{.}lr@{.}lr@{.}lr@{.}lr@{.}lr@{.}lr@{.}lr@{.}lr@{.}l|}
\hline
K& \multicolumn{2}{c}{$r^2$} & \multicolumn{2}{c}{$r^4$} & \multicolumn{2}{c}{$\zeta_d$} & \multicolumn{2}{c}{$F^2_{dd}$} & \multicolumn{2}{c}{$F^4_{dd}$} & \multicolumn{2}{c}{$\zeta_{2p}$} & \multicolumn{2}{c}{$F^2_{pd}$}  & \multicolumn{2}{c}{$G^1_{pd}$} & \multicolumn{2}{c}{$G^3_{pd}$} \vline \\
\hline
$2p^{6}$ $3d^{1\phantom{0}}$& \multicolumn{2}{c}{$*$}& \multicolumn{2}{c}{$*$}& $0$&$000$& \phantom{0}$0$&$000$& \phantom{0}$0$&$000$& \multicolumn{2}{c}{}& \multicolumn{2}{c}{}& \multicolumn{2}{c}{}& \multicolumn{2}{c}{} \vline \\
$2p^{5}$ $3d^{2\phantom{0}}$& \multicolumn{2}{c}{$*$}& \multicolumn{2}{c}{$*$}& $0$&$002$& \phantom{0}$2$&$723$& \phantom{0}$1$&$622$& \phantom{0}$1$&$880$& $1$&$045$& $0$&$583$& $0$&$329$ \\
$2p^{5}$ $3d^{1\phantom{0}}$& $1$&$687$& $6$&$860$& $0$&$005$&  \phantom{0}$0$&$000$& \phantom{0}$0$&$000$& \phantom{0}$1$&$880$& $2$&$202$& $1$&$318$& $0$&$745$ \\
\hline\end{tabular}
\\
\\
\\
\begin{tabular}{|l|r@{.}lr@{.}lr@{.}lr@{.}lr@{.}lr@{.}lr@{.}lr@{.}lr@{.}l|}
\hline
Ca& \multicolumn{2}{c}{$r^2$} & \multicolumn{2}{c}{$r^4$} & \multicolumn{2}{c}{$\zeta_d$} & \multicolumn{2}{c}{$F^2_{dd}$} & \multicolumn{2}{c}{$F^4_{dd}$} & \multicolumn{2}{c}{$\zeta_{2p}$} & \multicolumn{2}{c}{$F^2_{pd}$}  & \multicolumn{2}{c}{$G^1_{pd}$} & \multicolumn{2}{c}{$G^3_{pd}$} \vline \\
\hline
$2p^{6}$ $3d^{2\phantom{0}}$& \multicolumn{2}{c}{$*$}& \multicolumn{2}{c}{$*$}& $0$&$001$& \phantom{0}$2$&$433$& \phantom{0}$1$&$452$& \multicolumn{2}{c}{}& \multicolumn{2}{c}{}& \multicolumn{2}{c}{}& \multicolumn{2}{c}{} \vline \\
$2p^{5}$ $3d^{3\phantom{0}}$& \multicolumn{2}{c}{$*$}& \multicolumn{2}{c}{$*$}& $0$&$005$& \phantom{0}$4$&$205$& \phantom{0}$2$&$529$& \phantom{0}$2$&$405$& $1$&$970$& $1$&$204$& $0$&$681$ \\
$2p^{6}$ $3d^{1\phantom{0}}$& \multicolumn{2}{c}{$*$}& \multicolumn{2}{c}{$*$}& $0$&$004$& \phantom{0}$0$&$000$& \phantom{0}$0$&$000$& \multicolumn{2}{c}{}& \multicolumn{2}{c}{}& \multicolumn{2}{c}{}& \multicolumn{2}{c}{} \vline \\
$2p^{5}$ $3d^{2\phantom{0}}$& $1$&$202$& $3$&$577$& $0$&$008$& \phantom{0}$6$&$258$& \phantom{0}$3$&$851$& \phantom{0}$2$&$405$& $2$&$917$& $1$&$851$& $1$&$048$ \\
$2p^{5}$ $3d^{1\phantom{0}}$& $0$&$781$& $1$&$306$& $0$&$011$& \phantom{0}$0$&$000$& \phantom{0}$0$&$000$& \phantom{0}$2$&$404$& $3$&$793$& $2$&$511$& $1$&$423$ \\
\hline\end{tabular}
\\
\\
\\
\begin{tabular}{|l|r@{.}lr@{.}lr@{.}lr@{.}lr@{.}lr@{.}lr@{.}lr@{.}lr@{.}l|}
\hline
Sc& \multicolumn{2}{c}{$r^2$} & \multicolumn{2}{c}{$r^4$} & \multicolumn{2}{c}{$\zeta_d$} & \multicolumn{2}{c}{$F^2_{dd}$} & \multicolumn{2}{c}{$F^4_{dd}$} & \multicolumn{2}{c}{$\zeta_{2p}$} & \multicolumn{2}{c}{$F^2_{pd}$}  & \multicolumn{2}{c}{$G^1_{pd}$} & \multicolumn{2}{c}{$G^3_{pd}$} \vline \\
\hline
$2p^{6}$ $3d^{3\phantom{0}}$& \multicolumn{2}{c}{$*$}& \multicolumn{2}{c}{$*$}& $0$&$005$& \phantom{0}$3$&$855$& \phantom{0}$2$&$311$& \multicolumn{2}{c}{}& \multicolumn{2}{c}{}& \multicolumn{2}{c}{}& \multicolumn{2}{c}{} \vline \\
$2p^{5}$ $3d^{4\phantom{0}}$& \multicolumn{2}{c}{$*$}& \multicolumn{2}{c}{$*$}& $0$&$010$& \phantom{0}$5$&$326$& \phantom{0}$3$&$224$& \phantom{0}$3$&$034$& $2$&$696$& $1$&$733$& $0$&$982$ \\
$2p^{6}$ $3d^{2\phantom{0}}$& $1$&$356$& $4$&$559$& $0$&$007$& \phantom{0}$5$&$851$& \phantom{0}$3$&$590$& \multicolumn{2}{c}{}& \multicolumn{2}{c}{}& \multicolumn{2}{c}{}& \multicolumn{2}{c}{} \vline \\
$2p^{5}$ $3d^{3\phantom{0}}$& $0$&$943$& $2$&$247$& $0$&$013$& \phantom{0}$7$&$116$& \phantom{0}$4$&$385$& \phantom{0}$3$&$033$& $3$&$532$& $2$&$331$& $1$&$321$ \\
$2p^{6}$ $3d^{1\phantom{0}}$& $0$&$844$& $1$&$535$& $0$&$010$& \phantom{0}$0$&$000$& \phantom{0}$0$&$000$& \multicolumn{2}{c}{}& \multicolumn{2}{c}{}& \multicolumn{2}{c}{}& \multicolumn{2}{c}{} \vline \\
$2p^{5}$ $3d^{2\phantom{0}}$& $0$&$658$& $0$&$946$& $0$&$017$& \phantom{0}$8$&$530$& \phantom{0}$5$&$321$& \phantom{0}$3$&$032$& $4$&$332$& $2$&$950$& $1$&$674$ \\
$2p^{5}$ $3d^{1\phantom{0}}$& $0$&$516$& $0$&$535$& $0$&$020$& \phantom{0}$0$&$000$& \phantom{0}$0$&$000$& \phantom{0}$3$&$032$& $5$&$095$& $3$&$582$& $2$&$035$ \\
\hline\end{tabular}
\\
\\
\\
\begin{tabular}{|l|r@{.}lr@{.}lr@{.}lr@{.}lr@{.}lr@{.}lr@{.}lr@{.}lr@{.}l|}
\hline
Ti& \multicolumn{2}{c}{$r^2$} & \multicolumn{2}{c}{$r^4$} & \multicolumn{2}{c}{$\zeta_d$} & \multicolumn{2}{c}{$F^2_{dd}$} & \multicolumn{2}{c}{$F^4_{dd}$} & \multicolumn{2}{c}{$\zeta_{2p}$} & \multicolumn{2}{c}{$F^2_{pd}$}  & \multicolumn{2}{c}{$G^1_{pd}$} & \multicolumn{2}{c}{$G^3_{pd}$} \vline \\
\hline
$2p^{6}$ $3d^{4\phantom{0}}$& \multicolumn{2}{c}{$*$}& \multicolumn{2}{c}{$*$}& $0$&$009$& \phantom{0}$5$&$002$& \phantom{0}$3$&$019$& \multicolumn{2}{c}{}& \multicolumn{2}{c}{}& \multicolumn{2}{c}{}& \multicolumn{2}{c}{} \vline \\
$2p^{5}$ $3d^{5\phantom{0}}$& $1$&$326$& $6$&$323$& $0$&$015$& \phantom{0}$6$&$261$& \phantom{0}$3$&$806$& \phantom{0}$3$&$778$& $3$&$329$& $2$&$216$& $1$&$257$ \\
$2p^{6}$ $3d^{3\phantom{0}}$& $1$&$029$& $2$&$692$& $0$&$012$& \phantom{0}$6$&$780$& \phantom{0}$4$&$167$& \multicolumn{2}{c}{}& \multicolumn{2}{c}{}& \multicolumn{2}{c}{}& \multicolumn{2}{c}{} \vline \\
$2p^{5}$ $3d^{4\phantom{0}}$& $0$&$775$& $1$&$546$& $0$&$019$& \phantom{0}$7$&$888$& \phantom{0}$4$&$865$& \phantom{0}$3$&$777$& $4$&$098$& $2$&$783$& $1$&$578$ \\
$2p^{6}$ $3d^{2\phantom{0}}$& $0$&$699$& $1$&$075$& $0$&$016$& \phantom{0}$8$&$243$& \phantom{0}$5$&$132$& \multicolumn{2}{c}{}& \multicolumn{2}{c}{}& \multicolumn{2}{c}{}& \multicolumn{2}{c}{} \vline \\
$2p^{5}$ $3d^{3\phantom{0}}$& $0$&$566$& $0$&$713$& $0$&$023$& \phantom{0}$9$&$213$& \phantom{0}$5$&$744$& \phantom{0}$3$&$776$& $4$&$849$& $3$&$376$& $1$&$917$ \\
$2p^{6}$ $3d^{1\phantom{0}}$& $0$&$538$& $0$&$587$& $0$&$019$& \phantom{0}$0$&$000$& \phantom{0}$0$&$000$& \multicolumn{2}{c}{}& \multicolumn{2}{c}{}& \multicolumn{2}{c}{}& \multicolumn{2}{c}{} \vline \\
\hline\end{tabular}
\\
\\
\\
\begin{tabular}{|l|r@{.}lr@{.}lr@{.}lr@{.}lr@{.}lr@{.}lr@{.}lr@{.}lr@{.}l|}
\hline
V& \multicolumn{2}{c}{$r^2$} & \multicolumn{2}{c}{$r^4$} & \multicolumn{2}{c}{$\zeta_d$} & \multicolumn{2}{c}{$F^2_{dd}$} & \multicolumn{2}{c}{$F^4_{dd}$} & \multicolumn{2}{c}{$\zeta_{2p}$} & \multicolumn{2}{c}{$F^2_{pd}$}  & \multicolumn{2}{c}{$G^1_{pd}$} & \multicolumn{2}{c}{$G^3_{pd}$} \vline \\
\hline
$2p^{6}$ $3d^{5\phantom{0}}$& $1$&$447$& $7$&$507$& $0$&$014$& \phantom{0}$5$&$970$& \phantom{0}$3$&$620$& \multicolumn{2}{c}{}& \multicolumn{2}{c}{}& \multicolumn{2}{c}{}& \multicolumn{2}{c}{} \vline \\
$2p^{5}$ $3d^{6\phantom{0}}$& $1$&$044$& $3$&$981$& $0$&$022$& \phantom{0}$7$&$095$& \phantom{0}$4$&$324$& \phantom{0}$4$&$652$& $3$&$911$& $2$&$673$& $1$&$517$ \\
$2p^{6}$ $3d^{4\phantom{0}}$& $0$&$829$& $1$&$786$& $0$&$018$& \phantom{0}$7$&$599$& \phantom{0}$4$&$676$& \multicolumn{2}{c}{}& \multicolumn{2}{c}{}& \multicolumn{2}{c}{}& \multicolumn{2}{c}{} \vline \\
$2p^{5}$ $3d^{5\phantom{0}}$& $0$&$655$& $1$&$124$& $0$&$026$& \phantom{0}$8$&$612$& \phantom{0}$5$&$314$& \phantom{0}$4$&$651$& $4$&$634$& $3$&$218$& $1$&$827$ \\
$2p^{6}$ $3d^{3\phantom{0}}$& $0$&$595$& $0$&$793$& $0$&$022$& \phantom{0}$8$&$961$& \phantom{0}$5$&$576$& \multicolumn{2}{c}{}& \multicolumn{2}{c}{}& \multicolumn{2}{c}{}& \multicolumn{2}{c}{} \vline \\
$2p^{5}$ $3d^{4\phantom{0}}$& $0$&$494$& $0$&$553$& $0$&$031$& \phantom{0}$9$&$875$& \phantom{0}$6$&$152$& \phantom{0}$4$&$650$& $5$&$351$& $3$&$792$& $2$&$154$ \\
$2p^{6}$ $3d^{2\phantom{0}}$& $0$&$470$& $0$&$456$& $0$&$027$& $10$&$126$& \phantom{0}$6$&$353$& \multicolumn{2}{c}{}& \multicolumn{2}{c}{}& \multicolumn{2}{c}{}& \multicolumn{2}{c}{} \vline \\
\hline\end{tabular}
\\
\\
\\
\begin{tabular}{|l|r@{.}lr@{.}lr@{.}lr@{.}lr@{.}lr@{.}lr@{.}lr@{.}lr@{.}l|}
\hline
Cr& \multicolumn{2}{c}{$r^2$} & \multicolumn{2}{c}{$r^4$} & \multicolumn{2}{c}{$\zeta_d$} & \multicolumn{2}{c}{$F^2_{dd}$} & \multicolumn{2}{c}{$F^4_{dd}$} & \multicolumn{2}{c}{$\zeta_{2p}$} & \multicolumn{2}{c}{$F^2_{pd}$}  & \multicolumn{2}{c}{$G^1_{pd}$} & \multicolumn{2}{c}{$G^3_{pd}$} \vline \\
\hline
$2p^{6}$ $3d^{6\phantom{0}}$& $1$&$121$& $4$&$612$& $0$&$021$& \phantom{0}$6$&$833$& \phantom{0}$4$&$155$& \multicolumn{2}{c}{}& \multicolumn{2}{c}{}& \multicolumn{2}{c}{}& \multicolumn{2}{c}{} \vline \\
$2p^{5}$ $3d^{7\phantom{0}}$& $0$&$856$& $2$&$709$& $0$&$030$& \phantom{0}$7$&$866$& \phantom{0}$4$&$803$& \phantom{0}$5$&$671$& $4$&$460$& $3$&$112$& $1$&$767$ \\
$2p^{6}$ $3d^{5\phantom{0}}$& $0$&$692$& $1$&$267$& $0$&$025$& \phantom{0}$8$&$355$& \phantom{0}$5$&$146$& \multicolumn{2}{c}{}& \multicolumn{2}{c}{}& \multicolumn{2}{c}{}& \multicolumn{2}{c}{} \vline \\
$2p^{5}$ $3d^{6\phantom{0}}$& $0$&$565$& $0$&$851$& $0$&$035$& \phantom{0}$9$&$303$& \phantom{0}$5$&$742$& \phantom{0}$5$&$669$& $5$&$151$& $3$&$641$& $2$&$068$ \\
$2p^{6}$ $3d^{4\phantom{0}}$& $0$&$515$& $0$&$605$& $0$&$030$& \phantom{0}$9$&$648$& \phantom{0}$6$&$001$& \multicolumn{2}{c}{}& \multicolumn{2}{c}{}& \multicolumn{2}{c}{}& \multicolumn{2}{c}{} \vline \\
$2p^{5}$ $3d^{5\phantom{0}}$& $0$&$437$& $0$&$439$& $0$&$041$& $10$&$521$& \phantom{0}$6$&$551$& \phantom{0}$5$&$668$& $5$&$840$& $4$&$201$& $2$&$387$ \\
$2p^{6}$ $3d^{3\phantom{0}}$& $0$&$416$& $0$&$362$& $0$&$035$& $10$&$776$& \phantom{0}$6$&$754$& \multicolumn{2}{c}{}& \multicolumn{2}{c}{}& \multicolumn{2}{c}{}& \multicolumn{2}{c}{} \vline \\
\hline\end{tabular}
\\
\\
\\
\begin{tabular}{|l|r@{.}lr@{.}lr@{.}lr@{.}lr@{.}lr@{.}lr@{.}lr@{.}lr@{.}l|}
\hline
Mn& \multicolumn{2}{c}{$r^2$} & \multicolumn{2}{c}{$r^4$} & \multicolumn{2}{c}{$\zeta_d$} & \multicolumn{2}{c}{$F^2_{dd}$} & \multicolumn{2}{c}{$F^4_{dd}$} & \multicolumn{2}{c}{$\zeta_{2p}$} & \multicolumn{2}{c}{$F^2_{pd}$}  & \multicolumn{2}{c}{$G^1_{pd}$} & \multicolumn{2}{c}{$G^3_{pd}$} \vline \\
\hline
$2p^{6}$ $3d^{7\phantom{0}}$& \multicolumn{2}{c}{$*$}& \multicolumn{2}{c}{$*$}& $0$&$032$& \phantom{0}$5$&$312$& \phantom{0}$3$&$268$& \multicolumn{2}{c}{}& \multicolumn{2}{c}{}& \multicolumn{2}{c}{}& \multicolumn{2}{c}{} \vline \\
$2p^{5}$ $3d^{8\phantom{0}}$& $0$&$721$& $1$&$945$& $0$&$040$& \phantom{0}$8$&$594$& \phantom{0}$5$&$255$& \phantom{0}$6$&$849$& $4$&$987$& $3$&$539$& $2$&$010$ \\
$2p^{6}$ $3d^{6\phantom{0}}$& $0$&$592$& $0$&$942$& $0$&$035$& \phantom{0}$9$&$072$& \phantom{0}$5$&$590$& \multicolumn{2}{c}{}& \multicolumn{2}{c}{}& \multicolumn{2}{c}{}& \multicolumn{2}{c}{} \vline \\
$2p^{5}$ $3d^{7\phantom{0}}$& $0$&$495$& $0$&$662$& $0$&$046$& \phantom{0}$9$&$971$& \phantom{0}$6$&$156$& \phantom{0}$6$&$847$& $5$&$652$& $4$&$056$& $2$&$304$ \\
$2p^{6}$ $3d^{5\phantom{0}}$& $0$&$452$& $0$&$475$& $0$&$040$& $10$&$315$& \phantom{0}$6$&$413$& \multicolumn{2}{c}{}& \multicolumn{2}{c}{}& \multicolumn{2}{c}{}& \multicolumn{2}{c}{} \vline \\
$2p^{5}$ $3d^{6\phantom{0}}$& $0$&$390$& $0$&$355$& $0$&$053$& $11$&$154$& \phantom{0}$6$&$942$& \phantom{0}$6$&$846$& $6$&$320$& $4$&$603$& $2$&$617$ \\
$2p^{6}$ $3d^{4\phantom{0}}$& $0$&$371$& $0$&$293$& $0$&$046$& $11$&$414$& \phantom{0}$7$&$147$& \multicolumn{2}{c}{}& \multicolumn{2}{c}{}& \multicolumn{2}{c}{}& \multicolumn{2}{c}{} \vline \\
\hline\end{tabular}
\\
\\
\\
\begin{tabular}{|l|r@{.}lr@{.}lr@{.}lr@{.}lr@{.}lr@{.}lr@{.}lr@{.}lr@{.}l|}
\hline
Fe& \multicolumn{2}{c}{$r^2$} & \multicolumn{2}{c}{$r^4$} & \multicolumn{2}{c}{$\zeta_d$} & \multicolumn{2}{c}{$F^2_{dd}$} & \multicolumn{2}{c}{$F^4_{dd}$} & \multicolumn{2}{c}{$\zeta_{2p}$} & \multicolumn{2}{c}{$F^2_{pd}$}  & \multicolumn{2}{c}{$G^1_{pd}$} & \multicolumn{2}{c}{$G^3_{pd}$} \vline \\
\hline
$2p^{6}$ $3d^{8\phantom{0}}$& \multicolumn{2}{c}{$*$}& \multicolumn{2}{c}{$*$}& $0$&$021$& \phantom{0}$2$&$087$& \phantom{0}$1$&$275$& \multicolumn{2}{c}{}& \multicolumn{2}{c}{}& \multicolumn{2}{c}{}& \multicolumn{2}{c}{} \vline \\
$2p^{5}$ $3d^{9\phantom{0}}$& $0$&$620$& $1$&$452$& $0$&$053$& \phantom{0}$9$&$293$& \phantom{0}$5$&$688$& \phantom{0}$8$&$203$& $5$&$498$& $3$&$957$& $2$&$248$ \\
$2p^{6}$ $3d^{7\phantom{0}}$& $0$&$515$& $0$&$723$& $0$&$045$& \phantom{0}$9$&$761$& \phantom{0}$6$&$017$& \multicolumn{2}{c}{}& \multicolumn{2}{c}{}& \multicolumn{2}{c}{}& \multicolumn{2}{c}{} \vline \\
$2p^{5}$ $3d^{8\phantom{0}}$& $0$&$438$& $0$&$527$& $0$&$059$& $10$&$622$& \phantom{0}$6$&$559$& \phantom{0}$8$&$202$& $6$&$142$& $4$&$463$& $2$&$537$ \\
$2p^{6}$ $3d^{6\phantom{0}}$& $0$&$402$& $0$&$380$& $0$&$052$& $10$&$965$& \phantom{0}$6$&$815$& \multicolumn{2}{c}{}& \multicolumn{2}{c}{}& \multicolumn{2}{c}{}& \multicolumn{2}{c}{} \vline \\
$2p^{5}$ $3d^{7\phantom{0}}$& $0$&$351$& $0$&$292$& $0$&$067$& $11$&$778$& \phantom{0}$7$&$327$& \phantom{0}$8$&$200$& $6$&$792$& $5$&$000$& $2$&$843$ \\
$2p^{6}$ $3d^{5\phantom{0}}$& $0$&$334$& $0$&$240$& $0$&$059$& $12$&$042$& \phantom{0}$7$&$534$& \multicolumn{2}{c}{}& \multicolumn{2}{c}{}& \multicolumn{2}{c}{}& \multicolumn{2}{c}{} \vline \\
\hline\end{tabular}
\\
\\
\\
\begin{tabular}{|l|r@{.}lr@{.}lr@{.}lr@{.}lr@{.}lr@{.}lr@{.}lr@{.}lr@{.}l|}
\hline
Co& \multicolumn{2}{c}{$r^2$} & \multicolumn{2}{c}{$r^4$} & \multicolumn{2}{c}{$\zeta_d$} & \multicolumn{2}{c}{$F^2_{dd}$} & \multicolumn{2}{c}{$F^4_{dd}$} & \multicolumn{2}{c}{$\zeta_{2p}$} & \multicolumn{2}{c}{$F^2_{pd}$}  & \multicolumn{2}{c}{$G^1_{pd}$} & \multicolumn{2}{c}{$G^3_{pd}$} \vline \\
\hline
$2p^{6}$ $3d^{9\phantom{0}}$& \multicolumn{2}{c}{$*$}& \multicolumn{2}{c}{$*$}& $0$&$027$& \phantom{0}$2$&$196$& \phantom{0}$1$&$348$& \multicolumn{2}{c}{}& \multicolumn{2}{c}{}& \multicolumn{2}{c}{}& \multicolumn{2}{c}{} \vline \\
$2p^{5}$ $3d^{10}$& $0$&$541$& $1$&$117$& $0$&$067$& \phantom{0}$9$&$969$& \phantom{0}$6$&$108$& \phantom{0}$9$&$752$& $5$&$996$& $4$&$367$& $2$&$482$ \\
$2p^{6}$ $3d^{8\phantom{0}}$& $0$&$453$& $0$&$570$& $0$&$059$& $10$&$430$& \phantom{0}$6$&$431$& \multicolumn{2}{c}{}& \multicolumn{2}{c}{}& \multicolumn{2}{c}{}& \multicolumn{2}{c}{} \vline \\
$2p^{5}$ $3d^{9\phantom{0}}$& $0$&$392$& $0$&$427$& $0$&$075$& $11$&$261$& \phantom{0}$6$&$954$& \phantom{0}$9$&$750$& $6$&$624$& $4$&$865$& $2$&$766$ \\
$2p^{6}$ $3d^{7\phantom{0}}$& $0$&$360$& $0$&$310$& $0$&$066$& $11$&$604$& \phantom{0}$7$&$209$& \multicolumn{2}{c}{}& \multicolumn{2}{c}{}& \multicolumn{2}{c}{}& \multicolumn{2}{c}{} \vline \\
$2p^{5}$ $3d^{8\phantom{0}}$& $0$&$318$& $0$&$243$& $0$&$083$& $12$&$395$& \phantom{0}$7$&$707$& \phantom{0}$9$&$748$& $7$&$259$& $5$&$394$& $3$&$068$ \\
$2p^{6}$ $3d^{6\phantom{0}}$& $0$&$302$& $0$&$200$& $0$&$074$& $12$&$662$& \phantom{0}$7$&$916$& \multicolumn{2}{c}{}& \multicolumn{2}{c}{}& \multicolumn{2}{c}{}& \multicolumn{2}{c}{} \vline \\
\hline\end{tabular}
\\
\\
\\
\begin{tabular}{|l|r@{.}lr@{.}lr@{.}lr@{.}lr@{.}lr@{.}lr@{.}lr@{.}lr@{.}l|}
\hline
Ni& \multicolumn{2}{c}{$r^2$} & \multicolumn{2}{c}{$r^4$} & \multicolumn{2}{c}{$\zeta_d$} & \multicolumn{2}{c}{$F^2_{dd}$} & \multicolumn{2}{c}{$F^4_{dd}$} & \multicolumn{2}{c}{$\zeta_{2p}$} & \multicolumn{2}{c}{$F^2_{pd}$}  & \multicolumn{2}{c}{$G^1_{pd}$} & \multicolumn{2}{c}{$G^3_{pd}$} \vline \\
\hline
$2p^{6}$ $3d^{9\phantom{0}}$& $0$&$404$& $0$&$458$& $0$&$074$& $11$&$084$& \phantom{0}$6$&$835$& \multicolumn{2}{c}{}& \multicolumn{2}{c}{}& \multicolumn{2}{c}{}& \multicolumn{2}{c}{} \vline \\
$2p^{5}$ $3d^{10}$& $0$&$353$& $0$&$351$& $0$&$093$& $11$&$890$& \phantom{0}$7$&$343$& \phantom{0}$11$&$509$& $7$&$098$& $5$&$262$& $2$&$993$ \\
$2p^{6}$ $3d^{8\phantom{0}}$& $0$&$325$& $0$&$256$& $0$&$083$& $12$&$233$& \phantom{0}$7$&$597$& \multicolumn{2}{c}{}& \multicolumn{2}{c}{}& \multicolumn{2}{c}{}& \multicolumn{2}{c}{} \vline \\
$2p^{5}$ $3d^{9\phantom{0}}$& $0$&$289$& $0$&$204$& $0$&$102$& $13$&$005$& \phantom{0}$8$&$084$& \phantom{0}$11$&$507$& $7$&$720$& $5$&$783$& $3$&$290$ \\
$2p^{6}$ $3d^{7\phantom{0}}$& $0$&$275$& $0$&$168$& $0$&$091$& $13$&$276$& \phantom{0}$8$&$294$& \multicolumn{2}{c}{}& \multicolumn{2}{c}{}& \multicolumn{2}{c}{}& \multicolumn{2}{c}{} \vline \\
$2p^{5}$ $3d^{8\phantom{0}}$& $0$&$248$& $0$&$137$& $0$&$112$& $14$&$021$& \phantom{0}$8$&$763$& \phantom{0}$11$&$506$& $8$&$349$& $6$&$329$& $3$&$602$ \\
$2p^{6}$ $3d^{6\phantom{0}}$& $0$&$240$& $0$&$120$& $0$&$101$& $14$&$244$& \phantom{0}$8$&$944$& \multicolumn{2}{c}{}& \multicolumn{2}{c}{}& \multicolumn{2}{c}{}& \multicolumn{2}{c}{} \vline \\
$2p^{5}$ $3d^{7\phantom{0}}$& $0$&$218$& $0$&$100$& $0$&$122$& $14$&$965$& \phantom{0}$9$&$399$& \phantom{0}$11$&$505$& $8$&$984$& $6$&$898$& $3$&$929$ \\
\hline\end{tabular}
\\
\\
\\
\begin{tabular}{|l|r@{.}lr@{.}lr@{.}lr@{.}lr@{.}lr@{.}lr@{.}lr@{.}lr@{.}l|}
\hline
Cu& \multicolumn{2}{c}{$r^2$} & \multicolumn{2}{c}{$r^4$} & \multicolumn{2}{c}{$\zeta_d$} & \multicolumn{2}{c}{$F^2_{dd}$} & \multicolumn{2}{c}{$F^4_{dd}$} & \multicolumn{2}{c}{$\zeta_{2p}$} & \multicolumn{2}{c}{$F^2_{pd}$}  & \multicolumn{2}{c}{$G^1_{pd}$} & \multicolumn{2}{c}{$G^3_{pd}$} \vline \\
\hline
$2p^{6}$ $3d^{9\phantom{0}}$& $0$&$295$& $0$&$214$& $0$&$102$& $12$&$854$& \phantom{0}$7$&$980$& \multicolumn{2}{c}{}& \multicolumn{2}{c}{}& \multicolumn{2}{c}{}& \multicolumn{2}{c}{} \vline \\
$2p^{5}$ $3d^{10}$& $0$&$265$& $0$&$173$& $0$&$124$& $13$&$611$& \phantom{0}$8$&$457$& \phantom{0}$13$&$498$& $8$&$177$& $6$&$169$& $3$&$510$ \\
$2p^{6}$ $3d^{8\phantom{0}}$& $0$&$252$& $0$&$143$& $0$&$112$& $13$&$885$& \phantom{0}$8$&$669$& \multicolumn{2}{c}{}& \multicolumn{2}{c}{}& \multicolumn{2}{c}{}& \multicolumn{2}{c}{} \vline \\
$2p^{5}$ $3d^{9\phantom{0}}$& $0$&$228$& $0$&$118$& $0$&$135$& $14$&$617$& \phantom{0}$9$&$130$& \phantom{0}$13$&$496$& $8$&$796$& $6$&$708$& $3$&$818$ \\
$2p^{6}$ $3d^{7\phantom{0}}$& $0$&$221$& $0$&$103$& $0$&$123$& $14$&$845$& \phantom{0}$9$&$313$& \multicolumn{2}{c}{}& \multicolumn{2}{c}{}& \multicolumn{2}{c}{}& \multicolumn{2}{c}{} \vline \\
$2p^{5}$ $3d^{8\phantom{0}}$& $0$&$202$& $0$&$087$& $0$&$147$& $15$&$556$& \phantom{0}$9$&$762$& \phantom{0}$13$&$495$& $9$&$421$& $7$&$270$& $4$&$141$ \\
\hline\end{tabular}
\\
\\
\\
\begin{tabular}{|l|r@{.}lr@{.}lr@{.}lr@{.}lr@{.}lr@{.}lr@{.}lr@{.}lr@{.}l|}
\hline
Zn& \multicolumn{2}{c}{$r^2$} & \multicolumn{2}{c}{$r^4$} & \multicolumn{2}{c}{$\zeta_d$} & \multicolumn{2}{c}{$F^2_{dd}$} & \multicolumn{2}{c}{$F^4_{dd}$} & \multicolumn{2}{c}{$\zeta_{2p}$} & \multicolumn{2}{c}{$F^2_{pd}$}  & \multicolumn{2}{c}{$G^1_{pd}$} & \multicolumn{2}{c}{$G^3_{pd}$} \vline \\
\hline
$2p^{6}$ $3d^{9\phantom{0}}$& $0$&$232$& $0$&$122$& $0$&$136$& $14$&$489$& \phantom{0}$9$&$041$& \multicolumn{2}{c}{}& \multicolumn{2}{c}{}& \multicolumn{2}{c}{}& \multicolumn{2}{c}{} \vline \\
$2p^{5}$ $3d^{10}$& $0$&$211$& $0$&$102$& $0$&$162$& $15$&$210$& \phantom{0}$9$&$495$& \phantom{0}$15$&$738$& $9$&$240$& $7$&$084$& $4$&$033$ \\
$2p^{6}$ $3d^{8\phantom{0}}$& $0$&$204$& $0$&$089$& $0$&$147$& $15$&$443$& \phantom{0}$9$&$681$& \multicolumn{2}{c}{}& \multicolumn{2}{c}{}& \multicolumn{2}{c}{}& \multicolumn{2}{c}{} \vline \\
$2p^{5}$ $3d^{9\phantom{0}}$& $0$&$187$& $0$&$076$& $0$&$175$& $16$&$145$& $10$&$124$& \phantom{0}$15$&$737$& $9$&$857$& $7$&$639$& $4$&$352$ \\
\hline\end{tabular}
\\
\\
\\
\\
\\
\\
\\
\\
\begin{huge}$4d$ elements\end{huge}
\\
\\
\\
\begin{tabular}{|l|r@{ }lr@{ }lr@{.}lr@{.}lr@{.}lr@{.}lr@{.}lr@{.}lr@{.}l|}
\hline
Rb& \multicolumn{2}{c}{$r^2$} & \multicolumn{2}{c}{$r^4$} & \multicolumn{2}{c}{$\zeta_d$} & \multicolumn{2}{c}{$F^2_{dd}$} & \multicolumn{2}{c}{$F^4_{dd}$} & \multicolumn{2}{c}{$\zeta_{2p}$} & \multicolumn{2}{c}{$F^2_{pd}$}  & \multicolumn{2}{c}{$G^1_{pd}$} & \multicolumn{2}{c}{$G^3_{pd}$} \vline \\
\hline
$2p^{6}$ $4d^{1\phantom{0}}$& \phantom{0}& $*$ \phantom{0}& \phantom{0}& $*$ \phantom{0}& $0$&$001$& \phantom{0}$0$&$000$& \phantom{0}$0$&$000$& \multicolumn{2}{c}{}& \multicolumn{2}{c}{}& \multicolumn{2}{c}{}& \multicolumn{2}{c}{} \vline \\
$2p^{5}$ $4d^{2\phantom{0}}$& \multicolumn{2}{c}{$*$}& \multicolumn{2}{c}{$*$}& $0$&$009$& \phantom{0}$2$&$530$& \phantom{0}$1$&$553$& \phantom{0}$40$&$394$& $0$&$232$& $0$&$171$& $0$&$099$ \\
$2p^{5}$ $4d^{1\phantom{0}}$& \multicolumn{2}{c}{$*$}& \multicolumn{2}{c}{$*$}& $0$&$017$& \phantom{0}$0$&$000$& \phantom{0}$0$&$000$& \phantom{0}$40$&$393$& $0$&$440$& $0$&$333$& $0$&$192$ \\
\hline\end{tabular}
\\
\\
\\
\begin{tabular}{|l|r@{.}lr@{.}lr@{.}lr@{.}lr@{.}lr@{.}lr@{.}lr@{.}lr@{.}l|}
\hline
Sr& \multicolumn{2}{c}{$r^2$} & \multicolumn{2}{c}{$r^4$} & \multicolumn{2}{c}{$\zeta_d$} & \multicolumn{2}{c}{$F^2_{dd}$} & \multicolumn{2}{c}{$F^4_{dd}$} & \multicolumn{2}{c}{$\zeta_{2p}$} & \multicolumn{2}{c}{$F^2_{pd}$}  & \multicolumn{2}{c}{$G^1_{pd}$} & \multicolumn{2}{c}{$G^3_{pd}$} \vline \\
\hline
$2p^{6}$ $4d^{2\phantom{0}}$& \multicolumn{2}{c}{$*$}& \multicolumn{2}{c}{$*$}& $0$&$008$& \phantom{0}$2$&$459$& \phantom{0}$1$&$510$& \multicolumn{2}{c}{}& \multicolumn{2}{c}{}& \multicolumn{2}{c}{}& \multicolumn{2}{c}{} \vline \\
$2p^{5}$ $4d^{3\phantom{0}}$& \multicolumn{2}{c}{$*$}& \multicolumn{2}{c}{$*$}& $0$&$020$& \phantom{0}$3$&$585$& \phantom{0}$2$&$230$& \phantom{0}$45$&$518$& $0$&$459$& $0$&$350$& $0$&$203$ \\
$2p^{6}$ $4d^{1\phantom{0}}$& \multicolumn{2}{c}{$*$}& \multicolumn{2}{c}{$*$}& $0$&$016$& \phantom{0}$0$&$000$& \phantom{0}$0$&$000$& \multicolumn{2}{c}{}& \multicolumn{2}{c}{}& \multicolumn{2}{c}{}& \multicolumn{2}{c}{} \vline \\
$2p^{5}$ $4d^{2\phantom{0}}$& $2$&$173$& $9$&$473$& $0$&$029$& \phantom{0}$4$&$784$& \phantom{0}$3$&$052$& \phantom{0}$45$&$518$& $0$&$656$& $0$&$508$& $0$&$294$ \\
$2p^{5}$ $4d^{1\phantom{0}}$& $1$&$579$& $4$&$467$& $0$&$038$& \phantom{0}$0$&$000$& \phantom{0}$0$&$000$& \phantom{0}$45$&$517$& $0$&$857$& $0$&$672$& $0$&$389$ \\
\hline\end{tabular}
\\
\\
\\
\begin{tabular}{|l|r@{.}lr@{.}lr@{.}lr@{.}lr@{.}lr@{.}lr@{.}lr@{.}lr@{.}l|}
\hline
Y& \multicolumn{2}{c}{$r^2$} & \multicolumn{2}{c}{$r^4$} & \multicolumn{2}{c}{$\zeta_d$} & \multicolumn{2}{c}{$F^2_{dd}$} & \multicolumn{2}{c}{$F^4_{dd}$} & \multicolumn{2}{c}{$\zeta_{2p}$} & \multicolumn{2}{c}{$F^2_{pd}$}  & \multicolumn{2}{c}{$G^1_{pd}$} & \multicolumn{2}{c}{$G^3_{pd}$} \vline \\
\hline
$2p^{6}$ $4d^{3\phantom{0}}$& \multicolumn{2}{c}{$*$}& \multicolumn{2}{c}{$*$}& $0$&$019$& \phantom{0}$3$&$504$& \phantom{0}$2$&$177$& \multicolumn{2}{c}{}& \multicolumn{2}{c}{}& \multicolumn{2}{c}{}& \multicolumn{2}{c}{} \vline \\
$2p^{5}$ $4d^{4\phantom{0}}$& \multicolumn{2}{c}{$*$}& \multicolumn{2}{c}{$*$}& $0$&$033$& \phantom{0}$4$&$415$& \phantom{0}$2$&$774$& \phantom{0}$51$&$132$& $0$&$666$& $0$&$519$& $0$&$301$ \\
$2p^{6}$ $4d^{2\phantom{0}}$& \multicolumn{2}{c}{$*$}& \multicolumn{2}{c}{$*$}& $0$&$028$& \phantom{0}$4$&$696$& \phantom{0}$2$&$992$& \multicolumn{2}{c}{}& \multicolumn{2}{c}{}& \multicolumn{2}{c}{}& \multicolumn{2}{c}{} \vline \\
$2p^{5}$ $4d^{3\phantom{0}}$& $1$&$689$& $5$&$745$& $0$&$042$& \phantom{0}$5$&$456$& \phantom{0}$3$&$497$& \phantom{0}$51$&$131$& $0$&$855$& $0$&$673$& $0$&$390$ \\
$2p^{6}$ $4d^{1\phantom{0}}$& $1$&$620$& $4$&$708$& $0$&$038$& \phantom{0}$0$&$000$& \phantom{0}$0$&$000$& \multicolumn{2}{c}{}& \multicolumn{2}{c}{}& \multicolumn{2}{c}{}& \multicolumn{2}{c}{} \vline \\
$2p^{5}$ $4d^{2\phantom{0}}$& $1$&$306$& $3$&$079$& $0$&$052$& \phantom{0}$6$&$257$& \phantom{0}$4$&$067$& \phantom{0}$51$&$131$& $1$&$047$& $0$&$832$& $0$&$482$ \\
$2p^{5}$ $4d^{1\phantom{0}}$& $1$&$089$& $2$&$006$& $0$&$062$& \phantom{0}$0$&$000$& \phantom{0}$0$&$000$& \phantom{0}$51$&$130$& $1$&$238$& $0$&$992$& $0$&$575$ \\
\hline\end{tabular}
\\
\\
\\
\begin{tabular}{|l|r@{.}lr@{.}lr@{.}lr@{.}lr@{.}lr@{.}lr@{.}lr@{.}lr@{.}l|}
\hline
Zr& \multicolumn{2}{c}{$r^2$} & \multicolumn{2}{c}{$r^4$} & \multicolumn{2}{c}{$\zeta_d$} & \multicolumn{2}{c}{$F^2_{dd}$} & \multicolumn{2}{c}{$F^4_{dd}$} & \multicolumn{2}{c}{$\zeta_{2p}$} & \multicolumn{2}{c}{$F^2_{pd}$}  & \multicolumn{2}{c}{$G^1_{pd}$} & \multicolumn{2}{c}{$G^3_{pd}$} \vline \\
\hline
$2p^{6}$ $4d^{4\phantom{0}}$& \multicolumn{2}{c}{$*$}& \multicolumn{2}{c}{$*$}& $0$&$032$& \phantom{0}$4$&$338$& \phantom{0}$2$&$723$& \multicolumn{2}{c}{}& \multicolumn{2}{c}{}& \multicolumn{2}{c}{}& \multicolumn{2}{c}{} \vline \\
$2p^{5}$ $4d^{5\phantom{0}}$& $1$&$949$& $9$&$637$& $0$&$047$& \phantom{0}$5$&$119$& \phantom{0}$3$&$238$& \phantom{0}$57$&$267$& $0$&$863$& $0$&$683$& $0$&$396$ \\
$2p^{6}$ $4d^{3\phantom{0}}$& $1$&$734$& $6$&$062$& $0$&$042$& \phantom{0}$5$&$379$& \phantom{0}$3$&$444$& \multicolumn{2}{c}{}& \multicolumn{2}{c}{}& \multicolumn{2}{c}{}& \multicolumn{2}{c}{} \vline \\
$2p^{5}$ $4d^{4\phantom{0}}$& $1$&$381$& $3$&$846$& $0$&$058$& \phantom{0}$6$&$051$& \phantom{0}$3$&$892$& \phantom{0}$57$&$267$& $1$&$046$& $0$&$833$& $0$&$483$ \\
$2p^{6}$ $4d^{2\phantom{0}}$& $1$&$333$& $3$&$213$& $0$&$052$& \phantom{0}$6$&$188$& \phantom{0}$4$&$020$& \multicolumn{2}{c}{}& \multicolumn{2}{c}{}& \multicolumn{2}{c}{}& \multicolumn{2}{c}{} \vline \\
$2p^{5}$ $4d^{3\phantom{0}}$& $1$&$111$& $2$&$240$& $0$&$068$& \phantom{0}$6$&$786$& \phantom{0}$4$&$418$& \phantom{0}$57$&$266$& $1$&$232$& $0$&$989$& $0$&$574$ \\
$2p^{6}$ $4d^{1\phantom{0}}$& $1$&$107$& $2$&$076$& $0$&$062$& \phantom{0}$0$&$000$& \phantom{0}$0$&$000$& \multicolumn{2}{c}{}& \multicolumn{2}{c}{}& \multicolumn{2}{c}{}& \multicolumn{2}{c}{} \vline \\
\hline\end{tabular}
\\
\\
\\
\begin{tabular}{|l|r@{.}lr@{.}lr@{.}lr@{.}lr@{.}lr@{.}lr@{.}lr@{.}lr@{.}l|}
\hline
Nb& \multicolumn{2}{c}{$r^2$} & \multicolumn{2}{c}{$r^4$} & \multicolumn{2}{c}{$\zeta_d$} & \multicolumn{2}{c}{$F^2_{dd}$} & \multicolumn{2}{c}{$F^4_{dd}$} & \multicolumn{2}{c}{$\zeta_{2p}$} & \multicolumn{2}{c}{$F^2_{pd}$}  & \multicolumn{2}{c}{$G^1_{pd}$} & \multicolumn{2}{c}{$G^3_{pd}$} \vline \\
\hline
$2p^{6}$ $4d^{5\phantom{0}}$& \multicolumn{2}{c}{$*$}& \multicolumn{2}{c}{$*$}& $0$&$047$& \phantom{0}$5$&$047$& \phantom{0}$3$&$190$& \multicolumn{2}{c}{}& \multicolumn{2}{c}{}& \multicolumn{2}{c}{}& \multicolumn{2}{c}{} \vline \\
$2p^{5}$ $4d^{6\phantom{0}}$& $1$&$558$& $6$&$078$& $0$&$064$& \phantom{0}$5$&$745$& \phantom{0}$3$&$653$& \phantom{0}$63$&$959$& $1$&$052$& $0$&$842$& $0$&$489$ \\
$2p^{6}$ $4d^{4\phantom{0}}$& $1$&$411$& $4$&$020$& $0$&$057$& \phantom{0}$5$&$983$& \phantom{0}$3$&$844$& \multicolumn{2}{c}{}& \multicolumn{2}{c}{}& \multicolumn{2}{c}{}& \multicolumn{2}{c}{} \vline \\
$2p^{5}$ $4d^{5\phantom{0}}$& $1$&$165$& $2$&$738$& $0$&$075$& \phantom{0}$6$&$600$& \phantom{0}$4$&$256$& \phantom{0}$63$&$958$& $1$&$232$& $0$&$991$& $0$&$575$ \\
$2p^{6}$ $4d^{3\phantom{0}}$& $1$&$130$& $2$&$320$& $0$&$068$& \phantom{0}$6$&$724$& \phantom{0}$4$&$375$& \multicolumn{2}{c}{}& \multicolumn{2}{c}{}& \multicolumn{2}{c}{}& \multicolumn{2}{c}{} \vline \\
$2p^{5}$ $4d^{4\phantom{0}}$& $0$&$963$& $1$&$691$& $0$&$087$& \phantom{0}$7$&$288$& \phantom{0}$4$&$750$& \phantom{0}$63$&$958$& $1$&$414$& $1$&$144$& $0$&$665$ \\
$2p^{6}$ $4d^{2\phantom{0}}$& $0$&$960$& $1$&$573$& $0$&$079$& \phantom{0}$7$&$338$& \phantom{0}$4$&$820$& \multicolumn{2}{c}{}& \multicolumn{2}{c}{}& \multicolumn{2}{c}{}& \multicolumn{2}{c}{} \vline \\
\hline\end{tabular}
\\
\\
\\
\begin{tabular}{|l|r@{.}lr@{.}lr@{.}lr@{.}lr@{.}lr@{.}lr@{.}lr@{.}lr@{.}l|}
\hline
Mo& \multicolumn{2}{c}{$r^2$} & \multicolumn{2}{c}{$r^4$} & \multicolumn{2}{c}{$\zeta_d$} & \multicolumn{2}{c}{$F^2_{dd}$} & \multicolumn{2}{c}{$F^4_{dd}$} & \multicolumn{2}{c}{$\zeta_{2p}$} & \multicolumn{2}{c}{$F^2_{pd}$}  & \multicolumn{2}{c}{$G^1_{pd}$} & \multicolumn{2}{c}{$G^3_{pd}$} \vline \\
\hline
$2p^{6}$ $4d^{6\phantom{0}}$& $0$&$000$& $0$&$000$& $0$&$000$& \phantom{0}$5$&$679$& \phantom{0}$3$&$608$& \multicolumn{2}{c}{}& \multicolumn{2}{c}{}& \multicolumn{2}{c}{}& \multicolumn{2}{c}{} \vline \\
$2p^{5}$ $4d^{7\phantom{0}}$& $1$&$291$& $4$&$126$& $0$&$083$& \phantom{0}$6$&$320$& \phantom{0}$4$&$034$& \phantom{0}$71$&$243$& $1$&$237$& $0$&$998$& $0$&$580$ \\
$2p^{6}$ $4d^{5\phantom{0}}$& $1$&$186$& $2$&$843$& $0$&$075$& \phantom{0}$6$&$538$& \phantom{0}$4$&$213$& \multicolumn{2}{c}{}& \multicolumn{2}{c}{}& \multicolumn{2}{c}{}& \multicolumn{2}{c}{} \vline \\
$2p^{5}$ $4d^{6\phantom{0}}$& $1$&$004$& $2$&$034$& $0$&$095$& \phantom{0}$7$&$118$& \phantom{0}$4$&$599$& \phantom{0}$71$&$242$& $1$&$414$& $1$&$147$& $0$&$666$ \\
$2p^{6}$ $4d^{4\phantom{0}}$& $0$&$977$& $1$&$743$& $0$&$087$& \phantom{0}$7$&$231$& \phantom{0}$4$&$711$& \multicolumn{2}{c}{}& \multicolumn{2}{c}{}& \multicolumn{2}{c}{}& \multicolumn{2}{c}{} \vline \\
$2p^{5}$ $4d^{5\phantom{0}}$& $0$&$847$& $1$&$313$& $0$&$108$& \phantom{0}$7$&$769$& \phantom{0}$5$&$069$& \phantom{0}$71$&$242$& $1$&$593$& $1$&$299$& $0$&$755$ \\
$2p^{6}$ $4d^{3\phantom{0}}$& $0$&$845$& $1$&$225$& $0$&$099$& \phantom{0}$7$&$815$& \phantom{0}$5$&$135$& \multicolumn{2}{c}{}& \multicolumn{2}{c}{}& \multicolumn{2}{c}{}& \multicolumn{2}{c}{} \vline \\
\hline\end{tabular}
\\
\\
\\
\begin{tabular}{|l|r@{.}lr@{.}lr@{.}lr@{.}lr@{.}lr@{.}lr@{.}lr@{.}lr@{.}l|}
\hline
Tc& \multicolumn{2}{c}{$r^2$} & \multicolumn{2}{c}{$r^4$} & \multicolumn{2}{c}{$\zeta_d$} & \multicolumn{2}{c}{$F^2_{dd}$} & \multicolumn{2}{c}{$F^4_{dd}$} & \multicolumn{2}{c}{$\zeta_{2p}$} & \multicolumn{2}{c}{$F^2_{pd}$}  & \multicolumn{2}{c}{$G^1_{pd}$} & \multicolumn{2}{c}{$G^3_{pd}$} \vline \\
\hline
$2p^{6}$ $4d^{7\phantom{0}}$& $1$&$316$& $4$&$299$& $0$&$083$& \phantom{0}$6$&$259$& \phantom{0}$3$&$992$& \multicolumn{2}{c}{}& \multicolumn{2}{c}{}& \multicolumn{2}{c}{}& \multicolumn{2}{c}{} \vline \\
$2p^{5}$ $4d^{8\phantom{0}}$& $1$&$098$& $2$&$951$& $0$&$104$& \phantom{0}$6$&$860$& \phantom{0}$4$&$392$& \phantom{0}$79$&$157$& $1$&$418$& $1$&$153$& $0$&$671$ \\
$2p^{6}$ $4d^{6\phantom{0}}$& $1$&$019$& $2$&$102$& $0$&$095$& \phantom{0}$7$&$061$& \phantom{0}$4$&$560$& \multicolumn{2}{c}{}& \multicolumn{2}{c}{}& \multicolumn{2}{c}{}& \multicolumn{2}{c}{} \vline \\
$2p^{5}$ $4d^{7\phantom{0}}$& $0$&$879$& $1$&$560$& $0$&$118$& \phantom{0}$7$&$613$& \phantom{0}$4$&$927$& \phantom{0}$79$&$156$& $1$&$592$& $1$&$300$& $0$&$756$ \\
$2p^{6}$ $4d^{5\phantom{0}}$& $0$&$858$& $1$&$350$& $0$&$108$& \phantom{0}$7$&$717$& \phantom{0}$5$&$033$& \multicolumn{2}{c}{}& \multicolumn{2}{c}{}& \multicolumn{2}{c}{}& \multicolumn{2}{c}{} \vline \\
$2p^{5}$ $4d^{6\phantom{0}}$& $0$&$753$& $1$&$044$& $0$&$132$& \phantom{0}$8$&$236$& \phantom{0}$5$&$378$& \phantom{0}$79$&$155$& $1$&$769$& $1$&$452$& $0$&$845$ \\
$2p^{6}$ $4d^{4\phantom{0}}$& $0$&$751$& $0$&$975$& $0$&$121$& \phantom{0}$8$&$278$& \phantom{0}$5$&$441$& \multicolumn{2}{c}{}& \multicolumn{2}{c}{}& \multicolumn{2}{c}{}& \multicolumn{2}{c}{} \vline \\
\hline\end{tabular}
\\
\\
\\
\begin{tabular}{|l|r@{.}lr@{.}lr@{.}lr@{.}lr@{.}lr@{.}lr@{.}lr@{.}lr@{.}l|}
\hline
Ru& \multicolumn{2}{c}{$r^2$} & \multicolumn{2}{c}{$r^4$} & \multicolumn{2}{c}{$\zeta_d$} & \multicolumn{2}{c}{$F^2_{dd}$} & \multicolumn{2}{c}{$F^4_{dd}$} & \multicolumn{2}{c}{$\zeta_{2p}$} & \multicolumn{2}{c}{$F^2_{pd}$}  & \multicolumn{2}{c}{$G^1_{pd}$} & \multicolumn{2}{c}{$G^3_{pd}$} \vline \\
\hline
$2p^{6}$ $4d^{8\phantom{0}}$& $1$&$116$& $3$&$060$& $0$&$104$& \phantom{0}$6$&$802$& \phantom{0}$4$&$352$& \multicolumn{2}{c}{}& \multicolumn{2}{c}{}& \multicolumn{2}{c}{}& \multicolumn{2}{c}{} \vline \\
$2p^{5}$ $4d^{9\phantom{0}}$& $0$&$952$& $2$&$193$& $0$&$129$& \phantom{0}$7$&$373$& \phantom{0}$4$&$732$& \phantom{0}$87$&$739$& $1$&$596$& $1$&$306$& $0$&$761$ \\
$2p^{6}$ $4d^{7\phantom{0}}$& $0$&$890$& $1$&$605$& $0$&$118$& \phantom{0}$7$&$560$& \phantom{0}$4$&$890$& \multicolumn{2}{c}{}& \multicolumn{2}{c}{}& \multicolumn{2}{c}{}& \multicolumn{2}{c}{} \vline \\
$2p^{5}$ $4d^{8\phantom{0}}$& $0$&$779$& $1$&$226$& $0$&$143$& \phantom{0}$8$&$090$& \phantom{0}$5$&$243$& \phantom{0}$87$&$738$& $1$&$769$& $1$&$453$& $0$&$846$ \\
$2p^{6}$ $4d^{6\phantom{0}}$& $0$&$762$& $1$&$068$& $0$&$132$& \phantom{0}$8$&$187$& \phantom{0}$5$&$344$& \multicolumn{2}{c}{}& \multicolumn{2}{c}{}& \multicolumn{2}{c}{}& \multicolumn{2}{c}{} \vline \\
$2p^{5}$ $4d^{7\phantom{0}}$& $0$&$676$& $0$&$844$& $0$&$158$& \phantom{0}$8$&$690$& \phantom{0}$5$&$678$& \phantom{0}$87$&$737$& $1$&$944$& $1$&$604$& $0$&$934$ \\
$2p^{6}$ $4d^{5\phantom{0}}$& $0$&$674$& $0$&$790$& $0$&$146$& \phantom{0}$8$&$729$& \phantom{0}$5$&$739$& \multicolumn{2}{c}{}& \multicolumn{2}{c}{}& \multicolumn{2}{c}{}& \multicolumn{2}{c}{} \vline \\
\hline\end{tabular}
\\
\\
\\
\begin{tabular}{|l|r@{.}lr@{.}lr@{.}lr@{.}lr@{.}lr@{.}lr@{.}lr@{.}lr@{.}l|}
\hline
Rh& \multicolumn{2}{c}{$r^2$} & \multicolumn{2}{c}{$r^4$} & \multicolumn{2}{c}{$\zeta_d$} & \multicolumn{2}{c}{$F^2_{dd}$} & \multicolumn{2}{c}{$F^4_{dd}$} & \multicolumn{2}{c}{$\zeta_{2p}$} & \multicolumn{2}{c}{$F^2_{pd}$}  & \multicolumn{2}{c}{$G^1_{pd}$} & \multicolumn{2}{c}{$G^3_{pd}$} \vline \\
\hline
$2p^{6}$ $4d^{9\phantom{0}}$& $0$&$965$& $2$&$265$& $0$&$129$& \phantom{0}$7$&$320$& \phantom{0}$4$&$695$& \multicolumn{2}{c}{}& \multicolumn{2}{c}{}& \multicolumn{2}{c}{}& \multicolumn{2}{c}{} \vline \\
$2p^{5}$ $4d^{10}$& $0$&$836$& $1$&$678$& $0$&$156$& \phantom{0}$7$&$866$& \phantom{0}$5$&$059$& \phantom{0}$97$&$031$& $1$&$772$& $1$&$458$& $0$&$850$ \\
$2p^{6}$ $4d^{8\phantom{0}}$& $0$&$787$& $1$&$257$& $0$&$144$& \phantom{0}$8$&$041$& \phantom{0}$5$&$209$& \multicolumn{2}{c}{}& \multicolumn{2}{c}{}& \multicolumn{2}{c}{}& \multicolumn{2}{c}{} \vline \\
$2p^{5}$ $4d^{9\phantom{0}}$& $0$&$697$& $0$&$982$& $0$&$172$& \phantom{0}$8$&$554$& \phantom{0}$5$&$551$& \phantom{0}$97$&$030$& $1$&$943$& $1$&$605$& $0$&$935$ \\
$2p^{6}$ $4d^{7\phantom{0}}$& $0$&$683$& $0$&$861$& $0$&$159$& \phantom{0}$8$&$645$& \phantom{0}$5$&$647$& \multicolumn{2}{c}{}& \multicolumn{2}{c}{}& \multicolumn{2}{c}{}& \multicolumn{2}{c}{} \vline \\
$2p^{5}$ $4d^{8\phantom{0}}$& $0$&$612$& $0$&$692$& $0$&$188$& \phantom{0}$9$&$135$& \phantom{0}$5$&$973$& \phantom{0}$97$&$029$& $2$&$117$& $1$&$755$& $1$&$023$ \\
$2p^{6}$ $4d^{6\phantom{0}}$& $0$&$610$& $0$&$649$& $0$&$175$& \phantom{0}$9$&$172$& \phantom{0}$6$&$031$& \multicolumn{2}{c}{}& \multicolumn{2}{c}{}& \multicolumn{2}{c}{}& \multicolumn{2}{c}{} \vline \\
\hline\end{tabular}
\\
\\
\\
\begin{tabular}{|l|r@{.}lr@{.}lr@{.}lr@{.}lr@{.}lr@{.}lr@{.}lr@{.}lr@{.}l|}
\hline
Pd& \multicolumn{2}{c}{$r^2$} & \multicolumn{2}{c}{$r^4$} & \multicolumn{2}{c}{$\zeta_d$} & \multicolumn{2}{c}{$F^2_{dd}$} & \multicolumn{2}{c}{$F^4_{dd}$} & \multicolumn{2}{c}{$\zeta_{2p}$} & \multicolumn{2}{c}{$F^2_{pd}$}  & \multicolumn{2}{c}{$G^1_{pd}$} & \multicolumn{2}{c}{$G^3_{pd}$} \vline \\
\hline
$2p^{6}$ $4d^{9\phantom{0}}$& $0$&$704$& $1$&$005$& $0$&$173$& \phantom{0}$8$&$508$& \phantom{0}$5$&$518$& \multicolumn{2}{c}{}& \multicolumn{2}{c}{}& \multicolumn{2}{c}{}& \multicolumn{2}{c}{} \vline \\
$2p^{5}$ $4d^{10}$& $0$&$629$& $0$&$800$& $0$&$204$& \phantom{0}$9$&$007$& \phantom{0}$5$&$850$& $107$&$075$& $2$&$117$& $1$&$756$& $1$&$025$ \\
$2p^{6}$ $4d^{8\phantom{0}}$& $0$&$617$& $0$&$705$& $0$&$189$& \phantom{0}$9$&$093$& \phantom{0}$5$&$943$& \multicolumn{2}{c}{}& \multicolumn{2}{c}{}& \multicolumn{2}{c}{}& \multicolumn{2}{c}{} \vline \\
$2p^{5}$ $4d^{9\phantom{0}}$& $0$&$556$& $0$&$575$& $0$&$221$& \phantom{0}$9$&$572$& \phantom{0}$6$&$261$& $107$&$074$& $2$&$290$& $1$&$907$& $1$&$113$ \\
$2p^{6}$ $4d^{7\phantom{0}}$& $0$&$555$& $0$&$540$& $0$&$206$& \phantom{0}$9$&$607$& \phantom{0}$6$&$319$& \multicolumn{2}{c}{}& \multicolumn{2}{c}{}& \multicolumn{2}{c}{}& \multicolumn{2}{c}{} \vline \\
$2p^{5}$ $4d^{8\phantom{0}}$& $0$&$504$& $0$&$447$& $0$&$239$& $10$&$072$& \phantom{0}$6$&$628$& $107$&$074$& $2$&$466$& $2$&$061$& $1$&$203$ \\
$2p^{6}$ $4d^{6\phantom{0}}$& $0$&$507$& $0$&$434$& $0$&$223$& $10$&$069$& \phantom{0}$6$&$659$& \multicolumn{2}{c}{}& \multicolumn{2}{c}{}& \multicolumn{2}{c}{}& \multicolumn{2}{c}{} \vline \\
$2p^{5}$ $4d^{7\phantom{0}}$& $0$&$464$& $0$&$364$& $0$&$258$& $10$&$522$& \phantom{0}$6$&$960$& $107$&$075$& $2$&$644$& $2$&$219$& $1$&$295$ \\
\hline\end{tabular}
\\
\\
\\
\begin{tabular}{|l|r@{.}lr@{.}lr@{.}lr@{.}lr@{.}lr@{.}lr@{.}lr@{.}lr@{.}l|}
\hline
Ag& \multicolumn{2}{c}{$r^2$} & \multicolumn{2}{c}{$r^4$} & \multicolumn{2}{c}{$\zeta_d$} & \multicolumn{2}{c}{$F^2_{dd}$} & \multicolumn{2}{c}{$F^4_{dd}$} & \multicolumn{2}{c}{$\zeta_{2p}$} & \multicolumn{2}{c}{$F^2_{pd}$}  & \multicolumn{2}{c}{$G^1_{pd}$} & \multicolumn{2}{c}{$G^3_{pd}$} \vline \\
\hline
$2p^{6}$ $4d^{9\phantom{0}}$& $0$&$561$& $0$&$585$& $0$&$223$& \phantom{0}$9$&$533$& \phantom{0}$6$&$233$& \multicolumn{2}{c}{}& \multicolumn{2}{c}{}& \multicolumn{2}{c}{}& \multicolumn{2}{c}{} \vline \\
$2p^{5}$ $4d^{10}$& $0$&$509$& $0$&$483$& $0$&$258$& $10$&$002$& \phantom{0}$6$&$546$& $117$&$915$& $2$&$461$& $2$&$057$& $1$&$201$ \\
$2p^{6}$ $4d^{8\phantom{0}}$& $0$&$508$& $0$&$454$& $0$&$241$& $10$&$036$& \phantom{0}$6$&$602$& \multicolumn{2}{c}{}& \multicolumn{2}{c}{}& \multicolumn{2}{c}{}& \multicolumn{2}{c}{} \vline \\
$2p^{5}$ $4d^{9\phantom{0}}$& $0$&$464$& $0$&$380$& $0$&$278$& $10$&$493$& \phantom{0}$6$&$906$& $117$&$915$& $2$&$636$& $2$&$210$& $1$&$291$ \\
$2p^{6}$ $4d^{7\phantom{0}}$& $0$&$467$& $0$&$368$& $0$&$260$& $10$&$490$& \phantom{0}$6$&$936$& \multicolumn{2}{c}{}& \multicolumn{2}{c}{}& \multicolumn{2}{c}{}& \multicolumn{2}{c}{} \vline \\
$2p^{5}$ $4d^{8\phantom{0}}$& $0$&$429$& $0$&$312$& $0$&$298$& $10$&$938$& \phantom{0}$7$&$234$& $117$&$916$& $2$&$813$& $2$&$368$& $1$&$383$ \\
\hline\end{tabular}
\\
\\
\\
\begin{tabular}{|l|r@{.}lr@{.}lr@{.}lr@{.}lr@{.}lr@{.}lr@{.}lr@{.}lr@{.}l|}
\hline
Cd& \multicolumn{2}{c}{$r^2$} & \multicolumn{2}{c}{$r^4$} & \multicolumn{2}{c}{$\zeta_d$} & \multicolumn{2}{c}{$F^2_{dd}$} & \multicolumn{2}{c}{$F^4_{dd}$} & \multicolumn{2}{c}{$\zeta_{2p}$} & \multicolumn{2}{c}{$F^2_{pd}$}  & \multicolumn{2}{c}{$G^1_{pd}$} & \multicolumn{2}{c}{$G^3_{pd}$} \vline \\
\hline
$2p^{6}$ $4d^{9\phantom{0}}$& $0$&$467$& $0$&$385$& $0$&$280$& $10$&$459$& \phantom{0}$6$&$881$& \multicolumn{2}{c}{}& \multicolumn{2}{c}{}& \multicolumn{2}{c}{}& \multicolumn{2}{c}{} \vline \\
$2p^{5}$ $4d^{10}$& $0$&$429$& $0$&$326$& $0$&$320$& $10$&$910$& \phantom{0}$7$&$181$& $129$&$599$& $2$&$804$& $2$&$359$& $1$&$379$ \\
$2p^{6}$ $4d^{8\phantom{0}}$& $0$&$431$& $0$&$315$& $0$&$300$& $10$&$907$& \phantom{0}$7$&$211$& \multicolumn{2}{c}{}& \multicolumn{2}{c}{}& \multicolumn{2}{c}{}& \multicolumn{2}{c}{} \vline \\
$2p^{5}$ $4d^{9\phantom{0}}$& $0$&$397$& $0$&$269$& $0$&$342$& $11$&$350$& \phantom{0}$7$&$505$& $129$&$600$& $2$&$981$& $2$&$516$& $1$&$471$ \\
\hline\end{tabular}
\end{fontsize}

\chapter{Source-code of A$_{k,m}$}
\label{ApendixAkm}

The program A$_{k,m}$ can calculate the Madelung potential of an infinite point charge distribution. The Madelung potential is expanded on spherical harmonics and can be used as crystal field for a cluster calculation. The Madelung sum does not converge absolutely. In order to find the sum an Ewald summation has been used. The program has been tested on a linux platform with the Gnu f77 compiler and the Intel 8.1 ifort compiler. Both run fine, the Intel compiler is about two times faster. A short explanation to the program can be found in the comments fields of the source-code.

The power of the program lies not within the exactness of the potential or crystal-fields calculated. A point charge model is very crude. However if one has a system with some odd symmetry and one wants to know quickly which parameters of the crystal field are zero and which are not, this program gives a very fast answer. In other words, it is a good starting-point for fitting.

An example input file for a cubic rocksalt with 1 \AA \:between the different atoms and a charge of plus and minus one at each site is printed below. One only has to input the charges within the unit cell. Charges on the boundaries of the unit cell should be reduced in charge corresponding to the number of times they reappear in the crystal structure, or as shown below only taken into account once. The first three lines are the vectors spanning the unit cell. The next line is a multiplication factor for the x,y and z positions of the charges. The fifth line contains one number, the amount of charge at the origin. The sixth line contains the number of charges within the unit cell. The next n lines contain the charge and its position (q, x, y, z).

\begin{verbatim}
2.000 0.000 0.0000
1.000 1.000 0.0000
1.000 0.000 1.0000
1.000 1.000 1.0000
 1.0
1
-1.0 1.0 0.0 0.0
\end{verbatim}

Saving this file as in.dat and run Akm will give the file out.dat. For the give example this file is plotted below. One can nicely see that most of the $A_{k,m}$ values are zero, as one should expect for a cubic symmetry.

\begin{verbatim}
 Calculation of the Madelung potential
 Expanded on renormalized spherical harmonics
 V=A_{k,m} r^k C_{k,m}
 
 input crystal structure:
 Latice parameters:
a=  2.000000  0.000000  0.000000
b=  1.000000  1.000000  0.000000
c=  1.000000  0.000000  1.000000
 Positions of charges:
   q         x         y         z
  1.000000  0.000000  0.000000  0.000000
 -1.000000  1.000000  0.000000  0.000000
 
Ewald sum calculated with G=  0.772609  1.545218 and  3.090435
 Time used to calculate   0.109999999403954       sec.
 
 Potential found (A_{k,m} parameters)
A_{0, 0}=(   25.16407095,    0.00000000)
A_{1, 0}=(    0.00000000,    0.00000000)
A_{1, 1}=(    0.00000000,    0.00000000)
A_{2, 0}=(    0.00000000,    0.00000000)
A_{2, 1}=(    0.00000000,    0.00000000)
A_{2, 2}=(    0.00000000,    0.00000000)
A_{3, 0}=(    0.00000000,    0.00000000)
A_{3, 1}=(    0.00000000,    0.00000000)
A_{3, 2}=(    0.00000000,    0.00000000)
A_{3, 3}=(    0.00000000,    0.00000000)
A_{4, 0}=(   51.52982716,    0.00000000)
A_{4, 1}=(    0.00000000,    0.00000000)
A_{4, 2}=(    0.00000000,    0.00000000)
A_{4, 3}=(    0.00000000,    0.00000000)
A_{4, 4}=(   30.79496183,    0.00000000)
A_{5, 0}=(    0.00000000,    0.00000000)
A_{5, 1}=(    0.00000000,    0.00000000)
A_{5, 2}=(    0.00000000,    0.00000000)
A_{5, 3}=(    0.00000000,    0.00000000)
A_{5, 4}=(    0.00000000,    0.00000000)
A_{5, 5}=(    0.00000000,    0.00000000)
A_{6, 0}=(   14.24830629,    0.00000000)
A_{6, 1}=(    0.00000000,    0.00000000)
A_{6, 2}=(    0.00000000,    0.00000000)
A_{6, 3}=(    0.00000000,    0.00000000)
A_{6, 4}=(  -26.65614024,    0.00000000)
A_{6, 5}=(    0.00000000,    0.00000000)
A_{6, 6}=(    0.00000000,    0.00000000)
 
 input for Tanaka's programme
 
 QAk={ang
     ,0, 0,  25.164071*rk,   0.000000*rk
     ,1,-1,   0.000000*rk,   0.000000*rk
     ,1, 0,   0.000000*rk,   0.000000*rk
     ,1, 1,   0.000000*rk,   0.000000*rk
     ,2,-2,   0.000000*rk,   0.000000*rk
     ,2,-1,   0.000000*rk,   0.000000*rk
     ,2, 0,   0.000000*rk,   0.000000*rk
     ,2, 1,   0.000000*rk,   0.000000*rk
     ,2, 2,   0.000000*rk,   0.000000*rk
     ,3,-3,   0.000000*rk,   0.000000*rk
     ,3,-2,   0.000000*rk,   0.000000*rk
     ,3,-1,   0.000000*rk,   0.000000*rk
     ,3, 0,   0.000000*rk,   0.000000*rk
     ,3, 1,   0.000000*rk,   0.000000*rk
     ,3, 2,   0.000000*rk,   0.000000*rk
     ,3, 3,   0.000000*rk,   0.000000*rk
     ,4,-4,  30.794962*rk,   0.000000*rk
     ,4,-3,   0.000000*rk,   0.000000*rk
     ,4,-2,   0.000000*rk,   0.000000*rk
     ,4,-1,   0.000000*rk,   0.000000*rk
     ,4, 0,  51.529827*rk,   0.000000*rk
     ,4, 1,   0.000000*rk,   0.000000*rk
     ,4, 2,   0.000000*rk,   0.000000*rk
     ,4, 3,   0.000000*rk,   0.000000*rk
     ,4, 4,  30.794962*rk,   0.000000*rk
     ,5,-5,   0.000000*rk,   0.000000*rk
     ,5,-4,   0.000000*rk,   0.000000*rk
     ,5,-3,   0.000000*rk,   0.000000*rk
     ,5,-2,   0.000000*rk,   0.000000*rk
     ,5,-1,   0.000000*rk,   0.000000*rk
     ,5, 0,   0.000000*rk,   0.000000*rk
     ,5, 1,   0.000000*rk,   0.000000*rk
     ,5, 2,   0.000000*rk,   0.000000*rk
     ,5, 3,   0.000000*rk,   0.000000*rk
     ,5, 4,   0.000000*rk,   0.000000*rk
     ,5, 5,   0.000000*rk,   0.000000*rk
     ,6,-6,   0.000000*rk,   0.000000*rk
     ,6,-5,   0.000000*rk,   0.000000*rk
     ,6,-4, -26.656140*rk,   0.000000*rk
     ,6,-3,   0.000000*rk,   0.000000*rk
     ,6,-2,   0.000000*rk,   0.000000*rk
     ,6,-1,   0.000000*rk,   0.000000*rk
     ,6, 0,  14.248306*rk,   0.000000*rk
     ,6, 1,   0.000000*rk,   0.000000*rk
     ,6, 2,   0.000000*rk,   0.000000*rk
     ,6, 3,   0.000000*rk,   0.000000*rk
     ,6, 4, -26.656140*rk,   0.000000*rk
     ,6, 5,   0.000000*rk,   0.000000*rk
     ,6, 6,   0.000000*rk,   0.000000*rk
                                        };
\end{verbatim}

The source-code of the program is printed below.

\begin{small}
\begin{verbatim}
C
C A program to calculate the Madelung potential at a crystal
C lattice site, expanded on spherical harmonics up to 6-th order
C Programmed by M.W. Haverkort and M. Grueninger 
C               06-03 at Koeln University.
C
C Input is read from in.dat
C Output is written to out.dat
C working info is written to standard output
C
C The input file should be named in.dat and have the form
C a(x)   a(y)   a(z)
C b(x)   b(y)   b(z)
C c(x)   c(y)   c(z)
C factorX factorY factorZ
C charge at origin
C number N of other charges
C q(1)   x(1)   y(1)   z(1)
C ...
C q(N)   x(N)   y(N)   z(N)
C
C The positions x(i) will be multiplied by the number factorX.
C For orthogonal unit-cells one can input the crystal structure 
C in relative coordinates.
C
C The Ewald sums is expanded on spherical Harmonics.
C The radial part of the Ewald sum is expanded on a Taylor series:
C V(r)=\sum_{k=0}^{k=\inf}\sum_{m=-k}^{m=k} 
C                               A_{k,m} r^k C_m^k(\theta\phi)
C
C The coefficients A_{k,m} are found with the relation:
C A_{k,m}=\frac{1}{\sqrt{(k-m)!(k+m)!}} d_z^k-m(-d_x+i d_y)^m V(r)
C for m=>0
C for m<0 we use A_{k,m}=-1^m A_{k,m}^*
C
C Convergence, or G independency is checked by calculating 
C the Ewald sum for 3 different values of G.
C If the sums did converge they should all be the same.
C The G values taken are one time the G for which computation time
C is minimum, one time Gopt*2 and one time Gopt/2
C
      program Akm
C
      integer Mq
      parameter (Mq=100)
C Mq=maximum number of charges in unit cell
C warning parameters also defined in subroutines
C should be the same in all subroutine before compiling
      real*8 a(3),b(3),c(3),q(Mq,0:3),G,T
      complex*16 V(0:6,-6:6),Vd(0:6,-6:6),Vh(0:6,-6:6)
      integer Nq,Ok
      call getinput(a,b,c,q,Nq)
      call calculateG(G,a,b,c)
      call ewaldsum(V,a,b,c,q,Nq,G)
      G=G/2D0
      call ewaldsum(Vh,a,b,c,q,Nq,G)
      G=G*4D0
      call ewaldsum(Vd,a,b,c,q,Nq,G)
      G=G/2D0
      call converged(V,Vh,Vd,Ok)
      call writeoutput(Vh,V,Vd,a,b,c,q,Nq,G,Ok)
      end
C
C Check convergence. If the sum converged O.K. the sum is G
C independent. This program calculates the sum for optimum G, 2*G 
C and G/2 This subroutine checks if these are within 8 digits the 
C same, if so the subroutine returns OK=1 if not OK=0
C
      subroutine converged(V,Vh,Vd,Ok)
      complex*16 V(0:6,-6:6),Vd(0:6,-6:6),Vh(0:6,-6:6)
      real*8 acc
      integer Ok,k,m
      acc=10D-8
      OK=1
      DO k=0,6
       DO m=-k,k
        If (Dabs(Dreal(V(k,m))-Dreal(Vd(k,m))).GE.acc) then 
         OK=0
         write(*,*)'No convergence for real part of V',k,',',m
        Endif
        If (Dabs(Dimag(V(k,m))-Dimag(Vd(k,m))).GE.acc) then 
         OK=0
         write(*,*)'No convergence for imag part of V',k,',',m
        Endif
       Enddo
      Enddo
      end

C
C Calculate G such that the sum goes over a minimum of points
C Nsumpoints=Na*Nb*Nc+Nga*Ngb*Ngc
C
      subroutine calculateG(G,a,b,c)
      real*8 PI
      parameter (PI=3.1415926535D0)      
      real*8 a(3),b(3),c(3),G,ga(3),gb(3),gc(3),Na,Nb,Nc,
     &       Nga,Ngb,Ngc,x(3),Vunit
      CALL reciprocallatice(ga,gb,gc,a,b,c)
      CALL Cros(x,a,b)
      Vunit=dabs(c(1)*x(1)+c(2)*x(2)+c(3)*x(3))
      CALL Cros(x,c,b)
      Na=int(1D0+10D0*DSQRT((x(1)**2+x(2)**2+x(3)**2))/
     &       Vunit)
      CALL Cros(x,a,c)
      Nb=int(1D0+10D0*DSQRT((x(1)**2+x(2)**2+x(3)**2))/
     &       Vunit)
      CALL Cros(x,a,b)
      Nc=int(1D0+10D0*DSQRT((x(1)**2+x(2)**2+x(3)**2))/
     &       Vunit)
      CALL Cros(x,ga,gb)
      Vunit=dabs(gc(1)*x(1)+gc(2)*x(2)+gc(3)*x(3))
      CALL Cros(x,gc,gb)
      Nga=int(1D0+15D0*DSQRT((x(1)**2+x(2)**2+x(3)**2))/
     &       Vunit)
      CALL Cros(x,ga,gc)
      Ngb=int(1D0+15D0*DSQRT((x(1)**2+x(2)**2+x(3)**2))/
     &       Vunit)
      CALL Cros(x,ga,gb)
      Ngc=int(1D0+15D0*DSQRT((x(1)**2+x(2)**2+x(3)**2))/
     &       Vunit)
      G=(Na*Nb*Nc/(Nga*Ngb*Ngc))**(1D0/6D0)
      end      
C
C
C
      subroutine writeoutput(Vh,V,Vd,a,b,c,q,Nq,G,Ok)
      integer Mq
      parameter (Mq=100)
      real*8 a(3),b(3),c(3),q(Mq,0:3),G,T
      complex*16 V(0:6,-6:6),Vh(0:6,-6:6),Vd(0:6,-6:6)
      integer i,k,m,Nq,Ok
      CALL CLOCK(T)
      write(*,*)'start of writeoutput at',T,'sec.'
      open(10,file='out.dat')
      IF (Ok.Eq.0) then
       write(10,*)'Warning !!!!!!!!!!!'
       write(10,*)'The system did not converge completely (10^-8)'
       write(10,*)'check your results and G dependence'
       write(10,*)'Small differences could be due to number loss'
       write(10,*)'If it is not number loss you are looking at'
       write(10,*)'Please contact Maurits Haverkort'
       write(10,*)'haverkort@ph2.uni-koeln.de'
       write(10,*)''
       write(10,*)''
      endif
 100  format(A,3F10.6)
      write(10,*)'Calculation of the Madelung potential'
      write(10,*)'Expanded on renormalized spherical harmonics'
      write(10,*)'V=A_{k,m} r^k C_{k,m}'
      write(10,*)
      write(10,*)'input crystal structure:'
      write(10,*)'Latice parameters:'
      write(10,100)'a=',a
      write(10,100)'b=',b
      write(10,100)'c=',c
      write(10,*)'Positions of charges:'
      write(10,*)'  q         x         y         z'
 101  format(4F10.6)
      DO i=1,Nq
       write(10,101)q(i,0),q(i,1),q(i,2),q(i,3)
      ENDDO
      write(10,*)
 102  format(A,2F10.6,A,F10.6)
      write(10,102)'Ewald sum calculated with G=',G/2D0,G,
     &           ' and',G*2D0
      write(10,*)'Time used to calculate ',T,' sec.'
      write(10,*)
      write(10,*)'Potential found (A_{k,m} parameters)'
 103  format(A,I1,A,I2,A,'(',F14.8,',',F14.8,')')
 104  format(A,'(',F14.8,',',F14.8,')')
      DO k=0,6
       DO m=0,k
        IF(OK.EQ.0)THEN
         write(10,103)'A_{',k,',',m,'}=',Vh(k,m)
         write(10,104)'         ',V(k,m)
         write(10,104)'         ',Vd(k,m)
        ELSE
         write(10,103)'A_{',k,',',m,'}=',V(k,m)
        ENDIF
       ENDDO
      ENDDO
 105  format(A,',',I1,',',I2,',',F11.6,'*rk,',F11.6,'*rk')
      write(10,*)
      write(10,*)'input for Tanaka''s programme'
      write(10,*)
      write(10,  *)'QAk={ang'
      DO k=0,6
       DO m=-k,k
        write(10,105)'     ',k,m,V(k,m)
       ENDDO
      ENDDO
      write(10,*)'                                       };'
      close(10)
      CALL CLOCK(T)
      write(*,*) 'wrote output to out.dat at',T,'sec.'
      end
C
C Read input parameters from in.dat
C The input file should be named in.dat and have the form
C a(x)   a(y)   a(z)
C b(x)   b(y)   b(z)
C c(x)   c(y)   c(z)
C factorX factorY factorZ
C charge at origin
C number N of other charges
C q(1)   x(1)   y(1)   z(1)
C ...
C q(N)   x(N)   y(N)   z(N)
C
      subroutine getinput(a,b,c,q,Nq)
      integer Mq
      parameter (Mq=100)
      real*8 a(3),b(3),c(3),q(Mq,0:3),T,xfac,yfac,zfac
      integer i,Nq
      CALL CLOCK(T)
      write(*,*)'start of getinput at',T,'sec.'
      open(10,file='in.dat')
      read(10,*) a(1),a(2),a(3)
      read(10,*) b(1),b(2),b(3)
      read(10,*) c(1),c(2),c(3)
      read(10,*) xfac,yfac,zfac
      read(10,*) q(1,0)
      q(1,1)=0D0
      q(1,2)=0D0
      q(1,2)=0D0
      read(10,*) Nq
      Nq=Nq+1
      IF (Nq.GT.Mq) THEN
       write(*,*) 'Error in getinput'
       write(*,*) 'Nq to big you must increase Mq to at least',Nq
      ENDIF
      DO i=2,Nq
       read(10,*) q(i,0),q(i,1),q(i,2),q(i,3)
       q(i,1)=q(i,1)*xfac
       q(i,2)=q(i,2)*yfac
       q(i,3)=q(i,3)*zfac
      ENDDO
      close(10)
      write(*,*)'parameters are read.'
      write(*,*)'a=',a
      write(*,*)'b=',b
      write(*,*)'c=',c
      write(*,*)'the prefactors are',xfac,yfac,zfac
      write(*,*)'The number of charges and their positions are:'
      DO i=1,Nq
       write(*,*) q(i,0),q(i,1),q(i,2),q(i,3)
      ENDDO
      CALL CLOCK(T)
      write(*,*)'End of getinput at',T,'sec.'
      end
C
C Current time in seconds after start of programme
C
      subroutine CLOCK(CPU)
      REAL*8 CPU
      REAL*4 DUM, T(2)
      DUM=ETIME(T)
      CPU=DBLE(T(1))
      end
C
C sum over the lattice with use of Ewald summation
C V is output a,b,c are the lattice parameters
C q gives the position of the charges
C Nq the number of the charges
C G defines the truncation frequency
C The result should not depend on G
C the convergence rate of the fourier transformed
C and normal sum do depend on G
C
      subroutine EWALDSUM(V,a,b,c,q,Nq,G)
C
      integer Mq
      real*8 PI,eps0,JtoeV,e,Anstr
      parameter (Mq=100,PI=3.1415926535D0,eps0=8.854187817D-12,
     &      JtoeV=6.2414509745D18,e=1.602176462D-19,Anstr=1D10)
      real*8 a(3),b(3),c(3),ga(3),gb(3),gc(3),q(Mq,0:3),G,x(3)
     &      ,r,Ve,derfctemp,dexptemp,temp,T,Vunit,kmmfac,kpmfac
      complex*16 V(0:6,-6:6),Vtemp(0:6,-6:6),sum,ctemp,xiy,ixy,iz
      integer i,j,k,m,Nq,Na,Nb,Nc,ia,ib,ic
C initialize the potential to be 0.
      DO k=0,6
       DO m=-6,6
        V(k,m)=(0D0,0D0)
        Vtemp(k,m)=(0D0,0D0)
       ENDDO
      ENDDO
      CALL reciprocallatice(ga,gb,gc,a,b,c)
C do the normal space part of the summation
C calculate Na,Nb,Nc such that erfc<1D-17
C That happens when G*Na*|a_n|>6
C The radius of the sphere included has to be 6/G
C but for higher orders factors of r^k play also a role
C sum up to 10/G and the convergence is depended on G
C but for G=1 better than 1D-17 for all cases
      write(*,*) 'start normal space of the Ewald sum'
      write(*,*) 'the radius of the sphere over which is summed'
     &           ,10/G
      CALL Cros(x,a,b)
      Vunit=dabs(c(1)*x(1)+c(2)*x(2)+c(3)*x(3))
      Ve=Vunit
      CALL Cros(x,c,b)
      Na=int(1D0+10D0/G*DSQRT((x(1)**2+x(2)**2+x(3)**2))/
     &       Vunit)
      CALL Cros(x,a,c)
      Nb=int(1D0+10D0/G*DSQRT((x(1)**2+x(2)**2+x(3)**2))/
     &       Vunit)
      CALL Cros(x,a,b)
      Nc=int(1D0+10D0/G*DSQRT((x(1)**2+x(2)**2+x(3)**2))/
     &       Vunit)
      write(*,*) 'this fits in a parallelepiped with Na,Nb,Nc='
     &           ,Na,Nb,Nc
      write(*,*) 'with the basis vectors:'
      write(*,*) 'a:',a
      write(*,*) 'b:',b
      write(*,*) 'c:',c
C do the normal space summation
      DO ia=-Na,Na
       DO ib=-Nb,Nb
        DO ic=-Nc,Nc
         DO j=1,Nq
          IF(.NOT.((ia.EQ.0).AND.(ib.EQ.0).AND.(ic.EQ.0)
     &       .AND.(j.EQ.1)))THEN
           x(1)=ia*a(1)+ib*b(1)+ic*c(1)+q(j,1)
           x(2)=ia*a(2)+ib*b(2)+ic*c(2)+q(j,2)
           x(3)=ia*a(3)+ib*b(3)+ic*c(3)+q(j,3)
           r=dsqrt(x(1)**2+x(2)**2+x(3)**2)
           xiy=dcmplx(x(1),-x(2))
           derfctemp=derfc(G*r)
           dexptemp=dexp(-G**2*r**2)/dsqrt(PI)
           V(0,0)=V(0,0)+q(j,0)*derfctemp/r
           V(1,0)=V(1,0)+x(3)*q(j,0)*(2D0*r*G*dexptemp+derfctemp)/r**3
           V(1,1)=V(1,1)-xiy *q(j,0)*(2D0*r*G*dexptemp+derfctemp)/r**3
           V(2,0)=V(2,0)-q(j,0)*(2D0*r*G*dexptemp*(-3D0*x(3)**2+r**2
     &           *(1D0-2D0*x(3)**2*G**2))+(r**2-3D0*x(3)**2)*derfctemp)
     &           /r**5
           V(2,1)=V(2,1)-xiy*x(3)*q(j,0)*((6D0*r*G+4D0*r**3*G**3)
     &           *dexptemp+3D0*derfctemp)/r**5
           V(2,2)=V(2,2)+xiy**2*q(j,0)*((6D0*r*G+4D0*r**3*G**3)*dexptemp
     &           +3D0*derfctemp)/r**5
           V(3,0)=V(3,0)+x(3)*q(j,0)*(2D0*r*G*dexptemp*(15D0*x(3)**2
     &           +r**2*(-9D0+2D0*G**2*(5D0*x(3)**2+r**2*(-3D0+2D0
     &           *x(3)**2*G**2))))
     &           +(-9D0*r**2+15D0*x(3)**2)*derfctemp)/r**7
           V(3,1)=V(3,1)+xiy*q(j,0)*(dexptemp*(6D0*r*(r**2-5D0*x(3)**2)
     &           *G+4D0*r**3*(r**2-5D0*x(3)**2)*G**3-8D0*r**5*x(3)**2
     &           *G**5)+3D0*(r**2-5D0*x(3)**2)*derfctemp)/r**7
           V(3,2)=V(3,2)+xiy**2*x(3)*q(j,0)*(dexptemp*(30D0*r*G+20D0
     &           *r**3*G**3+8D0*r**5*G**5)+15D0*derfctemp)/r**7
           V(3,3)=V(3,3)-xiy**3*q(j,0)*(dexptemp*(30D0*r*G+20D0
     &           *r**3*G**3+8D0*r**5*G**5)+15D0*derfctemp)/r**7
           V(4,0)=V(4,0)+q(j,0)*(dexptemp*2D0*r*G*(105D0*x(3)**4+10D0
     &           *r**2*x(3)**2*(-9D0+7D0*x(3)**2*G**2)+r**4*(9D0-60D0
     &           *x(3)**2*G**2+28D0*x(3)**4*G**4)+r**6*(6D0*G**2-24D0
     &           *x(3)**2*G**4+8D0*x(3)**4*G**6))+3D0*(3D0*r**4-30D0
     &           *r**2*x(3)**2+35D0*x(3)**4)*derfctemp)/r**9
           V(4,1)=V(4,1)+xiy*x(3)*q(j,0)*(-2D0*r*G*dexptemp*(105D0
     &           *x(3)**2+r**2*(-45D0+2D0*G**2*(35D0*x(3)**2+r**2
     &           *(-15D0+2D0*G**2*(7D0*x(3)**2+r**2*(-3D0+2D0*x(3)**2
     &           *G**2))))))+15D0*(3D0*r**2-7D0*x(3)**2)*derfctemp)/r**9
           V(4,2)=V(4,2)+xiy**2*q(j,0)*(2D0*r*G*dexptemp*(105*x(3)**2
     &           +r**2*(-15D0-10D0*(r**2-7D0*x(3)**2)*G**2-4D0*r**2
     &           *(r**2-7D0*x(3)**2)*G**4+8D0*r**4*x(3)**2*G**6))
     &           -15D0*(r**2-7D0*x(3)**2)*derfctemp)/r**9
           V(4,3)=V(4,3)-xiy**3*x(3)*q(j,0)*(2D0*r*G*dexptemp*(105D0
     &           +70D0*r**2*G**2+28D0*r**4*G**4+8D0*r**6*G**6)
     &           +105D0*derfctemp)/r**9
           V(4,4)=V(4,4)+xiy**4*q(j,0)*(2D0*r*G*dexptemp*(105D0
     &           +70D0*r**2*G**2+28D0*r**4*G**4+8D0*r**6*G**6)
     &           +105D0*derfctemp)/r**9
           V(5,0)=V(5,0)+x(3)*q(j,0)*(2D0*r*G*dexptemp*(945D0*x(3)**4
     &           +210D0*r**2*x(3)**2*(-5D0+3D0*x(3)**2*G**2)+4D0*r**8
     &           *G**4*(15D0-20D0*x(3)**2*G**2+4D0*x(3)**4*G**4)+2D0
     &           *r**6*G**2*(75D0-140D0*x(3)**2*G**2+36D0*x(3)**4*G**4)
     &           +r**4*(225D0-700D0*x(3)**2*G**2+252D0*x(3)**4*G**4))+15D0
     &           *(15D0*r**4-70D0*r**2*x(3)**2+63D0*x(3)**4)*derfctemp)
     &           /r**11
           V(5,1)=V(5,1)-xiy*q(j,0)*(dexptemp*2D0*r*G*(945D0*x(3)**4
     &           +630D0*r**2*x(3)**2*(-1D0+x(3)**2*G**2)+4D0*r**8*G**4
     &           *(3D0-12D0*x(3)**2*G**2+4D0*x(3)**4*G**4)+6D0*r**6*G**2
     &           *(5D0-28D0*x(3)**2*G**2+12D0*x(3)**4*G**4)+3D0*r**4
     &           *(15D0-140D0*x(3)**2*G**2+84D0*x(3)**4*G**4))+45D0
     &           *(r**4-14D0*r**2*x(3)**2+21D0*x(3)**4)*derfctemp)/r**11
           V(5,2)=V(5,2)+xiy**2*x(3)*q(j,0)*(2D0*r*G*dexptemp*(945D0
     &           *x(3)**2+8D0*r**8*G**6*(-3D0+2D0*x(3)**2*G**2)+315D0
     &           *r**2*(-1D0+2D0*x(3)**2*G**2)+12D0*r**6*G**4*(-7D0+6D0
     &           *x(3)**2*G**2)+42D0*r**4*G**2*(-5D0+6D0*x(3)**2*G**2))
     &           -315D0*(r**2-3D0*x(3)**2)*derfctemp)/r**11
           V(5,3)=V(5,3)+xiy**3*q(j,0)*(dexptemp*(210D0*r*(r**2-9D0
     &           *x(3)**2)*G+140D0*r**3*(r**2-9D0*x(3)**2)*G**3+56D0
     &           *r**5*(r**2-9D0*x(3)**2)*G**5+16D0*r**7*(r**2-9D0
     &           *x(3)**2)*G**7-32*r**9*x(3)**2*G**9)+105*(r**2-9D0
     &           *x(3)**2)*derfctemp)/r**11
           V(5,4)=V(5,4)+xiy**4*x(3)*q(j,0)*(2D0*r*G*dexptemp*(945D0
     &           +630D0*r**2*G**2+252D0*r**4*G**4+72D0*r**6*G**6
     &           +16D0*r**8*G**8)+945D0*derfctemp)/r**11
           V(5,5)=V(5,5)-xiy**5*q(j,0)*(2D0*r*G*dexptemp*(945D0
     &           +630D0*r**2*G**2+252D0*r**4*G**4+72D0*r**6*G**6
     &           +16D0*r**8*G**8)+945D0*derfctemp)/r**11
           V(6,0)=V(6,0)+q(j,0)*(2D0*r*G*dexptemp*(10395D0*x(3)**6+315D0
     &           *r**2*x(3)**4*(-45D0+22D0*x(3)**2*G**2)+63D0*r**4
     &           *x(3)**2*(75D0-150D0*x(3)**2*G**2+44D0*x(3)**4*G**4)
     &           +4D0*r**10*G**4*(-15D0+90D0*x(3)**2*G**2-60D0*x(3)**4
     &           *G**4+8D0*x(3)**6*G**6)+2D0*r**8*G**2*(-75D0+630D0
     &           *x(3)**2*G**2-540D0*x(3)**4*G**4+88D0*x(3)**6*G**6)
     &           +9D0*r**6*(-25D0+350D0*x(3)**2*G**2-420D0*x(3)**4*G**4
     &           +88D0*x(3)**6*G**6))-45D0*(5D0*r**6-105D0*r**4*x(3)**2
     &           +315D0*r**2*x(3)**4-231D0*x(3)**6)*derfctemp)/r**13
           V(6,1)=V(6,1)-xiy*x(3)*q(j,0)*(2D0*r*G*dexptemp*(10395D0
     &           *x(3)**4+630D0*r**2*x(3)**2*(-15D0+11D0*x(3)**2*G**2)
     &           +8D0*r**10*G**6*(15D0-20D0*x(3)**2*G**2+4D0*x(3)**4
     &           *G**4)+4D0*r**8*G**4*(105D0-180D0*x(3)**2*G**2+44D0
     &           *x(3)**4*G**4)+63D0*r**4*(25D0-100D0*x(3)**2*G**2+44D0
     &           *x(3)**4*G**4)+6D0*r**6*G**2*(175D0-420D0*x(3)**2*G**2
     &           +132D0*x(3)**4*G**4))+315D0*(5D0*r**4-30D0*r**2*x(3)**2
     &           +33D0*x(3)**4)*derfctemp)/r**13
           V(6,2)=V(6,2)+xiy**2*q(j,0)*(dexptemp*((630D0*r*G+420D0*r**3
     &           *G**3+168D0*r**5*G**5+48D0*r**7*G**7)*(r**4-18D0*r**2
     &           *x(3)**2+33D0*x(3)**4)-32D0*r**9*x(3)**2*(6D0*r**2-11D0
     &           *x(3)**2)*G**9+64D0*r**11*x(3)**4*G**11)+315D0*(r**4
     &           -18D0*r**2*x(3)**2+33D0*x(3)**4)*derfctemp)/r**13
           V(6,3)=V(6,3)-xiy**3*x(3)*q(j,0)*(2D0*r*G*dexptemp*(10395D0
     &           *x(3)**2+r**2*(-2835D0+2D0*G**2*(3465D0*x(3)**2+r**2
     &           *(-945D0+2D0*G**2*(693D0*x(3)**2+r**2*(-189D0+2D0*G**2
     &           *(99D0*x(3)**2+r**2*(-27D0+2D0*G**2*(11D0*x(3)**2+r**2
     &           *(-3D0+2D0*x(3)**2*G**2))))))))))-945D0*(3D0*r**2-11D0
     &           *x(3)**2)*derfctemp)/r**13
           V(6,4)=V(6,4)+xiy**4*q(j,0)*(2D0*r*G*dexptemp*(10395*x(3)**2
     &           +r**2*(-945D0+2D0*G**2*(3465D0*x(3)**2+r**2*(-315D0
     &           +2D0*G**2*(693D0*x(3)**2+r**2*(-63D0+2D0*G**2*(99D0
     &           *x(3)**2+r**2*(-9D0+2D0*G**2*(11D0*x(3)**2+r**2*(-1D0
     &           +2D0*x(3)**2*G**2))))))))))-945D0*(r**2-11D0*x(3)**2)
     &           *derfctemp)/r**13
           V(6,5)=V(6,5)-xiy**5*x(3)*q(j,0)*(2D0*r*G*dexptemp*(10395D0
     &           +2D0*r**2*G**2*(3465D0+1386D0*r**2*G**2+396D0*r**4*G**4
     &           +88D0*r**6*G**6+16D0*r**8*G**8))+10395D0*derfctemp)
     &           /r**13
           V(6,6)=V(6,6)+xiy**6*q(j,0)*(2D0*r*G*dexptemp*(10395D0
     &           +2D0*r**2*G**2*(3465D0+1386D0*r**2*G**2+396D0*r**4*G**4
     &           +88D0*r**6*G**6+16D0*r**8*G**8))+10395D0*derfctemp)
     &           /r**13
          ENDIF
         ENDDO
        ENDDO
       ENDDO
      ENDDO   
      CALL CLOCK(T)
      write(*,*) 'end of the normal space summation at',T,'sec.'
C do the summation over k space
C calculate boundaries for Na,Nb and Nc in sum over ga,gb,gc
C radius over which has to be integrated r:
C such that e^-r^2/4G^2<1D-12
C r=12G
C but higher orders of the potential make problems, see normal
C space sum. Sum up to r=15G
      write(*,*) 'start k-space of the Ewald sum'
      write(*,*) 'the radius of the sphere over which is summed',15*G
      CALL Cros(x,ga,gb)
      Vunit=dabs(gc(1)*x(1)+gc(2)*x(2)+gc(3)*x(3))
      CALL Cros(x,gc,gb)
      Na=int(1D0+15D0*G*DSQRT((x(1)**2+x(2)**2+x(3)**2))/
     &       Vunit)
      CALL Cros(x,ga,gc)
      Nb=int(1D0+15D0*G*DSQRT((x(1)**2+x(2)**2+x(3)**2))/
     &       Vunit)
      CALL Cros(x,ga,gb)
      Nc=int(1D0+15D0*G*DSQRT((x(1)**2+x(2)**2+x(3)**2))/
     &       Vunit)
      write(*,*) 'this fits in a parallelepiped with Na,Nb,Nc=',Na,Nb,Nc
      write(*,*) 'with the basis vectors:'
      write(*,*) 'a:',ga
      write(*,*) 'b:',gb
      write(*,*) 'c:',gc
C do the k-space summation
      DO ia=-Na,Na
       DO ib=-Nb,Nb
        DO ic=-Nc,Nc
         IF(.NOT.((ia.EQ.0).AND.(ib.EQ.0).AND.(ic.EQ.0)))THEN
          x(1)=ia*ga(1)+ib*gb(1)+ic*gc(1)
          x(2)=ia*ga(2)+ib*gb(2)+ic*gc(2)
          x(3)=ia*ga(3)+ib*gb(3)+ic*gc(3)
          iz=Dcmplx(0D0,x(3))
          ixy=Dcmplx(x(2),x(1))
          r=dsqrt(x(1)**2+x(2)**2+x(3)**2)
          sum=0D0
          DO j=1,Nq
           sum=sum+q(j,0)*cdexp(-dcmplx(0D0,1D0)*
     &         (x(1)*q(j,1)+x(2)*q(j,2)+x(3)*q(j,3)))
          ENDDO
          ctemp=4*PI*DEXP(-r**2/(4D0*G**2))*sum/(Ve*r**2)
          Vtemp(0,0)=Vtemp(0, 0)+ctemp
          Vtemp(1,0)=Vtemp(1, 0)+ctemp*iz
          Vtemp(1,1)=Vtemp(1, 1)-ctemp*ixy
          Vtemp(2,0)=Vtemp(2, 0)+ctemp*iz**2
          Vtemp(2,1)=Vtemp(2, 1)-ctemp*iz*ixy
          Vtemp(2,2)=Vtemp(2, 2)+ctemp*ixy**2
          Vtemp(3,0)=Vtemp(3, 0)+ctemp*iz**3
          Vtemp(3,1)=Vtemp(3, 1)-ctemp*iz**2*ixy
          Vtemp(3,2)=Vtemp(3, 2)+ctemp*iz*ixy**2
          Vtemp(3,3)=Vtemp(3, 3)-ctemp*ixy**3
          Vtemp(4,0)=Vtemp(4, 0)+ctemp*iz**4
          Vtemp(4,1)=Vtemp(4, 1)-ctemp*iz**3*ixy
          Vtemp(4,2)=Vtemp(4, 2)+ctemp*iz**2*ixy**2
          Vtemp(4,3)=Vtemp(4, 3)-ctemp*iz*ixy**3
          Vtemp(4,4)=Vtemp(4, 4)+ctemp*ixy**4
          Vtemp(5,0)=Vtemp(5, 0)+ctemp*iz**5
          Vtemp(5,1)=Vtemp(5, 1)-ctemp*iz**4*ixy
          Vtemp(5,2)=Vtemp(5, 2)+ctemp*iz**3*ixy**2
          Vtemp(5,3)=Vtemp(5, 3)-ctemp*iz**2*ixy**3
          Vtemp(5,4)=Vtemp(5, 4)+ctemp*iz*ixy**4
          Vtemp(5,5)=Vtemp(5, 5)-ctemp*ixy**5
          Vtemp(6,0)=Vtemp(6, 0)+ctemp*iz**6
          Vtemp(6,1)=Vtemp(6, 1)-ctemp*iz**5*ixy
          Vtemp(6,2)=Vtemp(6, 2)+ctemp*iz**4*ixy**2
          Vtemp(6,3)=Vtemp(6, 3)-ctemp*iz**3*ixy**3
          Vtemp(6,4)=Vtemp(6, 4)+ctemp*iz**2*ixy**4
          Vtemp(6,5)=Vtemp(6, 5)-ctemp*iz*ixy**5
          Vtemp(6,6)=Vtemp(6, 6)+ctemp*ixy**6
         ENDIF
        ENDDO
       ENDDO
      ENDDO
C correct for charge at origin
      Vtemp(0,0)=Vtemp(0,0)-2D0  *G   *q(1,0)/Dsqrt(PI)
      Vtemp(2,0)=Vtemp(2,0)+4D0  *G**3*q(1,0)/(3D0*Dsqrt(PI))
      Vtemp(4,0)=Vtemp(4,0)-24D0 *G**5*q(1,0)/(5D0*Dsqrt(PI))
      Vtemp(6,0)=Vtemp(6,0)+240D0*G**7*q(1,0)/(7D0*Dsqrt(PI))
C include prefactor \frac{1}{(\sqrt{k-m)!(k+m)!}}
      DO k=0,6
       DO m=0,k
        kmmfac=1
        kpmfac=1
        DO i=1,k-m
         kmmfac=kmmfac*i
        ENDDO
        DO i=1,k+m
         kpmfac=kpmfac*i
        ENDDO
        V(k,m)=(1D0/Dsqrt(kmmfac*kpmfac))*(Vtemp(k,m)+V(k,m))
       ENDDO
      ENDDO
      DO k=0,6
       DO m=-k,-1
        V(k,m)=(-1)**m*Dcmplx(Dreal(V(k,-m)),-Dimag(V(k,-m)))
       ENDDO
      ENDDO
      CALL CLOCK(T)
      write(*,*) 'end of the k-space summation at,',T,'sec'
C from natural units to eV and A
      DO k=0,6
       DO m=-k,k
        V(k,m)=-e**2*JtoeV*Anstr/(4*PI*eps0)*V(k,m)
       ENDDO
      ENDDO
      end
C
C calculate the reciprocal lattice
C
      subroutine reciprocallatice(ga,gb,gc,a,b,c)
      real*8 PI
      parameter (PI=3.141592653589793285D0)
      real*8 a(3),b(3),c(3),ga(3),gb(3),gc(3),V
      CALL CROS(ga,b,c)
      CALL CROS(gb,c,a)
      CALL CROS(gc,a,b)
      CALL DOT(V,ga,a)
      ga(1)=2*PI*ga(1)/V
      ga(2)=2*PI*ga(2)/V
      ga(3)=2*PI*ga(3)/V
      gb(1)=2*PI*gb(1)/V
      gb(2)=2*PI*gb(2)/V
      gb(3)=2*PI*gb(3)/V
      gc(1)=2*PI*gc(1)/V
      gc(2)=2*PI*gc(2)/V
      gc(3)=2*PI*gc(3)/V
      end
C
C Calculate the dot product of two vectors
C
      subroutine DOT(d,a,b)
      real*8 d,a(3),b(3)
      d=a(1)*b(1)+a(2)*b(2)+a(3)*b(3)
      end
C
C Calculate the cross product of two vectors
C
      subroutine CROS(c,a,b)
      real*8 c(3),a(3),b(3)
      c(1)=a(2)*b(3)-a(3)*b(2)
      c(2)=a(3)*b(1)-a(1)*b(3)
      c(3)=a(1)*b(2)-a(2)*b(1)
      end
\end{verbatim}
\end{small}

\chapter{Manual for XTLS}
\label{ApendixXTLSManual}

The program XTLS, written by A. Tanaka, is a very powerful programme to solve any kind of model Hamiltonian. It can find the lowest eigen-states and operator values, as well as calculate spectra. It uses a Lancsos routine to find the lowest eigenstates of the Hamiltonian. With the use of Greens functions one can calculate absorption and electron emission spectra, as well as resonant processes. In this appendix we will explain how to set up an TMO$_6$ cluster Hamiltonian and how to calculate absorption and electron emission spectra.

When starting a new project or calculating a new material it is necessary to create three directories; xcards, xobjs, and xwrk. The xcards directory will contain the input files. The xobjs directory will contain the results of the calculation. The xwrk directory is used by the program to save temporary files, and a log file of the entire calculation containing some hints if errors are found.

The input file for a ground-state calculation consist of five parts. In the first part user variables and functions can be defined. Useful if one wants to check the dependence of the results as a function of some parameter. The second part contains the definition of the basis. Only the configurations in use have to be specified, the program will create the full basis for each configuration. In the third pard some global parameters can be specified. In the forth part one has to specify what one wants to calculate. Only the ground-state, or also an absorption or photo emission spectrum. In the fifth part the Hamiltonian has to be specified.

Lets start with a simple input file, calculating the energy levels of a single $d$ electron in $O_h$ symmetry.
\begin{verbatim}
XCRD: //example 01
(
)
CNFG:
      3d
  #i1  1
PARA:
EXEC:
  Mode=nop;
  Mag={3d};
  Ninit=10;
OPRT:
  10Dq(#i1 3d)=1.0;
XEND:
STOP:
\end{verbatim}
The input file, xcard starts with the commando XCRD: to lett the program know the file starts. Comments can be written behind //. 

The first part, written between brackets () contains user variables and is empty in the first example. 

The second part contains the definition of the configurations in use. It starts with CNFG: then there is one line with the orbitals used. The name of an orbital can be a number or letter, followed by the orbital quantum number of that orbital. 1s, 2p and 3d are good names for an orbital, but also Ld for the binding oxygen orbitals in a $TMO_6$ cluster is a good name, or Ap, Bp, Cp, Dp, Ep, and Fp are good names when there are 6 oxygen $2p$ orbitals belonging to different atoms. After the first line specifying the orbitals in use there can be many lines with different configurations. Each configuration has a name, starting with \#. It is common practise to take the first letter of the configuration name to be i for initial states, f for final states and m for intermediate states. After the name of the configuration one has to specify how much electrons are in each orbital. In the case of example 1 there is one orbital defined, a $3d$ orbital. There is one configuration defined with the name \#i1 and this configuration has one electron in the $3d$ shell.

The third part is used to define global parameters. In the case of example 1 it is empty.

The forth part tells the pc what has to be calculated. With the use of the Mode command one can specify if the computer will calculate an absorption spectra, Mode=xas; an photo emission spectra, Mode=pes; or only an energy level diagram of the initial-state, Mode=nop; The program can calculate the expectation value of many operators, like L,S, and T it calculates occupation numbers and gives out a density matrix. If one has many orbitals this output can become pretty large and hard to read, especially if one is only interested in one orbital. With the command Mag one can decide for which orbitals the operator information should be given in all its detail. In example one there is only one orbital, the 3d orbital and for this orbital all information should be printed in the output file. The computer also needs to know how much states it should compute. The command Ninit tells how many initial states should be included. The more initial states included the slower the programm becomes. But one should be careful not to exclude important initial states. In general all initial states that are within an energy range that can be thermal populated should be included in the calculation.

In the fifth part the operator is defined. In this case we only added a crystal field of $O_h$ symmetry. For that one can use the command 10Dq(\#configuration orbital)=value;. Where \#configuration and orbital should be replaced by the configuration and orbital the crystal field acts on.

The input file should end with XEND: and STOP: to tell the pc that everything is ready.

If we now want to include electron electron repulsion we need to include the command Rk in the operator. The operator Rk defines the values of the Slater integrals and for d-d interactions it expects the parameters $F^2$ and $F^4$. For a $d^8$ system the input file would look like:
\begin{verbatim}
XCRD: //example 02
(
)
CNFG:
      3d
  #i1  8
PARA:
EXEC:
  Mode=nop;
  Mag={3d};
  Ninit=45;
OPRT:
  10Dq(#i1 3d)=1.0;
  Rk(#i1 3d 3d)={10.000,6.250};
XEND:
STOP:
\end{verbatim}

We want to do cluster calculations and include the effect of covalency with the oxygens. Lets start with a simple cluster and project out the non-bonding oxygen orbitals. From the 6 oxygens, each with 6 2p orbitals only 10 orbitals are bonding and they have d orbital symmetry, as explained in chapter \ref{ChapterCluster}.
\begin{verbatim}
XCRD: //example 03
(
)
CNFG:
      3d Ld
  #i1  8 10
  #i2  9  9
  #i3 10  8
PARA:
  U(3d,3d)=6.0;
  Dlt(#i1 #i2 3d Ld)=5.0;
EXEC:
  Mode=nop;
  Mag={3d,Ld};
  Ninit=190;
OPRT:
  10Dq(#i1 3d)=1.0;
  10Dq(#i2 3d)=1.0;
  Rk(#i1 3d 3d)={10.000,6.250};
  VOh(#i1 #i2 3d Ld)={2.0,1.0};
  VOh(#i2 #i3 3d Ld)={2.0,1.0};
XEND:
STOP:
\end{verbatim}

In example 3 we included three configurations. A $d^8$ configuration, a $d^{9}\underline{L}$ and a $d^{10}\underline{L}^2$ configuration named \#i1, \#i2 and \#i3 respectively. In order to define the on-site energies of these configurations we defined a $U$ and a $\Delta$ as defined in the introduction \cite{Zaanen85}. The on-site energy of configuration \#i1 is zero by definition. The on-site energy of configuration \#i2 is $\Delta$ and the on-site energy of configuration \#i3 is $2\Delta+U$. We also added the operator $VOh$, which defines the hybridization between the $3d$ and $Ld$ orbitals. There are two parameters for the operator $VOh$. These parameters are the hybridization strength between the $e_{g}$ orbitals, $V_{e_g}=-\sqrt{3}pd\sigma$ and the hybridization between strength between the $t_{2g}$ orbitals, $V_{t_{2g}}=2pd\pi$. In example 3 we took $V_{e_g}=2.0$ eV and $V_{t_{2g}}=1.0$ eV.

We do not only want to calculate the ground-state of a cluster, but want to calculate spectra. In order to calculate spectra we need to add a final state. A final state is added when the final state configurations are defined and an extra $OPRT:$ command is added in which the final state Hamilton operator is defined. One also needs to define the dipole matrix elements. We normally set the dipole strength between the initial state and the final state equal to 1.0 as we are interested in relative intensities only.

In example 4 we show the input file for $2p$ XAS and in example 5 the input file for $3d$ valence band photo emission.

\begin{verbatim}
XCRD: //example 04
(
)
CNFG:
      2p 3d Ld
  #i1  6  8 10
  #i2  6  9  9
  #i3  6 10  8
  #f1  5  9 10
  #f2  5 10  9
PARA:
  U(3d,3d)=6.0;
  U(2p,3d)=8.0;
  Dlt(#i1 #i2 3d Ld)=5.0;
EXEC:
  Mode=xas;
  Mag={3d,Ld};
  Ninit=3;
OPRT:
  10Dq(#i1 3d)=1.0;
  10Dq(#i2 3d)=1.0;
  Rk(#i1 3d 3d)={10.000,6.250};
  VOh(#i1 #i2 3d Ld)={2.0,1.0};
  VOh(#i2 #i3 3d Ld)={2.0,1.0};
OPRT:
  10Dq(#f1 3d)=1.0;
  Rk(#f1 2p 3d)={6.000,5.000,4.000};
  VOh(#f1 #f2 3d Ld)={2.0,1.0};
  Zta{#f1 2p}=10.0;
  Zta{#f2 2p}=10.0;
OPRT:
  Dk(#i1 #f1 2p 3d)=1.0;
  Dk(#i2 #f2 2p 3d)=1.0;
XEND:
STOP:
\end{verbatim}

\begin{verbatim}
XCRD: //example 05
(
)
CNFG:
       3d Ld
  #i1   8 10
  #i2   9  9
  #i3  10  8
  #f1   7 10
  #f2   8  9
  #f3   9  8
  #f4  10  7
PARA:
  U(3d,3d)=6.0;
  U(2p,3d)=8.0;
  Dlt(#i1 #i2 3d Ld)=5.0;
EXEC:
  Mode=pes;
  Mag={3d,Ld};
  Ninit=3;
OPRT:
  10Dq(#i1 3d)=1.0;
  10Dq(#i2 3d)=1.0;
  Rk(#i1 3d 3d)={10.000,6.250};
  VOh(#i1 #i2 3d Ld)={2.0,1.0};
  VOh(#i2 #i3 3d Ld)={2.0,1.0};
OPRT:
  10Dq(#f1 3d)=1.0;
  10Dq(#f2 3d)=1.0;
  10Dq(#f3 3d)=1.0;
  10Dq(#f4 3d)=1.0;
  Rk(#f1 3d 3d)={10.000, 6.250};
  Rk(#f2 3d 3d)={10.000, 6.250};
  VOh(#f1 #f2 3d Ld)={2.0,1.0};
  VOh(#f2 #f3 3d Ld)={2.0,1.0};
  VOh(#f3 #f4 3d Ld)={2.0,1.0};
OPRT:
  Dk
XEND:
STOP:
\end{verbatim}

Sofar we have only discussed systems in cubic symmetry, without spin-orbit coupling, without a magnetic field and without magnetic exchange interaction. These interactions can be added with the use of the following commands.\\
Spin-orbit coupling:
\begin{verbatim}
  Zta(#i1 3d)=0.1;
\end{verbatim}
A magnetic field of strength B in the direction k,l,m where k,l,m are the directional cosines:
\begin{verbatim}
  Ba(#i1 3d)={B,k,l,m};
\end{verbatim}
An exchange field of strength H in the direction k,l,m where k,l,m are the directional cosines:
\begin{verbatim}
  Ha(#i1 3d)={H,x,y,z}
\end{verbatim}
A crystal field of arbitrary symmetry can be included with the command Akm for real crystal fields and CAkm for complex crystal fields. These commands expect the expansion coefficients of the crystal field expanded on spherical harmonics as discussed in chapter \ref{ChapterCluster}. Only nonzero $B_{k,m}$ values have to be listed. The real and imaginary part of $B_{k,m}$ are placed after each other, separated by a comma.
\begin{verbatim}
  CAkm(#i1 3d)={k1,m1,ReBk1m1,ImBk1m1,
                k2,m2,ReBk2m2,ImBk2m2};
\end{verbatim}
Hybridization of arbitrary symmetry can also be included with the use of potentials expanded on spherical harmonics. The hybridization between the $3d_{x^2-y^2}$ orbital and the $Ld_{x^2-y^2}$ orbital is called $V_{x^2-y^2,x^2-y^2}$ More general the hybridization between the $3d_{\tau}$ orbital and the $Ld_{\tau'}$ orbital is called $V_{\tau,\tau'}$. This potential $V$ can be expanded on spherical harmonics and gives parameters $B_{k,m}$. The way how to expand this is explained within chapter \ref{ChapterCluster}. If we have configuration \#i1 to be a $3d^n$ configuration and the configuration \#i2 is a $3d^{n+1}\underline{Ld}$ configuration then the potential V expanded on spherical harmonics can be added in the following way:
\begin{verbatim}
   CAkm(#i1 #i2 3d Ld)={k1,m1,ReBk1m1,ImBk1m1,
                        k2,m2,ReBk2m2,ImBk2m2};
\end{verbatim}

\backmatter

\chapter{Abstract}

The class of transition metal compounds shows an enormous richness of physical properties \cite{Tsuda91, Imada98}, such as metal-insulator transitions, colossal magneto-resistance, super-conductivity, magneto-optics and spin-depend transport. The theoretical description of these materials is still a challenge. Traditional methods using the independent electron approximation most of the time fail on even the simplest predictions. For example, many of the transition metal compounds, with NiO as the classical example, should be a metal according to band-structure calculations, but are in reality excellent insulators. 

The single band Mott-Hubbard model \cite{Mott49, Hubbard63} explains very nicely why many correlated materials are insulating. But even the Mott-Hubbard model has some problems in understanding the band-gap found for many of the transition metal compounds \cite{Sawatzky84}. With the recognition that transition metal compounds can be of the charge-transfer type or the Mott-Hubbard type \cite{Zaanen85}, depending on the ratio of $U$ and $\Delta$, also the band-gap can be understood. Whereby $U$ is defined as the repulsive Coulomb energy of two electrons on the same transition metal site and $\Delta$ is defined as the energy it costs to bring an electron from an oxygen site to a transition metal site.

The single band Mott-Hubbard model is however, even when charge transfer effects are included, inadequate in describing the full richness found in many of the transition metal compounds \cite{Birgeneau00, Tokura00, Orenstein00}. It now becomes more and more clear that in order to describe transition metal compounds the charge, orbital, spin and lattice degrees of freedom should all be taken into account. Especially the orbital degrees of freedom have not been considered to the full extend until recently. In the manganates, for example, orbital and charge ordering of the Mn ions play an important role for the colossal magneto-resistance of these materials \cite{Ramirez97, Khomskii97, Mizokawa95, Mizokawa96, Mizokawa97}. An other example would be the metal-insulator transition in V$_{2}$O$_{3}$ \cite{Park00, Ezhov99, Mila00}. The orbital occupation of the V ion changes drastically at the phase transition \cite{Park00}. This change in orbital occupation will change the local spin-spin correlations which in-turn will change the effective band-width. This indicates that not only electron-electron Coulomb repulsion in a single band must be considered, but a full multi-band theory including all interactions must be considered in order to understand this prototypical Mott-Hubbard system.

With the recognition that the local orbital occupation plays an important role in many of the transition metal compounds there is a need for experimental techniques that can measure the orbital occupation. This technique is soft x-ray absorption spectroscopy (XAS). For transition metal atoms one measures the local transition of a $2p$ core electron into the $3d$ valence shell. In chapter \ref{ChapterNiO} to chapter \ref{ChapterMnO} we used soft x-ray absorption spectroscopy to measure orbital occupations, crystal fields, and spin directions in thin films. In chapter \ref{ChapterCobaltates} to chapter \ref{ChapterVO2} we used soft x-ray absorption spectroscopy to gain insight in the importance of spin and orbital degrees of freedom in bulk transition metal compounds.

In chapter \ref{ChapterNiO} we present linear dichroism in the Ni $L_{2,3}$ x-ray absorption spectra of a monolayer NiO(001) on Ag(001) capped with MgO (001). The dichroic signal appears to be very similar to the magnetic linear dichroism observed for thicker antiferromagnetic NiO films. A detailed experimental and theoretical analysis reveals, however, that the dichroism is caused by crystal field effects. We present a practical experimental method for identifying the independent magnetic and crystal field contributions to the linear dichroic signal in spectra of NiO films with arbitrary thickness and lattice strain.

In chapter \ref{ChapterCoO} we first used XAS to study the properties of CoO bulk, as well as thin films. We confirm that the Co ion in CoO has a free orbital momentum in cubic symmetry. We confirm that spin-orbit coupling is very important for understanding the properties of CoO and show that it is not reduced from the Hartree-Fock value for a free Co$^{2+}$ ion. With the use of cluster calculations we can get a full consistent understanding of the XAS spectra and the polarization dependence of CoO. For CoO thin films we used XAS to show that we can control the orbital momentum and spin direction with the use of strain in thin CoO films. This finding opens up great opportunities for the use of exchange-bias, where people put an antiferromagnet adjoined to a ferromagnet in order to shift the magnetization hysteresis loop in one of the magnetic field directions \cite{Meiklejohn56, Meiklejohn57}. With the use of strain in the antiferromagnet one can chose if the system will be exchange-biased in the plane of the thin film, or perpendicular to the thin film surface. It also has great implications for the understanding of the exchange-bias phenomenon sofar. The exchange-bias effect takes place at the interface between the ferromagnet and antiferromagnet \cite{Nogues99, Berkowitz99}. At the interface there will be strain in the antiferromagnet and one can not assume that the antiferromagnet has the same spin structure at the interface as it has in the bulk.

In chapter \ref{ChapterMnO} we show how one can orientate spins in antiferromagnetic thin films with low magnetocrystalline anisotropy ($d^3$, $d^5$ and $d^8$ systems in $O_h$ symmetry) via the exchange coupling to adjacent antiferromagnetic films with high magnetocrystaline anisotropy ($d^6$ and $d^7$ systems in nearly $O_h$ symmetry). We have grown MnO thin films on CoO thin films with different predetermined spin orientation. With the use of Mn $L_{2,3}$ soft x-ray absorption spectroscopy we show that the Mn spin 'follows' the Co spin direction.

In chapter \ref{ChapterCobaltates} we study the spin-state problem within the cobaltates. Normally one is used to discuss the spin direction, or the magnetic spin angular momentum, $S_z$. Within the cobaltates, $d^6$ compounds, there is however a discussion about the size of the spin, or the spin angular momentum, $S^2=S(S+1)$. $S^2$ can be 0, 2, or 6 (S=0,1,2), referred to as a low-spin state, an intermediate-spin state and a high-spin state. Within the literature there is a lot of confusion about the spin state as deduced from magnetic, neutron and x-ray diffraction measurements in the newly synthesized layered cobalt perovskits \cite{Martin97, Maignan99, Yamaura99a, Yamaura99b, Loureiro01, Vogt00, Suard00, Fauth01, Burley03, Mitchell03, Moritomo00, Respaud01, Kusuya01, Frontera02, Fauth02, Taskin03, Soda03, Loureiro00, Knee03, Wu00, Kwon00, Wang01, Wu01, Wu02, Wu03}. These measurements determine the size of the spin angular momentum ($S^2$) from the maximum size of the magnetic spin momentum $S_z$. XAS is directly sensitive to the expectation value of $S^2$. We carried out a test experiment using a relatively simple model compound, namely Sr$_2$CoO$_3$Cl, in which there are no spin state transitions present and in which there is only one kind of Co$^{3+}$ ion coordination \cite{Loureiro00,Knee03}. Important is that this coordination is identical to the pyramidal CoO$_{5}$ present in the heavily debated layered perovskites \cite{Martin97, Maignan99, Yamaura99a, Yamaura99b, Loureiro01, Vogt00, Suard00, Fauth01, Burley03, Mitchell03, Moritomo00, Respaud01, Kusuya01, Frontera02, Fauth02, Taskin03, Soda03}. Using a \textit{spectroscopic} tool, that is soft x-ray absorption spectroscopy (XAS), we demonstrate that pyramidal Co$^{3+}$ ions are not in the often claimed intermediate-spin state but unambiguously in a high-spin state. This outcome suggests that the spin states and their temperature dependence in layered cobalt perovskites may be rather different in nature from those proposed in the recent literature.

In chapter \ref{ChapterLaTiO3} we study LaTiO$_{3}$. There has been a strong debate about the role of orbital degrees of freedom within the titanates and in LaTiO$_{3}$ especially \cite{Cwik03, Goral83, Meijer99, Mizokawa96, Keimer00, Khaliullin00, Mochizuki03, Pavarini04, Solovyev04, Craco04}. With the use of spin resolved circular polarized photo emission spectroscopy we confirmed that the orbital momentum in LaTiO$_{3}$ is indeed quenched \cite{Keimer00}. With the use of XAS we show that this is due to a relative large crystal field in the order of 120 to 300 meV. For a realistic description of materials one should not forget that there is a strong coupling between the orbitals and the lattice. 

In chapter \ref{ChapterVO2} we look at the metal-insulator transition in VO$_{2}$. With the use of XAS we show that the metal-insulator transition within this material is accompanied by a change in orbital occupation. The orbital occupation changes from almost isotropic in the metallic phase to the almost completely $\sigma$-polarized in the insulating phase, in close agreement with the two-site cluster model \cite{Tanaka04}. This very strong orbital polarization leads in fact to a change of the electronic structure of VO$_{2}$ from a 3-dimensional to effectively a 1-dimensional system \cite{Khomskii05}. The V ions in the chain along the c-axis are then very susceptible to a Peierls transition. In this respect, the MIT in VO$_{2}$ can indeed be regarded as a Peierls transition \cite{Wentzcovitch94}. However, to achieve the required dramatic change of the orbital occupation one also need the condition that strong electron correlations bring this narrow band system close to the Mott regime \cite{Rice94}. The MIT in VO$_{2}$ may therefore be labelled as a "collaborative" Mott-Peierls transition.

\chapter{Zusammenfassung}

Die Material Klasse der \"Uber\-gangs\-metall\-oxide zeichnet sich durch einen enormen Reich\-tum an physikalischen Effecten aus \cite{Tsuda91, Imada98}; dazu geh\"oren Metall-Isolator-\"Uber\-g\"ange, kolossaler Magnetwiderstand, Supraleitung, Magnetooptik und spinabh\"angiger Transport. Die theoretische Beschreibung dieser Materialien ist noch immer eine Herausforderung: Traditionelle Methoden, die auf der N\"aherung unabh\"angiger Elektronen basieren, scheitern oftmals schon bei einfachen Vorhersagen. Zum Beispiel sollten viele \"Uber\-gangs\-metall\-oxide, NiO ist das klassische Beispiel, laut klassischen Bandstrukturrechnungen Metalle sein, in Wirklichkeit sind sie jedoch Isolatoren.

Das Ein-Band-Hub\-bard\-mo\-dell \cite{Mott49, Hubbard63} beschreibt sehr gut, warum viele korrelierte Materialien isolierend sind; aber selbst mit dem Mott-Hub\-bard\-mo\-dell hat man Probleme, die Band\-l\"ucke in vielen \"Uber\-gangs\-metall\-ver\-bin\-dung zu verstehen \cite{Sawatzky84}. Mit der Erkenntnis, daß \"Uber\-gangs\-metall\-ver\-bin\-dungen, abh\"angig von Ver\-h\"alt\-nis $U$ zu $\Delta$, entweder vom La\-dungs\-trans\-fer- oder Mott-Hubbard-Typ sind, kann man auch die Bandl\"ucke verstehen \cite{Zaanen85}. Hierbei ist $U$ definiert als die abstoßende Coulomb-Energie zweier Elektronen desselben \"Uber\-gangs\-metall-Ions und $\Delta$ ist die Energie, die man braucht, um ein Elektron vom Sauerstoff auf das Metall zu \"ubertragen.

Das Ein-Band-Hubbardmodell scheitert jedoch, selbst wenn La\-dungs\-trans\-fer-Effekte ber\"ucksichtig werden, an der vollst\"andigen Beschreibung der Effekte einer Vielzahl von \"Uber\-gangs\-metall\-ver\-bin\-dungen  \cite{Birgeneau00, Tokura00, Orenstein00}. Es wird zunehmend klarer, daß Ladungs-, Spin- und Gitterfreiheitsgrade gleichermaßen ber\"ucksichtigt werden m\"ussen, um \"Uber\-gangs\-metall\-ver\-bin\-dungen korrekt zu beschreiben. Be\-son\-ders die orbitalen Freiheitsgrade wurden bis vor kurzem nicht vollst\"andig ber\"ucksichtigt. In Manganaten spielt beispielsweise die orbitale und Ladungsordnung der Manganionen eine große Rolle f\"ur den kolossalen Magnetwiderstand dieser Materialen \cite{Ramirez97, Khomskii97, Mizokawa95, Mizokawa96, Mizokawa97}. Ein anderes Beispiel ist der Metall-Isolator-\"Ubergang in $\rm V_2O_3$ \cite{Park00, Ezhov99, Mila00}: Die orbitale Besetzung des Vanadiumions \"andert sich stark am Phasen\"ubergang \cite{Park00}. Diese Änderung beeinflußt die lokale Spin-Spin-Wechselwirkung und so die effektive Bandbreite. Das deutet darauf hin, daß nicht nur die Elektron-Elektron-Coulombabstoßung in einem einzelnen Band ber\"ucksichtigt werden muß, sondern vielmehr eine Theorie nötig ist, die Mehr-Band-Effekte einschließlich aller Wechselwirkungen umfaßt, um dieses prototypische Mott-Hubbard-System zu verstehen.

When mann ber\"ucksichtigt, daß die lokale orbitale Besetzung eine große Rolle in vielen \"Uber\-gangs\-metall\-ver\-bin\-dungen spielt, wird eine experimentelle Methode benötigt, mit der die orbitale Besetzung gemessen werden kann. Dies erlaubt die Rönt\-gen\-ab\-sop\-tions\-spek\-tros\-ko\-pie. Bei \"Uber\-gangs\-metall\-ato\-men mißt man den lokalen \"Uber\-gang eines kernnahen $2p$-Elektrons in die $3d$-Valenzschale. In Kapitel \ref{ChapterNiO} bis \ref{ChapterMnO} wird die Rönt\-gen\-ab\-sop\-tions\-spek\-tros\-ko\-pie (XAS) benutzt, um die orbitale Besetzung, Kristallfelder und Spinrichtung in d\"unnen Filmen zu messen. In Kapitel \ref{ChapterCobaltates} bis \ref{ChapterVO2} wird versucht, mittels ihr ein Einblick in die Bedeutung der Spin- und orbitalen Freiheitsgrade in \"Uber\-gangs\-metall\-oxiden zu gewinnen.

In Kapitel \ref{ChapterNiO} zeigen wir den linearen Dichroismus in Nickel $L_{2,3}$-Absorption\-spektren einer Lage NiO(001) auf Ag(001), die mit MnO(001) abgedeckt ist. Das dichroische Signal scheint dem magnetischen, linearen Dichroismius dickerer, antiferromagnetischer NiO-Filme zu \"ahneln. Eine detailierte experimentelle und theoretischen Analyse zeigt jedoch, daß der Dichroismus durch Kristallfeldeffekte verursacht wird. Wir zeigen eine praktische Methode, die es erlaubt, die von einander unabh\"anigen magnetischen und Kristallfeld-Anteile des linearem dichroischen Signal im Spektrum von NiO-Filmen beliebiger Dicke und Verspannung zu identifizieren.

In Kapitel \ref{ChapterCoO} benutzen wir zuerst die XAS, um die Eigenschaften von CoO sowohl im Volumen als auch in d\"unnen Filmen zu studieren. Wir best\"atigen, daß die Cobaltionen in CoO in kubischer Symmetrie ein freies orbitales Moment haben. Weiter zeigen wir, daß die Spin-Bahn-Wechselwirkung sehr wichtig f\"ur das Verst\"andnis der CoO-Eigenschaften ist und, daß sie nicht gegen\"uber dem Hartree-Fock-Wert eines freien Co$^{2+}$-Ions reduziert ist. Mittels Cluster-Rechnung kann man ein vollkommen konsistentes Verst\"andnis der XAS-Spektren und der Polarisationsabh\"anigkeit von CoO erhalten. F\"ur d\"unne Filme von CoO haben wir XAS benutzt, um zu zeigen, daß wir das orbitale Moment und die Spinrichtung durch Verspannung kontrollieren können. Diese Entdeckung eröffnet weite Möglichkeiten f\"ur die Verwendung der Austauschanisotropie (exchange bias), bei der ein Antiferromagnet mit einem Ferromagneten verbunden wird, um die Magnetisierungs-Hysteresekurve in eine der Magnetfeldrichtungen zu verschieben \cite{Meiklejohn56, Meiklejohn57}. Mittels Verzerrung kann der Antiferromagnet daf\"ur sorgen, daß das System entweder in der Filmebene oder orthogonal dazu die Austauschanisotropie zeigt. Dies hat auch große Auswirkungen auf das Verst\"andnis der bisher bekannten Austauschanisotropie-Ph\"anomene. Der Austauschanisotrope-Effekt findet an der Grenzfl\"ache zwischen Antiferro- und Ferromagneten statt \cite{Nogues99, Berkowitz99}; an dieser wird der Antiferromagnet verzerrt, so daß man nicht erwarten kann, daß der Antiferromagnet an der Grenzfl\"ache dieselbe Spinstruktur hat wie im Volumen.

In Kapitel \ref{ChapterMnO} zeigen wir, wie man den Spin in antiferromagnetischen d\"unnen Filmen mit niedrigen magnetokristallinen Anisotropien ($d^3$-, $d^5$- und $d^8$-Sy\-ste\-men in $O_h$-Symmetrie) mittels Austauschkopplung an einen angrenzenden antiferromagnetischen Films mit starker magnetokristallinen Anisotrope ($d^6$- und $d^7$-Systeme in fast $O_h$-Symmetrie) ausrichten kann. Wir haben d\"unne MnO-Filme auf d\"unnen CoO-Filmen mit unterschiedlicher Spinorientierung gewachsen; mit Hilfe von XAS an der Mn-$L_{2,3}$-Kante können wir zeigen, daß der Mn-Spin der Richtung des Co-Spins folgt.

In Kapitel \ref{ChapterCobaltates}  studieren wir das Problem des Spinzustandes in den Cobaltaten. Normalerweise ist man es gewöhnt, die Spinrichtung oder das magnetische Spinmoment $S_z$ zu diskutieren. Bei den Cobaltaten ($d^6$-Verbindungen), wird jedoch \"uber die Größe des betrags der Spins, $S^2=S(S+1)$ diskutiert. $S^2$ kann 0, 2 oder 6 ($S=0,1,2$) sein und dementsprechend spricht man von einem Niedrig-, intermedi\"aren oder Hochspinsystem. In der Literatur herrscht einige Verwirrung \"uber den aus Magnetisierungs-, Neutronen- und Röntgenbeugungsmessungen hergeleiteten Spinzustands in neusynthetisierten Schicht-Cobaltperowskiten \cite{Martin97, Maignan99, Yamaura99a, Yamaura99b, Loureiro01, Vogt00, Suard00, Fauth01, Burley03, Mitchell03, Moritomo00, Respaud01, Kusuya01, Frontera02, Fauth02, Taskin03, Soda03, Loureiro00, Knee03, Wu00, Kwon00, Wang01, Wu01, Wu02, Wu03}. In diesen Messungen wurde die Größe des Spindrehmomentes $S^2$ aus dem maximalen Erwartungswert f\"ur $S_z$ bestimmt. XAS kann hingegen direkt den Erwartungswert von $S^2$ bestimmen. Wir haben ein Testexperiment mit einer relativ einfachen Modellverbindung, n\"amlich $\rm Sr_2CoO_3Cl$, durchgef\"uhrt, in der es keine Spin\"uberg\"ange gibt und in der die Co$^{3+}$-Ionen in nur einer Koordination vorliegen \cite{Loureiro00,Knee03}. Wichtig ist, daß diese Koordination die selbe ist wie in dem pyramidialen CoO$_5$ in den stark diskutierten Schichtperowskiten \cite{Martin97, Maignan99, Yamaura99a, Yamaura99b, Loureiro01, Vogt00, Suard00, Fauth01, Burley03, Mitchell03, Moritomo00, Respaud01, Kusuya01, Frontera02, Fauth02, Taskin03, Soda03}. Durch die Verwendung einer \textit{spektroskopischen} Technik wie der Rönt\-gen\-ab\-sorp\-tions\-spek\-tros\-ko\-pie konnten wir zeigen, daß die pyramidialen Co$^{3+}$-Ionen nicht, wie oft behaupted im intermedi\"aren Spinzustand, sondern zweifelsfrei im Hochspinzustand sind. Dieses Ergebnis l\"aßt darauf schließen, daß die Ursache des Spinzustands und dessen Temperaturabh\"anigkeit sich stark von der in der j\"ungeren Literatur diskutierten unterscheidet.

In Kapitel \ref{ChapterLaTiO3} studieren wir LaTiO$_3$. Es gab eine starke Diskussion \"uber die Rolle des orbitalen Freiheitsgrads innerhalb der Titanate und LaTiO$_3$ im besonderen \cite{Cwik03, Goral83, Meijer99, Mizokawa96, Keimer00, Khaliullin00, Mochizuki03, Pavarini04, Solovyev04, Craco04}. Mittels spinaufgel\"oster, zirklar polarisierter Pho\-to\-emissions\-spek\-tros\-ko\-pie konnten wir best\"atigen, daß das orbitale Moment in LaTiO$_3$ tat\-s\"ach\-lich ausgelöscht (gequenched) ist. Mit Hilfe der XAS zeigen wir, daß dies durch ein relativ starkes Kristallfeld im Bereich von 120 bis 300 meV verursacht wird. F\"ur eine realistische Beschreibung dieser Materialien mu{\ss} man ber\"ucksichtigen, daß es eine starke Kopplung zwischen den Orbitalen und dem Gitter gibt.

In Kapitel \ref{ChapterVO2} untersuchen wir den Metall-Isolator-\"Ubergang in VO$_2$. Mittels XAS zeigen wir, daß zusammen mit dem Metall-Isolator-\"Ubergang sich auch die orbitale Besetzung dieses Materials \"andert. Die orbitale Besetzung geht dabei von fast isotrop in der Metallphase zu fast vollst\"andig $\sigma$-polarisiert in der Isolatorphase \"uber, in guter \"Uber\-ein\-stim\-mung mit dem Doppelclustermodell \cite{Tanaka04}. Diese sehr starke orbitale Polarisation f\"uhrt tats\"achlich zu einer \"Anderung der elektronischen Struktur des VO$_2$ von einem dreidimensionalen zu einem effektiv eindimensionalen System \cite{Khomskii05}. Die Vanadiumionen in der Kette entlang der $c$-Achse sind dann sehr empfindlich auf den Peierls-\"Ubergang. In dieser Beziehung kann der Metall-Isolator-\"Ubergang in VO$_2$ tats\"achlich als Peierls-\"Ubergang betrachtet werden \cite{Wentzcovitch94}. Um jedoch diese dramatische Änderung in der orbitalen Besetzung zu erreichen, ist es zus\"atzlich n\"otig, daß durch starke Elektronenkorrelation das schmalbandige System in die N\"ahe des Mottbereiches gebracht wird. Der Metall-Isolator-\"Ubergang in VO$_2$ kann deshalb als "gemeinsamer" Mott-Peierls-\"Ubergang bezeichnet werden.

\makeatletter
\g@addto@macro\@openbib@code{\setlength\parsep{0pt}}
\makeatother

\chapter{Acknowledgements}

This thesis could not have been written without the help, input and tremendous amount of work of many people. Physics is teamwork and the thesis presented here is a result of that. 

First of all I would like to thank Hao Tjeng. Your input and enthusiasm for physics is amazing. You came up with so many extremely interesting physical subjects that I had no time to work on all of them. There are still many things to be done. I like the discussions we had and am still amazed by the speed of your argumentations. Your attitude towards physics, which clearly intermixes theoretical and experimental physics, is something I learned to value very high. By combining theory and experiment the most interesting questions come up and can be answered.

Second I would like to thank Arata Tanaka. The last four years I have been working with your program and I am still amazed by the amount of options it has, the amount of different problems it can solve and the large amount of information it gives about the initial states of the calculated system. The program is extremely fast and the input file format is very simple. I also would like to thank you for the help given via e-mail on how to write correct input files for you program and for your explanation on some general physics.

The thin film research could not have been done without growing thin films. Dear Szili Csiszar, thanks for the extreme care taken when growing the films of NiO, CoO and MnO. I realize that growing these films with the correct stoichiometry, correct valence and correct structure is not a simple task. Without these films a large part of this thesis would not be there. The two months you spend in Cologne were very nice and I liked the discussions on our CoO and MnO research.

The input of Zhiwei Hu takes a special place within this thesis. There is not a single sample measured without him. When I first met you I had some difficulties in understanding you, but after some time I got used to your language and everything became fine. I greatly value your work and enjoy working together. I look forward on working together on the projects you're still working on.

I would like to thank Lucy Hamdan for all the preparations done in the background.

Dear Sergey Streltsov, I greatly enjoyed your stays here in Cologne. The L(S)DA(+U) calculations you did on LaTiO$_{3}$ and VO$_{2}$ in order to obtain crystal-fields and the occupation matrices were essential for the interpretation of the XAS spectra of these materials. I am happy you learned me how to do LDA calculations (L(S)DA+(U) for you), although we never came much further than the H atom.

Hua Wu, I enjoyed our discussions about LDA calculations, which certainly will continue. The discussions we had and the calculations you did on the cobaltates were very helpful. Especially that you made us aware of the magnetic ordering dependence of the total energy within the cobaltates was very helpful.

Tobias Burnus, thanks for helping me out with computer and latex questions and for translating the abstract into German for me. 

Thomas Koethe, Jan Gegner, Hua Wu, Ronny Sutarto, Marlies Engels, Holger Ott, Cristian Sch\"u\ss ler-Langeheine and Tobias Burnus, many thanks for proof reading my PhD. thesis and all the valuable comments.

Justina Schlappa, Thomas Koethe, Dilek Madenci and Jan Gegner, thanks for the nice time in our room. I enjoined the discussions we had. Justina thanks for the plants and the posters in our room. Dilek, thanks for the singing$\setminus$humming.

Cristian Sch\"u\ss ler-Langeheine, thanks for the tours through Berlin.

Daniel Komskii, thanks for the explanations and nice discussions about all different problems within correlated materials.

Jonas Wheinen, thanks for keeping the network and printers alive.

Thomas Lorenz, thanks for explaining me about the cgs units in magnetism and for answering all the other questions I asked. Our discussions about the cobaltates will certainly continue.

Jurg Bayer and Alexander Goesling, I enjoyed organizing part of the publicity of the 'Tag der offenen t\"ur' together, even when it took much more time than planned.

Markus Grueninger, when you walked into my room with the simple question how I normally calculated the crystal fields I started the answer with; "We do not really calculate them, but fit them to experiment". You stressed me to have a better look at theory. The result of our work is the program presented in appendix \ref{ApendixAkm}. This work has changed my attitude towards crystal-fields. For high symmetry systems, I still think a simple fit to experiment is a very good option. However, for low symmetry systems it is absolutely necessary to get a crystal field from \textit{ab initio} calculations.

Reinhard R\"uckamp, your optical data on $d$--$d$ excitations are very good. The discussion afterwards and the time spent to understand these data were very nice.

All the people on their bikes, thanks for waiting for me after every mountain we had to climb since I lagged behind.

I also would like to thank all the people from downstairs, for the nice time spent together with a beer whenever there was time to celebrate something.

Vladimir Katayev, thanks for the good discussions we had on LaCoO$_{3}$.  It is a pity you left Cologne. The questions you answered me, even when you had left our institute, were most valuable.

Salvatore Altieri, thanks for the nice time spent in Taiwan and in Cologne. Our work on NiO thin films on Ag and the discussion we had afterwards on magnetic anisotropy in such films have been very stimulating.

Holger Roth, Matthias Cwik and Thomas Lorenz, many thanks for the bulk LaTiO$_{3}$ samples. W. Reichelt, many thanks for the VO$_{2}$ samples. I. Bonn and C. Felser, many thanks for the Sr$_{2}$CoO$_{3}$Cl samples. Without your sample preparation we would not have been able to measure these materials, and this thesis could not have been written.

Thanks to Julio Cezar, Nick Brookes and all other people at beamline ID08 of the ESRF. The stability of your beamline is absolutely amazing. The XMCD measurements we did on CoO were only possible due to this extreme stability and reproducibility. Giacomo Ghiringhelli thanks for the nice discussions we had during the last measurement cycle. It is a pity that our spin-resolved circular polarized photo emission experiments did not work out.

Thanks to Hui-Huang Hsiej, Hong-Ji Lin, C.T. Chen and all the other people who keep the dragon beamline at the NSRRC in Taiwan functioning. Most of the spectra in this thesis are measured at your beamline.

Frank N\"ubel und Helmut Volke, danke f\"ur die matte Referenz und die sch\"one Diskussionen beim Mahmoud.

Lieve Marlies, bedankt voor gewoon degene die je bent.

\chapter{Erkl\"arung}

Ich versichere, dass ich die von mir vorgelegte Dissertation selbst\"andig angefertigt, die benutzten Quellen und Hilfsmittel vollst\"andig angegeben und die Stellen der Arbeit - einschlie\ss lich Tabellen, Karten und Abbildungen -, die anderen Werken im Wortlaut oder dem Sinn nach entnommen sind, in jedem Einzelfall als Entlehnung kenntlich gemacht habe; dass diese Dissertation noch keiner Fakult\"at oder Universit\"at zur Pr\"ufung vorgelegen hat; dass sie - abgesehen von unten angegebenen Teilpublikationen - noch nicht ver\"offentlicht worden ist sowie, dass ich eine solche Ver\"offentlichung vor Abschluss des Promotionsverfahrens nicht vornehmen werde. Die Bestimmungen dieser  Promotionsordnung sind mir bekannt. Die von mir vorgelegte Dissertation ist von Prof. Dr. L. H. Tjeng betreut worden.

$\qquad$\\

$\qquad$\\

\begin{center}
$\qquad$\\
Maurits Haverkort
\end{center}

\chapter{Publications}

\textit{Magnetic versus crystal field linear dichroism in NiO thin films.}\\
M. W. Haverkort, S. I. Csiszar, Z. Hu, S. Altieri, A. Tanaka, H. H. Hsieh, H.-J. Lin, C. T. Chen, T. Hibma, and L. H. Tjeng,\\
Physical Review B \textbf{69}, 020408 (2004); cond-mat/0310634.

$\qquad$\\
\textit{Growth and properties of strained VO$_{x}$ thin films with controlled stoichiometry.}\\
A. D. Rata, A. R. Chezan, T. Hibma, M. W. Haverkort, L. H. Tjeng, H. H. Hsieh, H.-J. Lin, and C. T. Chen,\\
Physical Review B \textbf{69}, 075404 (2004); cond-mat/0301263.

$\qquad$\\
\textit{A different look at the spin state of Co$^{3+}$ ions in CoO$_{5}$ pyramidal coordination.}\\
Z. Hu, Hua Wu, M. W. Haverkort, H. H. Hsieh, H.-J. Lin, T. Lorenz, J. Baier, A. Reichl, I. Bonn, C. Felser, A. Tanaka, C. T. Chen, and L. H. Tjeng,\\
Physical Review Letters \textbf{92}, 207402 (2004); cond-mat/0310138.

$\qquad$\\
\textit{Determination of the orbital moment and crystal field splitting in LaTiO$_{3}$.}\\
M. W. Haverkort, Z. Hu, A. Tanaka, G. Ghiringhelli, H. Roth, M. Cwik, T. Lorenz, C. Schuessler-Langeheine, S. V. Streltsov, A. S. Mylnikova, V. I. Anisimov, C. de Nadai, N. B. Brookes, H. H. Hsieh, H.-J. Lin, C. T. Chen, T. Mizokawa, Y. Taguchi, Y. Tokura, D. I. Khomskii, and L. H. Tjeng,\\
Physical Review Letters \textbf{94}, 056401 (2005); cond-mat/0405516.

$\qquad$\\
\textit{Controlling orbital moment and spin orientation in CoO layers by strain.}\\
S. I. Csiszar, M. W. Haverkort, Z. Hu, A. Tanaka, H. H. Hsieh, H.-J. Lin, C. T. Chen, T. Hibma, and L. H. Tjeng,\\
Submitted to Physical Review Letters; cond-mat/0504519.

$\qquad$\\
\textit{Aligning spins in antiferromagnetic films using antiferromagnets.}\\
S. I. Csiszar, M. W. Haverkort, T. Burnus, Z. Hu, A. Tanaka, H. H. Hsieh, H.-J. Lin, C. T. Chen, T. Hibma, and L. H. Tjeng\\
Submitted to Physical Review Letters; cond-mat/0504520.

$\qquad$\\
\textit{Optical study of orbital excitations in transition-metal oxides.}\\
R. R\"uckamp, A. G\"ossling, E. Benckiser, M. W. Haverkort, H. Roth, T. Lorenz, A. Freimuth, L. Jongen, A. M\"oller, G. Meyer, P. Reutler, B. B\"uchner, A. Revcolevschi, S.-W. Cheong, C. Sekar, G. Krabbes, and M. Gr\"uninger,\\
Submitted to New Journal of Physics; cond-mat/0503405.

$\qquad$\\
\textit{Zero-field incommensurate spin-Peierls phase with interchain frustration\\ in TiOCl.}\\
R. R\"uckamp, J. Baier, M. Kriener, M. W. Haverkort, T. Lorenz, G. S. Uhrig, L. Jongen, A. M\"oller, G. Meyer, and M. Gr\"uninger\\
Submitted to Physical Review Letters; cond-mat/0503409.

$\qquad$\\
\textit{Nature of magnetism in Ca$_{3}$Co$_{2}$O$_{6}$.}\\
Hua Wu, M. W. Haverkort, D. I. Khomskii, and L. H. Tjeng \\
Submitted to Physical Review Letters; cond-mat/0504490.

$\qquad$\\
\textit{Orbital-assisted metal-insulator transition in VO$_{2}$.}\\
M. W. Haverkort, Z. Hu, A. Tanaka, W. Reichelt, S. V. Streltsov, M. A. Korotin, V. I. Anisimov, H. H. Hsieh, H.-J. Lin, C. T. Chen, D. I. Khomskii, and L. H. Tjeng\\
To be submitted to Phys. Rev. Lett.

\chapter{Curiculum vitae}

\begin{tabular}{ll}
  \multicolumn{2}{l}{\textbf{Pers\"onliche Daten}}\\
  Name                      & Maurits Wim Haverkort \\
  Geburtsdatum              & 13 Januar 1976 \\
  Geburtsort                & Swifterbant (Niederlande)\\
  Staatsangeh\"origkeit     & Niederlande \\
  Personenstand             & ledige \\
                            &  \\
  \multicolumn{2}{l}{\textbf{Studium}}\\
  Juli 1994                 & Abitur nach Sechs Jahren VWO \\
  August 1994 - Juli 1995   & Sch\"uleraustausch im USA \\
  August 1995 - Januar 2002 & Rijksuniversiteit Groningen \\
                            & Master Degree in Physik \\
  Februar 2002              & Beginn der Doktorarbeit am II. Physikalischen  \\
                            & Institut, Universit\"at zu K\"oln \\
\end{tabular}


\begin{thebibliography}{200}
\bibitem{SOremark}
Spin-orbit interaction in the $3d$ shell will not split the $^{3}A_{2}$ ground
  state of a Ni $3d^{8}$ ion in $O_{H}$ symmetry. Spin-orbit coupling does
  split the $^{3}$B$_{1}$ state for a Ni ion in $D_{4h}$, but this splitting is
  negligibly small, of order $10^{-4}$ eV.
\end{thebibliography}
\end {document}